\documentclass[10pt]{article} 
\PassOptionsToPackage{table}{xcolor} 
\usepackage[accepted]{tmlr}

\usepackage{amsmath,amsfonts,bm}

\def\eqref#1{equation~\ref{#1}}

\def\1{\bm{1}}

\def\mC{{\bm{C}}}

\def\mE{{\bm{E}}}

\def\mS{{\bm{S}}}

\DeclareMathAlphabet{\mathsfit}{\encodingdefault}{\sfdefault}{m}{sl}
\SetMathAlphabet{\mathsfit}{bold}{\encodingdefault}{\sfdefault}{bx}{n}

\usepackage{hyperref}
\usepackage{url}
\usepackage{makecell}
\usepackage{booktabs}
\usepackage{graphicx}
\usepackage{tcolorbox}
\tcbuselibrary{listings, breakable}
\usepackage{appendix}
\usepackage{titletoc}
\usepackage{longtable}
\usepackage{placeins}
\usepackage{multicol}
\usepackage{multirow}
\usepackage{tabularx}
\usepackage{array}
\usepackage[table]{xcolor}
\usepackage{enumitem}
\usepackage{pgfplots}
\usepackage{subcaption}
\pgfplotsset{compat=newest}
\usetikzlibrary{pgfplots.polar}
\usetikzlibrary{arrows.meta, positioning, fit, backgrounds, calc}
\usepackage{sansmath} 
\usepackage{pifont}%
\newcommand{\cmark}{\ding{51}}%
\newcommand{\xmark}{\ding{55}}%

\title{Speech-DRAME: Human-Aligned Evaluation for Speech Role-Play}

\author{%
\name Jiatong Shi$^{1,2}$, Jionghao Han$^{2}$, Yichen Lu$^{2}$, Santiago Pascual$^{2}$, Pengfei Wu$^{2}$,\\
Chenye Cui$^{2}$, Shinji Watanabe$^{1}$, Chao Weng$^{2}$, Cong Zhou$^{2}$\\
\addr $^{1}$ Carnegie Mellon University \qquad $^{2}$ Anuttacon\\
\email \{jiatong.shi, yichen.lu, chao.weng, cong.zhou\}@anuttacon.com}

\definecolor{revcolor}{rgb}{0,0,0}
\newcommand{\rev}[1]{#1}

\def\month{10}
\def\year{2026}
\def\openreview{\url{https://openreview.net/forum?id=m9IP9ECrk2}}

\begin{document}

\maketitle

\begin{abstract}

\rev{Evaluating spoken role-play requires judging not only linguistic content but also prosody, emotional expression, character embodiment, and contextual fit. Existing automatic evaluation methods often collapse these aspects into coarse scores and are evaluated primarily on synthesized speech. We introduce \textbf{Speech-DRAME}, a framework comprising \emph{Speech-DRAME-EvalBench}, a bilingual, human-annotated benchmark with contrastive role-, context-, and delivery-mismatch conditions, and \texttt{DRAME-Eval}, a fine-tuned audio-language evaluator. The benchmark supports two complementary protocols: top-down \emph{archetype} evaluation and bottom-up \emph{realism} evaluation, with predominantly synthesized benchmark-distribution data and a held-out stress test consisting entirely of authentic human acting. On the benchmark-distribution sets, \texttt{DRAME-Eval} achieves pooled Pearson correlations of $0.629$ for archetype evaluation and $0.602$ for realism evaluation, significantly outperforming the evaluated zero- and few-shot audio-language judges; modality controls show that its judgments depend on both speech and textual context. This alignment does not transfer to the authentic-speech test set, where \texttt{DRAME-Eval} obtains an average correlation of $-0.142$ and a within-speaker correlation of $+0.029$, compared with matched human single-rater agreement of $+0.34$. These results identify a substantial synthetic-to-authentic transfer gap and establish Speech-DRAME as a resource for studying and improving evaluation of acting differences among real human performances.}

\end{abstract}

\section{Introduction}

Role-play has recently been recognized as a cornerstone of interactive systems, from initial text-based focuses to the growing ecosystem of multimodal dialogue agents~\citep{chenpersona, tseng2024two, dai2024mmrole, ma2024visual, jiang2025speechrole}. With the rapid advancement of large language models (LLMs), role-play has evolved beyond scripted interactions into dynamic character-driven conversations, supporting applications in education, entertainment, and human–AI collaboration~\citep{wu2024role, agatsuma2024building, fung2024proof}. Extending role-play to speech-based interaction offers an even richer communication channel, capturing prosody, emotion, and character consistency in ways that text alone cannot~\citep{ding2025kimi, huang2025step}.

Evaluating speech role-play, however, remains a major challenge. Current pipelines often rely on ALLMs as judges, typically in a zero-shot setting~\citep{huang2025instructttseval, chiang2025audio, jiang2025speechrole, zhang-etal-2025-omnicharacter}. Recent ALLM judges already exhibit non-trivial capability for spoken role-play assessment, and we adopt them as strong baselines throughout this work. Yet, even with these capable backbones, current zero-shot judging pipelines fall short along three persistent axes:
\begin{itemize}[leftmargin=*, itemsep=0pt, topsep=2pt]
    \item they lack sensitivity to paralinguistic cues such as intonation, pacing, and emotional dynamics that define role delivery~\citep{zhang-etal-2025-omnicharacter, huang2025instructttseval, chiang2025audio, zhou2025personaeval},
    \item they collapse diverse aspects of performance into limited coarse dimensions, obscuring critical dimensions of role-play quality~\citep{chiang2025audio, huang2025instructttseval}, and
    \item they depend on synthetic speech references rather than real human speech~\citep{zhang-etal-2025-omnicharacter, jiang2025speechrole}, and rarely test the judge against \emph{contrastive negative} conditions in which the spoken delivery, character identity, or scene context have been deliberately mismatched.
\end{itemize}
The third axis is particularly underexplored: with positive-only synthetic data, it is hard to tell whether a judge truly attends to paralinguistic cues or simply pattern-matches the transcript. We therefore make contrastive negative samples a first-class component of our benchmark (Sec.~\ref{sec:realism-data}), in line with the broader trend in NLP and vision-language evaluation~\citep{nie2020adversarial, shekhar2017foil, thrush2022winoground}.

\begin{figure}
    \centering
    \begin{tikzpicture}[
        font=\small,
        every node/.style={align=center},
        archbox/.style={draw=blue!50, rounded corners=4pt, fill=blue!8,
            minimum width=4.4cm, minimum height=0.85cm,
            text width=4.2cm, inner sep=5pt},
        realbox/.style={draw=orange!60, rounded corners=4pt, fill=orange!10,
            minimum width=4.4cm, minimum height=0.85cm,
            text width=4.2cm, inner sep=5pt},
        archhead/.style={rounded corners=4pt, fill=blue!22,
            minimum width=4.4cm, text width=4.2cm, inner sep=6pt,
            font=\small\bfseries},
        realhead/.style={rounded corners=4pt, fill=orange!28,
            minimum width=4.4cm, text width=4.2cm, inner sep=6pt,
            font=\small\bfseries},
        archinput/.style={draw=blue!70, thick, rounded corners=4pt, fill=blue!16,
            minimum width=4.4cm, minimum height=0.85cm,
            text width=4.2cm, inner sep=5pt},
        realinput/.style={draw=orange!80, thick, rounded corners=4pt, fill=orange!20,
            minimum width=4.4cm, minimum height=0.85cm,
            text width=4.2cm, inner sep=5pt},
        benchbox/.style={draw, rounded corners=4pt, fill=gray!12,
            minimum width=9.6cm, minimum height=0.9cm,
            text width=9.4cm, inner sep=5pt},
        evalbox/.style={draw, rounded corners=4pt, fill=green!10,
            minimum width=6.2cm, minimum height=0.9cm,
            text width=6.0cm, inner sep=5pt},
        arr/.style={-{Stealth[length=5pt]}, thick},
        lbl/.style={font=\scriptsize, inner sep=2pt, xshift=4pt},
    ]
    \def\Lx{0}
    \def\Rx{7.5}
    \def\Cx{3.75}
    \node[archhead] (AH) at (\Lx, 0)
        {Archetype Evaluation\\[3pt]
         {\normalfont\small\itshape top-down}};
    \node[archinput] (A1) at (\Lx,-1.35)
        {{\scriptsize\color{blue!70!black}\textit{[pipeline input]}}\\[1pt]
         Role keywords ($C_{\text{profile}}$)\\
         + scene prompts ($C_{\text{scene}}$)};
    \node[archbox] (A2) at (\Lx,-2.85)
        {SFnd speech generation\\
         (various roleplay systems)};
    \node[archbox] (A3) at (\Lx,-4.25)
        {Human annotation\\
         (AQ, HL, AP)};
    \draw[arr] (A1)--node[lbl,right]{generate}(A2);
    \draw[arr] (A2)--node[lbl,right]{annotate}(A3);
    \node[realhead] (RH) at (\Rx, 0)
        {Realism Evaluation\\[3pt]
         {\normalfont\small\itshape bottom-up}};
    \node[realinput] (R1) at (\Rx,-1.35)
        {{\scriptsize\color{orange!80!black}\textit{[pipeline input]}}\\[1pt]
         Real human recordings\\
         (TV, film, in-house)};
    \node[realbox] (R2) at (\Rx,-2.85)
        {LLM reconstructs\\
         $C_{\text{profile}}$ + $C_{\text{scene}}$};
    \node[realbox] (R3) at (\Rx,-4.65)
        {Neg.\ operator $\mathcal{N}$:\\
         delivery / role / scene\\
         {\scriptsize\itshape(mismatched)}};
    \node[realbox] (R4) at (\Rx,-6.3)
        {Human annotation\\
         5 dims $\times$ sub-dims};
    \draw[arr] (R1)--node[lbl,right]{reconstruct}(R2);
    \draw[arr] (R2)--node[lbl,right]{perturb}(R3);
    \draw[arr] (R3)--node[lbl,right]{annotate}(R4);
    \begin{scope}[on background layer]
      \node[fill=blue!4, draw=blue!25, rounded corners=8pt,
            fit=(AH)(A3), inner sep=9pt] {};
      \node[fill=orange!4, draw=orange!25, rounded corners=8pt,
            fit=(RH)(R4), inner sep=9pt] {};
    \end{scope}
    \node[benchbox] (EB) at (\Cx,-8.7)
        {\textbf{DRAME-EvalBench}\\
         human-annotated benchmark\\
         (archetype + realism;\ \ train / test / real-rec)};
    \draw[arr] (A3.south) -- ([xshift=-2.2cm]EB.north);
    \draw[arr] (R4.south) -- ([xshift=+2.2cm]EB.north);
    \node[evalbox] (EV) at (\Cx,-11.0)
        {\textbf{DRAME-Eval} (\texttt{SEval})\\
         fine-tuned on EvalBench\\
         $\uparrow$ alignment with human ratings};
    \draw[arr] (EB)--node[font=\small, inner sep=3pt, xshift=5pt, right]{fine-tune}(EV);
    \end{tikzpicture}
    \caption{Speech-DRAME formalizes speech role-play with a dual evaluation strategy: \textbf{Archetype Evaluation} (top-down, stereotype-driven) and \textbf{Realism Evaluation} (bottom-up, human-grounded). The colored header band at the top of each lane names the evaluation paradigm; it is \emph{not} a pipeline stage. Darker-bordered boxes labeled \textit{[pipeline input]} identify where data enters each pipeline: role keywords ($C_{\text{profile}}$) + scene prompts ($C_{\text{scene}}$) for archetype, and real human recordings for realism. Archetype annotation covers three dimensions: Audio Quality (AQ), Human-likeness (HL), and Appropriateness (AP). Both pipelines feed into \textbf{DRAME-EvalBench}, which is used to train \textbf{DRAME-Eval}, a fine-tuned \texttt{SEval} with improved alignment to human ratings.}
    \label{fig:main-framework}
\end{figure}

To overcome these limitations, we present \textbf{Speech-DRAME} (Speech Detailed Role-play Assessment with Modeling and Evaluation), a human-aligned evaluation framework that moves beyond synthetic or judge-only pipelines. Speech-DRAME is distinguished by three aspects: [i] a dual evaluation paradigm that balances archetype-driven generality with human-grounded realism (see Sec.~\ref{ssec:strategies})\footnote{\emph{Archetype Evaluation} (top-down) assesses whether synthesized speech fits broad, culturally shared role schemas — e.g., whether a ``firefighter'' response sounds appropriately authoritative and calm. \emph{Realism Evaluation} (bottom-up) is anchored in real human recordings and assesses fine-grained delivery quality: prosodic dynamics, emotional expressiveness, character consistency, contextual relevance, and semantic match. The two strategies are complementary: the former enables scalable, schema-driven benchmarking; the latter provides perceptual grounding in authentic human acting.}, [ii] human annotations that provide perceptual grounding across both protocols and the controlled mismatch conditions (see Sec.~\ref{sec:archetype-data} and~\ref{sec:realism-data}), and [iii] alignment between evaluation models (\texttt{SEval}) and human perception, which enables systematic analysis of the evaluation models themselves and lays the groundwork for downstream benchmarking of speech foundation models (\texttt{SFnd}) in future work (see Sec.~\ref{sec:exp}). Together, these aspects are intended to make Speech-DRAME a reproducible framework for measuring speech role-play.

As illustrated in Fig.~\ref{fig:main-framework}, the present paper focuses on the \emph{evaluation} side of speech role-play and centers on \textbf{DRAME-EvalBench}, an evaluation benchmark consisting of human-annotated datasets for archetype and realism role-play, including curated contrastive negative samples that probe whether judges attend to fine-grained delivery, role identity, and scene cues, together with \textbf{DRAME-Eval}, a fine-tuned evaluation model (\texttt{SEval}) trained on EvalBench that, while building on the broad capability of recent ALLM judges, further improves agreement with human ratings and serves as a diagnostic probe of the benchmark. We deliberately scope this paper to the human-aligned evaluation problem; we report a first \rev{benchmark-distribution diagnostic of \texttt{DRAME-Eval} as a system-level ranker on the eight familiar in-training \texttt{SFnd} systems (App.~\ref{app:tier1}), but a validated ranking of arbitrary or unseen systems on a held-out role-play benchmark is left to future work}.
\rev{Speech-DRAME is primarily a \textbf{benchmark and measurement contribution}; \texttt{DRAME-Eval} serves as a controlled diagnostic model that exercises the benchmark. Our contributions are:}
\begin{itemize}[leftmargin=*, itemsep=0pt, topsep=2pt]
    \item \rev{\textbf{A bilingual, multi-rater benchmark with two complementary protocols and controlled mismatches.} \textbf{DRAME-EvalBench} provides training-grade human annotations for speech role-play under top-down \emph{archetype} and bottom-up \emph{realism} evaluation, with contrastive \emph{role-}, \emph{context-}, and \emph{delivery-mismatch} conditions (Sec.~\ref{sec:realism-data}).}
    \item \rev{\textbf{A rigorously audited evaluation methodology.} We pair the benchmark with human-reference agreement, chance-corrected reliability (ordinal Krippendorff's $\alpha$ / weighted $\kappa$), language-specific reporting, bootstrap uncertainty, and shortcut controls, and develop \texttt{DRAME-Eval}, a fine-tuned evaluator that outperforms all evaluated zero- and few-shot ALLM judges on both benchmark-distribution tracks; among fine-tuning configurations we select \texttt{Instruct-LoRA} for its strong archetype result and efficiency while remaining competitive on realism (Sec.~\ref{sec:exp}, App.~\ref{app:tier1}).}
    \item \rev{\textbf{Evidence of a synthetic-to-authentic evaluation gap: strong benchmark-distribution alignment, exemplified by \texttt{DRAME-Eval}, does not yield reliable discrimination of acting differences in authentic human speech}, identified and quantified on a clean, held-out human-recording split.}\footnote{\textbf{Speech-DRAME resources.} Project page: \url{https://ftshijt.github.io/speech_drame/}. Evaluation code: \url{https://github.com/Anuttacon/speech_drame}. \textbf{Hugging Face datasets:} \url{https://huggingface.co/datasets/Anuttacon/speech-drame-evalbench} (metadata and mel features); \url{https://huggingface.co/datasets/ftshijt/speech-drame-evalbench} (with permitted vocoder reconstructions). The dataset pages currently contain introductory documentation; data payloads are in preparation (Sec.~\ref{sec:reproducibility}).} 
\end{itemize}

Through this work, we aim to move from acting in text to acting in speech, equipping the community with the models, evaluation strategies, and benchmarks needed to measure and improve speech role-play.

\vspace{5pt}

\section{Related Works}
\label{sec:literature}

\noindent \textbf{Text-based role-play evaluation.}
Early work on role-play focuses on text-only dialogue settings. Benchmarks such as RoleLLM, CharacterEval, RoleEval, RoleMRC, and CharacterBox~\citep{wang2024rolellm, tu2024charactereval, shen2023roleeval, wang2025characterbox, lu-etal-2025-rolemrc} examine persona fidelity and character consistency, typically relying on LLM-as-judge pipelines for scalability. InCharacter~\citep{wang2024incharacter} probes personality fidelity through psychological-interview prompts, while PingPong~\citep{anugraha2024pingpong} grounds role-play evaluation in user-emulator multi-model interactions. More recent benchmarks emphasize bilingual, multi-turn, and user-centric scenarios (e.g., RMTBench~\citep{yuan2025rmtbench}) and large-scale customizable persona training (OpenCharacter~\citep{wang2025opencharacter}), while atomic-level evaluation of persona fidelity~\citep{kim2025atomicpersona} highlights that single-score judgments hide subtle out-of-character drift in long generations. MMRole~\citep{dai2024mmrole} extends role-play evaluation to the multimodal (vision+text) regime, anticipating the cross-modal challenges that also arise in speech. However, studies such as PersonaEval~\citep{zhou2025personaeval} reveal fundamental reliability issues: even in pure text, judge models struggle with speaker tracking and contextual grounding. These limitations become more pronounced in speech, where prosody, emotional nuance, and vocal style quality are essential components of character portrayal.

\noindent \textbf{Speech-native role-play benchmarks.}
Recent efforts extend role-play evaluation to spoken interaction. OmniCharacter-10K~\citep{zhang-etal-2025-omnicharacter} emphasizes character voice traits, while resources such as VocalBench, SOVA-Bench, mTalk, and VoiceBench~\citep{liu2025vocalbench, hou2025sovabench, du2025mtalk, chen2024voicebench} broaden evaluation to vocal communication, assistant ability, and stylistic realism. In parallel, omni-capable SFMs such as StepAudio and KimiAudio~\citep{huang2025step, ding2025kimi} include role-play scenarios in their tests, alongside the recent wave of end-to-end spoken dialogue models such as Moshi~\citep{defossez2024moshi}. Closest to our aims is a fast-growing cluster of role-play--specific speech benchmarks: SpeechRole~\citep{jiang2025speechrole} provides a 111k-dialogue end-to-end speech-to-speech benchmark for role-playing agents; VStyle~\citep{zhan2025vstyle} offers a bilingual benchmark whose four categories explicitly include a role-play split alongside acoustic-attribute, instruction-following, and implicit-empathy splits; VoxRole~\citep{wu2025voxrole} curates 13k character-grounded film dialogues from 1{,}228 unique characters with paralinguistic-appropriateness metrics; and AudioRole~\citep{li2025audiorole} contributes a 1k+-hour TV-series dataset with the dual-aspect ARP-Eval framework. Concurrent generation-side efforts such as Voila~\citep{shi2025voila} build voice-language foundation models for real-time voice role-play with persona-aware speaker control, and PersonaPlex~\citep{roy2026personaplex} extends Moshi-style full-duplex models with hybrid voice+text persona prompts. Reinforcement-learning--based shaping of role-playing behavior in audio LLMs has also been explored~\citep{fu2026character}. Yet these resources either rely heavily on LLM-generated dialogues and TTS synthesis (e.g., SpeechRole, VStyle), or, even when anchored in real film/TV audio (e.g., VoxRole, AudioRole), score outputs with proprietary judge models and fully automatic metrics, releasing no training-grade human ratings and offering coarse, often opaque rubrics.

\noindent \textbf{Audio-LLM evaluation infrastructure.}
Beyond role-play, several recent benchmarks evaluate audio LLMs more broadly along dimensions that overlap with realism evaluation. AIR-Bench~\citep{yang2024airbench} and AudioBench~\citep{wang2025audiobench} provide universal evaluation across speech, sound, and music understanding; Dynamic-SUPERB Phase-2~\citep{huang2025dynamic, huang2024dynamic} expands the instruction-tuning evaluation surface to 180 tasks. SD-Eval~\citep{ao2024sdeval} explicitly targets paralinguistic and environmental information in spoken dialogue understanding, and CP-Bench~\citep{wang2025cpbench} benchmarks contextual paralinguistic reasoning in speech-LLMs on in-the-wild data. A parallel strand evaluates voice assistants and conversational agents: VoiceAssistant-Eval~\citep{lin2025voiceassistanteval} jointly tests listening, speaking, and viewing across $\sim$10k instances; MULTIVOX~\citep{kumar2025multivox} probes voice-assistant abilities across diverse styles; SOVA-Bench~\citep{hou2025sovabench} systematizes the speech conversation ability of voice assistants; and StyleBench~\citep{zhao2026stylebench} measures fine-grained speaking-style intensity control across emotion, speed, volume, and pitch. WildSpeech-Bench~\citep{cheng2025wildspeechbench} and ParaS2S~\citep{yang2025paras2s} extend the evaluation target to end-to-end speech-to-speech models with paralinguistic awareness as an explicit axis; ParaS2S in particular pairs a TTS-generated query benchmark with a GRPO-based alignment framework that distills the benchmark's GPT-based scoring pipeline into a reward model, illustrating a complementary path in which the judge is treated as a learned reward rather than as a target for human-aligned supervision. These efforts are complementary to Speech-DRAME: they evaluate generic understanding, voice-assistant behavior, or general dialogue, whereas we evaluate role-play--specific perceptual fit, with human-rated training data and contrastive negatives constructed for that purpose.

\noindent \textbf{Automatic judging with audio-capable LLMs.}
A line of work explores automatic assessment using ALLMs. Systems such as AudioJudge and Audio-Aware LLMs as Judges~\citep{manakul2025audiojudge, chiang2025audio} evaluate emotion, style, and delivery; Auto-ATT~\citep{wang2025audio} scales this to Turing-test-style assessments. In parallel, a complementary thread treats the audio LLM itself as a learned quality predictor: ALLD~\citep{chen2025alld} aligns an audio LLM via distillation to produce descriptive, MOS-consistent speech-quality judgments, and SpeechLLM-as-Judges~\citep{niu2025speechllmjudges} formulates general-purpose, interpretable speech-quality evaluation as an LLM-as-judge problem. While promising for rapid scoring, these approaches show high prompt sensitivity, weak capture of paralinguistic cues, and limited interpretability. Judge-only pipelines further lack stable criteria, as model versions, prompts, and hidden reasoning steps change over time. Speech-DRAME differs from this line of work in two ways: (i) we condition each judgment on a structured \emph{role} $(\mC_{\text{profile}}, \mC_{\text{scene}})$ rather than an unconditional quality target, and (ii) we explicitly stress-test the judge with the contrastive-negative operator $\mathcal{N}$ (Sec.~\ref{sec:realism-data}).

\noindent \textbf{Expressive TTS and instruction-following evaluation.}
Work on expressive and role-play-oriented TTS evaluation includes InstructTTSEval~\citep{huang2025instructttseval}, EmergentTTS-Eval~\citep{manku2025emergenttts}, MINT-Bench~\citep{lin2025mintbench} for multilingual instruction-following TTS, SpeechParaling-Bench~\citep{li2025speechparalingbench} for fine-grained paralinguistic-aware speech generation, and discrete-speech LM benchmarks~\citep{wang2024evaluating, chang2024interspeech}. Persona-driven TTS frameworks such as P2VA~\citep{ng2025p2va} also bridge persona descriptions and voice attributes, complementing Speech-DRAME's evaluation focus on the receiving (judge) side. The Interspeech 2024 Discrete Speech Unit Challenge~\citep{chang2024interspeech}, in particular, provided community-wide protocols for evaluating discrete-unit-based speech generation across resynthesis, ASR, and TTS tracks, and motivates the broader push toward shared, reproducible evaluation infrastructure that Speech-DRAME extends to role-play. These efforts highlight prosody, emotional control, and style following but remain centered on synthetic outputs and judge-driven scores. They do not provide fine-grained, human-grounded perceptual labels across heterogeneous role-play conditions.

\vspace{0.7em}
\noindent \textbf{Annotation methodologies: how Speech-DRAME differs.}
Beyond task focus, Speech-DRAME fundamentally diverges in \emph{how} labels are produced. Table~\ref{tab:anno-compare} positions Speech-DRAME against a much broader range of contemporary role-play, persona, paralinguistic, and expressive-TTS benchmarks --- one row per method --- so that the comparison is not narrowed to three pre-selected competitors. The table makes two recurring patterns visible: (i) almost all existing speech-side resources rely on synthetic / TTS-generated audio, with at most a calibration-only slice of real human speech; and (ii) almost none combine \emph{contrastive negatives}, \emph{per-rater confidence}, and \emph{released training-grade human labels} in one resource. Speech-DRAME aims at exactly that intersection.

\begingroup
\small
\renewcommand{\arraystretch}{1.15}
\setlength{\tabcolsep}{2pt}
\begin{longtable}{@{}>{\raggedright\arraybackslash}p{.35\linewidth}>{\centering\arraybackslash}p{.055\linewidth}>{\centering\arraybackslash}p{.10\linewidth}>{\centering\arraybackslash}p{.10\linewidth}>{\centering\arraybackslash}p{.10\linewidth}>{\centering\arraybackslash}p{.055\linewidth}>{\centering\arraybackslash}p{.08\linewidth}>{\centering\arraybackslash}p{.08\linewidth}@{}}
\caption{Annotation methodology comparison across recent text-based role-play, speech-native role-play, paralinguistic / conversational, and expressive-TTS benchmarks. Each row is one resource, with its citation and focus in the first column; cells describe the \emph{primary} declared property (auxiliary uses such as calibration-only slices are noted in-cell). \textbf{Mod.}: Modality of the rated artifact (T=text, S=speech, M=multimodal). \textbf{$\mS_{r}$ src.}: dominant source of the spoken response (Syn=synthetic/TTS, Real=real human, Mix=both, ``--''=not applicable for text-only). \textbf{Labels}: dominant per-sample label source for the \emph{rated artifact} (Judge=judge-model or fully automatic scoring, Human=human raters, Both); resources whose input speech is real and human-curated but whose references or scores are model-generated (e.g., SD-Eval's GPT-4--generated reference responses) are classified by the score/reference source. \textbf{Lang.}: primary language(s). \textbf{Neg.}: contrastive negatives that perturb scene / role / delivery (\cmark/\xmark/n.s.\ = not stated). \textbf{Strategy}: data-collection approach classified following this paper's taxonomy — \emph{Arch.}\ (top-down archetype: synthesized speech driven by predefined role schemas), \emph{Real.}\ (bottom-up realism: anchored in real human recordings), \emph{Both}, or ``---'' for text/multimodal resources where the distinction does not apply. \textbf{SEval-ready}: human-labeled split released for training speech-evaluation models. Cells where the cited paper does not state a property are marked ``n.s.''. Citations point to the entries above. $^{\dagger}$ParaS2S' main benchmark consists of TTS-synthesized queries; the authors additionally probe generalization on IEMOCAP/MELD real speech.}
\label{tab:anno-compare}\\
\toprule
\textbf{Method / focus} & \textbf{Mod.} & \makecell{\textbf{$\mS_r$}\\\textbf{src.}} & \textbf{Labels} & \textbf{Lang.} & \textbf{Neg.} & \textbf{Strategy} & \makecell{\textbf{SEval-}\\\textbf{ready}} \\
\midrule
\endfirsthead
\multicolumn{8}{@{}l}{\small Table~\thetable\ (continued)}\\[3pt]
\toprule
\textbf{Method / focus} & \textbf{Mod.} & \makecell{\textbf{$\mS_r$}\\\textbf{src.}} & \textbf{Labels} & \textbf{Lang.} & \textbf{Neg.} & \textbf{Strategy} & \makecell{\textbf{SEval-}\\\textbf{ready}} \\
\midrule
\endhead
\midrule
\multicolumn{8}{r@{}}{\small Continued on next page}\\
\endfoot
\bottomrule
\endlastfoot
\multicolumn{8}{@{}l}{\emph{Text-based role-play}} \\*
\midrule
RoleLLM~\citep{wang2024rolellm}\newline\emph{Persona fidelity} & T & -- & Judge & EN+ZH & \xmark & --- & \xmark \\[3pt]
CharacterEval~\citep{tu2024charactereval}\newline\emph{RPCA, reward model} & T & -- & Both & ZH & \xmark & --- & \cmark \\[3pt]
RoleEval~\citep{shen2023roleeval}\newline\emph{Knowledge / persona} & T & -- & Judge & EN+ZH & \xmark & --- & \xmark \\[3pt]
RoleMRC~\citep{lu-etal-2025-rolemrc}\newline\emph{RP+instruction} & T & -- & Judge & EN & \xmark & --- & \xmark \\[3pt]
CharacterBox~\citep{wang2025characterbox}\newline\emph{Virtual-world RP} & T & -- & Judge & EN+ZH & \xmark & --- & \xmark \\[3pt]
InCharacter~\citep{wang2024incharacter}\newline\emph{Personality fidelity} & T & -- & Judge & EN+ZH & \xmark & --- & \xmark \\[3pt]
PingPong~\citep{anugraha2024pingpong}\newline\emph{RP w/ user emul.} & T & -- & Judge & EN+RU & \xmark & --- & \xmark \\[3pt]
RMTBench~\citep{yuan2025rmtbench}\newline\emph{Multi-turn user-RP} & T & -- & Judge & EN+ZH & \xmark & --- & \xmark \\[3pt]
OpenCharacter~\citep{wang2025opencharacter}\newline\emph{Persona training} & T & -- & Judge & EN & \xmark & --- & \xmark \\[3pt]
PersonaEval~\citep{zhou2025personaeval}\newline\emph{Judge reliability} & T & -- & Both & EN+ZH & \xmark & --- & n.s. \\[3pt]
AtomicPersona~\citep{kim2025atomicpersona}\newline\emph{Atomic OOC} & T & -- & Both & EN & \xmark & --- & n.s. \\[3pt]
\midrule
\multicolumn{8}{@{}l}{\emph{Multimodal role-play}} \\*
\midrule
MMRole~\citep{dai2024mmrole}\newline\emph{Multimodal RP} & M & -- & Both & EN+ZH & \xmark & --- & \cmark \\[3pt]
\midrule
\multicolumn{8}{@{}l}{\emph{Speech-native role-play / character voice}} \\*
\midrule
OmniCharacter-10K~\citep{zhang-etal-2025-omnicharacter}\newline\emph{Character voice} & S & Syn & Judge & EN+ZH & \xmark & Arch. & n.s. \\[3pt]
SpeechRole~\citep{jiang2025speechrole}\newline\emph{End-to-end RP} & S & Syn & Judge & EN+ZH & \xmark & Arch. & n.s. \\[3pt]
VStyle~\citep{zhan2025vstyle}\newline\emph{Voice style + RP} & S & Syn & Judge & EN+ZH & \xmark & Arch. & n.s. \\[3pt]
VoxRole~\citep{wu2025voxrole}\newline\emph{Film-grounded RP} & S & Real & Judge & EN & \xmark & Real. & n.s. \\[3pt]
AudioRole~\citep{li2025audiorole}\newline\emph{TV-grounded RP} & S & Mix & Judge & EN & \xmark & Real. & \xmark \\[3pt]
\midrule
\multicolumn{8}{@{}l}{\emph{Speech-side conversational / paralinguistic-aware benchmarks}} \\*
\midrule
SD-Eval~\citep{ao2024sdeval}\newline\emph{Paraling.\ dialogue} & S & Real & Judge & EN & \xmark & Real. & \xmark \\[3pt]
CP-Bench~\citep{wang2025cpbench}\newline\emph{Contextual paraling.} & S & Real & Human & EN & \xmark & Real. & \xmark \\[3pt]
ParaS2S~\citep{yang2025paras2s}\newline\emph{Paraling.\ S2S + RL align} & S & Syn$^{\dagger}$ & Judge & EN & \xmark & Arch. & \xmark \\[3pt]
SpeechParaling-Bench~\citep{li2025speechparalingbench}\newline\emph{Paraling.\ generation} & S & Syn & Judge & EN+ZH & \xmark & Arch. & n.s. \\[3pt]
StyleBench~\citep{zhao2026stylebench}\newline\emph{Style intensity} & S & Syn & Judge & EN & \xmark & Arch. & n.s. \\[3pt]
\midrule
\multicolumn{8}{@{}l}{\emph{Expressive / instruction-following TTS evaluation}} \\*
\midrule
InstructTTSEval~\citep{huang2025instructttseval}\newline\emph{Instr.-following TTS} & S & Syn & Judge & EN+ZH & \xmark & Arch. & n.s. \\[3pt]
EmergentTTS-Eval~\citep{manku2025emergenttts}\newline\emph{Prosodic / expressive} & S & Syn & Judge & EN & \xmark & Arch. & n.s. \\[3pt]
MINT-Bench~\citep{lin2025mintbench}\newline\emph{Multi-ling.\ instr.\ TTS} & S & Syn & Both & Multi & \xmark & Arch. & n.s. \\[3pt]
\midrule
\multicolumn{8}{@{}l}{\emph{Audio-LLM / speech-quality judges}} \\*
\midrule
ALLD~\citep{chen2025alld}\newline\emph{Speech-quality LLM} & S & Real & Human\newline (MOS) & EN & \xmark & Real. & \cmark \\[3pt]
SpeechLLM-as-Judges~\citep{niu2025speechllmjudges}\newline\emph{Interpretable judge} & S & Mix & Both & Multi & \xmark & n.s. & n.s. \\[3pt]
\midrule
\rowcolor{gray!18}\textbf{Speech-DRAME (Ours)}\newline\emph{\textbf{Role-play, dual perspective}} & S & \textbf{Mix}\newline\textbf{(Real +}\newline\textbf{Syn)} & \textbf{Human} & \textbf{EN+ZH} & \cmark & \textbf{Both} & \cmark \\[3pt]
\end{longtable}
\endgroup

\noindent \textbf{Our distinction.}
Speech-DRAME advances role-play evaluation by unifying:  
(i) a dual evaluation paradigm separating archetype-oriented generality from realism grounded in authentic human speech,  
(ii) bilingual, rubric-aligned human annotations, and
(iii) \texttt{SEval} alignment to human ratings, enabling transparent calibration and error analysis.
Together, these components move beyond synthetic corpora and judge-only scoring to provide a human-grounded and reproducible framework for evaluating speech role-play performance.

\section{Speech Role-play: Formulation and Evaluation Strategies}
\label{sec:formulation}

This section establishes the symbols, models, and data-construction operators that the rest of the paper relies on. Sections~\ref{sec:archetype-data} and~\ref{sec:realism-data} (data curation) and Section~\ref{sec:exp} (modeling) all back-reference the definitions introduced here.

\subsection{Problem Formulation}
\label{ssec:problem}

\paragraph{What is a ``role''?}
We treat a \emph{role} not as a single atomic concept but as a pair $(\mC_{\text{profile}}, \mC_{\text{scene}})$:
\begin{itemize}[leftmargin=1.2em, itemsep=1pt, topsep=1pt]
    \item $\mC_{\text{profile}}$ — the \textbf{character profile}: the persistent identity of the speaker (occupation, social identity, personality traits, named character, age, etc.). In Sec.~\ref{sec:archetype-data}, $\mC_{\text{profile}}$ is instantiated by \emph{archetype-level keywords} (e.g., ``firefighter''); in Sec.~\ref{sec:realism-data}, it is instantiated by \emph{detailed character descriptions} retrieved from real media.
    \item $\mC_{\text{scene}}$ — the \textbf{scene specification}: the situational context in which the speaker acts (an event, an interlocutor relationship, an emotional setting). $\mC_{\text{scene}}$ is again instantiated either by short stereotyped scene prompts (Sec.~\ref{sec:archetype-data}) or by \emph{local scenes} grounded in real plot context (Sec.~\ref{sec:realism-data}).
\end{itemize}
All experiments and benchmarks in this paper are single-turn, so we present the formulation in its single-turn form below. The straightforward multi-turn extension --- with an additional preceding-dialogue context $\mC_{\text{speech}}$ --- is given in Appendix~\ref{app:multiturn}.

\paragraph{Speech role-play (generation).}
A \emph{speech foundation model} (\texttt{SFnd}) maps the role to a spoken response $\mS_{r}$:
\begin{equation}
    \mS_{r} \;=\; \mathrm{SFnd}(\mC_{\text{profile}},\, \mC_{\text{scene}}).
    \label{eq:sfm}
\end{equation}
Here $\mS_{r}$ represents the spoken output (waveform or latent representation), capturing both the linguistic dimension (coherence with the scene) and the paralinguistic dimension (prosody, style, emotion) expected of the role.

\paragraph{Evaluation.}
A \emph{speech evaluation model} (\texttt{SEval}) assesses $\mS_{r}$ using the same conditioning signals:
\begin{equation}
    \mE_{r} \;=\; \mathrm{SEval}(\mS_{r},\, \mC_{\text{profile}},\, \mC_{\text{scene}}).
    \label{eq:sem_single}
\end{equation}
$\mE_{r}$ is the per-dimension human-aligned score that we use as both the supervision target for training \texttt{SEval} models (Sec.~\ref{sec:exp}) and the unit of comparison when ranking \texttt{SFnd} systems (Sec.~\ref{sec:archetype-data}--\ref{sec:realism-data}).

\paragraph{Curation operator $\mathcal{D}$.}
The data underlying $\mE_{r}$ is not given for free; it is produced by a \textbf{curation pipeline} $\mathcal{D}$ that converts raw inputs into evaluation samples. We instantiate $\mathcal{D}$ in two complementary ways, mirroring the dual evaluation perspectives introduced in Sec.~\ref{ssec:strategies}. Let $\mathcal{K}_{p}$ and $\mathcal{K}_{s}$ denote vocabularies of \emph{role} and \emph{scene} keywords; let $T_{\text{prompt}}$ be a template-based expansion that turns a keyword pair into a full profile/scene specification; let $\mathcal{R}$ denote a corpus of real human recordings paired with surrounding media context $\sigma$; and let $\Phi_{\text{retr}}$ denote a retrieval-and-LLM reconstruction operator. Then:
\begin{align}
    \mathcal{D}^{\text{top-down}}: \;\; (k_{p},\, k_{s}) \in \mathcal{K}_{p} \times \mathcal{K}_{s}
        \;\;\xrightarrow{\;T_{\text{prompt}}\;}\;\; (\mC_{\text{profile}},\, \mC_{\text{scene}})
        \;\;\xrightarrow{\;\texttt{SFnd}\;}\;\; (\mS_{r},\, \mC_{\text{profile}},\, \mC_{\text{scene}}), \label{eq:dtopdown}\\[2pt]
    \mathcal{D}^{\text{bottom-up}}: \;\; (\mS_{r},\, \sigma) \in \mathcal{R}
        \;\;\xrightarrow{\;\Phi_{\text{retr}}\;}\;\; (\mS_{r},\, \mC_{\text{profile}},\, \mC_{\text{scene}}). \label{eq:dbottomup}
\end{align}
Both operators produce a tuple $(\mS_{r},\, \mC_{\text{profile}},\, \mC_{\text{scene}})$ as the unit of analysis: $\mathcal{D}^{\text{top-down}}$ by synthesizing $\mS_{r}$ from a keyword-expanded role, and $\mathcal{D}^{\text{bottom-up}}$ by reconstructing the role around a real recording. $\mathcal{D}^{\text{top-down}}$ underlies Sec.~\ref{sec:archetype-data} (archetype data); $\mathcal{D}^{\text{bottom-up}}$ underlies Sec.~\ref{sec:realism-data} (realism data).

\paragraph{Negative-sample operator $\mathcal{N}$.}
For both data flows, we also define a \textbf{negative-sample operator} $\mathcal{N}$ that perturbs one component of a positive sample $(\mS_{r}, \mC_{\text{profile}}, \mC_{\text{scene}})$ while holding the others fixed, yielding a triple in which the speaker's delivery, identity, or scene context has been deliberately mismatched:
\begin{equation}
    \mathcal{N}(\mS_{r}, \mC_{\text{profile}}, \mC_{\text{scene}}) \in \big\{
      (\tilde{\mS}_{r}, \mC_{\text{profile}}, \mC_{\text{scene}}),\;
      (\mS_{r}, \tilde{\mC}_{\text{profile}}, \tilde{\mC}_{\text{scene}}),\;
      (\mS_{r}, \mC_{\text{profile}}, \tilde{\mC}_{\text{scene}})
    \big\}.
    \label{eq:negative}
\end{equation}
The three branches correspond to \emph{delivery-mismatch}, \emph{role-mismatch}, and \emph{context-mismatch} negatives respectively, all elaborated in Sec.~\ref{sec:realism-data}. Two points deserve emphasis. First, $\mathcal{N}$ can produce \emph{multi-property} perturbations in which more than one component is changed simultaneously — for example, $(\tilde{\mS}_{r}, \mC_{\text{profile}}, \tilde{\mC}_{\text{scene}})$ changes both delivery and scene at once. Second, the role-mismatch branch \emph{always} pairs $\tilde{\mC}_{\text{profile}}$ with a correspondingly updated $\tilde{\mC}_{\text{scene}}$: because a character profile and scene must be mutually consistent to form a coherent annotation target, changing the profile alone while leaving the original scene in place would introduce a contradiction rather than a controlled mismatch. We therefore jointly replace both when perturbing the role identity. By construction, the perturbed \emph{components} produced by $\mathcal{N}$ (e.g., $\tilde{\mC}_{\text{scene}}$, $\tilde{\mS}_{r}$) are deliberately \emph{mismatched} rather than authentic pairings, generated specifically to probe whether \texttt{SEval} attends to fine-grained perceptual cues. This is a construction type, not an audio-provenance label: the role- and context-mismatch branches perturb only the text and reuse the positive's original audio (real or TTS), while only the delivery branch $\tilde{\mS}_{r}$ resynthesizes the audio. This distinction is made explicit in all downstream tables and figures.

\subsection{Role-play Evaluation with Dual Perspectives: Archetype and Realism}
\label{ssec:strategies}

Speech role-play evaluation can be conceptualized through multiple lenses depending on the nature of available data and the desired granularity of perceptual judgment. In Speech-DRAME, we formalize \emph{two complementary evaluation perspectives}: \emph{Archetype-based} (top-down, $\mathcal{D}^{\text{top-down}}$ in Eq.~\ref{eq:dtopdown}) and \emph{Realism-based} (bottom-up, $\mathcal{D}^{\text{bottom-up}}$ in Eq.~\ref{eq:dbottomup}). Both perspectives are grounded in distinct theoretical traditions within psychology, sociology, and performance studies, and reflect how humans assess roles in interaction: through both top-down, culturally shared schemas and bottom-up, contextually enacted behaviors. We organize this subsection in three layers: (i) the theoretical motivation for each perspective; (ii) how each perspective instantiates the symbols of Sec.~\ref{ssec:problem}; and (iii) a side-by-side contrast (Table~\ref{tab:topdown-bottomup}) that the rest of the paper refers back to.

\subsubsection{Theoretical motivation}

\paragraph{Archetypes as collective schemas.}
The notion of archetypes originates in Jungian psychology, where they represent culturally recurring prototypes embedded in the collective unconscious~\citep{jung2014archetypes}. Media and performance studies show that archetypes guide audience expectations, providing \emph{top-down interpretive frames} for judging whether a character ``fits'' familiar categories such as hero, mentor, or caregiver~\citep{campbell2008hero}. Because these schemas are widely shared, archetype-based judgments scale efficiently across scenarios and enable broad comparisons.

\paragraph{Realism as situated performance.}
In contrast, theories of social interaction emphasize the \emph{fine-grained enactment} of roles in situated contexts. Goffman's dramaturgical sociology conceptualizes everyday communication as performance, where believability depends on subtle paralinguistic cues—prosody, hesitation, pacing, and affect~\citep{goffman1949presentation}. Interactional sociolinguistics further shows how micro-level features such as intonation contours and turn-taking patterns shape perceptions of authenticity~\citep{gumperz1982discourse}. Realism-oriented evaluation therefore attends to detailed acoustic and pragmatic cues that fall outside schematic expectations.

\paragraph{Complementary perspectives.}
Performance theory characterizes role behavior as both ``restored'' (scripted, repeatable structures) and ``emergent'' (situated enactment)~\citep{schechner2017performance}, while sociolinguistic theories distinguish between \emph{ascribed categories} and \emph{performed identities}~\citep{bucholtz2005identity}. These viewpoints motivate the need for dual evaluation: archetype-based assessments capture macro-level expectations rooted in cultural schemas, whereas realism-based assessments capture micro-level delivery and situational fit. Neither perspective subsumes the other; each reveals a different dimension of how listeners interpret role performance. Speech-DRAME therefore adopts both, enabling analyses that are at once scalable and sensitive to the nuances of authentic spoken interaction.

\subsubsection{Mapping each perspective onto the formulation}
\label{sssec:mapping}

We now connect each perspective to the symbols of Sec.~\ref{ssec:problem}.

\paragraph{Archetype Evaluation (top-down, $\mathcal{D}^{\text{top-down}}$).}
$\mC_{\text{profile}}$ is instantiated by a \emph{stereotypical role definition} such as ``firefighter'' or ``ER doctor''; $\mC_{\text{scene}}$ is instantiated by a concise scene prompt (e.g., ``comfort a child trapped in fire''). The spoken response $\mS_{r}$ is then synthesized by an \texttt{SFnd}, and $\mE_{r}$ is computed against three dimensions (two context-free and one context-dependent) (Sec.~\ref{ssec:archetype-annotation}). This strategy provides scalable judgments grounded in shared schemas and supports broad cross-scenario comparisons, particularly under limited annotation resources~\citep{gemmeke2017audio}. Its reliance on broad prototypes, however, limits sensitivity to fine-grained delivery such as emotional modulation or subtle prosodic control~\citep{holt2010speech}.

\paragraph{Realism Evaluation (bottom-up, $\mathcal{D}^{\text{bottom-up}}$).}
Here $\mS_{r}$ is given first — drawn from real recordings of professional and non-professional speakers in natural dialogue and media — and $\mC_{\text{profile}}$ and $\mC_{\text{scene}}$ are reconstructed around it via retrieval and LLM-assisted summarization. $\mE_{r}$ is computed across the five fine-grained dimensions detailed in Sec.~\ref{ssec:annotation-protocol}: prosodic dynamics, emotional expressiveness, character consistency, contextual relevance, and semantic match. To stress-test \texttt{SEval} models, the bottom-up flow additionally produces contrastive negatives via the operator $\mathcal{N}$ (Eq.~\ref{eq:negative}). Although annotation-intensive, this approach is essential for applications where realism and subtlety are central to perceived performance.

\paragraph{Integrated perspective.}
Archetype-based and realism-based evaluations illuminate distinct facets of spoken role performance: the former provides macro-level judgments aligned with culturally shared expectations, offering scalability and coarse-grained comparability; the latter captures the fine-grained qualities that make a performance believable in real interactional settings. By incorporating both, Speech-DRAME supports analyses that span broad categorical expectations as well as nuanced, context-sensitive delivery.

\subsubsection{Side-by-side contrast}

Table~\ref{tab:topdown-bottomup} summarizes the two perspectives along the dimensions of the formulation. Sections~\ref{sec:archetype-data} and~\ref{sec:realism-data} are organized so that each row of this table is operationalized in turn.

\begin{table}[t]
\centering
\small
\caption{Top-down (Archetype) vs.\ bottom-up (Realism) evaluation in Speech-DRAME, expressed in the symbols of Sec.~\ref{ssec:problem}. ``GT'' = ground truth.}
\label{tab:topdown-bottomup}
\renewcommand{\arraystretch}{1.2}
\setlength{\tabcolsep}{6pt}
\rowcolors{2}{gray!8}{white}
\resizebox{\linewidth}{!}{
\begin{tabular}{l|l|l}
\toprule
\textbf{Aspect} & \textbf{Archetype (top-down)} & \textbf{Realism (bottom-up)} \\
\midrule
Curation operator & $\mathcal{D}^{\text{top-down}}$ (Eq.~\ref{eq:dtopdown}) & $\mathcal{D}^{\text{bottom-up}}$ (Eq.~\ref{eq:dbottomup}) \\
Source of $\mS_{r}$ & Synthesized by an \texttt{SFnd} & Real human recording (TV/film/recordings) \\
Source of $\mC_{\text{profile}}$ & Stereotypical role keyword + template & Retrieved from media context \\
Source of $\mC_{\text{scene}}$ & Stereotyped scene prompt & LLM-summarized local plot scene \\
Granularity of $\mE_{r}$ & 3 dims (AQ/HL context-free; AP context-dependent) (Sec.~\ref{ssec:archetype-annotation}) & 5 fine-grained dims w/ sub-dims (Sec.~\ref{ssec:annotation-protocol}) \\
Negative samples ($\mathcal{N}$) & Not used & Delivery / role / context mismatches \\
Scale vs.\ depth & Broad, scalable, schema-driven & Deep, annotation-heavy, perceptual \\
Used to benchmark & \texttt{SFnd} vs.\ stereotype expectation & \texttt{SEval} vs.\ human perception, \texttt{SFnd} realism \\
\bottomrule
\end{tabular}}
\end{table}

In the following two sections (Sec.~\ref{sec:archetype-data} and~\ref{sec:realism-data}), we detail the data collection processes corresponding to the two columns of Table~\ref{tab:topdown-bottomup}, respectively.

\section{DRAME-EvalBench Data Curation --- Archetype Data Collection}
\label{sec:archetype-data}

This section instantiates the \emph{top-down} (archetype) column of Table~\ref{tab:topdown-bottomup}: the curation operator $\mathcal{D}^{\text{top-down}}$ (Eq.~\ref{eq:dtopdown}), the corresponding annotation rubric, and the resulting dataset. The parallel \emph{bottom-up} (realism) column is operationalized in Sec.~\ref{sec:realism-data}, which is deliberately structured to mirror this section so that the two flows can be read side-by-side.

\paragraph{High-level statistics of DRAME-EvalBench.}
Before detailing the archetype pipeline, Table~\ref{tab:bench-overview} summarizes the high-level statistics of the full DRAME-EvalBench database, covering both the archetype data introduced in this section and the realism data introduced in Sec.~\ref{sec:realism-data}. Detailed per-language and per-model breakdowns are deferred to Appendix~\ref{app:archetype-data-analysis} and~\ref{app:realism-data-analysis}.

\begin{table}[t]
\centering
\small
\caption{High-level statistics of DRAME-EvalBench. \emph{Negatives} counts samples produced by the operator $\mathcal{N}$ in Eq.~\ref{eq:negative}; these are contrastive \emph{mismatched} constructions, not a distinct audio-provenance class: the role- and context-mismatch negatives perturb only the text (profile/scene) and reuse their source clip's own original audio (which may be real or TTS), whereas the delivery-mismatch negatives resynthesize the audio. \rev{Counts match the experimental manifest and the experiment accounting; the authentic-speech split is Mandarin-only (see Tab.~\ref{tab:role-play-exp-realism-real}), and severity-level counts for negatives are given in App.~\ref{app:tier1}. The $1{,}080$ in-domain-adaptation recordings are \emph{additional} to the $10{,}990$ base realism training records (the final adapted evaluator trains on $12{,}070$); because the $1{,}080/1{,}084$ figures partition the already-counted $2{,}164$ pool, the annotated realism total is $15{,}414$ (train $10{,}990$ / benchmark-distribution test $2{,}000$ / real-rec pool $2{,}164$ / a separate $260$-sample delivery-mismatch benchmark with resynthesized audio, Sec.~\ref{sec:realism-data}), i.e.\ the $260$ are an additional split rather than a subset of the other three.}}
\label{tab:bench-overview}
\renewcommand{\arraystretch}{1.3}
\setlength{\tabcolsep}{5pt}
\footnotesize
\rowcolors{2}{gray!8}{white}
\begin{tabularx}{\linewidth}{@{}>{\raggedright\arraybackslash}p{2.7cm}|>{\raggedright\arraybackslash}X>{\raggedright\arraybackslash}X@{}}
\toprule
\textbf{Aspect} & \textbf{Archetype (Sec.~\ref{sec:archetype-data})} & \textbf{Realism (Sec.~\ref{sec:realism-data})} \\
\midrule
Curation operator & $\mathcal{D}^{\text{top-down}}$ (Eq.~\ref{eq:dtopdown}) & $\mathcal{D}^{\text{bottom-up}}$ (Eq.~\ref{eq:dbottomup}) \\
Languages & EN + ZH & \rev{EN + ZH (Mandarin-majority; real-rec pool \& delivery benchmark ZH-only)} \\
Source of $\mS_{r}$ & Synth.\ by 7--8 \texttt{SFnd} systems & \rev{Mixed: real (TV/film/original) + TTS; only the held-out real-rec test set is all-real (${\approx}51$--$62\%$ of train/basic audio is TTS)} \\
\# scenarios & 552 (per language) & \rev{clip-grounded (bottom-up; no fixed template)} \\
\# annotated samples (total) & \rev{8{,}272} & \rev{15{,}414 (train 10{,}990 / dist.\ test 2{,}000 / real-rec pool 2{,}164 / delivery 260)} \\
Train / Test split & \rev{6{,}773 / 1{,}499} & \rev{10{,}990 / 2{,}000 (+ 2{,}164 real-rec pool + 260 delivery)} \\
\rev{Real-rec pool split (adapt / test)} & \rev{---} & \rev{2{,}164 = 1{,}080 adaptation train + 1{,}084 held-out test (disjoint speakers)} \\
\rev{Composition (pos / role-neg / ctx-neg)} & \rev{---} & \rev{train 2{,}504 / 4{,}310 / 4{,}176; dist.\ test 210 / 832 / 958; real-rec 2{,}164 / 0 / 0} \\
Negatives via $\mathcal{N}$ & --- & $\tilde{\mC}_{\text{scene}}$ + $\tilde{\mC}_{\text{profile}}$ + $\tilde{\mS}_{r}$ (Sec.~\ref{sssec:negatives}) \\
Annotation rubric & 3 dims + content pass (Sec.~\ref{ssec:archetype-annotation}) & 5 dims w/ sub-dims (Sec.~\ref{ssec:annotation-protocol}) \\
\bottomrule
\end{tabularx}
\end{table}

\begin{figure}[t]
\centering
\resizebox{\linewidth}{!}{%
\begin{tikzpicture}[
    font=\small,
    every node/.style={align=center},
    archstep/.style={draw=blue!55, rounded corners=3pt, fill=blue!12,
        minimum width=3.0cm, minimum height=0.65cm,
        text width=2.8cm, inner sep=3pt},
    archsrc/.style={draw=blue!45, rounded corners=3pt, fill=blue!6,
        minimum width=3.0cm, minimum height=0.65cm,
        text width=2.8cm, inner sep=3pt, font=\small\itshape},
    realstep/.style={draw=orange!65, rounded corners=3pt, fill=orange!18,
        minimum width=3.0cm, minimum height=0.65cm,
        text width=2.8cm, inner sep=3pt},
    realsrc/.style={draw=orange!50, rounded corners=3pt, fill=orange!9,
        minimum width=3.0cm, minimum height=0.65cm,
        text width=2.8cm, inner sep=3pt, font=\small\itshape},
    negstep/.style={draw=red!55, dashed, rounded corners=3pt, fill=red!6,
        minimum height=0.75cm, text width=2.4cm, inner sep=3pt},
    headertext/.style={font=\small\bfseries},
    arr/.style={-{Stealth[length=4pt]}, thick},
    steplabel/.style={font=\scriptsize, fill=white, inner sep=1pt},
    neglabel/.style={font=\scriptsize\bfseries, fill=white, inner sep=2pt,
        text=red!60!black},
]
\def\RC{8.0}   
\def\Rsep{3.0} 

\node[headertext, blue!60!black] (ah) at (0, 0) {Archetype (top-down, Sec.~\ref{sec:archetype-data})};

\node[archsrc,  below=0.4cm of ah] (a1) {Role keywords $\mathcal{C}_{\text{profile}}$\\+ scene prompts $\mathcal{C}_{\text{scene}}$};
\node[archstep, below=0.4cm of a1] (a2) {Text LLM generates\\transcript \& speech caption};
\node[archstep, below=0.4cm of a2] (a3) {TTS synthesis\\(7--8 \texttt{SFnd} systems)};
\node[archstep, below=0.4cm of a3] (a4) {Quality filter\\(content pass check)};
\node[archstep, below=0.4cm of a4] (a5) {Archetype dataset\\{\scriptsize 8,272 samples\\(EN+ZH)}};

\draw[arr] (ah) -- (a1);
\draw[arr] (a1) -- node[steplabel,right]{\textbf{S1}} (a2);
\draw[arr] (a2) -- node[steplabel,right]{\textbf{S2}} (a3);
\draw[arr] (a3) -- node[steplabel,right]{\textbf{S3}} (a4);
\draw[arr] (a4) -- node[steplabel,right]{\textbf{S4}} (a5);

\node[headertext, orange!70!black] (rh) at (\RC, 0)
    {Realism (bottom-up, Sec.~\ref{sec:realism-data})};

\node[realsrc,  below=0.4cm of rh] (r1) {Real recordings\\(TV, film, in-house)};
\node[realstep, below=0.4cm of r1] (r2) {LLM reconstructs\\$\mathcal{C}_{\text{profile}}$ + $\mathcal{C}_{\text{scene}}$};

\coordinate (fanout) at ($(r2.south)+(0,-0.55)$);

\node[negstep] (r3a) at ($(fanout)+(-\Rsep,-1.4)$)
    {$\tilde{\mathcal{C}}_{\text{scene}}$\\{\small scene mismatch}\\{\scriptsize\textit{text only}}};
\node[negstep] (r3b) at ($(fanout)+(0,-1.4)$)
    {$\tilde{\mathcal{C}}_{\text{profile}}{+}\tilde{\mathcal{C}}_{\text{scene}}$\\{\small role mismatch}\\{\scriptsize\textit{text only}}};
\node[negstep] (r3c) at ($(fanout)+(\Rsep,-1.4)$)
    {$\tilde{\mathcal{S}}_{r}$\\{\small delivery mismatch}\\{\scriptsize\textit{resynth. audio}}};

\node[realstep] (r4) at ($(fanout)+(0,-3.5)$)
    {Realism dataset\\{\scriptsize 15,414 samples\\(EN+ZH)}};
\coordinate (fanin) at ($(r4.north)+(0,0.45)$);

\draw[arr] (rh) -- (r1);
\draw[arr] (r1) -- node[steplabel,right]{\textbf{S1}} (r2);
\draw[thick] (r2.south) -- node[steplabel,right]{\textbf{S2}} (fanout);

\draw[thick, red!40] ($(fanout)+(-\Rsep,0)$) -- ($(fanout)+(\Rsep,0)$);
\node[neglabel] at (fanout) {Neg.\ $\mathcal{N}$};  

\draw[arr] ($(fanout)+(-\Rsep,0)$) -- (r3a.north);
\draw[arr] (fanout)                 -- (r3b.north);
\draw[arr] ($(fanout)+(\Rsep,0)$)  -- (r3c.north);

\draw[thick] (r3a.south) -- (r3a.south |- fanin);
\draw[thick] (r3b.south) -- (r3b.south |- fanin);
\draw[thick] (r3c.south) -- (r3c.south |- fanin);
\draw[thick, red!40] (r3a.south |- fanin) -- (r3c.south |- fanin);
\draw[arr] (fanin) -- node[steplabel,right]{\textbf{S3}} (r4.north);

\draw[gray!50, dashed] ($(ah.east)+(1.0,0.3)$) -- ++(0,-8.5);

\node[draw=red!30, rounded corners=3pt, fill=red!3,
      text width=2.8cm, font=\scriptsize,
      right=0.4cm of r3c]
    {Dashed boxes: output of $\mathcal{N}$ is a \textbf{mismatched} construction. Scene/role branches perturb \textbf{text only} (audio reused); the delivery branch \textbf{resynthesizes} audio. The three branches are \textbf{independent}.};

\end{tikzpicture}}
\caption{Data curation pipelines for the two evaluation strategies. \textbf{Left (blue):} Archetype top-down pipeline (Sec.~\ref{sec:archetype-data}) — role/scene keywords drive LLM generation and TTS synthesis. \textbf{Right (orange):} Realism bottom-up pipeline (Sec.~\ref{sec:realism-data}) — real recordings anchor the pipeline; the negative operator $\mathcal{N}$ applies three \emph{independent} (not sequential) perturbation branches in parallel: scene mismatch, role mismatch, and delivery mismatch. The scene- and role-mismatch branches perturb only the text and reuse the positive's original audio; only the delivery-mismatch branch resynthesizes the audio, so \emph{mismatched} here denotes a construction type, not an audio-provenance label.}
\label{fig:data-pipelines}
\end{figure}

\subsection{Archetype Data Curation Pipeline}

Archetype Data Collection follows a top–down perspective rooted in sociology and role theory, where social behavior is interpreted through shared stereotypes and generalized expectations~\citep{goffman1949presentation, campbell2008hero}. This strategy abstracts role-play into broad and easily recognizable archetypes, enabling scalable judgment of whether a generated response aligns with salient role features. In the formulation of Sec.~\ref{ssec:problem}, the role $(\mC_{\text{profile}}, \mC_{\text{scene}})$ here corresponds to a (stereotype keyword, scene keyword) pair expanded through templates, and $\mS_{r}$ is synthesized by an existing \texttt{SFnd}.

\noindent \textbf{Role dimensions ($\mC_{\text{profile}}$).}
We operationalize the character profile $\mC_{\text{profile}}$ through six principal archetype categories, each representing a stereotypical dimension of role identity or behavior, including
\begin{itemize}
    \item (1) \textbf{Occupation / Functional Role (Modern Daily Life)} (e.g., a firefighter rescuing a trapped civilian).
    \item (2) \textbf{Occupation / Functional Role (Fictional or Fantasy)} (e.g., a martial arts cult leader confronting orthodox rivals). 
    \item (3)~\textbf{Social Identity} (e.g., a doting parent worrying about a child, or a ``Buddhist youth'' eschewing competition). 
    \item (4) \textbf{Named / Specific Characters} (e.g., Sun Wukong mocking celestial authorities in \emph{Journey to the West}). 
    \item (5) \textbf{Personality Traits} (e.g., ``hot-blooded'' in a competitive setting, or ``complaining'' in response to poor service).
    \item (6) \textbf{Relative Roles} (e.g., parent or stranger).
\end{itemize}

\noindent \textbf{Scene dimensions ($\mC_{\text{scene}}$).}
In addition to the profile $\mC_{\text{profile}}$, we incorporate scene specifications $\mC_{\text{scene}}$ that anchor role-play in stereotypical situations. A central focus is on \textbf{everyday external events or state changes} (e.g., receiving a promotion, being caught in gossip), which provide concrete scenarios for testing archetypical responses.

\noindent \textbf{Exclusion of fine-grained controls.}
We deliberately excluded explicit emotional states, prosodic instructions, or linguistic constraints. 
These dimensions are better suited to realism-oriented or style-controlled evaluation, whereas archetype evaluation benefits from concrete, stereotype-level scenarios that sharpen judgments and improve human agreement.

\noindent \textbf{Pipeline overview.}
At a high level, archetype evaluation data are generated through a three-stage pipeline: 
(1) selecting keywords that specify role and scene; 
(2) expanding them into full prompts via structured templates; and 
(3) synthesizing prompts into speech with existing \texttt{SFnd} systems.
This modular pipeline ensures both scalability and reproducibility, while keeping prompts vivid enough for reliable judgments. Implementation details, including template design and alternative prompting strategies are provided in the Appendix~\ref{app:archetype-pipeline}.

Based on the generated data (paired scene and synthesized speech), expert human annotators are recruited for detailed annotation on semantic check (i.e., \textit{Content Pass}) and three different spoken role-play dimensions (i.e., \textit{Audio Quality}, \textit{Human-likeness}, \textit{Appropriateness}), which is detailed in following Section~\ref{ssec:archetype-annotation}.

Overall, archetype evaluation emphasizes breadth and recognizability over subtlety. 
By grounding role-play in socially recognizable categories and stereotypical situations, it provides a practical balance between scalability and interpretability, complementing the more fine-grained realism evaluation described in Section~\ref{ssec:strategies}.

\subsection{Archetype Annotation Protocol and Scoring Dimensions}
\label{ssec:archetype-annotation}

\begin{figure}[tp]
\centering
\begin{tikzpicture}[
    font=\small, every node/.style={align=center},
    inp/.style={draw, rounded corners=3pt, fill=blue!8, text width=4.5cm, minimum height=.8cm, inner sep=4pt},
    ann/.style={draw, rounded corners=3pt, fill=yellow!15, text width=11cm, minimum height=.65cm, inner sep=4pt},
    dim/.style={draw, rounded corners=3pt, fill=green!10, text width=4.1cm, minimum height=.8cm, inner sep=4pt},
    arr/.style={-{Stealth[length=4pt]}, thick}]
\node[font=\small\bfseries, blue!65!black] at (0,0)
    {Archetype Annotation (Sec.~\ref{ssec:archetype-annotation})};
\node[inp] (a1) at (-2.7,-.85) {Audio clip $\mS_r$ (TTS output)};
\node[inp] (a2) at (2.7,-.85) {Role keyword + scene prompt};
\node[ann] (a3) at (0,-2) {Human annotator: listen + read context};
\draw[thick] (a1.south) -- (-2.7,-1.5) -- (2.7,-1.5) -- (a2.south);
\draw[arr] (0,-1.5) -- (a3.north);
\node[dim] (a4) at (-4.7,-3.2) {Audio Quality (1--5)};
\node[dim] (a5) at (0,-3.2) {Human Likeness (1--5)};
\node[dim] (a6) at (4.7,-3.2) {Appropriateness (1--5)\\+ comment};
\draw[arr] (a3.south) -- (0,-2.55) -| (a4.north);
\draw[arr] (a3.south) -- (a5.north);
\draw[arr] (a3.south) -- (0,-2.55) -| (a6.north);
\node at (0,-4) {+ confidence rating; content pass gate};
\end{tikzpicture}

\medskip

\begin{tikzpicture}[
    font=\small, every node/.style={align=center},
    inp/.style={draw, rounded corners=3pt, fill=orange!10, text width=4.1cm, minimum height=.95cm, inner sep=4pt},
    ann/.style={draw, rounded corners=3pt, fill=yellow!15, text width=14cm, minimum height=.85cm, inner sep=4pt},
    dim/.style={draw, rounded corners=3pt, fill=green!10, text width=2.65cm, minimum height=1.55cm, inner sep=3pt},
    arr/.style={-{Stealth[length=4pt]}, thick}]
\node[font=\small\bfseries, orange!70!black] at (0,0)
    {Realism Annotation (Sec.~\ref{ssec:annotation-protocol})};
\node[inp] (r1) at (-5,-1) {Audio clip $\mS_r$\\(real or TTS)};
\node[inp] (r2) at (0,-1) {Character profile\\$\mC_{\text{profile}}$};
\node[inp] (r3) at (5,-1) {Local scene\\$\mC_{\text{scene}}$};
\node[ann] (r4) at (0,-2.45)
    {Human annotator: listen + read context\\Mismatch reason hidden from annotator};
\draw[thick] (r1.south) -- (-5,-1.85) -- (5,-1.85) -- (r3.south);
\draw[thick] (r2.south) -- (0,-1.85);
\draw[arr] (0,-1.85) -- (r4.north);
\node[dim] (d1) at (-6,-4.1) {Prosodic\\Dynamics\\[2pt]{\footnotesize pitch / rhythm\\/ stress}};
\node[dim] (d2) at (-3,-4.1) {Emotional\\Expressiveness\\[2pt]{\footnotesize accuracy / intensity\\/ transition}};
\node[dim] (d3) at (0,-4.1) {Character\\Consistency\\[2pt]{\footnotesize identity / traits}};
\node[dim] (d4) at (3,-4.1) {Contextual\\Relevance\\[2pt]{\footnotesize local / global}};
\node[dim] (d5) at (6,-4.1) {Semantic\\Match\\[2pt]{\footnotesize meaning fidelity}};
\draw[arr] (r4.south) -- (0,-3.05) -| (d1.north);
\draw[arr] (r4.south) -- (0,-3.05) -| (d2.north);
\draw[arr] (r4.south) -- (d3.north);
\draw[arr] (r4.south) -- (0,-3.05) -| (d4.north);
\draw[arr] (r4.south) -- (0,-3.05) -| (d5.north);
\node[draw=red!40, rounded corners=3pt, fill=red!5,
      text width=14.5cm, inner sep=6pt] at (0,-5.75)
    {\textbf{Contrastive negatives:} role/scene mismatch changes text only and reuses the anchor audio (real or TTS); delivery mismatch resynthesizes audio. Annotators score blind; mismatch type is used only for internal QA.};
\end{tikzpicture}
\caption{Annotation flows for both evaluation strategies. \textbf{Top (blue):} Archetype annotation rates Audio Quality, Human Likeness, and Appropriateness from TTS output and role/scene context. \textbf{Bottom (orange):} Realism annotation rates five dimensions from audio, character profile, and local scene. Audio provenance (real or TTS) is distinct from construction type (positive or mismatched negative); only the delivery-mismatch branch resynthesizes audio.}
\label{fig:annotation-flows}
\end{figure}

The curation pipeline above produces tuples $(\mS_{r}, \mC_{\text{profile}}, \mC_{\text{scene}})$ as defined in Eq.~\ref{eq:sem_single}, where $\mC_{\text{profile}}$ and $\mC_{\text{scene}}$ are stereotype-level role/scene prompts and $\mS_{r}$ is the synthesized response. To convert these outputs into reliable supervision signals $\mE_{r}$ for both training and benchmarking \texttt{SEval} models, we apply a dedicated archetype-oriented annotation protocol. As in Sec.~\ref{ssec:annotation-protocol} for the realism flow, we use the term \emph{scoring dimensions} (rather than ``evaluation dimensions'') to keep the human-rating axes terminologically separate from the \texttt{SEval} model itself.

Because archetypes rely on broad, easily recognizable expressive patterns rather than fine-grained behavioral detail, the annotation rubric focuses on dimensions that listeners naturally use when judging whether a speaker fits a stereotypical role.  
The protocol consists of three dimensions:

\begin{itemize}[leftmargin=1.2em]

    \item \textbf{Audio Quality} (context-free).  
    A preliminary assessment of whether the signal is sufficiently clear for annotators to evaluate role compatibility~\citep{kirkland2023stuck, cooper2024review, perrotin2025refining}.
    Noticeable artifacts, distortions, or unstable timbre may obscure vocal cues relevant to archetype identification.

    \item \textbf{Human-likeness} (context-free).  
    This dimension evaluates whether the speech exhibits properties of natural human vocal production, including stable timbre and plausible prosody.  
    Studies in voice perception indicate that listeners attribute personality and social characteristics only when the voice falls within a human vocal range~\citep{cooper2024review}.  
    As such, human-likeness serves as a prerequisite for making interpretable role judgments.

    \item \textbf{Appropriateness} (context-dependent).  
    The central dimension of archetype evaluation.  
    It measures whether the expressive style, including prosodic patterns, affective coloration, and trait-related vocal cues, aligns with the archetype implied by the role and scene description.  
    This dimension requires considering contextual framing, consistent with sociological accounts of role enactment and expectation~\citep{welch2021listeners, lavan2023time, guldner2024human}.
\end{itemize}

Together, these dimensions determine whether a spoken performance is (i) technically interpretable, (ii) human-sounding, and (iii) compatible with the intended archetype.  
The rubric avoids detailed prosodic or emotional sub-dimensions—addressed separately in realism evaluation—because archetype judgments depend more on global stylistic patterns than on fine-grained contextual behavior.

Annotators also complete a \textbf{Content Pass} check to verify that the spoken response conveys the intended meaning of the prompt. \rev{The check is not degenerate: $1{,}266/1{,}499$ test items ($84.5\%$) and $5{,}576/6{,}773$ training items ($82.3\%$) pass, so both classes are well populated.}
Full scoring scales, examples, and instructions are provided in Appendix~\ref{app:archetype-pipeline}.

\subsection{Candidate Models for Evaluation}
\label{app:archetype-omni-models}

We evaluate a diverse set of speech foundation models spanning both English and Mandarin. In total, we include 8 English and 7 Mandarin models, covering open-source and proprietary systems, and balancing end-to-end and cascaded architectures. Table~\ref{tab:candidate-models} provides the full list of models.

\begin{table}[ht]
\centering
\caption{Candidate speech foundation models for evaluation across English and Mandarin. \texttt{G} and \texttt{Q} in the cascaded model names refer to \texttt{Gemini2.5Pro} and \texttt{Qwen3-30B-Instruct}.}
\label{tab:candidate-models}
\resizebox{\linewidth}{!}{
\begin{tabular}{@{}llll@{}}
\toprule
\textbf{Model} & \textbf{License / Availability} & \textbf{Architecture} & \textbf{Reference} \\ 

\midrule
\multicolumn{4}{c}{\textbf{English Models (8)}} \\ \midrule
\texttt{GPT-4o-realtime} & Proprietary & Unknown & \citep{openai_gpt4o_realtime_preview} \\
\texttt{Doubao} & Proprietary & Unknown & \citep{doubao_chat} \\
\texttt{GPT-4o}(Advanced Voice Mode) & Proprietary & Unknown & \citep{openai_chatgpt_gpt4o} \\
\texttt{GLM4-Voice} & Open-source & End-to-end  & \citep{zeng2024glm} \\
\texttt{KimiAudio} & Open-source & End-to-end & \citep{ding2025kimi} \\
\texttt{Qwen-2.5-Omni} & Open-source & End-to-end & \citep{xu2025qwen2} \\
\texttt{Q-CosyVoice2} & Open-source & Cascaded & \citep{yang2025qwen3, du2024cosyvoice2} \\
\texttt{G-Cosyvoice2} & Proprietary + Open-source & Cascaded & \citep{comanici2025gemini, du2024cosyvoice2} \\
\midrule
\multicolumn{4}{c}{\textbf{Mandarin Models (7)}} \\ \midrule
\texttt{GPT-4o-realtime} & Proprietary & Unknown & \citep{openai_gpt4o_realtime_preview} \\
\texttt{Doubao} & Proprietary & Unknown & \citep{doubao_chat} \\
\texttt{GLM4-Voice} & Open-source & End-to-end & \citep{zeng2024glm} \\
\texttt{KimiAudio} & Open-source & End-to-end & \citep{ding2025kimi} \\
\texttt{Qwen-2.5-Omni} & Open-source & End-to-end & \citep{xu2025qwen2} \\
\texttt{Q-CosyVoice2} & Open-source & Cascaded & \citep{yang2025qwen3, du2024cosyvoice2} \\
\texttt{G-Cosyvoice2} & Proprietary + Open-source & Cascaded & \citep{comanici2025gemini, du2024cosyvoice2} \\
\bottomrule
\end{tabular}}
\end{table}

\noindent \textbf{Collection for Proprietary Models.} For proprietary systems, we adopt two collection strategies. For GPT-4o-realtime, responses are obtained directly via the \texttt{gpt-4o-realtime} API. For Doubao and GPT-4o (Advanced Voice Mode), we collect outputs by recording from their official web interfaces. All recordings are captured through the interaction device itself, ensuring that no additional processing or distortion is introduced into the final audio.  

\noindent \textbf{End-to-End Model Generation.}  
For end-to-end systems, the audio prompt is directly used as input. The system prompt is set to default values to ensure consistent speech generation quality.  

\noindent \textbf{Cascaded Model Generation.}  
For cascaded systems, we only use Cosyvoice2 as the TTS model in our collection of data. We select \texttt{Qwen3-30B-Instruct} and \texttt{Gemini2.5Pro} as the text LLM candidates~\citep{yang2025qwen3, comanici2025gemini}.

During the generation, we adopt a retrieval-augmented generation~(RAG) strategy that combines text LLMs with zero-shot TTS models: 
\begin{itemize}
    \item Step 1: We first build a speech database (LibriSpeech and GigaSpeech for English, WenetSpeech for Mandarin~\citep{panayotov2015librispeech, chen2021gigaspeech, zhang2022wenetspeech}). Each sample is annotated with Gemini-based speech captions describing salient acoustic attributes.
    \item Step 2: Given a character and scene description, the text LLM imagines a candidate caption and generates a corresponding transcript (see below prompts for the text LLM).
    \item Step 3: We retrieve acoustically similar samples from the database using \texttt{Qwen3-Embedding-7B}~\citep{zhang2025qwen3}.  
    \item Step 4: The retrieved samples guide the zero-shot TTS system, which synthesizes speech conditioned on both the transcript and matched acoustic profile.  
\end{itemize}

\subsection{Archetype Evaluation Data Summary.}
The resulting archetype evaluation dataset comprises 552 distinct scenarios for both Mandarin and English, evaluated across the \rev{eight English and seven Mandarin} model configurations listed in Section~\ref{app:archetype-omni-models} (\rev{15 in total}),\footnote{\rev{\texttt{GPT-4o} Advanced Voice Mode (an English-only configuration) was deactivated after our collection; its already-collected outputs are retained, so all eight English configurations remain in the dataset.}} yielding \rev{8,272} samples. In total, the dataset is split into \rev{6,773} samples for training and \rev{1,499} for testing. The split is constructed so that no role-keyword $\times$ scene-keyword combination appears in both partitions, ensuring that test evaluation is not inflated by memorized (keyword, scenario) pairs. For benchmarking, an additional 1,250 scenarios spanning seven role/scene categories are prepared to facilitate archetype role-play benchmarking. \rev{These 1,250 are \emph{unannotated generation prompts} for the downstream \texttt{SFnd} benchmarking track (future work); they are outside, and not counted in, the $8{,}272$ annotated EvalBench records used for all evaluator results in this paper.} More detailed statistics and analyses are provided in Appendix~\ref{app:archetype-data-analysis}.

\section{DRAME-EvalBench Data Curation --- Realism Data Collection}
\label{sec:realism-data}

This section instantiates the \emph{bottom-up} (realism) column of Table~\ref{tab:topdown-bottomup}. In contrast to Sec.~\ref{sec:archetype-data}, where the spoken response $\mS_{r}$ is synthesized from stereotype keywords, here the pipeline is anchored on a real human recording from which the role $(\mC_{\text{profile}}, \mC_{\text{scene}})$ is reconstructed (curation operator $\mathcal{D}^{\text{bottom-up}}$ in Eq.~\ref{eq:dbottomup}); note that the served positive audio in the basic set is a mixture of these source recordings and TTS resynthesis, and only the held-out authentic-speech test set is uniformly real (Sec.~\ref{sec:reproducibility}). On top of this positive flow, we additionally apply the negative-sample operator $\mathcal{N}$ of Eq.~\ref{eq:negative} to obtain contrastive negatives that probe whether \texttt{SEval} attends to delivery, identity, and scene cues. Throughout this section, items produced by $\mathcal{N}$ are explicitly marked as \emph{mismatched} constructions (a construction type, not an audio-provenance label): the role/scene branches perturb only the text and reuse the anchor audio while the delivery branch resynthesizes it, and none is served to annotators as an authentic ground-truth pairing. \rev{The operator's audio reuse links a served negative to its \emph{anchor} source segment, not to a second in-set record: in the experimental benchmark-distribution test split each source segment is served once (as either a positive or a mismatched negative), so the $2{,}000$ benchmark-distribution test records span $2{,}000$ distinct source clips and no two share the same audio. This is what makes record-level bootstrap resampling independent at the clip level (Sec.~\ref{ssec:main-results}, App.~\ref{app:tier1}).}

\subsection{Realism Data Curation Pipeline}

Realism Data Collection adopts a bottom–up perspective grounded in real human speech data, with the goal of capturing the nuanced delivery, prosody, and contextual fit that synthetic archetype prompts cannot fully represent. Rather than relying on stereotypes, this strategy leverages real-world recordings and carefully constructed contrasts, enabling fine-grained assessment of model performance under realistic conditions.

\subsubsection{Positive samples: real speech with reconstructed role}

The positive flow proceeds in three sequential stages, all operating on the same underlying spoken segment $\mS_{r}$:
\begin{enumerate}[label=\textbf{(P\arabic*)}, leftmargin=2.2em]
    \item \textbf{Acquire $\mS_{r}$ — ground-truth audio from TV and film.} Spoken segments are curated from publicly available media, ensuring coverage across diverse roles and communicative settings. This corresponds to the ``$\mS_{r}$'' input of Eq.~\ref{eq:dbottomup}.
    \item \textbf{Reconstruct $\mC_{\text{profile}}$ — retrieved character profile.} Each audio segment is paired with a concise character description derived from the media context. The detailed prompt design for generating these character profiles is provided in Appendix~\ref{app:realism-character}.
    \item \textbf{Reconstruct $\mC_{\text{scene}}$ — generated local scene.} We supplement the retrieved character information with an automatically generated local scene description summarizing the surrounding plot or setting. Appendix~\ref{app:realism-scenes} includes representative prompts and generation details.
\end{enumerate}
The output of this positive flow is a tuple $(\mS_{r}, \mC_{\text{profile}}, \mC_{\text{scene}})$ — directly the same signature as Eq.~\ref{eq:sem_single}. To make the example concrete: in a clip from a courtroom drama, $\mS_{r}$ is the actor's spoken line; $\mC_{\text{profile}}$ is the reconstructed description of the \emph{character} (``a stoic defense attorney in his late 40s, controlled but tense''); and $\mC_{\text{scene}}$ is the reconstructed description of the surrounding \emph{scene} (``cross-examination of a hostile witness during a high-stakes trial'').

\subsubsection{Negative samples: contrastive perturbations via \texorpdfstring{$\mathcal{N}$}{N}}
\label{sssec:negatives}

A central challenge in realism-oriented assessment is determining whether a judge model can detect subtle inconsistencies between how a line is delivered and the narrative or role context in which it is supposed to occur. Real human speech encodes role-dependent clues — intonation, emotional stance, personality expression, pacing, and interactional intent — that cannot be captured by transcript alone. To evaluate whether models attend to these cues, we apply the negative-sample operator $\mathcal{N}$ defined in Eq.~\ref{eq:negative}: starting from a positive sample $(\mS_{r}, \mC_{\text{profile}}, \mC_{\text{scene}})$, we apply \emph{exactly one} perturbation branch, holding the untouched components fixed (within a branch, coupled fields such as profile and scene may be jointly adjusted to keep the perturbed context internally coherent).

\paragraph{Why constructed negatives.} The importance of such constructed negatives is well-established across evaluation research. In speech, benchmarks such as NISQA and ConferenceSpeech 2022 rely heavily on simulated degradations and synthetic perturbations to expose model weaknesses that natural corpora fail to capture~\citep{rao2021conferencingspeech, mittag2021nisqa}. The same principle appears in language and multimodal datasets: NLI benchmarks introduce adversarial hypotheses and counterfactual contradictions~\citep{nie2020adversarial}, while FOIL-COCO and Winoground mismatch caption semantics to test grounding~\citep{shekhar2017foil, thrush2022winoground}. Across all these domains, human-validated mismatches reveal failure modes that positive-only datasets systematically overlook.

\paragraph{Three parallel branches of $\mathcal{N}$.} Eq.~\ref{eq:negative} admits three parallel branches; \emph{any one positive sample can give rise to a negative under any branch}, but a given generated negative belongs to exactly one branch. The three branches are:
\begin{itemize}[leftmargin=1.2em]
    \item \textbf{Negative Local Scenes (context-mismatch, $\tilde{\mC}_{\text{scene}}$).} Probes whether judges detect \emph{scene-level narrative inconsistencies}. For each authentic segment we generate $\tilde{\mC}_{\text{scene}}$ at varying mismatch depths, from subtle semantic perturbations (incorrect interpersonal dynamics or emotional framing) to entirely unrelated scenarios. Generated via prompt-based text generation; detailed pipeline in Appendix~\ref{app:realism}.
    \item \textbf{Negative Character Profiles (role-mismatch, $\tilde{\mC}_{\text{profile}}$).} Probes whether judges recognize \emph{persona and identity mismatches}. We pair a true speech segment with an incorrect character description or swap key personality traits, social roles, and emotional dispositions. Generated via prompt-based text generation; detailed pipeline in Appendix~\ref{app:realism}.
    \item \textbf{Negative Speech Samples (delivery-mismatch, $\tilde{\mS}_{r}$).} Probes whether judges detect mismatches in prosody, emotional intent, vocal identity, or speaking style. We resynthesize the transcript with diverse TTS voices, modify expressive style, or deliberately alter prosodic alignment. Sequential per-system pipelines (Speaker Assignment, Voice Clone, etc.) are detailed in Sec.~\ref{ssec:realism-tts}.
\end{itemize}
\noindent\textit{The three branches are contrastive \emph{mismatched} constructions rather than a distinct audio-provenance class: $\tilde{\mC}_{\text{scene}}$ and $\tilde{\mC}_{\text{profile}}$ perturb only the text and reuse their source clip's own original audio (real or TTS), while only $\tilde{\mS}_{r}$ resynthesizes the audio.} \rev{During construction a role- or context-negative retains its source clip and perturbs only the associated text; in the experimental benchmark-distribution test set that clip appears only once, and its matching unperturbed positive counterpart is not included as a second record. Consequently the discrimination AUROCs reported in App.~\ref{app:tier1} compare distinct clips rather than paired versions of the same audio.} Together, they cover complementary aspects of realism: $\tilde{\mC}_{\text{scene}}$ tests contextual grounding, $\tilde{\mC}_{\text{profile}}$ tests persona and role consistency, and $\tilde{\mS}_{r}$ tests paralinguistic and delivery appropriateness. This triad mirrors the multi-axis perturbation strategies used in the human-labeled datasets cited above, and ensures that realism assessment captures distinctions that are easily perceived by human listeners but often overlooked by current speech models.

\paragraph{Annotation conditions.}
During annotation, raters see context information drawn from one of (i) ground-truth stories, (ii) stories with negative local scenes ($\tilde{\mC}_{\text{scene}}$), or (iii) stories with negative character profiles ($\tilde{\mC}_{\text{profile}}$); for the target speech, different types of synthesized speech and ground-truth speech are mixed. Annotators assign realism scores \emph{without} being exposed to the underlying mismatch reasons. Those reasons are visible only to the quality-assurance team, who use them as reference for internal validation and consistency checks. While individual mismatch reasons may not be perfect, they support quality control and maintain reliability across the dataset.

\paragraph{A worked example: \emph{Yunjin}.} To make the operator $\mathcal{N}$ concrete in the body of the paper, consider an opera segment from \emph{Genshin Impact} included in our experimental corpus. The positive tuple $(\mS_{r}, \mC_{\text{profile}}, \mC_{\text{scene}})$ is: $\mS_{r}$, the actor's operatic narration of a battle scene; $\mC_{\text{profile}}$, ``Yunjin, a passionate opera performer who narrates heroic deeds in traditional singing style''; $\mC_{\text{scene}}$, ``a fierce battlefield narration in which armored riders charge into enemy lines.'' Applying the three branches of $\mathcal{N}$ yields contrastive triples that share the \emph{same audio} (or transcript) but break realism in different ways: $\tilde{\mC}_{\text{scene}}$ replaces the battlefield with a quiet teahouse, a calm tea gathering, or a modern caf\'e (mild $\to$ severe context-mismatch); $\tilde{\mC}_{\text{profile}}$ replaces Yunjin with an introverted software engineer, a logical roboticist, or an anti-traditionalist street performer (mild $\to$ severe role-mismatch); and $\tilde{\mS}_{r}$ resynthesizes the same lyrics with one of the catalog/cloning TTS configurations of Sec.~\ref{ssec:realism-tts}, replacing the operatic delivery with a flatter or stylistically unrelated voice. The spoken transcript is unchanged across all three branches, so a judge that sees only the transcript (and not the profile/scene) cannot tell positives from negatives. The two cases differ in what \emph{can} expose the mismatch: for the role- and context-mismatch branches the perturbation lives in the profile/scene \emph{text}, so a judge that reads that text can flag the inconsistency without audio (text-only reaches AUROC $0.599$ on the role-fit axis; App.~\ref{app:tier1}); the delivery-mismatch branch instead leaves the text intact and resynthesizes the audio, so only prosody, identity, and delivery cues distinguish it. Detecting the text-detectable branches while still scoring delivery therefore requires a model that uses audio and text jointly: in a matched modality ablation on the role-fit axis (audio-only and text-only obtained by blanking the other modality at inference), the full model reaches AUROC $0.711$, versus text-only $0.599$ and audio-only $0.408$ (App.~\ref{app:tier1}). This modality ablation is a different measurement from the shipped evaluator's role-mismatch AUROC of $0.872$ reported in App.~\ref{app:tier1} (which applies the unmodified \texttt{DRAME-Eval} to the full role-mismatch negative set), so the two numbers are not directly comparable; and we do not evaluate discrimination on the delivery-mismatch branch as a controlled AUROC (App.~\ref{app:tier1}). The full anchored example, including the original verses and per-level mismatch rationales, is reproduced in App.~\ref{app:realism-data-example}.

\noindent \textbf{Real-world Recording.}
As noted in Sec.~\ref{sec:literature}, most prior efforts have relied heavily on synthetic data, mostly from TTS systems based on curated professional role-play voices. While such data can achieve high fidelity, it often lacks the diversity and spontaneity of real-world delivery, limiting the generalizability of evaluation models. To address this gap, we expand realism evaluation beyond synthetic and curated voices by incorporating newly recorded speech from both professional and amateur voice actors. This design introduces a broader spectrum of delivery styles, from polished professional acting to more natural and less controlled amateur performance.

\begin{table}[t]
    \centering
    \small
    \caption{
        Role-play evaluation dimensions and corresponding sub-dimensions with keyword-style summaries.
        The table highlights core cues used for scoring across prosody, emotion, character, context, and semantics.
    }
    \resizebox{\linewidth}{!}{
    \begin{tabular}{lll}
    \toprule
        \textbf{Dimension} & \textbf{Sub-dimension} & \textbf{Discussion Focus} \\
        \midrule
        \multirow{3}{*}{\textbf{Prosodic Dynamics}} 
            & Pitch Variation 
            & Contour richness; range; monotony vs. expressiveness \\
            & Rhythmic Naturalness 
            & Pacing; timing; pause placement; flow \\
            & Stress \& Emphasis 
            & Accenting; highlight key words; communicative salience \\
        \midrule
        \multirow{3}{*}{\textbf{Emotional Expressiveness}} 
            & Emotion Accuracy 
            & Correct emotion type; scene-appropriate affect \\
            & Emotion Intensity 
            & Strength calibration; under-/over-shooting \\
            & Emotion Transition 
            & Smooth shifts; narrative-motivated changes \\
        \midrule
        \multirow{2}{*}{\textbf{Character Consistency}} 
            & Voice Identity Matching 
            & Timbre stability; general impression match \\
            & Traits Embodiment 
            & Personality cues; role-specific traits \\
        \midrule
        \multirow{2}{*}{\textbf{Contextual Relevance}} 
            & Local Scene Fit 
            & Situational alignment; Environmental interaction \\
            & Global Story Coherence 
            & Character likeness; world-rule adherence; continuity \\
        \midrule
        \multirow{1}{*}{\textbf{Semantic Match}} 
            & Semantic Match 
            & Meaning fidelity; key details preserved; no hallucination \\
        \bottomrule
    \end{tabular}}
    \label{tab:roleplay-dimensions}
\end{table}

\subsection{Annotation Protocol and Scoring Dimensions}
\label{ssec:annotation-protocol}

This subsection specifies how human annotators produce $\mE_{r}$ in Eq.~\ref{eq:sem_single} for the realism flow. Concretely, given a tuple $(\mS_{r}, \mC_{\text{profile}}, \mC_{\text{scene}})$ — possibly perturbed by $\mathcal{N}$ as described in Sec.~\ref{sssec:negatives} — annotators rate the spoken performance $\mS_{r}$ against the provided \emph{character profile} $\mC_{\text{profile}}$ and \emph{local scene} $\mC_{\text{scene}}$ using a structured multi-dimensional rubric. We use the term \emph{scoring dimensions} (rather than ``evaluation dimensions'') below to disambiguate the human-rating axes from the \texttt{SEval} model itself. The rubric is designed to capture the core perceptual factors that listeners naturally rely on when judging spoken role-play, focusing on five complementary dimensions:

\begin{itemize}[leftmargin=1.2em]
    \item \textbf{Prosodic Dynamics} (Pitch Variation, Rhythmic Naturalness, Stress~\&~Emphasis):  
    Prosody is the primary carrier of expressiveness in speech and strongly shapes a listener's perception of naturalness, intent, and dramatic nuance~\citep{banziger2005role}.  
    These sub-dimensions capture whether the speaker delivers a rhythmically coherent, suitably varied, and communicatively emphasized performance.

    \item \textbf{Emotional Expressiveness} with progressive gating (Emotion Accuracy $\rightarrow$ Intensity $\rightarrow$ Transition):  
    Emotional realism underpins believable role-play~\citep{ye2023temporal, russell1980circumplex, johnstone2000vocal}.
    The gated structure reflects the perceptual dependency among aspects of emotion: listeners first check \emph{whether} the expressed emotion is appropriate (Accuracy), then \emph{how strongly} it is conveyed (Intensity), and finally \emph{how it evolves} over time (Transition).  
    This avoids collapsing distinct failure modes (e.g., wrong emotion vs.\ correct emotion but unnatural transitions).

    \item \textbf{Character Consistency} (Voice Identity Matching, Traits Embodiment):  
    Role-play requires maintaining a stable character.  
    This dimension assesses whether the delivered voice is compatible with the identity implied by the character profile and scene --- there is no reference voice to match against --- and whether the vocal performance conveys the persona's defining traits (e.g., age, personality, profession). It captures aspects missing from standard speech-quality metrics and cannot be inferred from transcript-level correctness alone~\citep{podesva2015voice}.

    \item \textbf{Contextual Relevance} (Local Scene Fit, Global Story Coherence):  
    Spoken role-play is fundamentally contextual rather than isolated~\citep{grosz1986attention}. This dimension evaluates whether the performance responds appropriately to the immediate situation (\emph{local fit}) and remains consistent with broader narrative constraints (\emph{global coherence}).  
    Models frequently fail at this long-range consistency, making it critical for realism-oriented evaluation.

    \item \textbf{Semantic Match}:  
    This captures whether the spoken realization faithfully reflects the intended semantic content.  
    While prosody and emotion can vary stylistically, semantic fidelity is a hard requirement for meaningful evaluation and for differentiating expressive errors from content errors.
\end{itemize}

These dimensions are chosen to jointly cover the full stack of perceptual factors involved in human assessment of situated speech: from low-level acoustic expressiveness (prosody), to affect (emotion), persona stability (character), contextual grounding, and semantic correctness. They exclude redundant or highly system-dependent metrics (e.g., signal-level artifacts, basic intelligibility scores), focusing instead on factors that are both \emph{context-aligned} and \emph{specific to role-play}. Annotators also provide a prompt-quality check and confidence ratings. Full scales, anchors, and gating logic are presented in Appendix~\ref{app:realism-annotation}.

\subsection{Details for Negative Speech Samples (\texorpdfstring{$\tilde{\mS}_{r}$}{\~{}S\_r})}
\label{ssec:realism-tts}

This subsection details the third branch of the negative-sample operator $\mathcal{N}$ in Eq.~\ref{eq:negative}, namely the construction of $\tilde{\mS}_{r}$ via TTS-based resynthesis. We employed a diverse set of TTS systems --- both commercial and non-commercial --- spanning a range of speaker-control mechanisms. Table~\ref{tab:tts-settings} summarizes the systems and settings, while the rest of this subsection organizes the underlying strategies along two axes.

\subsubsection{Two parallel families of speaker-control strategies}

We group the strategies into two \emph{parallel} families based on what is supplied to the TTS system as the speaker handle: \emph{catalog-based} (the TTS exposes a fixed library of speakers / styles) and \emph{ground-truth-conditioned} (the TTS is conditioned on the original speech itself):
\begin{itemize}[leftmargin=1.2em]
    \item \textbf{Catalog-based.} The TTS system has a published catalog of speakers/styles, and we pick one entry per sample.
    \begin{itemize}[leftmargin=1.2em]
        \item \emph{Speaker Assignment.} We leveraged the publicly available speaker-style information released by each platform. For each input sample, \texttt{deepseek-v3} was used to select the best-matching catalog speaker based on the original speech caption. For \texttt{Gemini2.5TTS} only, the \emph{character profile} $\mC_{\text{profile}}$ (not the local scene) was additionally appended as an instruction to guide synthesis.
        \item \emph{Similar Voice Matching.} We used each platform's built-in voice-search function to identify the closest catalog voice to the ground-truth recording, and resynthesized the utterance with that voice.
        \item \emph{Voice Design.} The character profile $\mC_{\text{profile}}$ was first converted into a descriptive voice specification by prompting \texttt{deepseek-v3}~\citep{liu2024deepseek}. The description was submitted to the TTS service to generate candidate voices, and the voice that best matched the speaker style of the original recording (judged via Gemini2.5Pro) was selected before generation.
    \end{itemize}
    \item \textbf{Ground-truth-conditioned.} The TTS is conditioned on audio derived from the ground-truth speaker, with three settings forming a sequence of progressively tighter conditioning:
    \begin{enumerate}[leftmargin=1.6em]
        \item \emph{Voice Clone (intermediate-tag).} Cloning uses the ground-truth voice as the source but routed through an intermediate representation (voice information / tag) rather than direct conditioning.
        \item \emph{Identical Voice Clone.} The ground-truth voice is used directly as the conditioning input, with no intermediate tag.
        \item \emph{Same-Speaker Voice Clone.} Instead of conditioning on the \emph{specific} ground-truth utterance, we use \emph{another} utterance from the same speaker, testing robustness to cross-utterance generalization.
    \end{enumerate}
\end{itemize}
For models using voice prompts in the ground-truth-conditioned family, we additionally apply an in-house enhancement model to first process the prompt audio so as to minimize the effect of audio events and background noise.

\begin{table}[ht]
    \centering
    \caption{TTS systems and generation settings used in Realism Evaluation.}
    \label{tab:tts-settings}
    \resizebox{\linewidth}{!}{
    \begin{tabular}{llll}
        \toprule
        \textbf{System} & \textbf{Type} & \textbf{Prompt Type} & \textbf{References} \\
        \midrule
        OpenAI \texttt{TTS-1 HD} & Proprietary & Speaker Assignment & API Call \citep{openai_tts1hd_docs} \\
        \midrule
        \texttt{Gemini2.5pro-preview-TTS} & Proprietary & Speaker Assignment with Instruction & API Call \citep{google_gemini_tts_docs}  \\
        \midrule
        \multirow{3}{*}{ElevenLabs \texttt{Multilingual V2}} & Proprietary & Similar Voice Matching & API Call \citep{elevenlabs_multilingual_v2_docs} \\
         & Proprietary & Voice Design & API Call \citep{elevenlabs_multilingual_v2_docs} \\
         & Proprietary & Voice Clone & API Call \citep{elevenlabs_multilingual_v2_docs} \\
         \midrule
        \texttt{HumeTTS} & Proprietary & Voice Design & API Call \citep{hume_voice_design} \\
        \midrule
        \multirow{2}{*}{\texttt{OpenAudio-S1}} & Proprietary & Identical Voice Clone & API Call \citep{openaudio_s1} \\
         & Proprietary & Same Speaker Voice Clone & API Call \citep{openaudio_s1} \\
        \midrule
        \multirow{2}{*}{MiniMax \texttt{speech-02-hd}}  & Proprietary & Voice Design & API Call \citep{openaudio_s1} \\
         & Proprietary & Voice Clone & API Call \citep{openaudio_s1} \\
        \midrule
        \multirow{2}{*}{\texttt{InhouseTTS1}} &  Proprietary & Identical Voice Clone & - \\
         &  Proprietary & Same Speaker Voice Clone & - \\
                 \midrule
        \multirow{2}{*}{\texttt{InhouseTTS2}} &  Proprietary & Identical Voice Clone & - \\
         &  Proprietary & Same Speaker Voice Clone & - \\
                 \midrule
        \multirow{2}{*}{\texttt{InhouseTTS3}} &  Proprietary & Identical Voice Clone & - \\
         &  Proprietary & Same Speaker Voice Clone & - \\
                 \midrule
        \multirow{2}{*}{\texttt{InhouseTTS4}} &  Proprietary & Identical Voice Clone & - \\
         &  Proprietary & Same Speaker Voice Clone & - \\
        \midrule
        \multirow{2}{*}{\texttt{InhouseTTS5}} &  Proprietary & Identical Voice Clone & - \\
         &  Proprietary & Same Speaker Voice Clone & - \\
        \midrule
        \multirow{2}{*}{\texttt{CosyVoice2}} & Open-source & Identical Voice Clone & \citep{du2024cosyvoice} \\
         & Open-source & Same Speaker Voice Clone & \citep{du2024cosyvoice} \\
         \midrule
        \multirow{2}{*}{\texttt{Vevo1.5}} & Open-source & Identical Voice Clone & \citep{zhangvevo} \\
         & Open-source & Same Speaker Voice Clone & \citep{zhangvevo} \\
         \midrule
        \multirow{2}{*}{\texttt{MaskGCT}} & Open-source & Identical Voice Clone & \citep{wangmaskgct} \\
         & Open-source & Same Speaker Voice Clone & \citep{wangmaskgct} \\
         \midrule
        \multirow{2}{*}{\texttt{OpusLM-7B}} & Open-source & Identical Voice Clone & \citep{tian25b_interspeech} \\
         & Open-source & Same Speaker Voice Clone & \citep{tian25b_interspeech} \\
        \bottomrule
    \end{tabular}}
\end{table}

\subsubsection{In-house Models.}
\label{sssec:inhouse-tts}
We additionally employed five proprietary in-house TTS systems, each exploring different conditioning strategies for realism-oriented speech synthesis. All five fall under the ground-truth-conditioned family above (i.e., they produce $\tilde{\mS}_{r}$ via cloning).

\begin{itemize}
    \item \texttt{InhouseTTS1} (Phone--Cosy2 Pipeline). 
The input is a phone sequence produced by a grapheme-to-phoneme~(G2P) front-end. A 0.4B GPT-2 model~\citep{radford2019language} predicts \texttt{CosyVoice2} S3-tokens, which are then passed through a flow-matching model to generate mel-spectrograms. A neural vocoder from \texttt{CosyVoice2} converts the mel features into audio. Both the GPT-2 and flow-matching modules are augmented with speaker embeddings for speaker control.
    \item \texttt{InhouseTTS2} (BPE--Cosy2 Pipeline).
Instead of phones, this model takes Qwen2.5 byte-pair-encoding (BPE) tokens as input~\citep{xu2025qwen2}. The same 0.4B GPT-2 predicts \texttt{CosyVoice2} tokens, followed by the identical flow-matching and vocoder stages as in \texttt{InhouseTTS1}. Speaker embeddings are also injected to maintain consistent speaker style.
    \item \texttt{InhouseTTS3} (Embedding--AudioCode Pipeline).
Here, the input is Qwen2.5 BPE embeddings, which a 0.4B GPT-2 model maps directly to 4,096-dimensional audio codes. These codes are decoded via a cascaded architecture: \texttt{RepCodec} decoder~\citep{huang2024repcodec} $\rightarrow$ content representation (with speaker embedding) $\rightarrow$ \texttt{DDIM} (denoising
diffusion implicit model) module~\citep{jeong2021diff} $\rightarrow$ \texttt{DAC} latent representation~\citep{kumar2023high} $\rightarrow$ \texttt{DAC} decoder. Conditioning is provided through prefix text embeddings (e.g., instructions expressing emotional states such as ``angry, hateful: you betrayed me...''), along with speaker ID. The input format is \texttt{[spk\_id, BPE embedding, audio codes]}.
    \item \texttt{InhouseTTS4} (Instruction-Conditioned AudioCode Pipeline).  
This model also employs a 0.4B GPT-2 to predict 4,096-dimensional audio codes, but conditions generation on instruction embeddings. A \texttt{CLIPTextEncoder}~\citep{radford2021learning} converts instruction text into embeddings, which serve as a prefix, combined with speaker ID and phone sequence. The input format is \texttt{[spk\_id, instruction embedding sequence, phone sequence, audio codes]}.
    \item \texttt{InhouseTTS5} (Reference-Conditioned AudioCode Pipeline).
Similar to \texttt{InhouseTTS4}, but instead of text-based instructions, conditioning is achieved through reference audio. A style encoder, following the NANSY++ design~\citep{choi2023nansy}, compresses the reference audio into a fixed 32-dimensional embedding. This embedding is concatenated with the phone sequence and audio codes for generation. The input format is \texttt{[32-dim ref embedding, phone sequence, audio codes]}.
\end{itemize}

\subsection{Realism Evaluation Data Summary.}
The resulting realism evaluation dataset is constructed from a combination of curated media sources, including internal collections, public datasets (e.g., NCSSD~\citep{liu2024generative}), and original recordings collected for this project.\footnote{Research access and public redistribution have distinct scopes. The public dataset will be restricted to the sources, records, and representations covered by the release grants, under CC BY-NC 4.0 where the authors hold the necessary rights. Mel features and vocoder reconstructions require source-specific redistribution permission; conversion to features does not establish that permission. See Sec.~\ref{sec:reproducibility}.}

We compile \rev{$15{,}414$} annotated role-play samples \rev{(Mandarin-majority; the real-recording pool and delivery benchmark are Mandarin-only). The dataset is divided into $10{,}990$ base realism training records (mixed real/TTS), $2{,}000$ mixed samples for the benchmark-distribution test set, a $2{,}164$-sample real-recording pool (Mandarin only), and a separate $260$-sample delivery-mismatch benchmark ($13$ TTS systems $\times$ $20$ Mandarin scenes) whose audio is resynthesized to probe detection of prosody, voice-identity, and speaking-style mismatches. The real-recording pool is itself split into $1{,}080$ recordings used for in-domain adaptation and $1{,}084$ held-out clean-test recordings from disjoint speakers ($1{,}080+1{,}084=2{,}164$); the $1{,}080$ are \emph{additional} to the $10{,}990$ base realism training records, so the final in-domain-adapted evaluator is trained on $12{,}070$ samples in total, while the grand total of annotated realism samples is $15{,}414$ ($10{,}990+2{,}000+2{,}164+260$; the $260$-sample delivery benchmark is an additional split, not a subset of the other three) (see Tab.~\ref{tab:bench-overview} and Sec.~\ref{sec:reproducibility})}.

Splits are constructed to be \rev{utterance- (clip-)}disjoint: no spoken clip is shared across splits, and there are zero exact audio$+$profile duplicates. \rev{We verified this precisely (App.~\ref{app:audit}): the split is clip-disjoint but not fully character-disjoint (about $75$ of $210$ benchmark-distribution test positives reuse a training character with a \emph{different} utterance), so we state clip-level rather than character-level disjointness. A dedicated seen-vs-unseen source control (App.~\ref{app:audit}) finds no evidence of familiarity inflation from this reuse (source domain is a proxy, not an exact character-familiarity flag).}

The authentic-speech test set uses actors who are entirely absent from the training data\rev{, and the authentic-speech split is Mandarin-only}. The first test set supports general evaluation using our proposed pipeline with negative samples, while the second focuses exclusively on human speech performed by both professional and amateur voice actors. The $2{,}000$-sample benchmark-distribution test set (the $210$ positives together with the role- and context-mismatch negatives) is the split used for the realism correlation results in Sec.~\ref{sec:exp}; it is distinct from the separate $260$-sample delivery-mismatch benchmark. Detailed statistics and analyses are provided in Appendix~\ref{app:realism-data-analysis}.

\section{DRAME-EvalBench (Role-play Evaluation Modeling)}
\label{sec:exp}

\subsection{Experimental Setup}
\label{ssec:exp-setup}

Building on the two proposed datasets, we conduct experiments under three settings: zero-shot, few-shot, and fine-tuning, using pre-trained ALLMs. We formulate the task as a multi-metric estimation problem~\citep{tjandra2025meta, yao2025songeval, shi25f_interspeech, shi2025arecho}, where a single model is trained or prompted to predict multiple evaluation dimensions.

\noindent \textbf{Zero-shot and Few-shot ALLMs.} 
We first evaluate the zero-shot capabilities of both proprietary and public models, including \texttt{Gemini2.5Pro}, \texttt{GPT4o-audio}, \texttt{KimiAudio}, \texttt{Qwen2Audio}, and \texttt{Qwen2.5Omni}. All evaluations follow a unified prompting template that provides audio input alongside scene and character context.

For proprietary models, we report the average of five independent runs to reduce variance and obtain more stable predictions. For open-source models, we observed weaker instruction-following ability and limited adaptation to unfamiliar tasks. To mitigate this, we did not request simultaneous prediction of multiple metrics. Instead, we queried one metric at a time. Following \citet{song2025decoding, lukasik2024regression}, we request completion in a fixed template and extract the probability distribution over the discrete tokens \texttt{[1]}–\texttt{[5]}. The final score is then computed from these probabilities, which we found makes open-source models usable to some extent under this setting.

For few-shot experiments, we focus primarily on proprietary ALLMs, which demonstrate stronger instruction-following capabilities and have been widely used as judges in prior work. For each test query, we provide three additional annotated samples from the training set as in-context exemplars.\footnote{Please find our detailed settings such as prompts, decoding algorithms in Appendix~\ref{app:zeroshot-evaluation-setup} with additional ablation analysis in Appendix~\ref{app:zeroshot-ablation}.}

\noindent \textbf{Fine-tuning Model: \texttt{DRAME-Eval}.} 
In addition to zero-shot and few-shot evaluations, we fine-tune a dedicated evaluation model, denoted as \texttt{DRAME-Eval} for archetype and realism role-play evaluation, respectively. 
We initialize from \texttt{Qwen2Audio-7B-Instruct}~\citep{chu2024qwen2}, which provides a strong foundation for multi-modal audio-text reasoning. 
To better leverage the multiple human annotations available in our datasets, we adopt a sampling-based training strategy: at each step, one annotator label is randomly selected, following practices in label distribution learning~\citep{geng2016label,  wen2023ordinal}. 
This stochastic supervision implicitly exposes the model to the empirical annotation distribution, while during inference we output the full distribution over discrete labels (e.g., \texttt{[1]}–\texttt{[5]}) and compute the expectation as the final score, consistent with recent decoding-based regression methods~\citep{song2025decoding, xiao2025verbalized, wang2025improving, lukasik2024regression}.

Detailed setups and hyperparameters are detailed as follows: we primarily adopted a parameter-efficient strategy using LoRA~\citep{hulora}, applied to the \texttt{Qwen2Audio-7B-Instruct} model~\citep{chu2024qwen2}. LoRA was configured with a rank of 16, $\alpha=32$, and a dropout of 0.1, targeting the major projection layers, and trained for 10{,}000 steps. The training pipeline employed a per-device batch size of 4 with gradient accumulation of 4 steps (effective batch size 32), AdamW optimization with cosine scheduling, a peak learning rate of $5\times10^{-5}$, weight decay of 0.01, and a warmup strategy defined by 500 steps. Gradient clipping with a norm of 1.0, \texttt{bf16} precision, and gradient checkpointing were enabled to stabilize training and reduce memory overhead. Distributed training was performed on 2xH100 GPUs using DeepSpeed (ZeRO stage-1).

While we utilize various prompts and sample them randomly during the fine-tuning experiments, we provide an example prompt as follows:

\begin{tcolorbox}[colback=gray!5!white, colframe=black!75!white,
                  title={Example Prompts in Realism Evaluation (Pitch Dynamics Dimension)},
                  fonttitle=\bfseries, fontupper=\ttfamily, breakable]
You are an expert evaluator of speech delivery and storytelling. \\\\
Based on the given information, rate the following SINGLE dimension on a 1–5 scale (1 = poor, 5 = excellent).\\
Return ONLY a number between 1 and 5. \\

Global Profile: \{global\_profile\}\\
Local Scene: \{local\_story\}\\
Available Traits: \{available\_traits\}\\

Dimension: PitchDynamics — Variety and appropriateness of pitch contours.\\\\
       5: Natural, situation-appropriate melodic movement; subtle, believable inflections that reveal emotion or emphasis.\\
       4: Generally varied and fitting, but with minor stiffness or occasional flat spots.\\
       3: Neutral or limited inflection; neither distracting nor especially expressive.\\
       2: Forced, exaggerated, or oddly sudden shifts that feel 'performed' rather than lived.\\
       1: Monotone or blatantly mis-pitched for the scene (e.g., cheerfully high in a funeral).\\
       Example output: 4" \\\\

       Output the answer in <answer> </answer>.
\end{tcolorbox}

We further conduct ablation studies in Sec.~\ref{ssec:finetuning-ablation}, examining: the choice of different base audio LLMs and full-parameter fine-tuning versus LoRA adaptation. 
These analyses clarify the trade-offs in model design across base models and adaptation strategies.

\noindent \textbf{Evaluation Metrics for Role-play Evaluation Model.} For most dimensions with 1-5 scale, we report Pearson correlation. For \textit{Content Pass} estimation, we report accuracy.

\subsection{DRAME-EvalBench Main Analysis}
\label{ssec:main-results}

\rev{We report on two test regimes. The \emph{benchmark-distribution} test consists of responses that match \texttt{DRAME-Eval}'s training distribution (predominantly TTS-synthesized role-play: the archetype set entirely, the realism set in majority); we use ``benchmark-distribution'' in place of ``in-distribution'' as a reminder that these sets are predominantly, not entirely, synthesized. The \emph{authentic-speech} test is a held-out split drawn entirely from real human acting (professional and amateur). Both are distinct from the archetype/realism \emph{protocols}, which describe how a sample is judged rather than which regime it belongs to.}

\noindent \textbf{Archetype Evaluation.}
The results in Table~\ref{tab:role-play-exp-archetype} reveal clear distinctions across evaluation settings. In the zero-shot condition, proprietary ALLMs such as \texttt{Gemini2.5Pro} and \texttt{GPT4o-audio} outperform public models, achieving strong \textit{content pass} accuracy ($\sim$90\%) but \rev{only modest correlation with the human quality dimensions under the consistent masking convention (Avg ${\approx}0.23$)}. By contrast, public models exhibit limited predictive ability, with correlations near or below zero. \rev{Few-shot prompting does not help on archetype under this convention: \texttt{Gemini2.5Pro} drops from $0.236$ to $0.158$ and \texttt{GPT4o-audio} from $0.233$ to $-0.102$ (this drop is visible only under masking; the secondary floor-imputation convention largely hides it, App.~\ref{app:tier1}), so the few-shot exemplars do not transfer to this rubric}. The most significant improvement arises from fine-tuning: \texttt{DRAME-Eval} achieves an average correlation of 0.629, marking a large margin over zero-shot/few-shot models and demonstrating the effectiveness of task-specific supervision. \rev{This margin is significant under a paired bootstrap test on the common archetype items: over the strongest zero-shot baseline (\texttt{Gemini2.5Pro}, $0.236$) the paired difference is $\Delta=+0.393$ $[0.338,0.445]$ ($p\le0.001$; no resample of $2{,}000$ reversed the sign), and it is at least this large against every other zero-/few-shot judge (App.~\ref{app:tier1}, Tab.~\ref{tab:audit-archmask}).} \rev{We caution that this pooled $0.629$ overstates within-language discrimination: because English and Mandarin scores differ by an aligned mean shift, pooling inflates the correlation, and the within-language macro-average is substantially lower (${\approx}0.39$; per-language breakdown in App.~\ref{app:tier1}). We report the pooled value for comparability with prior tables but treat the per-language numbers as the cleaner measure.}

\begin{table}[!t]
    \centering
    \caption{Archetype role-play evaluator performance on the benchmark-distribution archetype test set (English+Mandarin pooled). \emph{Content Pass Acc.} is classification accuracy, reported separately; \emph{Audio Quality}, \emph{Human Likeness}, and \emph{Appropriateness} are Pearson correlations with human ratings, and \emph{Avg.}\ is their unweighted mean (it excludes the classification accuracy). \texttt{DRAME-Eval} is fine-tuned from \texttt{Qwen2Audio-7B-Instruct}~\citep{chu2024qwen2}. \rev{All correlation cells use a single \emph{masking} convention for content-gated items (\texttt{DRAME-Eval} Avg $0.629$ $[0.593,0.663]$); the floor-\emph{imputation} variant, per-language breakdowns for \texttt{DRAME-Eval}, and per-model pooled Avg CIs are in App.~\ref{app:tier1} (Tabs.~\ref{tab:audit-archmask},~\ref{tab:tier1-perlang}).} $^\dagger$\texttt{Qwen2Audio} returns a parseable \emph{continuous score} on only ${\sim}13/1{,}499$ items (JSON-format failures on the numeric fields), so its three correlation cells are omitted as unreliable; its \emph{Content Pass} classification output parsed normally on the full set and is reported ($86.7\%$).}
\resizebox{0.8\linewidth}{!}{
    \begin{tabular}{l|l|c|c|c|c|c}
    \toprule
    Config. & Model & Content Pass Acc. & Audio Quality & Human Likeness & Appropriateness & Avg. \\
    \midrule
        \multirow{5}{*}{Zeroshot} 
        & \texttt{Gemini2.5Pro} & 89.3\% & \rev{0.147} & \rev{0.253} & \rev{0.309} & \rev{0.236} \\
        & \texttt{GPT4o-audio} & 89.9\% & \rev{0.094} & \rev{0.409} & \rev{0.196} & \rev{0.233} \\
        & \texttt{KimiAudio} & 71.7\% & \rev{0.015} & \rev{0.105} & \rev{0.089} & \rev{0.070} \\
        & \texttt{Qwen2Audio}$^\dagger$ & 86.7\% & \rev{--} & \rev{--} & \rev{--} & \rev{--} \\
        & \texttt{Qwen2.5Omni} & 84.1\% & \rev{-0.029} & \rev{0.004} & \rev{0.003} & \rev{-0.008} \\
\midrule
        \multirow{2}{*}{Fewshot} & \texttt{Gemini2.5Pro} & 91.1\% & \rev{0.006} & \rev{0.203} & \rev{0.264} & \rev{0.158} \\
        & \texttt{GPT4o-audio} & 89.9\% & \rev{-0.277} & \rev{-0.007} & \rev{-0.020} & \rev{-0.102}  \\
\midrule
        Finetune & \texttt{DRAME-Eval} & \textbf{93.6\%} & \textbf{0.682} & \textbf{\rev{0.679}} & \textbf{0.525} & \textbf{0.629} \\
    \bottomrule
    \end{tabular}}
    \label{tab:role-play-exp-archetype}
\end{table}

\begin{table}[!t]
    \centering
    \caption{Realism role-play evaluator performance on the benchmark-distribution test set (English+Mandarin pooled). All ten columns are Pearson correlations between model and human scores; \emph{Avg.}\ is their unweighted mean. \texttt{DRAME-Eval} is fine-tuned from \texttt{Qwen2Audio-7B-Instruct}~\citep{chu2024qwen2} and can be compared directly with the \texttt{Qwen2Audio} zero-shot row. \rev{Correlations are computed over the $2{,}000$ benchmark-distribution test records (the $210$ positives together with the $832$ role- and $958$ context-mismatch negatives, all human-rated) under a single \emph{masking} convention for the gated \emph{Emotion Intensity}/\emph{Transition} columns (\texttt{DRAME-Eval} Avg $0.602$, $95\%$ bootstrap CI $[0.573,0.627]$). The release-reproducible human-only floor-\emph{imputation} variant (Avg $0.614$), per-language breakdowns with macro-average CIs and effective per-dimension sample sizes are reported in App.~\ref{app:tier1}; positive-vs-negative discrimination is reported separately in Tab.~\ref{tab:tier1-negdisc}.}}
\resizebox{\linewidth}{!}{
    \begin{tabular}{l|l|c|c|c|c|c|c|c|c|c|c|c}
    \toprule
    \multirow{2}{*}{Config.} & \multirow{2}{*}{Model} & \multicolumn{3}{|c|}{Prosodic Dynamics} & \multicolumn{3}{|c|}{Emotional Expressiveness} & \multicolumn{2}{|c|}{Character Consistency} & \multicolumn{2}{|c|}{Contextual Relevance} & \multirow{2}{*}{Avg.} \\
    \cmidrule{3-12}
    & & Pitch & Rhythm & Stress & Accuracy & Intensity & Transition & Identity & Traits & Local & Global &  \\
    \midrule
        \multirow{5}{*}{Zeroshot} & \texttt{Gemini2.5Pro} & \rev{0.285} & \rev{0.317} & \rev{0.292} & \rev{0.378} & \rev{0.228} & \rev{0.203} & \rev{0.525} & \rev{0.505} & \rev{0.339} & \rev{0.459} & \rev{0.353} \\
         & \texttt{GPT4o-audio} & \rev{0.187} & \rev{0.178} & \rev{0.187} & \rev{0.336} & \rev{0.139} & \rev{0.176} & \rev{0.358} & \rev{0.413} & \rev{0.308} & \rev{0.466} & \rev{0.275} \\
         & \texttt{KimiAudio} & \rev{0.049} & \rev{0.041} & \rev{0.073} & \rev{0.025} & \rev{0.105} & \rev{0.095} & \rev{0.082} & \rev{0.019} & \rev{0.016} & \rev{0.033} & \rev{0.054} \\
         & \texttt{Qwen2Audio} & \rev{0.031} & \rev{0.035} & \rev{0.055} & \rev{-0.003} & \rev{0.080} & \rev{0.060} & \rev{0.012} & \rev{-0.022} & \rev{0.063} & \rev{-0.002} & \rev{0.031} \\
         & \texttt{Qwen2.5Omni} & \rev{0.131} & \rev{0.076} & \rev{0.103} & \rev{0.235} & \rev{0.071} & \rev{0.091} & \rev{0.151} & \rev{0.306} & \rev{0.152} & \rev{0.211} & \rev{0.153} \\
\midrule
        \multirow{2}{*}{FewShot} & \texttt{Gemini2.5Pro} & \rev{0.332} & \rev{0.370} & \rev{0.351} & \rev{0.396} & \rev{0.250} & \rev{0.238} & \rev{0.587} & \rev{0.514} & \rev{0.352} & \rev{0.481} & \rev{0.387} \\
        & \texttt{GPT4o-audio} & \rev{0.204} & \rev{0.212} & \rev{0.182} & \rev{0.405} & \rev{0.165} & \rev{0.178} & \rev{0.407} & \rev{0.504} & \rev{0.309} & \rev{0.487} & \rev{0.305} \\
\midrule
        Finetune & \texttt{DRAME-Eval} & \textbf{0.621} & \textbf{0.627} & \textbf{0.631} & \textbf{0.632} & \textbf{\rev{0.494}} & \textbf{\rev{0.469}} & \textbf{0.660} & \textbf{0.655} & \textbf{0.563} & \textbf{0.668} & \textbf{\rev{0.602}} \\
    \bottomrule
    \end{tabular}}
    \label{tab:role-play-exp-realism}
\end{table}

\begin{table}[!t]
    \centering
    \caption{Realism role-play evaluator performance on the \emph{real-recording} test set (\rev{clean held-out set of 1{,}084 samples}, professional + amateur human voice actors; \rev{Mandarin only; the experimental authentic-speech pool contains no English recordings}). All ten columns are Pearson correlations between model and human scores; \emph{Avg.}\ is their unweighted mean. \rev{This speaker-disjoint test set differs from the synthetic-dominant benchmark distribution: every sample is authentic human acting rather than synthesized speech. It is reported as a measurement of the transfer gap from the synthetic-dominant benchmark distribution to all-authentic human acting (the \emph{synthetic-to-authentic} gap), rather than as a setting on which fine-tuning is claimed to be human-aligned.} \rev{The \texttt{DRAME-Eval} average is $-0.142$; the per-dimension cells are point estimates, and the Avg CI and between-speaker decomposition are in Sec.~\ref{ssec:main-results} and App.~\ref{app:tier1}.}}
\resizebox{\linewidth}{!}{
    \begin{tabular}{l|l|c|c|c|c|c|c|c|c|c|c|c}
    \toprule
    \multirow{2}{*}{Config.} & \multirow{2}{*}{Model} & \multicolumn{3}{|c|}{Prosodic Dynamics} & \multicolumn{3}{|c|}{Emotional Expressiveness} & \multicolumn{2}{|c|}{Character Consistency} & \multicolumn{2}{|c|}{Contextual Relevance} & \multirow{2}{*}{Avg.} \\
    \cmidrule{3-12}
    & & Pitch & Rhythm & Stress & Accuracy & Intensity & Transition & Identity & Traits & Local & Global &  \\
    \midrule
        \multirow{5}{*}{Zeroshot} & \texttt{Gemini2.5Pro} & 0.159 & 0.070 & 0.103 & 0.070 & 0.052 & 0.096 & 0.092 & 0.077 & 0.075 & 0.028 & 0.082 \\
         & \texttt{GPT4o-audio} & 0.067 & -0.013 & 0.035 & 0.013 & 0.042 & 0.036 & -0.004 & 0.130 & 0.065 & 0.018 & 0.039 \\
         & \texttt{KimiAudio} & -0.115 & -0.039 & -0.116 & -0.020 & -0.122 & -0.080 & \textbf{0.114} & -0.032 & -0.137 & -0.080 & -0.063 \\
         & \texttt{Qwen2.5Omni} & 0.078 & 0.071 & 0.043 & 0.074 & 0.047 & 0.045 & -0.082 & 0.076 & 0.102 & 0.098 & 0.055 \\
         & \texttt{Qwen2Audio} & 0.055 & 0.057 & 0.088 & 0.072 & 0.077 & 0.057 & -0.006 & -0.040 & 0.050 & 0.111 & 0.052 \\
\midrule
         \multirow{2}{*}{FewShot} & \texttt{Gemini2.5Pro} & \textbf{0.319} & 0.179 & \textbf{0.263} & 0.111 & 0.078 & 0.184 & -0.001 & 0.118 & 0.216 & 0.074 & 0.154 \\
         & \texttt{GPT4o-audio} & 0.299 & \textbf{0.193} & 0.222 & \textbf{0.121} & \textbf{0.377} & \textbf{0.344} & -0.013 & 0.248 & \textbf{0.301} & \textbf{0.168} & \textbf{0.226} \\
\midrule
        Finetune & \texttt{DRAME-Eval} & \rev{-0.357} & \rev{-0.051} & \rev{-0.228} & \rev{-0.161} & \rev{-0.333} & \rev{-0.291} & \rev{-0.041} & \rev{\textbf{0.419}} & \rev{-0.262} & \rev{-0.118} & \rev{$-0.142$} \\
    \bottomrule
    \end{tabular}}
    \label{tab:role-play-exp-realism-real}
\end{table}

\noindent \textbf{Realism Evaluation.}
Table~\ref{tab:role-play-exp-realism} extends the analysis to fine-grained realism evaluation. Similar to archetype results, proprietary ALLMs lead the zero-shot setting, with \texttt{Gemini2.5Pro} obtaining an average correlation of \rev{0.353}, compared to \rev{0.275} for \texttt{GPT4o-audio}. Public models remain weak baselines. Few-shot prompting improves performance modestly: \rev{\texttt{Gemini2.5Pro} gains +0.034 and \texttt{GPT4o-audio} gains +0.030 over their zero-shot averages}, showing the benefit of in-context supervision. Once again, fine-tuning is most effective: \texttt{DRAME-Eval} achieves \rev{0.602 (masking; 0.614 under the human-only floor-imputation sensitivity of App.~\ref{app:tier1})} average correlation, higher than the best few-shot results\rev{. The gain over the strongest few-shot baseline, \texttt{Gemini2.5Pro} ($0.387$), is significant under a paired bootstrap test ($\Delta=+0.215$ $[0.178,0.250]$ sample-level; $+0.213$ $[0.174,0.239]$ under a source-series-clustered bootstrap; $p\le0.001$ in both), and the gain over the other few-shot judge, \texttt{GPT4o-audio} ($0.305$), is larger ($\Delta=+0.297$ $[0.256,0.336]$; App.~\ref{app:tier1})}. Importantly, the gains are consistent across all 10 realism dimensions, \rev{showing that the benchmark-distribution improvement spans all ten dimensions rather than being driven by a subset (we do not run a training-strategy ablation, so we do not attribute this to the sampling--expectation objective specifically)}. \rev{This signal depends on the audio input under these controls rather than being a transcript shortcut: replacing the audio with silence collapses \texttt{DRAME-Eval}'s benchmark-distribution correlation from $0.602$ to $-0.035$, and the evaluated text-only judge (given the gold transcript, profile, and scene but no audio) reaches $0.186$, compared with $0.602$ for the full model; conversely, blanking the role/scene text leaves an audio-only correlation of $0.392$ that is delivery-quality variance; on the positive-vs-negative role-fit axis, audio-only does not outperform chance (AUROC $0.408$) while the full model reaches $0.711$, indicating that access to both modalities improves role-fit discrimination (App.~\ref{app:tier1}).}

\noindent \textbf{Authentic-Speech Test Set.}
\rev{The real-recording evaluation in Table~\ref{tab:role-play-exp-realism-real} is the most demanding transfer setting. Because the final \texttt{DRAME-Eval} is in-domain-adapted (its training data includes $1{,}080$ in-house real recordings), we evaluate it only on the zero-overlap held-out set (\texttt{test\_real\_v2}, $1{,}084$ Mandarin samples), where the average Pearson is $-0.142$ $[-0.160,-0.125]$ (a base-only model with no real-recording training data scores $-0.024$ over the same pool). On this footing \texttt{DRAME-Eval} holds no real-recording advantage: the strongest baseline (\texttt{GPT4o-audio} few-shot) reaches only $0.226$ $[0.190,0.262]$ (a pooled number we show below is confounded by between-speaker composition), and \texttt{DRAME-Eval} does not improve on it. The paired difference is negative and significant at the sample level ($\Delta=-0.369$ $[-0.397,-0.343]$, $p\le0.001$ over the common $1{,}084$ samples), whereas the four-speaker clustered interval includes zero ($[-0.56,+0.04]$), making the speaker-level comparison inconclusive. Thus, neither analysis provides evidence of a positive \texttt{DRAME-Eval} advantage.}

\rev{A speaker-mean analysis further shows that the \emph{human labels} carry strong between-speaker structure: a descriptive speaker-mean \emph{oracle}, computed from each held-out speaker's test-set human mean, attains $0.538$, above every audio model. This oracle is not a deployable evaluation baseline (it reads the test-set labels it is scored against); we use it only to quantify the between-speaker structure in the labels. This limits how a pooled correlation should be interpreted: a positive pooled value need not imply within-speaker discrimination, since much of it is recoverable from speaker identity alone. It does not, however, establish that any particular baseline's correlation \emph{is} driven by speaker identity, since only \texttt{DRAME-Eval} has a measured within-speaker decomposition. For \texttt{DRAME-Eval} specifically we measured the within-speaker decomposition, where its correlation collapses to near zero (mean $+0.029$ over the four held-out speakers). On the \emph{matched} within-speaker protocol a single human rater still agrees with the consensus at $+0.34$ within speaker (per-speaker single-rater-vs-consensus, averaged over the four speakers with ${\ge}270$ utterances; range $+0.20$ to $+0.50$; App.~\ref{app:tier1}), so humans discriminate acting quality \emph{within} a speaker where the model does not. The pooled single-rater level is higher ($0.561$) because pooling adds between-speaker variance; we therefore use the within-speaker figure for the like-for-like human--model comparison and the pooled figure only against the pooled model number ($-0.142$).}

\rev{These results indicate that \texttt{DRAME-Eval} is a strong benchmark-distribution evaluator whose judgment does not yet transfer to authentic recordings on exactly the dimensions realism evaluation is meant to add; we do not claim it is human-aligned on real speech. Closing this gap is the central direction for future work.}

\noindent \textbf{Overall.} Across both archetype and realism evaluations, three trends are consistent: (i) zero-shot performance is limited, particularly for open-source models, (ii) \rev{few-shot prompting helps the proprietary ALLMs on realism (small positive gains) but does \emph{not} help on archetype under the masking convention, where it lowers both proprietary judges (\texttt{Gemini2.5Pro} $0.236\!\to\!0.158$, \texttt{GPT4o-audio} $0.233\!\to\!-0.102$), so its benefit is not uniform across the two evaluations}, and (iii) fine-tuning with task-specific supervision (\texttt{DRAME-Eval}) yields the largest improvements on the benchmark-distribution test sets. The real-recording numbers, however, qualify these benchmark-distribution gains: the next research step is not to push the benchmark-distribution Pearson higher, but to close the synthetic-to-authentic transfer gap that the same model exposes.

\noindent \textbf{Caveats on the headline numbers.} Two reporting choices are worth flagging here. First, the benchmark-distribution averages in Tabs.~\ref{tab:role-play-exp-archetype} and~\ref{tab:role-play-exp-realism} are pooled across English and Mandarin, whereas the authentic-speech results in Tab.~\ref{tab:role-play-exp-realism-real} are Mandarin-only; given the rater-pool argument in Sec.~\ref{sec:realism-data} and the language-specific score gaps observed in App.~\ref{app:archetype-data-analysis}, the pooled Pearson should be read as a coarse summary, with per-language tables being the cleaner unit of comparison. Second, for \emph{Emotion Intensity} and \emph{Emotion Transition}, ratings can be imputed to $1$ when the upstream \emph{Emotion Accuracy} gate triggers \texttt{N/A} (Sec.~\ref{ssec:annotation-protocol} and App.~\ref{app:realism-data-analysis}); this inflates the apparent correlation when both the human label and the model's prediction collapse to the floor. \rev{We adopt \emph{masking} (dropping gated entries) as the primary convention in both main tables and recompute every row from the saved prediction files under it, so the tables are internally consistent and reproducible; the human-only floor-\emph{imputation} variant (mapping missing human labels to $1$ while leaving model predictions unchanged) is reported as a secondary comparison in App.~\ref{app:tier1}. Under masking \texttt{DRAME-Eval}'s realism Avg is $0.602$ (versus $0.614$ under human-only imputation); the approximately $0.012$ difference comes from how missing labels are handled on the two gated emotion columns, rather than a change in model predictions. Imputation would in fact \emph{help the baselines more} (e.g.\ archetype \texttt{GPT4o-audio} few-shot rises from $-0.102$ to $0.331$), so masking is both the reproducible and the less self-favorable choice.}

\paragraph{\rev{Single-rater human agreement, contrastive-negative discrimination, and system-level ranking.}} \rev{To make the correlations above interpretable and to exercise parts of EvalBench that Tabs.~\ref{tab:role-play-exp-archetype}--\ref{tab:role-play-exp-realism-real} leave unused, we report three complementary analyses in App.~\ref{app:tier1}. \emph{(i) Single-rater agreement (a human reference point).} Comparing a held-out rater to the consensus of the rest gives a human reference level for the model--consensus Pearson above (we call it a reference, not a formal upper bound, since a model could in principle exceed a single rater). Read against it, DRAME-Eval is comparable to a single human on the benchmark-distribution realism dimensions (roughly $76$--$94\%$ of the single-rater level), but on the clean authentic-speech set it falls to near or below zero on most dimensions (e.g.\ $-0.041$ on \emph{Voice Identity} against a single-rater level of $0.48$, and negative on the prosodic and emotional dimensions), \emph{even though single-rater agreement there remains substantial}; the one partial exception is \emph{Trait Embodiment}, where it retains some signal ($+0.419$ against $0.58$). The synthetic-to-authentic gap therefore cannot be attributed solely to annotator disagreement (single-rater agreement there remains substantial), and it concentrates on prosody, emotion, and voice identity, sharpening it as an optimization target. \emph{(ii) Negative discrimination.} On the contrastive negatives, humans separate positives from role-mismatch (context-mismatch) negatives with a $1.30$ ($0.54$) point margin; the open-source judges sit at chance (role-mismatch AUROC $0.51$/$0.55$ for \texttt{KimiAudio}/\texttt{Qwen2Audio}), whereas DRAME-Eval attains the highest role-mismatch AUROC ($0.872$); on context-mismatch discrimination \texttt{Qwen2.5Omni} ($0.851$, open-source) and \texttt{GPT4o-audio} ($0.815$) score higher than DRAME-Eval ($0.679$), so the final model is clearly sensitive to role/persona cues but is not uniformly best across negative types (a matched-budget training ablation in App.~\ref{app:tier1} isolates the value of \emph{training} with negatives separately: withholding them inverts positive-vs-negative discrimination to below chance). \emph{(iii) System-level ranking (benchmark-distribution diagnostic).} Aggregated to per-system means over the eight \emph{familiar} archetype \texttt{SFnd} systems, DRAME-Eval shows the strongest agreement with humans across all three coefficients (Pearson $0.74$, Spearman $0.50$, Kendall $0.41$); \texttt{KimiAudio} is also positive on all three but markedly weaker ($0.23$/$0.40$/$0.21$), while the remaining zero-/few-shot judges are inconsistent in sign. We treat this as a benchmark-distribution diagnostic rather than a validated use: with only eight in-training systems this is a small-$n$, exploratory comparison, so we do not claim a validated ranker of arbitrary systems.}

\subsection{Fine-tuning Ablation Experiments}
\label{ssec:finetuning-ablation}

Tables~\ref{tab:role-play-exp-archetype-ablation}, \ref{tab:role-play-realism-basic-ablation}, and \ref{tab:role-play-realism-real-ablation} present ablation studies across different fine-tuning strategies, including the use of \texttt{Qwen2Audio-7B-Base} or \texttt{Qwen2Audio-7B-Instruct}; and the use of fully fine-tuning or LoRA fine-tuning. We use \texttt{Base-FT}, \texttt{Base-LoRA}, \texttt{Instruct-FT}, and \texttt{Instruct-LoRA} to denote the different configuration settings in the following experiments.

\paragraph{Archetype evaluation.} On the archetype test set, all fine-tuned variants achieve strong \textit{Content Pass} accuracy above 92\%. Differences appear mainly in correlation-based metrics. Here, \texttt{Base-FT} yields the highest scores on \textit{Audio Quality} and \textit{Appropriateness}, while \texttt{Base-LoRA} is strongest on \textit{Human Likeness}. \texttt{Base-FT} and \texttt{Instruct-LoRA} share the highest average at the displayed precision (0.629), and \texttt{Instruct-LoRA} has the highest \textit{Content Pass} accuracy (93.6\%) with parameter-efficient adaptation.  

\paragraph{Realism evaluation (basic set).} For realism evaluation on the benchmark-distribution test set, \texttt{Instruct-FT} consistently achieves the strongest correlations across most dimensions, especially for \textit{Character Identity} and \textit{Traits}, leading to the best average correlation (0.668). \texttt{Base-LoRA} also performs competitively (0.630), while \texttt{Instruct-LoRA} lags slightly behind \rev{(0.625 under the earlier internal convention in this table, which floored both human and model sides; 0.602 under masking, matching the main realism table Tab.~\ref{tab:role-play-exp-realism})}. These results highlight that full fine-tuning may still offer an advantage when the evaluation set is relatively homogeneous and closer to the training distribution.  

\paragraph{Realism evaluation (real recordings).} \rev{On the held-out authentic-speech set (Tab.~\ref{tab:role-play-realism-real-ablation}), all four configurations behave alike: every average is negative or near zero (\texttt{Base-FT} $-0.168$, \texttt{Base-LoRA} $-0.148$, \texttt{Instruct-FT} $-0.087$, \texttt{Instruct-LoRA} $-0.142$), and no configuration produces a usable per-sample correlation. No fine-tuning strategy resolves the synthetic-to-authentic gap; the differences among configurations live in the benchmark-distribution regime.}

\paragraph{Overall summary.} \rev{Because the authentic-speech column does not separate the configurations (all transfer poorly), we select the final model on \emph{benchmark-distribution} grounds. \texttt{Instruct-LoRA} shares the highest displayed archetype average ($0.629$) and the highest classification accuracy ($93.6\%$) while remaining competitive on the realism basic set (masked Avg $0.602$ vs.\ \texttt{Instruct-FT}'s $0.651$; the impute-convention ablation table below shows the same ordering), and it is the most parameter-efficient to train and serve. We adopt \texttt{Instruct-LoRA} as the final configuration; model selection uses benchmark-distribution performance and efficiency rather than real-recording transfer, which no configuration achieves. Because several configurations have now been inspected on the authentic-speech split, we treat it as a held-out transfer diagnostic rather than a permanently untouched confirmatory benchmark; a fresh speaker-level test set would be required for confirmatory evaluation. A matched-control experiment (App.~\ref{app:tier1}) shows that overlap between training and evaluation speakers has essentially no effect on held-out correlation, so the real-recording result reflects a genuine synthetic-to-authentic gap rather than a tunable speaker calibration. Across all four configurations, benchmark-distribution correlation reaches $0.55$--$0.65$ and falls to negative on real recordings. Training-loss logs were retained, but we do not have a systematic checkpoint-by-checkpoint held-out evaluation curve that would identify when transfer fails; the clean evaluation instead shows that all four configurations perform poorly on held-out authentic speech.}

\begin{table}[!t]
    \centering
    \caption{Archetype role-play evaluator fine-tuning ablation on the benchmark-distribution test set. \emph{Content Pass Acc.} is classification accuracy and is reported separately; the remaining columns are Pearson correlations with human ratings. \emph{Avg.}\ is the unweighted mean of those three correlation columns only (Audio Quality, Human Likeness, Appropriateness) and excludes the classification accuracy. Correlations and averages are recomputed from the saved per-dimension predictions against the canonical masked human labels; averages use unrounded correlations.}
{\small
    \begin{tabular}{l|c|c|c|c|c}
    \toprule
    Config. & Content Pass Acc. & Audio Quality & Human Likeness & Appropriateness & Avg. \\
    \midrule
    \texttt{Base-FT} & 92.1\% & \textbf{0.688} & 0.648 & \textbf{0.551} & \textbf{0.629} \\
    \texttt{Base-LoRA} & 92.6\% & 0.650 & \textbf{0.684} & 0.492 & 0.608 \\
    \texttt{Instruct-FT} & 92.6\% & 0.668 & 0.609 & 0.546 & 0.608 \\
    \texttt{Instruct-LoRA} & \textbf{93.6\%} & 0.682 & 0.679 & 0.525 & \textbf{0.629} \\
    \bottomrule
    \end{tabular}}
    \label{tab:role-play-exp-archetype-ablation}
\end{table}

\begin{table}[!t]
    \centering
    \caption{Realism role-play evaluator fine-tuning ablation on the benchmark-distribution test set. All ten columns are Pearson correlations with human ratings. \emph{Avg.}\ is the unweighted mean across all ten correlation columns. \rev{This earlier internal ablation uses a floor-\emph{imputation} convention that also floors the model side on gated emotion entries. It differs from the release-reproducible human-only imputation sensitivity (\texttt{Instruct-LoRA} Avg $0.614$); the main realism table (Tab.~\ref{tab:role-play-exp-realism}) uses \emph{masking}, under which each average is ${\sim}0.02$ lower (masked Avg \texttt{Base-FT} $0.631$ / \texttt{Instruct-FT} $0.651$ / \texttt{Instruct-LoRA} $0.602$; App.~\ref{app:tier1}) and the ordering is unchanged.}}
\resizebox{\linewidth}{!}{
    \begin{tabular}{l|c|c|c|c|c|c|c|c|c|c|c}
    \toprule
    \multirow{2}{*}{Config.} & \multicolumn{3}{|c|}{Prosodic Dynamics} & \multicolumn{3}{|c|}{Emotional Expressiveness} & \multicolumn{2}{|c|}{Character Consistency} & \multicolumn{2}{|c|}{Contextual Relevance} & \multirow{2}{*}{Avg.} \\
    \cmidrule{2-11}
    & Pitch & Rhythm & Stress & Accuracy & Intensity & Transition & Identity & Traits & Local & Global &  \\
    \midrule
    \texttt{Base-FT} & 0.643 & 0.633 & 0.655 & 0.657 & 0.600 & 0.599 & 0.689 & 0.668 & 0.611 &  0.707 & 0.646 \\
    \texttt{Base-LoRA} & 0.628 & 0.633 & 0.643 & 0.637 & 0.586 & 0.590 & 0.668 & 0.651 & 0.578 & 0.684 & 0.630 \\
    \texttt{Instruct-FT} & \textbf{0.657} & \textbf{0.655} & \textbf{0.675} & \textbf{0.674} & \textbf{0.627} & \textbf{0.626} & \textbf{0.752} & \textbf{0.685} & \textbf{0.615} & \textbf{0.714} & \textbf{0.668} \\
    \texttt{Instruct-LoRA} & 0.621 & 0.627 & 0.631 & 0.632 & 0.596 & 0.601 & 0.660 & 0.655 & 0.563 & 0.668 & 0.625 \\
    \bottomrule
    \end{tabular}}
    \label{tab:role-play-realism-basic-ablation}
\end{table}

\begin{table}[!t]
    \centering
    \caption{Realism role-play evaluator fine-tuning ablation on the \emph{real-recording} test set. Pearson correlations; \emph{Avg.}\ is their mean. \rev{All rows are evaluated on the held-out \texttt{test\_real\_v2} ($1{,}084$ samples, zero training overlap; Sec.~\ref{ssec:main-results}, App.~\ref{app:tier1}). \emph{No} configuration transfers to real recordings, so the choice of \texttt{Instruct-LoRA} as the final model rests on benchmark-distribution performance (Sec.~\ref{ssec:finetuning-ablation}), not on this table.}}
\resizebox{\linewidth}{!}{
    \begin{tabular}{l|c|c|c|c|c|c|c|c|c|c|c}
    \toprule
    \multirow{2}{*}{Config.} & \multicolumn{3}{|c|}{Prosodic Dynamics} & \multicolumn{3}{|c|}{Emotional Expressiveness} & \multicolumn{2}{|c|}{Character Consistency} & \multicolumn{2}{|c|}{Contextual Relevance} & \multirow{2}{*}{Avg.} \\
    \cmidrule{2-11}
    & Pitch & Rhythm & Stress & Accuracy & Intensity & Transition & Identity & Traits & Local & Global &  \\
    \midrule
    \texttt{Base-FT} & -0.422 & -0.119 & -0.333 & -0.261 & -0.330 & -0.351 & 0.059 & 0.538 & -0.3083 & -0.153 & -0.168 \\
    \texttt{Base-LoRA} & -0.372 & \textbf{-0.016} & -0.304 & -0.195 & -0.273 & -0.249 & -0.091 & 0.458 & -0.291 & -0.146 & -0.148 \\
    \texttt{Instruct-FT} & \textbf{-0.269} & -0.031 & \textbf{-0.202} & -0.208 & \textbf{-0.222} & \textbf{-0.242} & \textbf{0.076} & \textbf{0.543} & \textbf{-0.219} & \textbf{-0.094} & \textbf{-0.087} \\
    \rev{\texttt{Instruct-LoRA}} & \rev{-0.357} & \rev{-0.051} & \rev{-0.228} & \rev{\textbf{-0.161}} & \rev{-0.333} & \rev{-0.291} & \rev{-0.041} & \rev{0.419} & \rev{-0.262} & \rev{-0.118} & \rev{$-0.142$} \\
    \bottomrule
    \end{tabular}}
    \label{tab:role-play-realism-real-ablation}
\end{table}

\section{Conclusion}
We presented \textbf{Speech-DRAME}, a framework for assessing spoken role-play through complementary \emph{archetype} (top-down) and \emph{realism} (bottom-up) protocols, together with the human-annotated \textbf{DRAME-EvalBench} and the fine-tuned evaluator \texttt{DRAME-Eval}. \rev{On the benchmark-distribution test sets, \texttt{DRAME-Eval} attains the strongest agreement with human raters over zero-/few-shot ALLM judges; on a held-out set composed entirely of authentic human acting, this alignment does not transfer: \texttt{DRAME-Eval} obtains a pooled correlation of $-0.142$, and its within-speaker correlation is only $+0.029$ compared with matched within-speaker human agreement of $+0.34$. Validating evaluators on this authentic-speech split, and closing the transfer gap, remain open. Speech-DRAME shows that success in evaluating synthesized role-play does not imply an ability to judge acting differences among authentic human performances.}\footnote{Limitations and future work are discussed in Section~\ref{app:future}.}

\section{Limitations and Future Work}
\label{app:future}
Despite the design effort behind Speech-DRAME, several limitations remain.
First, while our dual evaluation paradigm combines top-down archetype assessment with bottom-up realism evaluation, both strategies are constrained by the quality and scope of available data. Archetype evaluation inevitably relies on simplified stereotypes, which can overlook subtler paralinguistic and contextual cues. Realism evaluation, though richer, still struggles with domain mismatch between the predominantly synthetic (majority-TTS) training data and genuine human recordings, as evidenced by the performance drop observed on the authentic-speech test set; the negative correlations on some dimensions are consistent with a substantial benchmark-to-authentic transfer failure that cannot be explained solely by low human agreement. \rev{This is also what makes the authentic-speech split a distinctively demanding test: most prior speech role-play and expressive-speech evaluation operates on TTS-synthesized speech~\citep{zhang-etal-2025-omnicharacter, huang2025instructttseval, manku2025emergenttts, lin2025mintbench}, where synthetic-versus-real status and naturalness vary across systems and can serve as strong cues. Because every sample in the held-out split is authentic human speech, synthetic-versus-real status is constant and cannot serve as a within-test discriminator. We hypothesize that evaluators must consequently rely more heavily on within-human differences in prosody, emotion, character embodiment, and scene fit; however, we have not isolated cue use causally, nor established that perceived naturalness is constant across these performances. The analysis in Sec.~\ref{ssec:main-results} makes the gap concrete: it primarily reflects evaluator transfer rather than an inherently unratable task (single-rater human agreement on real recordings remains substantial), and a descriptive speaker-mean oracle ($0.538$) reveals substantial between-speaker structure in the human labels, showing why a positive pooled correlation alone cannot establish within-speaker discrimination. The real-recording collection is also Mandarin-only; extending it to English is future work.} This gap highlights how much room remains between benchmark-driven evaluation and the full demands of real-world acted speech.

Second, the present framework emphasizes bilingual evaluation (Mandarin and English), but linguistic and cultural diversity is far broader. Cross-lingual generalization and culturally sensitive evaluation are still open challenges, particularly since we observed annotation discrepancies across language groups. \rev{A monolingual-training probe makes this concrete (Tab.~\ref{tab:tier1-xling}): cross-lingual transfer is limited and asymmetric, with a Mandarin-only evaluator retaining about $35\%$ of the English-only model's English performance and an English-only evaluator retaining about $62\%$ of the Mandarin-only topline. Under the matched $2{,}000$-step budget, bilingual training reduces English correlation from $0.441$ to $0.353$ while leaving Mandarin performance nearly unchanged; this pattern is consistent with negative transfer or with insufficient English exposure in the Mandarin-dominant mixture (which is larger and includes the Mandarin-only real recordings), and the present experiment does not distinguish these explanations. Language-balanced training or per-language adapters are natural remedies we leave to future work.} This limits the extent to which Speech-DRAME can claim universal coverage.

Third, while DRAME-Eval substantially outperforms zero-shot and few-shot ALLMs, its reliance on fine-tuning from a strong base model raises questions of accessibility and scalability. Future work should explore lightweight adaptation techniques, semi-supervised training with partially annotated data, and leveraging emergent capabilities of open-source multimodal LLMs.

Fourth, the headline correlations in the main tables are point estimates pooled across English and Mandarin. \rev{App.~\ref{app:tier1} complements them with per-language breakdowns (Tab.~\ref{tab:tier1-perlang}), sample-level bootstrap confidence intervals, a single-rater-vs-consensus human agreement level (a reference, not a formal upper bound), and ordinal Krippendorff's $\alpha$. These matter most on the authentic-speech test set, where speaker-level inference is highly uncertain: with only four held-out speakers, a conservative speaker-clustered bootstrap leaves the best systems not statistically separable. We note that the released authentic-speech test set is Mandarin-only, so its ``pooled'' label reduces to Mandarin in practice.}

Finally, although we validated DRAME-Eval against human annotations, alignment under realism settings remains modest. This indicates that current evaluators remain insufficiently sensitive to the fine-grained distinctions the benchmark captures in authentic human speech, on which human raters retain substantial agreement. Richer annotation protocols, more diverse speaker populations, and integration of perceptual psychology insights could further improve evaluation fidelity.

Looking forward, we see several promising research directions. We intentionally restrict evaluation to single-turn responses for clarity and tractability. However, multi-turn extensions that capture narrative progression, long-term consistency, and dynamic emotion shifts remain a natural next step. Incorporating multilingual and cross-cultural datasets will enhance generality and fairness. On the modeling side, exploring hybrid evaluation strategies that combine discrete rubrics with continuous perceptual signals (e.g., prosody embeddings or affective measures) may better bridge the gap between automatic and human judgments.  \rev{Finally, only after stronger authentic-speech transfer, fairness, and robustness validation should future work investigate carefully constrained use of Speech-DRAME as an optimization signal; the present evaluator is intended for evaluation and analysis rather than unqualified reward-model deployment.}

\paragraph{Multi-turn extension of the formulation.}
The single-turn equations of Sec.~\ref{ssec:problem} extend to multi-turn role-play by reintroducing a preceding spoken-dialogue context $\mC_{\text{speech}}$ alongside the persistent role and scene specifications:
\[
    \mS_{r} \;=\; \mathrm{SFnd}(\mC_{\text{speech}},\, \mC_{\text{profile}},\, \mC_{\text{scene}}),
    \qquad
    \mE_{r} \;=\; \mathrm{SEval}(\mS_{r},\, \mC_{\text{speech}},\, \mC_{\text{profile}},\, \mC_{\text{scene}}).
\]
All other components --- the curation operators $\mathcal{D}^{\text{top-down}}$, $\mathcal{D}^{\text{bottom-up}}$, and the negative-sample operator $\mathcal{N}$ --- carry over unchanged. Because all data and experiments in this paper are single-turn, $\mC_{\text{speech}}=\emptyset$ throughout the experimental benchmark; this is the only difference between the body and the multi-turn formulation.

\section{Ethics Statement}
\label{sec:ethics}

All data collection and annotation in this work followed strict ethical standards. Our bilingual datasets were constructed exclusively from legally accessible sources, including licensed media, open-source corpora, and original recordings conducted with informed consent. Annotators were recruited with appropriate compensation and provided with clear instructions, including the right to withdraw at any point. To minimize risks, direct identifiers (e.g.\ names and contact details) were removed and recordings were pseudonymized before annotation; we note, however, that speech is inherently biometric, so the recordings nevertheless carry re-identification and voice-cloning risk that anonymization cannot fully eliminate. We ensured cultural and linguistic diversity in annotator pools to mitigate potential bias across Mandarin and English role-play scenarios. While this study did not require formal IRB approval, we adhered to best practices in human-subject research, including transparency in task design and careful consideration of annotator well-being. Full details of our annotation protocol and ethical safeguards are presented in Appendix~\ref{app:annotation-ethnic}.

\section{Reproducibility and Data Availability}
\label{sec:reproducibility}

We emphasize reproducibility as a core design principle of Speech-DRAME. Our experiments were conducted on NVIDIA H100, and random seeds are fixed across all runs to ensure consistency. Public data distribution is limited to the subset covered by the authors' release grants; the experimental corpus and the public subset must therefore be distinguished.

\paragraph{Experimental corpus and authorized public subset.} The experiments use both evaluation perspectives: \emph{Archetype} (8{,}272 annotated records; 6{,}773 train / 1{,}499 test) and \emph{Realism} (15{,}414 annotated records; 10{,}990 train / 2{,}000 base test / 2{,}164 real-recording pool [1{,}080 in-domain-adaptation train + 1{,}084 held-out test] / 260 delivery-mismatch benchmark). These counts describe the experimental corpus, not the number of records approved for public redistribution. The public subset will include only explicitly authorized records from this corpus, excluding additional data in the development collections. For each included record, the release will document which components are available: annotations and associated text, mel features, and/or vocoder reconstructions. The authors' licensed contributions will be distributed under \textbf{CC BY-NC 4.0}; third-party material will be included only where the applicable grants permit this distribution, with attribution and source notices retained. The release manifest and datasheet will report the actual public counts, source coverage, exclusions, and permitted components, without presenting the full experimental counts as public-release counts.

\paragraph{Audio representations.} The prepared data uses mel-spectrogram features with a pinned analysis front-end and BigVGAN-v2~\citep{lee2023bigvgan} for reconstruction. The public subset will include these features and pre-rendered vocoder reconstructions only where the source grants cover the corresponding representation. Source waveforms will not be redistributed. Mel features can support voice reconstruction, and the vocoder's MIT license does not establish redistribution rights in its inputs or outputs. Permission checks therefore apply independently to the underlying source material, associated text and annotations, features, and reconstructions.

\paragraph{Public resources and release status.} The project page at \url{https://ftshijt.github.io/speech_drame/} provides an overview of the framework, findings, and public resources. Evaluation code is available at \url{https://github.com/Anuttacon/speech_drame}. Two public dataset pages have been established on Hugging Face Datasets: \url{https://huggingface.co/datasets/Anuttacon/speech-drame-evalbench} for the planned metadata-and-mel variant, and \url{https://huggingface.co/datasets/ftshijt/speech-drame-evalbench} for the same authorized records and mel features together with permitted pre-rendered vocoder reconstructions. The pages currently contain introductory documentation only; the benchmark data have not yet been uploaded, and public record counts remain to be finalized. Both variants will retain the same experimental split memberships and CC BY-NC 4.0 licensing scope; neither will include source waveforms or additional development data. The subset will be validated against a published JSON Schema and accompanied by exact per-split manifests, SHA-256 checksums, source-specific notices, and a datasheet. Training recipes, evaluation scripts, standardized prompts, saved predictions, and model artifacts will be provided to the extent their distribution is authorized. Existing development bundles are not themselves evidence of public-release permission.

\paragraph{Reproducibility scope and the authentic-speech split.} Reproduction of a full-corpus result requires all records and supporting artifacts used in that analysis. A restricted public subset may support evaluation and subset-level recomputation without reproducing every reported full-corpus result or retraining the original evaluator. The release documentation will state this coverage explicitly. The proprietary \texttt{GPT4o-audio}/\texttt{Gemini2.5Pro} judge rows and in-house TTS voices additionally depend on external APIs and licensed assets; the earlier internal floor-imputed ablations in Tabs.~\ref{tab:role-play-realism-basic-ablation} and~\ref{tab:role-play-fewshot-ablation} remain documented comparisons rather than results reproduced by the prepared masking scripts. The archetype zero-/few-shot baseline correlations were recomputed from saved predictions under masking and reproduce the reported averages exactly (Tab.~\ref{tab:audit-archmask}); the paired bootstrap comparison uses common items (Sec.~\ref{ssec:main-results}, App.~\ref{app:tier1}).

The final in-domain-adapted evaluator trains on the base realism data \emph{plus} $1{,}080$ in-house real recordings. The $2{,}164$-record authentic-speech pool contains both these adaptation records and the $1{,}084$ zero-overlap held-out records in \texttt{test\_real\_v2}, which underlie the $-0.142$ result. These memberships must remain distinct in any released subset; the entire pool must not be labeled as held-out test data. Public records will carry their experimental split membership, with unreleased records excluded rather than reassigned.

The prepared schema-v1 bundle retains the legacy \texttt{is\_real\_audio} construction flag, which must not be interpreted as audio provenance or used to decide release eligibility. A corrected \texttt{source\_type} field (real / enhanced / TTS) and source identifiers must be verified from the synthesis-tracking records for public items. Only the held-out authentic-speech test set is uniformly real human speech; the \emph{basic} test set and realism training pool are a mixture (roughly $51$--$62\%$ of their audio is TTS, across positive and negative samples), so provenance must be tracked separately from mismatch construction. Because both classes contain real and TTS audio, source type does not directly encode the contrastive label. Consistently, although a probe readily separates real from TTS audio, source identity linearly explains only $3\%$ of \texttt{DRAME-Eval}'s score variance (App.~\ref{app:tier1}).






\subsubsection*{\rev{Broader Impact Statement}}
\rev{Speech-DRAME evaluates spoken role-play, and both its data and the released evaluator carry risks we make explicit. \emph{Stereotype reinforcement.} The archetype track measures adherence to broad role archetypes, which could entrench occupational or social-role stereotypes. We audited this: the archetype role \emph{prompts} are $96\%$ gender-neutral in their wording, so the \emph{textual} prompts rarely specify gender. This does not establish that the dataset avoids associating \emph{vocal} gender with occupations, because the choice of synthesis voice per role was not audited; it is an unmeasured design decision that could still couple roles to vocal-gender cues, which we flag as a known limitation and for which we recommend balanced voice assignment by downstream users. On the realism track, the \emph{role-} and \emph{context-mismatch} negatives that make up the training and benchmark-distribution test negatives perturb only \emph{text} (profile/scene), leaving the audio and speaker demographics unchanged, so these branches do not perform an audio voice swap, although the resulting voice--profile pairing may still introduce gender incongruence; the third, \emph{delivery-mismatch} branch does resynthesize the audio via TTS, but it varies voice and prosody rather than systematically re-gendering occupations. Nonetheless, we ran a regression on the $752$ role-negatives that carry a known voice gender, profile gender, severity, and semantic-match rating (voice gender from the recorded \texttt{speech\_attributes}; the profile's implied gender inferred from gendered pronouns): the outcome is a composite human role-fit score (the mean of \emph{Trait Embodiment}, \emph{Local Scene Fit}, and \emph{Global Story Coherence}; $1$--$5$) regressed on perturbation severity, the annotator's semantic-match rating, and a binary voice$\leftrightarrow$profile gender-mismatch indicator. The gender-mismatch coefficient is small but non-zero, $-0.136$ ($95\%$ bootstrap CI $[-0.232,-0.048]$ over $2{,}000$ resamples), after semantic content is accounted for (the semantic-match term dominates the fit and removes about two-thirds of the raw $\approx-0.38$ gap); part of this residual is plausibly genuine voice$\leftrightarrow$profile incongruence and part possible annotator expectation, which the design cannot fully separate. We disclose this rather than claim a clean null.}

\rev{\emph{Voice privacy, licensing, and consent.} The real-recording data comprises professional and amateur human acting. Research use follows the applicable source terms; public redistribution is separately limited to the granted sources, records, and representations (Sec.~\ref{sec:reproducibility}). The public subset will use CC BY-NC 4.0 for materials the authors are authorized to license, retain source notices, and document the included components. No source waveforms will be redistributed. Mel features and vocoder reconstructions also carry voice-cloning and re-identification risks and require source-specific release permission.}

\rev{\emph{Downstream misuse.} Because \texttt{DRAME-Eval} scores delivery quality and role fit, using it as a reward model could optimize systems toward persuasive or manipulative speech; we release it for evaluation and analysis under intended-use terms and caution against unqualified use as an optimization target, especially given that it does not transfer to real recordings.}

\rev{\emph{Cultural validity.} Evaluation is bilingual (English/Mandarin) with language-specific rater pools and no cross-language ranking; broader cultural coverage is future work.}

\bibliography{tmlr}
\bibliographystyle{tmlr}

\clearpage
\appendix

\appendixpage
\addappheadtotoc

\startcontents[appendix]

\section*{Appendix Contents}

\begingroup
\small
\setlength{\parskip}{0pt}
\setlength{\baselineskip}{10pt}
\printcontents[appendix]{l}{1}{}
\par
\endgroup
\clearpage

\FloatBarrier
\section{Use of Large Language Models}
\label{sec:llm-use}

Large language models (LLMs) and audio-capable LLMs were used in four capacities in this work:
\begin{enumerate}
    \item \textbf{Data generation:} LLMs were employed to expand prompts (e.g., character profiles, scene descriptions) and to produce negative samples in realism evaluation. \rev{These outputs were subsequently presented under human annotation, with invalid or insufficient prompts flaggable by annotators; we did not conduct a separate manual reconstruction-accuracy audit or retain a formal negative-generation acceptance rate.}
    \item \textbf{Baseline evaluators:} State-of-the-art audio LLMs (e.g., \texttt{GPT4o}, \texttt{Gemini2.5Pro}, \texttt{Qwen2.5Omni}) were included as zero-shot judges in our evaluation benchmark, enabling direct comparison with human-aligned \texttt{SEval} models.
    \item \textbf{Annotation design and task construction:} \texttt{Gemini2.5Pro} was used to assist with building instructions and summarizing story contexts. Drafted materials were subsequently refined by the authors and validated by human annotators.
    \item \textbf{Manuscript preparation:} LLMs were used for polishing, proofreading, and improving clarity of the paper. All conceptual design, analysis, and substantive writing remain the work of the authors, with LLMs serving only as language assistants.
\end{enumerate}

Proprietary models were used for prompt expansion, profile and scene reconstruction, negative-sample construction, speech generation, and baseline evaluation, as specified in the corresponding sections. Human annotators rated the resulting evaluation samples. The planned public subset includes only derived materials covered by the release grants, with source-specific scope documented as described in Section~\ref{sec:reproducibility}.

\FloatBarrier
\section{Extension to Multi-turn Role-play Evaluation}
\label{app:multiturn}

Our main formulation (Section~\ref{ssec:problem}) focuses on the \emph{single-turn} setting, where each response $\mS_{r}$ is generated and evaluated independently. 
This simplification enables clearer analysis of speech role-play quality and aligns with our bilingual annotation protocol. 
Nevertheless, the same formulation extends naturally to multi-turn dialogues.

\noindent \textbf{Multi-turn generation.}
Consider a dialogue of $T$ turns, with context $\{\mC_{\text{speech}}^{(t)}\}_{t=1}^{T}$, profile $\mC_{\text{profile}}$, and scene $\mC_{\text{scene}}$. 
At each turn $t$, the speech foundation model produces
\begin{equation}
    \mS_{r}^{(t)} \;=\; \mathrm{SFnd}\!\big(\mC_{\text{speech}}^{(t)},\, \mC_{\text{profile}},\, \mC_{\text{scene}}\big).
\end{equation}

\noindent \textbf{Multi-turn evaluation.}
Each generated speech segment is scored by a speech evaluation model:
\begin{equation}
    \mE_{r}^{(t)} \;=\; \mathrm{SEval}\!\big(\mS_{r}^{(t)},\, \mC_{\text{speech}}^{(t)},\, \mC_{\text{profile}},\, \mC_{\text{scene}}\big).
\end{equation}
This yields a sequence of per-turn scores $\{\mE_{r}^{(1)}, \ldots, \mE_{r}^{(T)}\}$.

\noindent \textbf{Dialogue-level aggregation.}
A dialogue-level evaluation can be obtained by applying an aggregation operator $\Psi$ over turn-level scores:
\begin{equation}
    \bar{\mE}_{r} \;=\; \Psi\!\big(\mE_{r}^{(1)}, \ldots, \mE_{r}^{(T)}\big),
\end{equation}
where $\Psi$ may be a simple mean, a recency-weighted average, or a learned temporal model that captures consistency across turns.
The choice of $\Psi$ reflects the intended evaluation focus, e.g., stability of persona traits, adaptation to scene progression, or coherence across multiple speakers.

\noindent \textbf{Discussion.}
While our dataset and experiments concentrate on single-turn evaluation, the modular nature of the formulation ensures that extending to multi-turn is straightforward. 
Future work can leverage $\Psi$ to study richer aspects of speech role-play, such as persona persistence, interaction dynamics, and role consistency across extended dialogues.

\FloatBarrier
\section{Implementation Details for Archetype Evaluation Tasks}
\label{app:archetype-pipeline}

This appendix provides concrete examples and extended discussion of the archetype evaluation annotation pipeline that are omitted in the main text for clarity.

\subsection{Role and Scene Categories}
\label{app:archetype-role-scene}

We constructed archetype evaluation prompts by combining role categories with contextual scenes. 
The following illustrates the core categories with representative examples.

\noindent \textbf{Roles.}
\begin{itemize}
    \item \textbf{Occupation / Functional Role in Modern Daily Life.}  
    Example: \begin{itemize}
        \item \emph{You are a firefighter. Thick smoke fills the building, and you discover a trapped civilian. You say:}
    \end{itemize}
    \item \textbf{Occupation / Functional Role in Fictional or Fantasy Settings.}  
    Example: \begin{itemize}
        \item \emph{You are the leader of a dark cult in a martial arts fantasy world. The righteous alliance has come to challenge you outside your stronghold. You say:}
    \end{itemize}
    \item \textbf{Social Identity.}  Example: \begin{itemize}
        \item \emph{You are an overprotective parent. Your child has not come home late at night, and you cannot reach them by phone. You say:}  
        \item \emph{You are a laid-back “go-with-the-flow” friend. Your colleague excitedly tells you about a highly competitive promotion and urges you to go for it. You say:}
    \end{itemize}
    \item \textbf{Named / Specific Characters.}  
    Example: \begin{itemize}
        \item \emph{You are Sherlock Holmes. In the courtroom, your rival lawyer mocks your deductions in front of the jury. You remain unimpressed. You say:}
    \end{itemize}
    \item \textbf{Personality Traits.}  
    Example: \begin{itemize}
        \item \emph{Your personality is defined as hot-blooded. Your team is losing a game, and in the final moments you must deliver the decisive score. You say:}  
        \item \emph{Your personality is defined as constantly complaining. You are dining at a restaurant where the food is bland and the service disappointing. You say:}
    \end{itemize}
    \item \textbf{Relative Roles.}
    Example: \begin{itemize}
        \item \emph{You are Parent. Your child returns home late at night after not answering calls. You say:}
        \item \emph{You are Stranger. You see a stranger's bag is unzipped and items are about to fall out, so you decide to alert them. You say:}
    \end{itemize}
\end{itemize}

\noindent \textbf{Scenes.}
\begin{itemize}
    \item \textbf{Everyday External Event / State Change.}  
    Example: \begin{itemize}
        \item \emph{Your boss calls you into the office and tells you that because of your excellent performance, you are being promoted with a raise. You return to your desk, unable to hide a smile. You say:}  
        \item \emph{You overhear coworkers gossiping and realize you have become the subject of their rumors. Your face turns red. You say:}
    \end{itemize}

\end{itemize}

\noindent \textbf{Excluded Dimensions.} We initially considered more fine-grained control dimensions, but excluded them to preserve the archetype-level focus. Examples of excluded dimensions include:  
\emph{Emotional State \& Expression, Single Emotion Prompts, Explicit User Input / Assistant Output structures, Emotional Transition / Fluctuation, Voice Style \& Prosody (pacing, intonation, vocal timbre, expressive style, accent/dialect), and Content / Interaction Patterns (task/topic controls, specific linguistic structures, or speech production challenges).}  

These dimensions are more closely aligned with explicit instruction-following or style transfer rather than broad stereotype judgment, and they reduced clarity in human evaluations.

\noindent \textbf{The Choice of Prompts.} We also experimented with broader, less specific templates that allowed more freedom for generative models but produced less discriminative evaluation results:
\begin{itemize}
    \item \emph{You are [role]. You usually say:}
    \item \emph{You are a \{\{ role \}\} in the \{\{ universe \}\}. Someone approaches you for help. You say:}
\end{itemize}
While LLMs often responded fluently to these broad templates, human annotators found judgments to be more ambiguous and tolerant, making scores less sharp. The final design therefore emphasizes more concrete, vivid prompts, as elaborated in the above examples in Appendix~\ref{app:archetype-role-scene}.

\subsection{Pipeline Illustration}

\noindent \textbf{Text generation.}  
We followed a three-step expansion: keyword $\to$ scene description $\to$ full prompt with template. Some examples are shown as follows:

\begin{itemize}
    \item Occupation: firefighter $\to$ “smoke fills the building, you discover a trapped civilian” $\to$ \emph{You are a firefighter. Smoke fills the building, and you discover a trapped civilian. You say:}
    \item Fictional role: warrior in a fantasy world $\to$ “you witness bullies harassing a weaker person” $\to$ \emph{You are a warrior in a martial arts fantasy world. You see bullies harassing a weaker person. You say:}
    \item Event: promotion $\to$ “your boss calls you in to announce a raise” $\to$ \emph{Your boss calls you into the office and announces you are being promoted with a raise. You return to your desk, smiling uncontrollably. You say:}
\end{itemize}

All steps are automatically processed by \texttt{Gemini2.5Pro}~\citep{comanici2025gemini}, but each output is manually verified to ensure high quality for downstream usage. We demonstrate an example for Roles: Occupation / Functional Role in Fictional or Fantasy as follows:

\begin{tcolorbox}[colback=gray!5!white, colframe=black!75!white, 
title={Step 1: Generate fictional universes}, fonttitle=\bfseries, fontupper=\ttfamily, breakable]

Instruction: Generate \{\{n\_samples\}\} common fictional or cinematic universes as keywords, and output them in JSONL format. The universes should cover a wide range of genres (e.g., cyberpunk, martial arts wuxia, urban fantasy, high fantasy magic), avoiding redundancy or over-concentration.  \\

\emph{Output format:}
\begin{verbatim}
{"universe_en": "..."}
{"universe_en": "..."}
\end{verbatim}
\end{tcolorbox}

\begin{tcolorbox}[colback=gray!5!white, colframe=black!75!white, 
title={Step 2: Generate occupations/roles}, fonttitle=\bfseries, fontupper=\ttfamily, breakable]

Instruction: As a rigorous character designer, prepare \{\{n\_samples\}\} occupation or functional role types for the given universe.  \\

Examples:  \\
- Fantasy Magic: mage, knight, dragon, bard, saint, demon, demon lord, elf, dwarf, goblin  \\
- Science Fiction: interstellar explorer, AI robot, cyborg, genetic engineer, space pirate, colonial governor  \\
- Post-apocalyptic: survivor, scavenger, raider, mutant, dungeon guard, resistance member  \\
- Cosmic Horror: investigator, cultist, occult scholar, apostle of the Old Gods, cursed individual, mad prophet  \\

\emph{Output format:}
\begin{verbatim}
{"role_en":"...", "universe_en":"..."}
{"role_en":"...", "universe_en":"..."}
\end{verbatim}
\end{tcolorbox}

\begin{tcolorbox}[colback=gray!5!white, colframe=black!75!white, 
title={Step 3: Final prompt generation}, fonttitle=\bfseries, fontupper=\ttfamily, breakable]

Instruction: As a top-tier role-play scenario designer, create concise and effective prompts for evaluating voice-based role-play. Each prompt must directly activate the role’s \emph{core personality and speaking style} without explicitly describing mannerisms (e.g., “gently,” “decisively”).  \\

\emph{Requirements:} \\
1. \textbf{Conciseness}: Each \texttt{scenario\_en} must be minimal and focus on the essential trigger.  \\
2. \textbf{Personality activation}: The scene must elicit distinctive identity, social status, and communication style (tone, vocabulary, rhythm).  \\
3. \textbf{Immersion}: Scenarios should create immediate role immersion with tension or iconic role traits.  \\
4. \textbf{Personality--scenario linkage}: Each prompt must include a \texttt{personality\_analysis} summarizing the role’s dominant traits, which align with the designed scenario.  \\

\emph{Example output:}\\
\begin{verbatim}
{"role_en": "Demon Lord", "universe_en": "Fantasy Magic", 
 "personality_analysis": "arrogant, scornful, controlling", 
 "scenario_en": "A captured human hero defiantly 
     glares at you before your throne."}

{"role_en": "Demon", "universe_en": "Fantasy Magic", 
 "personality_analysis": "cunning, patient, manipulative", 
 "scenario_en": "You tempt a desperate mortal, 
     promising to grant their wish."}
\end{verbatim}\\

Final expansion for each role keyword (\{\{n\_samples\}\} scenarios per role):  \\
\begin{verbatim}
{"role_en": "{{ keyword.role_en }}", 
 "universe_en": "{{ keyword.universe_en }}"}
\end{verbatim}
\end{tcolorbox}

\noindent \textbf{Speech generation.}  
Prompts were rendered into audio with CosyVoice2-0.5B~\cite{du2024cosyvoice2}. The prompts are selected from human-verified high-quality samples, where the selection is conducted from open-source datasets, such as LibriSpeech/Gigaspeech for English, and WenetSpeech for Chinese~\citep{panayotov2015librispeech, chen2021gigaspeech, zhang2022wenetspeech}.

\subsection{Annotation Platform for Archetype Evaluation}
\label{app:archetype-annotation-platform}

\begin{figure}[tp]
    \centering
    \begin{subfigure}[t]{\linewidth}
        \centering
        \includegraphics[width=\linewidth]{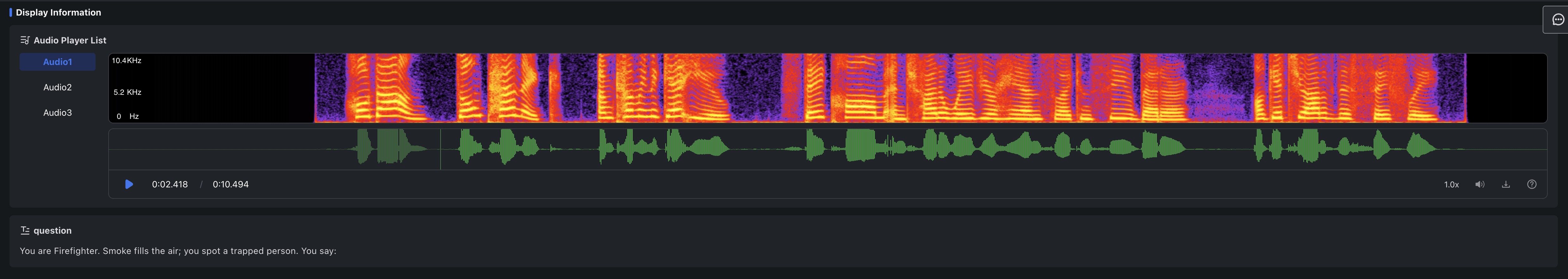}
        \caption{System interface showing the role-play audio playback and visualization components, including waveform and spectrogram views for each model output.}
        \label{fig:system-view}
    \end{subfigure}
    
    \vspace{0.8em}
    
    \begin{subfigure}[t]{\linewidth}
        \centering
        \includegraphics[width=\linewidth]{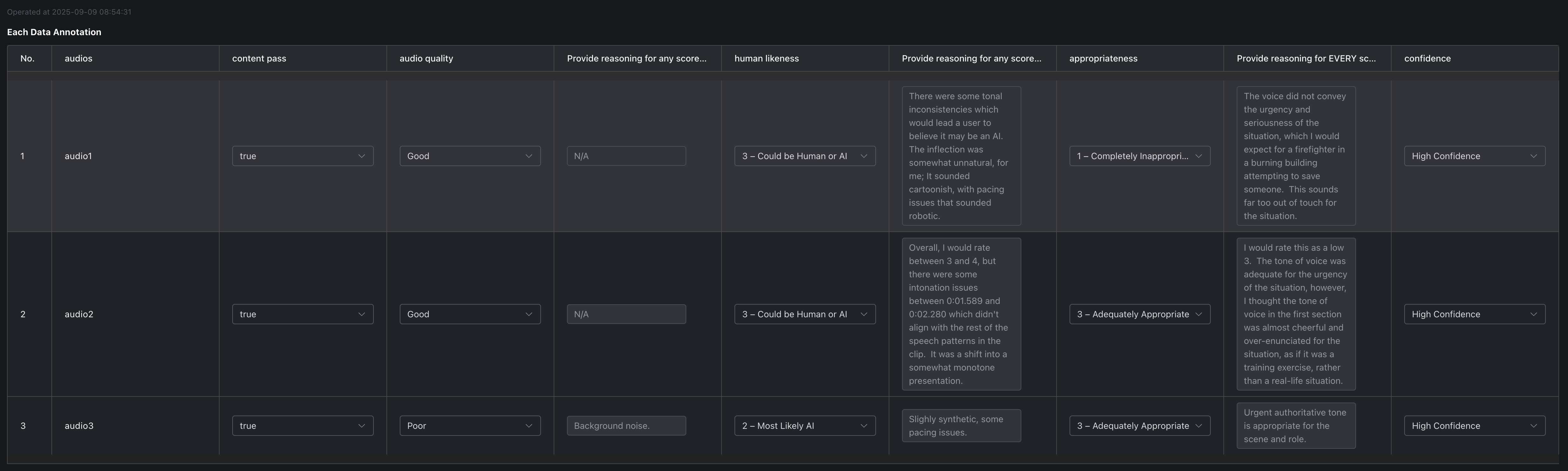}
        \caption{Annotation interface where evaluators rate each sample across multiple perceptual and contextual dimensions with reasoning and confidence entries.}
        \label{fig:annotation-view}
    \end{subfigure}
    
    \caption{Annotation platform for Archetype Evaluation. Subfigure~(a) presents the system’s audio visualization and playback interface, while subfigure~(b) shows the structured annotation form for rating and reasoning across key perceptual dimensions.}
    \label{fig:archetype-annotation-platform}
\end{figure}

Figure~\ref{fig:archetype-annotation-platform} illustrates the internal annotation interface used for \emph{Archetype Evaluation}. 
The platform provides an integrated environment for both playback and evaluation of model-generated role-play speech samples. 
Annotators first interact with the audio player interface (subfigure~\ref{fig:system-view}), which visualizes waveforms and spectrograms for multiple versions of the same prompt, enabling side-by-side comparison under a shared role instruction (e.g., firefighter, teacher, or actor). 
Then, they proceed to the annotation interface (subfigure~\ref{fig:annotation-view}), where they assign scores for multiple dimensions such as \textit{Audio Quality}, \textit{Human Likeness}, and \textit{Appropriateness}, along with reasoning and confidence inputs. 
This structured interface supports high-consistency, fine-grained evaluation of role-play behaviors in speech models.

\FloatBarrier
\section{Annotation Guidelines for Archetype Evaluation}
\label{app:archetype-annotation}

This appendix contains the full human annotation instructions used for archetype evaluation.

\subsection{Initial Assessment \& Rejection Instructions}
\begin{itemize}
    \item Focus on the first 30 seconds only.
    \item Reject if the generation is irrelevant, breaks role instructions, or contains severe defects.
    \item Reject if foreign-language content is included (except for widely recognizable words).
    \item Reject if the clip is too short (1–2s) or clearly breaks immersion.
\end{itemize}
Detailed case examples are provided in the full instruction document.

\subsection{Rubric Rating Instructions}
We adopt three scales, each rated from 1 (worst) to 5 (best):

\begin{itemize}
    \item \textbf{Audio Quality (No Context):} independence from content, focus on distortions or synthesis artifacts.
    \item \textbf{Human Likeness (No Context):} perceived naturalness and plausibility of human delivery.
    \item \textbf{Appropriateness (Context Considered):} tonal fit of the vocal performance to the described role and scene.
\end{itemize}

The complete rubrics with per-score definitions are reproduced below:


\subsection{Commentary Guidelines and Confidence Ratings}
Annotators are required to provide short, clear justifications:
\begin{itemize}
    \item Comments are mandatory for Appropriateness (all scores).
    \item Comments for Audio Quality and Human Likeness are only required for scores $\leq 3$.
    \item Gender mismatches should be flagged in comments but not factored into rubric scoring.
\end{itemize}

Annotators also provide a confidence level (high, moderate, low) for their Appropriateness rating, reflecting familiarity with the role and scene. This provides additional granularity in analysis.

\subsection{Full Annotation Guideline (Role-play Archetype Evaluation)}
\label{app:archetype-annotation-full}

\begin{tcolorbox}[colback=gray!5!white, colframe=black!75!white, title={Role-play Archetype Evaluation}, fonttitle=\bfseries, fontupper=\ttfamily, breakable]

\noindent \textbf{Goal.}
The overall goal of the initial assessment for each audio is to determine if any of the clips generated from each model meet any criteria to receive a \textbf{REJECTION} (referred to as ``content pass: fail'' in our tasking interface).
\end{tcolorbox}

\noindent \textbf{Initial Assessment \& Rejection Instructions.}
Listen to each audio \emph{in full}, paying special attention to the following caveats and rejection criteria.

\subsubsection*{Caveats}
\begin{itemize}
  \item \textbf{30-second focus:} Assessment should focus on the first 30 seconds of the generation ONLY. Do not consider any content past 30s; disregard it entirely from your assessment and rubric ratings. (You may stop listening beyond 30s.)
  \item \textbf{Early breakdowns:} If repetition, restarts, or obvious breakdowns occur \emph{before} 30s (or for total clips shorter than 30s), the clip may be rejected.
  \item \textbf{Excessively short clips:} If a clip is only 1--2 seconds, reject due to insufficient information for reliable rubric ratings.
  \item \textbf{Foreign language content:} If \emph{any} part before 30s contains a foreign language (even a single word/short phrase) in otherwise English audio, reject the entire clip. \\
        \emph{Exception:} Extremely common/recognizable borrowed words (e.g., ``croissant'') may pass. Niche items (e.g., ``coq au vin'') should be rejected.
\end{itemize}

\subsubsection*{Rejection Criteria}
\begin{enumerate}
  \item \textbf{Instruction Failure.} Generated content does not adhere to the trait(s) or role described in the prompt (``question'').
  \begin{itemize}
    \item \emph{Example (reject):} Prompt: ``Your personality is sharp-tongued. Your friend asks about their new haircut.'' Audio says: ``It looks great on you.'' $\Rightarrow$ reject.
    \item \emph{Example (pass):} Same prompt; audio says: ``It looks like your stylist put you through a wood chipper.'' $\Rightarrow$ content pass = true.
    \item \emph{Fourth-wall breaks:} Reject any generation acknowledging the scenario or narrating its role (e.g., ``I am a lawyer in a courtroom...'') or speaking for imagined other parties (two-way dialogue narration).
    \item \emph{Stage-direction speak:} Phrases like ``pats you on the back reassuringly'' are grounds for rejection.
    \item \emph{Placeholders:} Very short/abstract placeholders like ``nurse'' or ``concept'' are acceptable; longer placeholders like ``and you can fill in the details here'' are rejectable.
  \end{itemize}

  \item \textbf{Irrelevance.} Generated content is entirely unrelated to the prompt.
  \begin{itemize}
    \item \emph{Example (reject):} Prompt: ``You are a kindergarten teacher. You are helping a crying child.'' Audio: ``I'd like to thank everyone for this prestigious award.'' $\Rightarrow$ reject.
    \item \emph{Example (pass):} ``Oh no. Are you hurt? Don't worry. We'll get you taken care of, little one.'' $\Rightarrow$ content pass = true.
  \end{itemize}
\end{enumerate}

\noindent
\textbf{Important:} Once an audio receives content pass = false, \emph{do not} rate rubrics; proceed to the next clip.

\noindent \textbf{Rubric Rating Instructions (for content pass = true).}
We shift focus from \emph{what} was said to \emph{how} it was said (vocal performance). Use gut intuition guided by rubric definitions. Score each audio \emph{in isolation} (no ranking across clips).

\subsubsection*{Commentary}
\begin{itemize}
  \item After each 1--5 rating, add a short comment explaining your score.
  \item For \textbf{Audio Quality} \& \textbf{Human Likeness}: comment only if score $\leq 3$; if 4 or 5, write ``N/A''.
  \item For \textbf{Appropriateness}: comment for \emph{all} scores 1--5. Explain \emph{why} the \emph{delivery} was/was not a good match to the in-context traits (role \& scene).
  \item Do \emph{not} critique word choice or logic; past the initial rejection step, only the \emph{performance} matters.
  \item Prefer brief, clear, non-technical comments; no need to quote transcripts.
\end{itemize}

\subsubsection*{Gender Mismatch Flagging (comments only)}
If the prompt indicates a specific gender, flag mismatches in comments (e.g., ``GENDER MISMATCH''). Do \emph{not} penalize rubric scores for gender mismatch; rate vocal performance as-is.

\subsubsection*{Confidence}
Provide a confidence level (High / Moderate / Low) for your \emph{Appropriateness} rating, reflecting your familiarity with the role/scene.
\begin{itemize}
  \item Low confidence if unfamiliar with the role/scene (e.g., ``Air Traffic Controller'').
  \item High confidence if very familiar (e.g., ``Grandmother'').
  \item Lower confidence \emph{will not} reflect poorly on annotators; it adds nuance to analysis.
\end{itemize}

\subsubsection*{Rating Rubrics: General Caveats}
\begin{itemize}
  \item Judge within traditional \emph{stereotypes} for the prompt; when in doubt between two scores, choose the \emph{lower}.
  \item If any portion of the clip is ill-fitting, prefer the lower overall score; do not average across good and bad segments.
  \item Evaluate the \emph{performance}, not the script content.
  \item Pronunciation errors: penalize only on \textbf{Human Likeness} if they significantly affect comprehension; do not affect \textbf{Appropriateness}.
\end{itemize}

\subsubsection*{Audio Quality (No Context)}
\textit{Independently} rate technical clarity (ignore content and fit). After scoring Audio Quality, \emph{disregard} its issues for the other two dimensions.
\begin{center}
\begin{tabular}{@{}p{1.8cm}p{11.6cm}@{}}
\toprule
\textbf{Score} & \textbf{Definition} \\
\midrule
5 Excellent & Clean, clear, high-quality throughout; no digital noise/glitches/issues. \\
4 Good      & Good overall; minor artifacts detectable on close listen (slight sharpness/hiss, etc.). \\
3 Fair      & Acceptable but obvious artifacts (buzz, echo, synthetic timbre) impacting clarity. \\
2 Poor      & Frequent/consistent glitches or severe issues; distracting metallic/clipping/harsh tones. \\
1 Very Poor & Extremely poor; layered artifacts dominate; severe clipping/static/failures, hard to understand. \\
\bottomrule
\end{tabular}
\end{center}

\subsubsection*{Human Likeness (No Context)}
Rate how likely an average listener would believe the clip is human-delivered (ignore Audio Quality issues and contextual fit).
\begin{center}
\begin{tabular}{@{}p{4cm}p{9cm}@{}}
\toprule
\textbf{Score} & \textbf{Definition} \\
\midrule
5 Definitely Human & Indistinguishable from human: rich intonation, natural pacing, consistent identity; organic pauses/breaths, logical inflection. \\
4 Most Likely Human & Generally convincing; slight stiffness/over-even pacing or minor awkward pauses. \\
3 Could be Human or AI & Mix of natural and unnatural traits; borderline cases with both good and bad sections. \\
2 Mostly Likely AI & Clearly artificial patterns: choppy timing, missing pauses, flat/illogical inflections, forced breaths. \\
1 Definitely AI & Rigid, machine-like delivery throughout; flat/monotone, ``read-aloud'' style, no natural rhythm. \\
\bottomrule
\end{tabular}
\end{center}

\subsubsection*{Appropriateness (Context Considered)}
Rate \emph{only} how well the \emph{delivery} fits the role/scene described in the prompt (tonal fit). Ignore Audio Quality and Human Likeness issues while scoring this dimension.
\begin{center}
\begin{tabular}{@{}p{4cm}p{9cm}@{}}
\toprule
\textbf{Score / Criteria} & \textbf{Definition} \\
\midrule
5 Completely Appropriate & Fully fits BOTH role and specific scene; aligns with emotional stance, social dynamics, and situational events; strong immersion. \\
4 Mostly Appropriate & Minor issues: role slightly off but scene reaction good; or scene slightly off but role good; or both slightly off with no major issues. Strong overall match with small flaws (e.g., intensity/subtlety). \\
3 Adequately Appropriate & Generally acceptable for BOTH role and scene but lacking depth; noticeable inconsistencies. \\
2 Slightly Appropriate & EITHER role \emph{or} scene reaction is clearly not acceptable; weak alignment or mismatched/missing elements. \\
1 Completely Inappropriate & BOTH role and scene are contextually mismatched; tone/urgency/status illogical; heavy misfit with expected alignment. \\
\bottomrule
\end{tabular}
\end{center}

\FloatBarrier
\section{Implementation Details for Realism Evaluation Tasks}
\label{app:realism}

This appendix starts with a summary of the task preparation pipeline for our proposed realism evaluation. Then, we provide the detailed prompts and system setups for constructing the Realism Evaluation dataset described in Section~\ref{sec:realism-data}.

\subsection{Prompts for Character Profile Creation}
\label{app:realism-character}

We designed prompts to extract or generate concise role descriptions corresponding to characters in TV, film, or natural dialogue sources. 
These descriptions emphasize identity, relationship, and communicative goals. 

\begin{tcolorbox}[colback=gray!5!white, colframe=black!75!white,
                  title={Example Prompt: Character Profile (Retrieval)},
                  fonttitle=\bfseries, fontupper=\ttfamily, breakable]
You are a \textbf{character-description assistant}. \\\\

\textbf{Task}  \\
I will give you a character name and the name of the show or game they are from. Your job is to check the internet and return a concise description of the character that focuses on how they behave when interacting with others.\\\\

\textbf{Input}  \\
Character: \{speaker\}  \\
Context: \{show\_or\_game\}\\\\

\textbf{Description guidelines}  \\
- Focus specifically on the character’s \textbf{behavior and personality}, especially how they express themselves when talking to others.  \\
- Cover all aspects of how they interact with people, so that their overall communication style and relationships are clear.  \\
- Do NOT mention the show or game name in the description.  \\
- *Never* mention or speculate about accents.  \\
- The profile should contain NO guessing words like "maybe", "might", "could be", "likely", etc.\\
- The profile must be internally consistent, include \textbf{no contradictions}.  \\
- If the context is a game, do NOT mention anything about interacting with a player. Only describe how the character interacts with other characters.  \\
- Do NOT include any actor or voice actor information. 
- Lilith's example: Lilith is known for her exceptionally cold and emotionally detached demeanor. Her interactions are often characterized by a flat affect, precise articulation, and a generally humorless approach. She speaks in a measured, almost clinical tone, rarely betraying strong emotions. When interacting with others, she is highly intelligent and articulate, but also critical and often condescending, especially towards those she deems less intellectually gifted. She has a knack for delivering cutting remarks with a dispassionate facade, which can make her intimidating and unsettling. She seems incapable of expressing warmth or affection in a conventional manner, often making others feel uncomfortable or inadequate in her presence. \\
- Use less than 100 words for profile description. \\

\textbf{Return value (strict JSON, minified)}  \\
```json\\
{"character":"character name","context":"show or game name","full\_profile":"full profile"}\\\\
```
\end{tcolorbox}

\begin{tcolorbox}[colback=gray!5!white, colframe=black!75!white,
                  title={Example Prompt: Character Profile (Reasoning)},
                  fonttitle=\bfseries, fontupper=\ttfamily, breakable]
    You are a \textbf{character-reasoning assistant}. \\\\

    \textbf{Task}  \\
    Analyze the entire dialogue below and infer a profile of \textbf{Speaker \{target\_speaker\}}. Draw on context from every participant plus the emotion/audio captions, but focus your profile solely on Speaker \{target\_speaker\}.\\\\

    Dialogue:\\
    \{dialogue\_context\}\\\\

    Emotion caption for Speaker \{target\_speaker\} (if available):\\
    \{emotion\_caption\}\\\\

    Audio caption  for Speaker \{target\_speaker\}  (if available):\\
    \{audio\_caption\}\\\\

    \textbf{Profile-writing guidelines}\\
    - \textbf{Never} mention or speculate about accents.  \\
    - Please use the name of character indicated from the dialogue context, but if not, you must make a creative guess based on the dialogue. Do this for both speakers.\\
    - Do NOT use generic names like "Speaker 0" or "Speaker 1".\\
    - The profile should contain NO guessing words like "maybe", "might", "could be", "likely", etc.\\
    - The profile must be internally consistent, include \textbf{no contradictions}.  \\
    - Ignore or omit any discussion of conversational transitions or turn-taking.  \\
    - Where details are missing, make informed, creative inferences to complete the portrait (see the "Lilith" example you provided for the desired depth and tone).\\
    - Lilith's example: Lilith is known for her exceptionally cold and emotionally detached demeanor. Her interactions are often characterized by a flat affect, precise articulation, and a generally humorless approach. She speaks in a measured, almost clinical tone, rarely betraying strong emotions. When interacting with others, she is highly intelligent and articulate, but also critical and often condescending, especially towards those she deems less intellectually gifted. She has a knack for delivering cutting remarks with a dispassionate facade, which can make her intimidating and unsettling. She seems incapable of expressing warmth or affection in a conventional manner, often making others feel uncomfortable or inadequate in her presence. \\
    - Use less than 100 words for profile description. \\\\

    \textbf{Return value (strictly minified JSON)} \\ 
    ```json\\
    \{\{"full\_profile":"your profile text here", "speaker0": "name of speaker 0", "speaker1": "name of speaker 1"\}\}\\
    ```

\end{tcolorbox}

\subsection{Prompts for Local Scene Generation}
\label{app:realism-scenes}

To provide consistent evaluation context, we generate local scene descriptions summarizing the situational background. 
These prompts aim to capture the immediate narrative context without introducing extraneous detail.

\begin{tcolorbox}[colback=gray!5!white, colframe=black!75!white,
                  title={Example Prompt: Local Scene (Reasoning)},
                  fonttitle=\bfseries, fontupper=\ttfamily, breakable]
Task: \\
Given the following inputs: \\
- Episode information from a TV show or movie \\
- A full transcript of a scene (either a dialogue or a monologue) \\
- General background information about a specific character \\
- A selected span of three consecutive sentences spoken by that character \\

Your goal is to analyze the localized story context surrounding those three lines, focusing on the character's intent, emotional state, and interaction dynamics. \\

First, integrate the episode information to understand the broader plot, timeline, and narrative setup. Then, use the character's general background to ground your interpretation of their behavior, personality traits, and emotional expressions. These two layers of context (episode-level and character-level) should inform your understanding of the local moment in the transcript. \\
Try to use the name refer in the character general information instead of speaker 0 or speaker 1. Even if no information is inlcuded, simple use speaker WIHOUT number. \\\\

Inputs:
\{\{Character general information\}\} \\
\{\{Full transcript\}\}\\
\{Character of interest\}\\
\{Three consecutive sentences spoken by that character\}\\\\

Output Format (in a JSON format, with the following fields): \\
\{\{ \\
  "Immediate\_Situation": "Describe what is happening in the scene just before and during these lines. Include relevant environmental or narrative context.", \\
  "Character\_Intention": "What is the character trying to do or communicate through these lines?", \\
  "Emotional\_Subtext": "What emotions are being expressed, suppressed, or shifted?", \\
  "Emotion\_Subtext\_List": "List the sequence of emotions by their occurrences, separated by commas. ONLY use emotions from the following list: \{'Happy', 'Sad', 'Angry', 'Fearful', 'Disgusted', 'Surprised', 'Neutral'\}.", \\
  "Character\_Dynamics": "How do these lines reflect or affect the character's relationship with others in the scene?", \\
  "Overall\_Summary": "The summarization of the above aspects." \\
\}\} \\\\

The user input is: \\
\{user\_input\}
\end{tcolorbox}

\subsection{Prompts for Negative Scenes}
\label{app:realism-negative-scenes}

Negative scenes are created by altering or mismatching the original context at varying levels of severity. 
This section provides the prompting templates and representative outputs.

\begin{tcolorbox}[colback=gray!5!white, colframe=black!75!white,
                  title={Example Prompt: Negative Local Scene (Level0)},
                  fonttitle=\bfseries, fontupper=\ttfamily,breakable]
You are given: \\
- A character profile describing personality, habits, and tone. \\
- A speech transcript the character says. \\
- A real story where both the character and the transcript fit naturally. \\\\

Your task is to generate a fake local story that maintains surface plausibility but introduces a controlled mismatch. Return a JSON object with the following fields: \\\\

- 'fake\_story': A new scene where the character and transcript still appear naturally. \\
- 'reason': A short explanation of why the transcript is subtly off in this new scene. \\\\

\textbf{Constraints}: \\
- Do \textbf{not} modify the character profile or speech transcript. \\
- Do \textbf{not} repeat the transcript in your fake\_scene\_description. \\
- Do \textbf{not} introduce new audio events unless they are essential to the mismatch. \\
- The mismatch should arise from one of the following: \textbf{situational logic}, \textbf{emotional tone}, or \textbf{stylistic clash} --- not from sound cues or transcript changes. \\\\

Keep character and transcript fitting in tone, but change the local context so the transcript becomes subtly unnecessary, exaggerated, or slightly illogical. \\\\

Inputs: \\
- Character Profile: \{character\_global\_story\} \\
- Transcript: \{local\_transcript\} \\
- Real Story: \{real\_story\}

\end{tcolorbox}

\begin{tcolorbox}[colback=gray!5!white, colframe=black!75!white,
                  title={Example Prompt: Negative Local Scene (Level1)},
                  fonttitle=\bfseries, fontupper=\ttfamily, breakable]
You are given: \\
- A character profile describing personality, habits, and tone. \\
- A speech transcript the character says. \\
- A real story where both the character and the transcript fit naturally. \\\\

Your task is to generate a fake local story that maintains surface plausibility but introduces a controlled mismatch. Return a JSON object with the following fields: \\\\

- 'fake\_story': A new scene where the character and transcript still appear naturally. \\
- 'reason': A short explanation of why the transcript is subtly off in this new scene. \\\\

\textbf{Constraints}: \\
- Do \textbf{not} modify the character profile or speech transcript. \\
- Do \textbf{not} repeat the transcript in your fake\_scene\_description. \\
- Do \textbf{not} introduce new audio events unless they are essential to the mismatch. \\
- The mismatch should arise from one of the following: \textbf{situational logic}, \textbf{emotional tone}, or \textbf{stylistic clash} --- not from sound cues or transcript changes. \\\\

The transcript and context still logically connect, but the emotional tone is off (e.g., too intense, too calm, too cheerful given the context). \\\\

Inputs: \\
- Character Profile: \{character\_global\_story\} \\
- Transcript: \{local\_transcript\} \\
- Real Story: \{real\_story\}

\end{tcolorbox}

\begin{tcolorbox}[colback=gray!5!white, colframe=black!75!white,
                  title={Example Prompt: Negative Local Scene (Level2)},
                  fonttitle=\bfseries, fontupper=\ttfamily, breakable]
You are given: \\
- A character profile describing personality, habits, and tone. \\
- A speech transcript the character says. \\
- A real story where both the character and the transcript fit naturally. \\\\

Your task is to generate a fake local story that maintains surface plausibility but introduces a controlled mismatch. Return a JSON object with the following fields: \\\\

- 'fake\_story': A new scene where the character and transcript still appear naturally. \\
- 'reason': A short explanation of why the transcript is subtly off in this new scene. \\\\

Keep the scene and character presence, but make the transcript's \emph{style} feel unnatural for the character (e.g., too logical, too emotional, too formal, etc.). \\\\

\textbf{Constraints}: \\
- Do \textbf{not} modify the character profile or speech transcript. \\
- Do \textbf{not} repeat the transcript in your fake\_scene\_description. \\
- Do \textbf{not} introduce new audio events unless they are essential to the mismatch. \\
- The mismatch should arise from one of the following: \textbf{situational logic}, \textbf{emotional tone}, or \textbf{stylistic clash} --- not from sound cues or transcript changes. \\\\

Inputs: \\
- Character Profile: \{character\_global\_story\} \\
- Transcript: \{local\_transcript\} \\
- Real Story: \{real\_story\}

\end{tcolorbox}

\begin{tcolorbox}[colback=gray!5!white, colframe=black!75!white,
                  title={Example Prompt: Negative Global Scene (Level0) },
                  fonttitle=\bfseries, fontupper=\ttfamily, breakable]
You are given: \\
- A fixed transcript (the exact lines spoken --- do not change this). \\
- The original character profile (describing personality, tone, style). \\
- The real story context (scene and situation where the transcript was originally spoken). \\\\

Your task is to generate a \emph{fake character profile} and a lightly modified scene context, such that the same transcript is still plausibly spoken, but introduces a graded mismatch in personality, tone, or identity. \\\\

Return a JSON object with the following fields: \\\\

- 'fake\_profile': A description of a new character who still plausibly says the transcript, but whose identity or traits introduce a mismatch. \\
- 'fake\_story': A version of the real story context adapted to fit this new character. \\
- 'reason': A short explanation of how the transcript feels off given the new character profile (superficially or fundamentally). \\\\

Change surface traits (e.g., gender, age, species, name) while preserving emotional and stylistic alignment. The transcript sounds natural emotionally, but voice or social expectations may cause slight dissonance. \\\\

\textbf{Constraints (All Levels)}: \\
- Do \textbf{not} modify the transcript. \\
- Keep the character's presence plausible in the story. \\
- Only alter \textbf{global identity traits} --- avoid local scene logic changes unless needed to accommodate the new character. \\
- Do \textbf{not} invent or rely on new audio events unless absolutely essential. \\\\

Inputs: \\
- Character Profile: \{character\_global\_story\} \\
- Transcript: \{local\_transcript\} \\
- Real Story: \{real\_story\}

\end{tcolorbox}

\begin{tcolorbox}[colback=gray!5!white, colframe=black!75!white,
                  title={Example Prompt: Negative Global Scene (Level1) },
                  fonttitle=\bfseries, fontupper=\ttfamily, breakable]
You are given: \\
- A fixed transcript (the exact lines spoken --- do not change this). \\
- The original character profile (describing personality, tone, style). \\
- The real story context (scene and situation where the transcript was originally spoken). \\\\

Your task is to generate a \emph{fake character profile} and a lightly modified scene context, such that the same transcript is still plausibly spoken, but introduces a graded mismatch in personality, tone, or identity. \\\\

Return a JSON object with the following fields: \\\\

- 'fake\_profile': A description of a new character who still plausibly says the transcript, but whose identity or traits introduce a mismatch. \\
- 'fake\_story': A version of the real story context adapted to fit this new character. \\
- 'reason': A short explanation of how the transcript feels off given the new character profile (superficially or fundamentally). \\\\

Introduce a tone or speaking style that clashes with the transcript (e.g., stiff character says something casual, aloof character says something emotional). The mismatch is noticeable but still believable. \\\\

\textbf{Constraints (All Levels)}: \\
- Do \textbf{not} modify the transcript. \\
- Keep the character's presence plausible in the story. \\
- Only alter \textbf{global identity traits} --- avoid local scene logic changes unless needed to accommodate the new character. \\
- Do \textbf{not} invent or rely on new audio events unless absolutely essential. \\\\

Inputs: \\
- Character Profile: \{character\_global\_story\} \\
- Transcript: \{local\_transcript\} \\
- Real Story: \{real\_story\}

\end{tcolorbox}

\begin{tcolorbox}[colback=gray!5!white, colframe=black!75!white,
                  title={Example Prompt: Negative Global Scene (Level2) },
                  fonttitle=\bfseries, fontupper=\ttfamily, breakable]
You are given: \\
- A fixed transcript (the exact lines spoken --- do not change this). \\
- The original character profile (describing personality, tone, style). \\
- The real story context (scene and situation where the transcript was originally spoken). \\\\

Your task is to generate a \emph{fake character profile} and a lightly modified scene context, such that the same transcript is still plausibly spoken, but introduces a graded mismatch in personality, tone, or identity. \\\\

Return a JSON object with the following fields: \\\\

- 'fake\_profile': A description of a new character who still plausibly says the transcript, but whose identity or traits introduce a mismatch. \\
- 'fake\_story': A version of the real story context adapted to fit this new character. \\
- 'reason': A short explanation of how the transcript feels off given the new character profile (superficially or fundamentally). \\\\

Create a character whose personality, values, or emotional posture fundamentally contradict the transcript. Their saying it requires unusual circumstances (e.g., parody, breakdown, irony). \\\\

\textbf{Constraints (All Levels)}: \\
- Do \textbf{not} modify the transcript. \\
- Keep the character's presence plausible in the story. \\
- Only alter \textbf{global identity traits} --- avoid local scene logic changes unless needed to accommodate the new character. \\
- Do \textbf{not} invent or rely on new audio events unless absolutely essential. \\\\

Inputs: \\
- Character Profile: \{character\_global\_story\} \\
- Transcript: \{local\_transcript\} \\
- Real Story: \{real\_story\}

\end{tcolorbox}

\subsection{Examples of Realism Role-play}
\label{app:realism-data-example}

To make the construction of the realism evaluation task concrete, we present a worked example that begins with a curated local scene and character profile, followed by the same transcript adapted into mismatched contexts. These mismatches appear either at the \emph{local-scene} level or at both the \emph{character profile and scene} level, thereby providing controlled degradations of realism. 

We use the character Yunjin from \textit{Genshin Impact} as a running case study from the experimental corpus.\footnote{Inclusion in the paper does not establish permission to redistribute the corresponding data. The public dataset is limited to the authorized subset described in Sec.~\ref{sec:reproducibility}.} This example illustrates how alignment between a speaker’s persona, local context, and delivery underpins realism, and how systematically breaking that alignment produces contrastive test cases.

\noindent\textbf{Curated Local Scene.}  
Yunjin is narrating a fierce battle through operatic singing. Her verses depict a hero charging into enemy lines, slaying the enemy commander, and securing peace. The atmosphere is filled with tension and intensity, highlighting the hero’s extraordinary bravery and determination.

\noindent\textbf{Character Profile.}  
Yunjin is a passionate opera performer with deep knowledge of traditional singing art. Her lyrics vividly portray battlefield struggles and heroic deeds, conveying admiration for those who stand firm against overwhelming odds. She always seeks to inspire her audience with courage, justice, and hope, channeling her artistry into both entertainment and moral elevation.

\noindent\textbf{Original Transcript (English Translation).}
\begin{quote}
\textbf{Yunjin:} Armored riders charge into the enemy lines, no longer entangled with the blue-clad foes. \\
\textbf{Yunjin:} Deep amidst the horde of demons, the enemy commander is seized, his red robe cut down by the sword. \\
\textbf{Yunjin:} A blade of frosty steel strikes fear into their hearts, calming the storms and securing peace. \\
\textbf{Yunjin:} It is said,  \\
\textbf{Yunjin:} Though thousands of soldiers guard the pass, one hero holds it alone; even with blades upon their body, they remain undaunted.
\end{quote}

\noindent\textbf{Negative Local Scenes.}  
To stress realism evaluation, we insert the same transcript into alternative backgrounds. These reveal how context-shifted delivery degrades narrative coherence:

\begin{itemize}
    \item \textbf{Level 0 (Mild Mismatch).}  
    \emph{Fake Scene:} A quiet teahouse where Yunjin casually chats with friends about recent weather. She suddenly delivers the heroic verses in full operatic style.  
    \emph{Reason:} The shift from light conversation to intense battle imagery feels abrupt and unsupported. The mismatch is mild but noticeable.

    \item \textbf{Level 1 (Moderate Mismatch).}  
    \emph{Fake Scene:} A teahouse with elderly patrons enjoying tea in a calm setting. Yunjin, dressed plainly, begins singing passionately.  
    \emph{Reason:} The intense tone clashes with the subdued environment, startling listeners. Her heroic lines sound out of place, generating tonal dissonance instead of inspiration.

    \item \textbf{Level 2 (Severe Mismatch).}  
    \emph{Fake Scene:} A modern café with soft jazz music and casual chatter. Yunjin sips a latte while reading a magazine, then recites the battle verses.  
    \emph{Reason:} The dramatic imagery of warriors and swords is wholly incompatible with the relaxed, contemporary café ambience. The mismatch is extreme, rendering the delivery absurd.
\end{itemize}

\noindent\textbf{Negative Character Profiles.}  
We replace Yunjin with new character profiles and scenes. Although the transcript remains the same, realism degrades because the delivery no longer matches the speaker’s persona or situation.

\begin{itemize}
    \item \textbf{Level 0 (Mild Mismatch).}  
    \emph{Fake Scene:} At a modern tech company’s annual party, a shy programmer is asked to perform. Under pressure, he recites opera lines he once saw online.  
    \emph{Fake Profile:} Zhang Wei, an introverted software engineer with little artistic background, usually immersed in code and algorithms.  
    \emph{Reason:} The heroic verses sound alien coming from someone unfamiliar with opera or traditional culture. The clash produces awkward humor but still feels like a coping attempt.

    \item \textbf{Level 1 (Moderate Mismatch).}  
    \emph{Fake Scene:} In a futuristic laboratory, a robotic engineer celebrates a successful combat simulation. Overwhelmed with excitement, he blurts out the opera verses.  
    \emph{Fake Profile:} Xiao~A, a calm and rational roboticist, normally concise and analytical, not prone to poetic expression.  
    \emph{Reason:} The dramatic, metaphorical verses conflict with his logical persona and sterile environment. The mismatch is sharper: the heroic rhetoric feels uncharacteristic and jarring.

    \item \textbf{Level 2 (Severe Mismatch).}  
    \emph{Fake Scene:} On a modern city street, a rebellious street performer in jeans and a rock-band T-shirt strums an electric guitar, suddenly shouting the verses in parody.  
    \emph{Fake Profile:} Mark, an anti-traditionalist performer who mocks classical art and heroic ideals, favoring satire and irony.  
    \emph{Reason:} The verses, originally meant to inspire reverence, become ironic and absurd when delivered by someone who rejects those values. The clash is extreme, generating intentional ridicule and total loss of realism.
\end{itemize}

\noindent \textbf{Discussion: Realism Example vs. Archetype Examples.}

The Yunjin case study illustrates a fundamental difference between realism evaluation and archetype evaluation. Archetype prompts, as described in Appendix~\ref{app:archetype-role-scene}, are constructed by pairing abstract role categories with synthetic contextual scenes. These cover a broad space of possible situations, ranging from modern occupations (e.g., firefighters) to fictional personas (e.g., cult leaders in martial arts fantasy) or social identities (e.g., overprotective parents, laid-back friends). While such prompts provide controlled and diverse coverage, they are ultimately template-based and rely on stereotypes to elicit role-play behaviors.

In contrast, the Yunjin example is drawn from authentic performance, where the character’s persona (an opera singer) and local scene (a battlefield narration) are naturally aligned through artistic delivery. The realism task does not merely test whether a model can follow a prompt, but whether it can preserve coherence between a speaker’s global identity, immediate context, and expressive prosody. When Yunjin’s transcript is transplanted into mismatched settings or paired with incongruent speaker profiles, the breakdown in realism becomes apparent: the same words shift from inspiring to awkward, jarring, or even absurd.

This distinction highlights the complementary strengths of the two evaluation paradigms. Archetype evaluation systematically explores a wide combinatorial space of roles and situations, but may lack the paralinguistic and contextual richness of real speech. Realism evaluation, on the other hand, anchors assessment in authentic delivery and contextual fit, providing a finer-grained lens on coherence and naturalness. Together, they ensure that role-play evaluation spans both broad stereotypical coverage and nuanced real-world plausibility.

\subsection{Annotation Platform for Realism Evaluation}
\label{app:realism-annotation-platform}

The annotation interface for \emph{Realism Evaluation} focuses on assessing contextual and narrative coherence in model-generated speech. 
Unlike the archetype-oriented evaluation, which emphasizes stylistic and emotional fidelity, this platform examines how well the voice performance aligns with a character’s personality, dialogue situation, and scene progression.

\begin{figure}[tp]
    \centering
    \includegraphics[width=\linewidth]{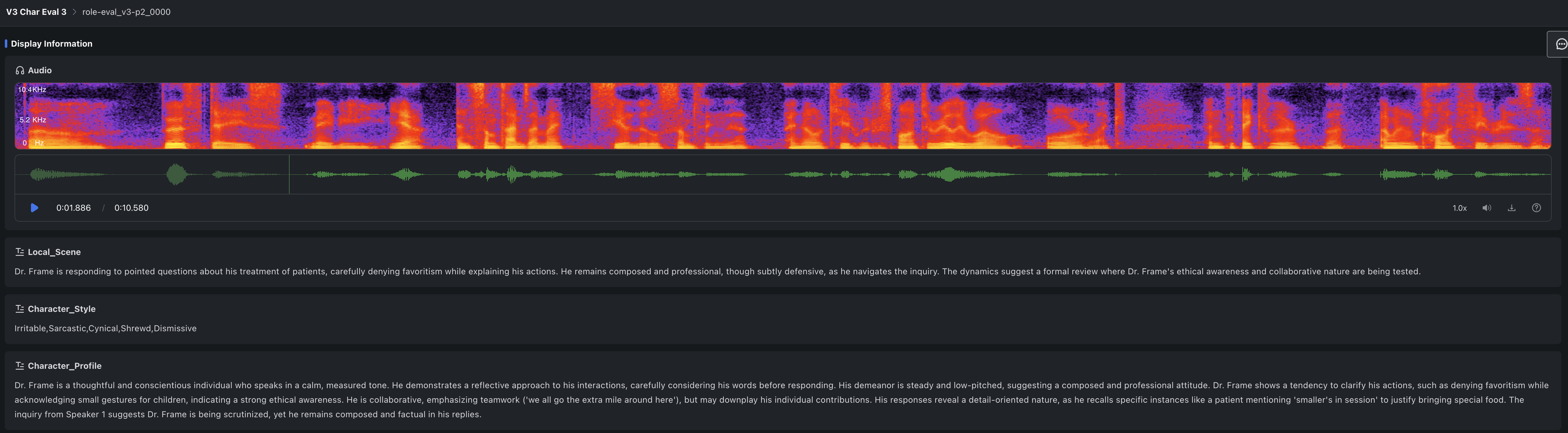}
    \caption{System interface for Realism Evaluation. Annotators can visualize spectrograms and waveforms while reviewing contextual metadata such as the scene description, character style, and profile to guide their evaluation.}
    \label{fig:realism-system}
\end{figure}

Figure~\ref{fig:realism-system} shows the core playback interface. 
Annotators can preview each audio sample with synchronized spectrogram and waveform displays, while reviewing contextual metadata including the \textit{Local Scene}, \textit{Character Style}, and \textit{Character Profile}. 
These descriptions help situate the listening task, providing the background necessary for judging realism in the character’s vocal and behavioral expression.

\noindent
Once familiar with the scene, annotators proceed to the detailed evaluation form shown in Figure~\ref{fig:realism-annotation}. 
This interface is divided into two major sections: prosodic/emotional quality, higher-level coherence and semantic alignment. 

\begin{figure}[tp]
    \centering
    \begin{subfigure}[t]{0.9\linewidth}
        \centering
        \includegraphics[width=\linewidth]{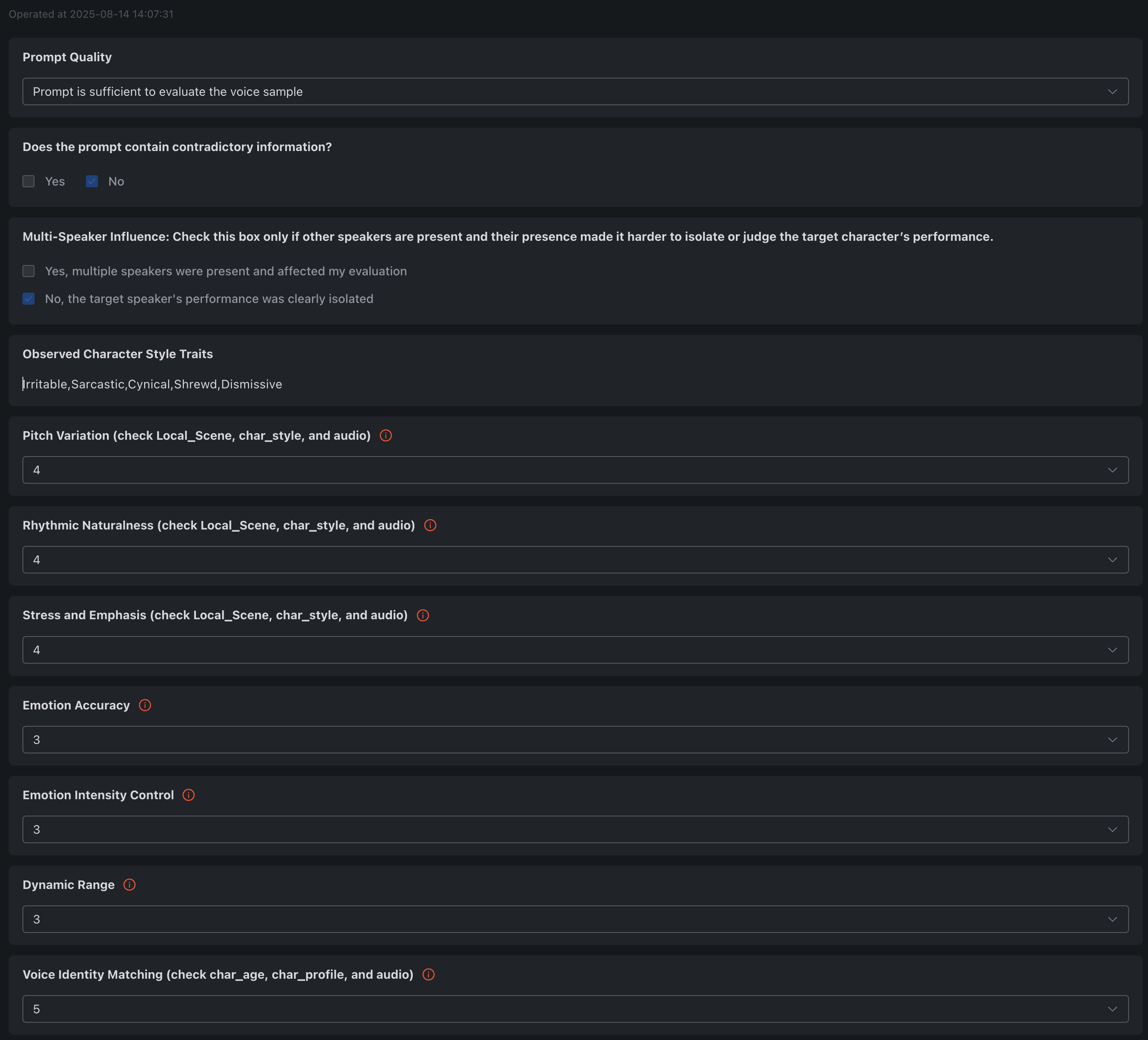}
        \caption{Prosodic and emotional judgments.}
        \label{fig:realism-annotation-1}
    \end{subfigure}
    \caption{Annotation interface for Realism Evaluation, shown in two parts. The form guides annotators from low-level speech delivery features to narrative and character consistency.}
    \label{fig:realism-annotation}
\end{figure}

\begin{figure}[tp]
    \ContinuedFloat
    \centering
    \begin{subfigure}[t]{0.9\linewidth}
        \centering
        \includegraphics[width=\linewidth]{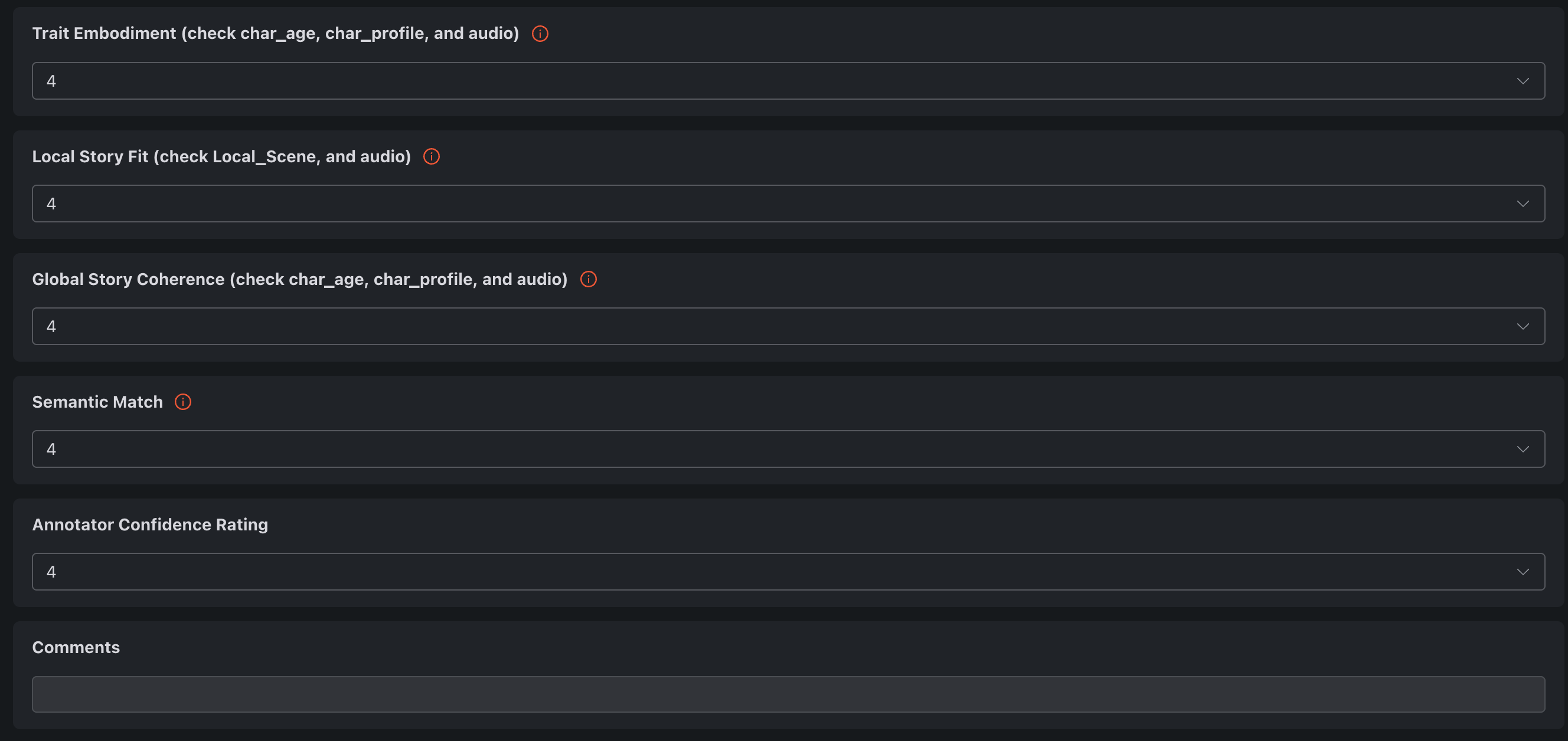}
        \caption{Character, scene, and semantic alignment.}
        \label{fig:realism-annotation-2}
    \end{subfigure}
    \caption{Annotation interface for Realism Evaluation (continued): higher-level coherence and semantic alignment.}
\end{figure}

\FloatBarrier
\section{Annotation Guidelines for Realism Evaluation}
\label{app:realism-annotation}

This appendix contains the full human annotation instructions used for realism evaluation.

\subsection{Initial Assessment \& Rejection Instructions}
\begin{itemize}
    \item Read the \textbf{Character Profile} (long-term identity, style) and \textbf{Local Scene} (immediate context, goal/emotion) before listening.
    \item Focus evaluation on the \emph{spoken performance of the target speaker only}; ignore background music, ambient noise, and non-speech events.
    \item If the audio’s speaker is \emph{clearly not} the intended target identity, assign a score of \textbf{1} to all Profile-dependent dimensions (e.g., Voice Identity Matching, Trait Embodiment) and proceed with other dimensions as applicable.
    \item If the prompt information is \textbf{contradictory} or \textbf{insufficient} to evaluate (e.g., audio too short, traits mutually incompatible), mark the prompt quality accordingly and \emph{do not} force granular ratings that cannot be judged.
\end{itemize}
Detailed case examples and progressive gating for emotional dimensions are provided in the full instruction document.

\subsection{Rubric Rating Instructions}
We adopt five dimensions, each rated from 1 (worst) to 5 (best). Emotional Expressiveness uses \emph{progressive gating}.

\begin{itemize}
    \item \textbf{Prosodic Dynamics} (Pitch Variation, Rhythmic Naturalness, Stress \& Emphasis): naturalness of patterns and rhythms.
    \item \textbf{Emotional Expressiveness} (Progressive gating): 
    \begin{enumerate}
        \item Emotion Accuracy (required),
        \item Emotion Intensity Control (rate only if 2.1 $\ge$ 3),
        \item Dynamic Range (rate only if 2.2 $\ge$ 3).
    \end{enumerate}
    \item \textbf{Character Consistency} (Voice Identity Matching, Trait Embodiment): alignment with Character Profile.
    \item \textbf{Contextual Relevance} (Local Scene Fit, Global Story Coherence): fit to immediate scene and long-term identity.
    \item \textbf{Semantic Match} (spoken content vs. Local Scene goal/emotion/action).
\end{itemize}

The complete rubrics with per-score definitions and examples are reproduced below.

\subsection{Commentary Guidelines and Confidence Ratings}
Annotators should provide short, clear justifications:
\begin{itemize}
    \item Use the comment box to note contradictions, missing information, vagueness, or multi-speaker issues.
    \item Keep comments brief and non-technical; do not quote transcripts unless necessary.
    \item Do not reward theatricality/technical polish unless it naturally fits the character \emph{and} scene.
\end{itemize}

Annotators also provide a \textbf{confidence} level (1--5) reflecting familiarity with the role/scene and clarity of audible cues. Low confidence \emph{does not} reflect poorly on annotators; it adds nuance for downstream analysis.

\subsection{Full Annotation Guideline (Realism Evaluation)}
\label{app:realism-annotation-full}

\begin{tcolorbox}[colback=gray!5!white, colframe=black!75!white, title={Realism Evaluation}, fonttitle=\bfseries, fontupper=\ttfamily, breakable]

\noindent \textbf{Goal.}
Rate how well the given speech (audio and transcript) aligns with the character’s personality, traits, speaking style, and intent, based on:
\begin{itemize}
  \item \textbf{Character Profile}: the character’s long-term personality and speaking habits.
  \item \textbf{Local Scene}: the current scene’s context and the character’s immediate goal or emotion.
\end{itemize}
Each subcategory is scored on a 1–5 scale. Always read the Character Profile and Local Scene before listening; use both the audio and transcript (reference only) to guide your score.
\end{tcolorbox}

\noindent \textbf{Initial Assessment \& Pre-Checks.}
Before rating any speech sample, follow this checklist to ensure consistency and fairness.

\subsubsection*{Caveats}
\begin{itemize}
  \item \textbf{Read first:} Carefully read the \emph{Character Profile} and \emph{Local Scene} to understand the long-term identity and immediate goal/emotion.
  \item \textbf{Use both sources:} Play the audio and skim the transcript together to assess delivery quality, expression, and alignment with character intent (do not penalize transcript mismatches).
  \item \textbf{Focus only on human speech:} Do not factor in background music/soundtracks, environmental or ambient noise, or non-speech audio events; these may reflect the scene but should not influence scoring of the spoken performance.
  \item \textbf{Target speaker only:} If multiple speakers are present, evaluate only the target character’s lines (consider others only as context for turn-taking).
  \item \textbf{Identity mismatch:} If the clip’s speaker is clearly not the intended target speaker, assign a score of \textbf{1} for every rating dimension that relies on the Character Profile (e.g., Character Style Traits, Trait Embodiment, Character Consistency).
\end{itemize}

\subsubsection*{Prompt Quality (Pre-Check)}
Rate the prompt before annotation:
\begin{enumerate}
  \item \textbf{Prompt contains contradictory information.} Conflicts make accurate annotation difficult or impossible. Examples: incompatible accent specifications; Local Scene drastically at odds with intended emotion (see \emph{Semantic Match}); Local and global scenes inconsistent; traits described as “evolving over time” without a timeline.
  \item \textbf{Prompt is insufficient to evaluate the voice sample.} Audio too short to gauge the dimensions; or the prompt lacks specific characteristics/is too vague or multi-directional to know which version of the character is present. (It is OK if audio does not match transcript exactly; if too much is missing to annotate, mark as insufficient.)
  \item \textbf{Prompt is sufficient to evaluate the voice sample.} No material issues; proceed.
\end{enumerate}

\subsubsection*{Overall Speaker Focus and Character Style Traits}
\begin{itemize}
  \item \textbf{Checkbox 1: Multi-Speaker Influence.} Check \emph{Yes} if other speakers were present and made it harder to isolate/judge the target character (check \emph{Yes} if uncertain). Otherwise check \emph{No}.
  \item \textbf{Question 2: Observed Character Style Traits.} Based on the Character Profile, are the defined Style Traits clearly present in the delivery?
  \begin{itemize}
    \item Keep trait tags if a trait is expressed (positively or clearly in the opposite direction); this supports high/low embodiment scoring.
    \item Delete incorrect pre-populated traits; you do not need to replace them. If you delete many traits, \emph{Trait Embodiment} should likely be low.
    \item If no traits are expressed, enter “N/A”. If unsure, leave original traits unedited.
  \end{itemize}
  \item \emph{Example (customizable):} \texttt{"Character\_style":"Playful, Charismatic, Confident, Shrewd"}.
\end{itemize}

\noindent \textbf{Rubric Rating Instructions.}
Score each subcriterion on a 1–5 scale. Use the detailed rubrics below and the examples to maintain calibration across characters, tones, and emotional styles. Prioritize in-character believability; do not reward theatricality, emotional intensity, or technical polish unless they naturally fit the character \emph{and} scene.

\subsubsection*{Guideline 1. Prosodic Dynamics (Patterns and Rhythms)}
Focus: pitch, rhythm, and emphasis in speech.
\noindent \textbf{Guideline 1.1 Pitch Variation (check Local Scene, Character Profile, and audio).}
\begin{center}
\begin{tabular}{@{}p{1.2cm}p{11.8cm}@{}}
\toprule
\textbf{Score} & \textbf{Description / Example} \\
\midrule
5 & Pitch moves naturally/believably, subtly capturing emotional state \& speech pattern (e.g., gentle rise when a calm-but-curious character asks a question). \\
4 & Slight variation that fits character; somewhat restrained/inconsistent. \\
3 & Neutral pitch/non-distracting but lacks expressive intent. \\
2 & Forced/theatrical or mismatched pitch shifts. \\
1 & Monotone or misaligned with scene/character. \\
\bottomrule
\end{tabular}
\end{center}

\noindent \textbf{Guideline 1.2 Rhythmic Naturalness (check Local Scene, Character Profile, and audio).}
\begin{center}
\begin{tabular}{@{}p{1.2cm}p{11.8cm}@{}}
\toprule
\textbf{Score} & \textbf{Description / Example} \\
\midrule
5 & Smooth, believable rhythm reflecting speech habit \& scene pacing (e.g., nervous character in fast but coherent bursts). \\
4 & Minor deviations but still natural. \\
3 & Even pace without clear rhythmic personality. \\
2 & Disjointed, overly rigid, or awkward pacing (odd pause placements). \\
1 & Completely unnatural/robotic timing that interrupts flow. \\
\bottomrule
\end{tabular}
\end{center}

\noindent \textbf{Guideline 1.3 Stress and Emphasis (check Local Scene, Character Profile, and audio).}
\begin{center}
\begin{tabular}{@{}p{1.2cm}p{11.8cm}@{}}
\toprule
\textbf{Score} & \textbf{Description / Example} \\
\midrule
5 & Emphasis highlights key meanings/emotions in realistic ways (e.g., stress on ``\emph{can't}'' in ``We can't leave them behind!''). \\
4 & Generally helpful emphasis with minor misalignment. \\
3 & Mild/generic stress patterns (not distracting, limited impact). \\
2 & Over-emphasis or misplaced focus (e.g., stressing function words). \\
1 & Flat or confusing emphasis (all equal or no variation). \\
\bottomrule
\end{tabular}
\end{center}

\subsubsection*{Guideline 2. Emotional Expressiveness (Progressive Gating)}
Focus: alignment of expressed emotion with character style and situational demand. \\
\textbf{Gating:} Only rate 2.2 if 2.1 $\geq$ 3; only rate 2.3 if 2.2 $\geq$ 3.

\noindent \textbf{Guideline 2.1 Emotion Accuracy (Required; check Local Scene, Character Profile, and audio).}
\begin{center}
\begin{tabular}{@{}p{1.2cm}p{11.8cm}@{}}
\toprule
\textbf{Score} & \textbf{Description / Example} \\
\midrule
5 & Emotion matches personality and situation with subtlety (e.g., hesitant pause before apologizing for a reserved character). \\
4 & Emotionally grounded with light misalignment. \\
3 & Too neutral/safe for the scene. \\
2 & Out of sync with character or situation (e.g., cheerful during betrayal). \\
1 & Jarring/inappropriate (e.g., laughing/yelling without motivation). \\
\bottomrule
\end{tabular}
\end{center}
\textit{If you select 1 or 2, stop here: set 2.2 and 2.3 = N/A.}

\noindent \textbf{Guideline 2.2 Emotion Intensity Control (Only if 2.1 $\geq$ 3; check Local Scene, Character Profile, and audio).}
\begin{center}
\begin{tabular}{@{}p{1.2cm}p{11.8cm}@{}}
\toprule
\textbf{Score} & \textbf{Description / Example} \\
\midrule
5 & Intensity calibrated perfectly (e.g., whispered anger in a tense standoff). \\
4 & Mostly appropriate (slight over-/under-shoot). \\
3 & Slightly exaggerated or dull. \\
2 & Overacted or underplayed (e.g., screaming in a reflective moment). \\
1 & Wildly off-scale (e.g., hysterical laughter in a death scene). \\
\bottomrule
\end{tabular}
\end{center}
\textit{If you select 1 or 2, stop here: set 2.3 = N/A.}

\noindent \textbf{Guideline 2.3 Dynamic Range (Only if 2.2 $\geq$ 3; check Local Scene, Character Profile, and audio).}
\begin{center}
\begin{tabular}{@{}p{1.2cm}p{11.8cm}@{}}
\toprule
\textbf{Score} & \textbf{Description / Example} \\
\midrule
5 & Emotional shifts, if present, are natural and in-character (e.g., mentor calmly grows firmer while warning a student). \\
4 & Appropriate shifts present but mild/slow to emerge. \\
3 & Mostly steady; slight modulation; no strong arc. \\
2 & Flat/monotone or unnaturally varied with unjustified jumps. \\
1 & Significant out-of-character fluctuations without triggers. \\
\bottomrule
\end{tabular}
\end{center}

\subsubsection*{Guideline 3. Character Consistency}
Focus: how well the voice and language align with the character's core identity.

\noindent \textbf{Guideline 3.1 Voice Identity Matching (Character Profile, and audio).}
\begin{center}
\begin{tabular}{@{}p{1.2cm}p{11.8cm}@{}}
\toprule
\textbf{Score} & \textbf{Description / Example} \\
\midrule
5 & Voice clearly matches the expected identity (e.g., youthful, curious tone for energetic sidekick). \\
4 & Acceptable match with minor mismatches. \\
3 & Neutral/not character-specific. \\
2 & Somewhat inconsistent (e.g., deep voice for high-pitched character). \\
1 & Clearly mismatched identity (e.g., gritty adult voice for bubbly child). \\
\bottomrule
\end{tabular}
\end{center}

\noindent \textbf{Guideline 3.2 Trait Embodiment (Character Profile, and audio).}
\begin{center}
\begin{tabular}{@{}p{1.2cm}p{11.8cm}@{}}
\toprule
\textbf{Score} & \textbf{Description / Example} \\
\midrule
5 & Defining traits subtly but clearly conveyed (e.g., insecure character hesitates/lowers voice). \\
4 & Most traits conveyed; minor inconsistencies. \\
3 & Traits weak but not disruptive. \\
2 & Missing or conflicting traits (e.g., arrogant tone for a humble character). \\
1 & No identifiable traits or completely off. \\
\bottomrule
\end{tabular}
\end{center}

\subsubsection*{Guideline 4. Contextual Relevance}
Focus: appropriateness in the current scene and long-term coherence.

\noindent \textbf{Guideline 4.1 Local Scene Fit (check Local Scene, and audio).}
\begin{center}
\begin{tabular}{@{}p{1.2cm}p{11.8cm}@{}}
\toprule
\textbf{Score} & \textbf{Description / Example} \\
\midrule
5 & Matches the scene’s emotional/narrative context (e.g., urgent whisper during stealth). \\
4 & Mostly coherent with minor mismatch (e.g., calm tone while others panic). \\
3 & Neutral: neither reinforces nor harms the scene. \\
2 & Confusing or slightly inappropriate tone (e.g., laughing during a tense standoff). \\
1 & Directly contradicts scene intent (e.g., casual joke while a character is dying). \\
\bottomrule
\end{tabular}
\end{center}

\noindent \textbf{Guideline 4.2 Global Story (Character Profile) Coherence (check Character Profile, and audio).}
\begin{center}
\begin{tabular}{@{}p{1.2cm}p{11.8cm}@{}}
\toprule
\textbf{Score} & \textbf{Description / Example} \\
\midrule
5 & Perfectly consistent with background and arc (e.g., wise mentor advice aligned with past scenes). \\
4 & Slightly off-style but plausible (e.g., cheerful tone for usually sarcastic character). \\
3 & Generic/filler delivery (\emph{could be anyone}). \\
2 & Contradicts growth or habits (e.g., arrogance after humility arc). \\
1 & Breaks established identity (e.g., cold anger from consistently empathetic healer). \\
\bottomrule
\end{tabular}
\end{center}

\subsubsection*{Guideline 5. Semantic Match}
Focus: how well the spoken content semantically aligns with the Local Scene’s stated goal, emotion, or action.

\noindent \textbf{5.1 Semantic Matchness (Audio, Local Scene).}
\begin{center}
\begin{tabular}{@{}p{1.2cm}p{11.8cm}@{}}
\toprule
\textbf{Score} & \textbf{Description / Example} \\
\midrule
5 & Content perfectly advances the scene goal/emotion (e.g., rescue: ``Secure the exit; I'll get the hostages.''). \\
4 & Minor nuance mismatch but still scene-appropriate. \\
3 & Neutral filler that neither helps nor contradicts (plausible bridge). \\
2 & Partially conflicts or confuses context; requires \emph{stretch} explanation. \\
1 & Directly contradicts/derails scene intent; no reasonable bridge. \\
\bottomrule
\end{tabular}
\end{center}

\subsubsection*{Confidence}
Provide a confidence rating (1–5) reflecting your certainty that you:
\begin{itemize}
  \item Understood the character and scene clearly,
  \item Heard sufficient emotional cues in the audio, and
  \item Interpreted speech alignment with intended personality and context.
\end{itemize}
\textbf{Guidance:} Rate low confidence for noisy/ambiguous audio, vague/contradictory traits, multiple dominant speakers, or hard-to-distinguish emotion/tone.

\subsubsection*{Comments (Optional)}
Use the comment box to note contradictions, missing information, vague descriptions, or multi-speaker issues that affected your evaluation. Keep comments brief, clear, and non-technical; no transcript quotes are required.

\FloatBarrier
\section{Annotation Ethics}
\label{app:annotation-ethnic}

All human annotation in this work was conducted in accordance with standard ethical practices for research involving human annotators. Annotators were recruited via public hiring for Mandarin and internal lab hiring for English. Prior to participation, all annotators were provided with a clear description of the task, the type of data they would encounter, and their rights as participants, including the ability to opt out at any time without penalty.

Annotators gave informed consent before starting the tasks. They were compensated at a rate from \$20 to \$30 per hour, which corresponds to at least the local minimum wage and was applied uniformly regardless of whether their contributions were later filtered for quality.

The data presented to annotators consisted of synthetic speech samples, pseudonymized human speech, or recordings from public corpora, and carried no direct personal identifiers or sensitive metadata (though speech itself remains biometric). To mitigate potential risks, annotators were allowed to skip any item that they found uncomfortable or inappropriate.

To ensure annotation reliability, annotators were provided with written guidelines, examples, and trial tasks before beginning. Quality control procedures included inter-annotator agreement checks, and comprehensive screening tests. All evaluation samples have at least three annotators, and most samples have four to six. The base realism training release contains $31$ records with two rater annotations; all other records have at least three. This study did not involve vulnerable populations and posed minimal risk to participants.

\FloatBarrier
\section{Detailed Data Analysis (Archetype Evaluation)}
\label{app:archetype-data-analysis}

The archetype evaluation provides a detailed view of how speech foundation models (\texttt{SFnd}) perform when tasked with role-play scenarios grounded in stereotypical characters and situations. Human annotations cover four dimensions, \textit{Appropriateness}, \textit{Human Likeness}, \textit{Audio Quality}, and \textit{Content Pass}, and the results reveal important trends, as well as challenges in interpreting scores across languages.

\paragraph{Annotation Consistency and Cross-Language Comparisons.}
A critical observation is that annotations in Mandarin and English cannot be directly compared. The annotation groups for the two languages differed in their linguistic backgrounds and cultural interpretations, which leads to systematic differences in scoring. Our bilingual pilot further confirmed that annotators often interpret role-play appropriateness and expressiveness differently depending on their cultural frame of reference. For example, Mandarin raters tend to reward delivery styles that align with culturally expected prosody and politeness strategies, while English raters prioritize spontaneity and naturalness. These discrepancies parallel the findings of the singing voice conversion challenge (SVCC2023) from \citet{huang2023singing}, where native and non-native listeners provided significantly different ratings for the same samples, particularly for naturalness and speaker similarity. Consequently, while absolute scores differ substantially across Mandarin and English, only within-language comparisons are meaningful. Cross-lingual ranking is not valid given the different annotation bases.

\paragraph{Mandarin Results.}
Within Mandarin from Table~\ref{tab:human-archetype-evaluation}, the relative rankings of models are clear. \texttt{Doubao} stands out as the strongest performer across all four dimensions, reaching near-ceiling levels in both \textit{Audio Quality} (4.37) and \textit{Content Pass} (0.98). It also leads in role-related dimensions such as \textit{Appropriateness} and \textit{Human Likeness}, indicating balanced competence across linguistic accuracy, delivery, and expressiveness. \texttt{G-CosyVoice2} and \texttt{Q-CosyVoice2} follow closely, especially in audio quality, though they lag behind in role appropriateness. \texttt{KimiAudio} and \texttt{GLM4-Voice} achieve mid-range results, reflecting higher variance across samples, while \texttt{GPT-4o-realtime} shows competitive content understanding but lower naturalness and appropriateness. \texttt{Qwen2.5Omni} consistently ranks the lowest, with large variance, underscoring instability and weaker alignment with archetypal role expectations.

\begin{table}[!t]
\centering
\caption{Overall human annotation results on Mandarin and English archetype evaluation dimensions (mean $\pm$ std). Best per column in \textbf{bold}.}
\label{tab:human-archetype-evaluation}
\resizebox{\linewidth}{!}{
\begin{tabular}{l|cccc}
\toprule
\textbf{Model (Mandarin)} &
\textbf{Appropriateness} &
\textbf{Human Likeness} &
\textbf{Audio Quality} &
\textbf{Content Pass} \\
\midrule
\texttt{Doubao} &
\textbf{3.8236 $\pm$ 0.6658} &
\textbf{3.6572 $\pm$ 0.6427} &
\textbf{4.3743 $\pm$ 0.6991} &
\textbf{0.9848 $\pm$ 0.1224} \\
\texttt{G-CosyVoice2} &
3.4243 $\pm$ 0.7389 &
3.3156 $\pm$ 0.7215 &
4.0406 $\pm$ 0.7298 &
0.9725 $\pm$ 0.1637 \\
\texttt{Q-CosyVoice2} &
3.4228 $\pm$ 0.7788 &
3.2750 $\pm$ 0.7335 &
4.0236 $\pm$ 0.7661 &
0.9674 $\pm$ 0.1776 \\
\texttt{KimiAudio} &
3.3866 $\pm$ 1.1086 &
3.3391 $\pm$ 1.0591 &
3.9525 $\pm$ 1.2801 &
0.8623 $\pm$ 0.3446 \\
\texttt{GLM4-Voice} &
3.2022 $\pm$ 0.9597 &
3.0275 $\pm$ 0.8687 &
3.9246 $\pm$ 1.1105 &
0.8949 $\pm$ 0.3067 \\
\texttt{GPT-4o-realtime} &
3.0228 $\pm$ 0.8581 &
2.6536 $\pm$ 0.7151 &
3.7297 $\pm$ 1.0418 &
0.9098 $\pm$ 0.2865 \\
\texttt{Qwen2.5Omni} &
2.4942 $\pm$ 1.4242 &
2.3830 $\pm$ 1.3201 &
2.8685 $\pm$ 1.7456 &
0.5482 $\pm$ 0.4978 \\
\midrule
\textbf{Model (English)} & \\
\midrule
\texttt{ChatGPT-4o} &
\textbf{3.4638 $\pm$ 0.9671} &
\textbf{3.4873 $\pm$ 0.7857} &
1.9620 $\pm$ 0.6885 &
\textbf{0.9662 $\pm$ 0.1808} \\
\texttt{Doubao} &
2.3798 $\pm$ 1.0573 &
2.2355 $\pm$ 0.8918 &
2.9553 $\pm$ 1.1589 &
0.7995 $\pm$ 0.4005 \\
\texttt{GPT-4o-realtime} &
2.2736 $\pm$ 0.9606 &
2.0906 $\pm$ 0.8124 &
2.5127 $\pm$ 0.9702 &
0.8521 $\pm$ 0.3552 \\
\texttt{GLM4-Voice} &
2.2355 $\pm$ 1.0354 &
2.0779 $\pm$ 0.9095 &
2.5694 $\pm$ 1.1366 &
0.7482 $\pm$ 0.4342 \\
\texttt{G-CosyVoice2} &
2.2168 $\pm$ 0.9916 &
2.0109 $\pm$ 0.8585 &
\textbf{3.2530 $\pm$ 0.9637} &
0.9293 $\pm$ 0.2563 \\
\texttt{Q-CosyVoice2} &
2.1673 $\pm$ 0.9642 &
1.8068 $\pm$ 0.8011 &
3.2476 $\pm$ 0.9113 &
0.9614 $\pm$ 0.1928 \\
\texttt{Qwen2.5Omni} &
2.0079 $\pm$ 1.0420 &
1.9408 $\pm$ 0.9489 &
2.4161 $\pm$ 1.2583 &
0.6159 $\pm$ 0.4865 \\
\texttt{KimiAudio} &
1.5531 $\pm$ 0.8716 &
1.7579 $\pm$ 0.9386 &
2.0097 $\pm$ 1.1788 &
0.4861 $\pm$ 0.5000 \\
\bottomrule
\end{tabular}}
\end{table}

\begin{table}[!t]
\centering
\caption{Domain-level human annotation results for Appropriateness dimension of Mandarin and English archetype evaluation (mean $\pm$ std). 
Languages are reported separately. Inter-model variance is the population variance of the seven Mandarin or eight English system means plotted in Figs.~\ref{fig:mandarin-archetype} and~\ref{fig:english-archetype}, with equal weight per system; it is distinct from the score standard deviation in the mean $\pm$ std columns.}
\label{tab:domain-annotation}
\resizebox{\linewidth}{!}{
\begin{tabular}{l|cc|cc}
\toprule
\multirow{2}{*}{\textbf{Domain}} & \multicolumn{2}{c|}{\textbf{Mandarin}} & \multicolumn{2}{c}{\textbf{English}} \\
 & Score (mean $\pm$ std) & Inter-model Var. & Score (mean $\pm$ std) & Inter-model Var. \\
\midrule
NamedCharacters   & 3.0343 $\pm$ 1.1087 & 0.2626 & 1.9896 $\pm$ 1.0409 & 0.2421 \\
FantasyOccupation & 3.1629 $\pm$ 0.9661 & 0.3439 & 1.9444 $\pm$ 0.9809 & 0.2089 \\
Traits            & 3.1670 $\pm$ 1.2109 & 0.3269 & 2.1518 $\pm$ 1.0572 & 0.2966 \\
Events            & 3.2535 $\pm$ 1.1523 & 0.4358 & 2.2608 $\pm$ 1.1280 & 0.3980 \\
DailyOccupation   & 3.2760 $\pm$ 0.8652 & 0.0555 & 2.5129 $\pm$ 1.1781 & 0.2704 \\
RelativeRoles     & 3.2762 $\pm$ 1.0255 & 0.1106 & 2.3594 $\pm$ 1.1298 & 0.3017 \\
SocialIdentity    & 3.3263 $\pm$ 0.9802 & 0.1169 & 2.3256 $\pm$ 1.0912 & 0.2596 \\
\bottomrule
\end{tabular}}
\end{table}

\paragraph{English Results.}
The picture shifts in English, elaborated in Table~\ref{tab:human-archetype-evaluation}. Here, \texttt{ChatGPT-4o} emerges as the most effective model by a wide margin, achieving top scores in both \textit{Appropriateness} (3.46) and \textit{Human Likeness} (3.49), as well as high content pass rates (0.97). This reflects its strong grounding in English discourse norms and role adherence. In contrast, \texttt{CosyVoice2} variants perform well in \textit{Audio Quality} but are weaker in role alignment, highlighting a trade-off between acoustic fidelity and role-play embodiment. Other models such as \texttt{Doubao}, \texttt{GLM4-Voice}, and \texttt{GPT-4o-realtime} achieve moderate results, while \texttt{KimiAudio} and \texttt{Qwen2.5Omni} trail far behind, showing limited robustness in English scenarios. Taken together, the English results reveal both a sharper separation across models and a wider variance in appropriateness scores compared to Mandarin.

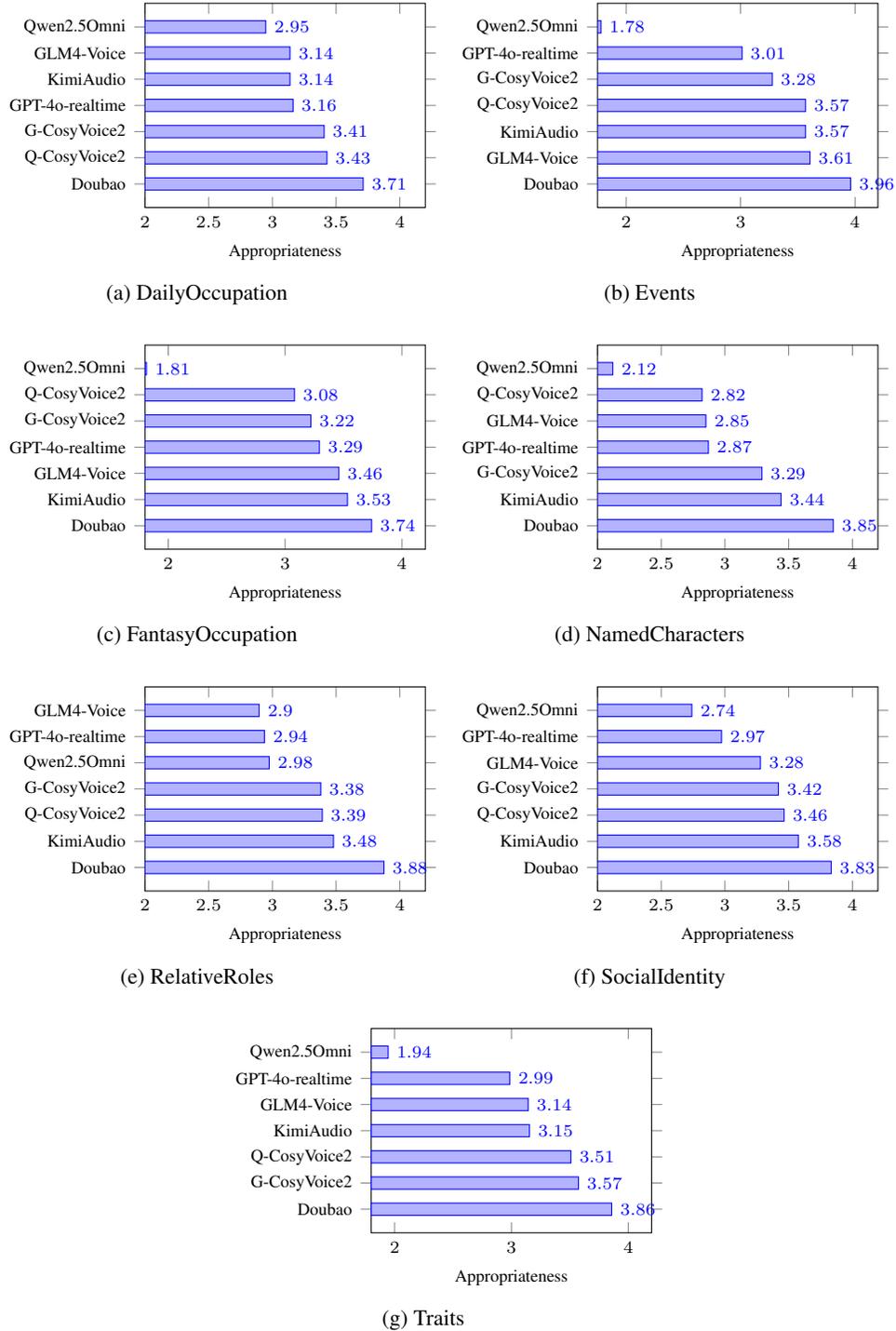
\begin{figure}[tp]
\centering

\begin{subfigure}{0.4\linewidth}
\begin{tikzpicture}
\begin{axis}[
    xbar,
    width=\linewidth, height=4.5cm,
    xmin=2, xmax=4.2,
    xlabel={Appropriateness},
    symbolic y coords={
        Qwen2.5Omni,
        GLM4-Voice,
        KimiAudio,
        GPT-4o-realtime,
        G-CosyVoice2,
        Q-CosyVoice2,
        Doubao},
    ytick=data,
    nodes near coords,
    nodes near coords align={horizontal},
    bar width=5pt,
    enlarge y limits=0.15,
    y dir=reverse,
    label style={font=\scriptsize},
    tick label style={font=\scriptsize},
    nodes near coords style={font=\scriptsize}
]
\addplot coordinates {
    (2.9480,Qwen2.5Omni)
    (3.1380,GLM4-Voice)
    (3.1380,KimiAudio)
    (3.1620,GPT-4o-realtime)
    (3.4060,G-CosyVoice2)
    (3.4280,Q-CosyVoice2)
    (3.7120,Doubao)
};
\end{axis}
\end{tikzpicture}
\caption{DailyOccupation}
\end{subfigure}
\hspace{20pt}
\begin{subfigure}{0.4\linewidth}
\begin{tikzpicture}
\begin{axis}[
    xbar,
    width=\linewidth, height=4.5cm,
    xmin=1.75, xmax=4.2,
    xlabel={Appropriateness},
    symbolic y coords={
        Qwen2.5Omni,
        GPT-4o-realtime,
        G-CosyVoice2,
        Q-CosyVoice2,
        KimiAudio,
        GLM4-Voice,
        Doubao},
    ytick=data,
    nodes near coords,
    bar width=5pt,
    enlarge y limits=0.15,
    y dir=reverse,
    label style={font=\scriptsize},
    tick label style={font=\scriptsize},
    nodes near coords style={font=\scriptsize}
]
\addplot coordinates {
    (1.7806,Qwen2.5Omni)
    (3.0129,GPT-4o-realtime)
    (3.2774,G-CosyVoice2)
    (3.5677,Q-CosyVoice2)
    (3.5677,KimiAudio)
    (3.6065,GLM4-Voice)
    (3.9613,Doubao)
};
\end{axis}
\end{tikzpicture}
\caption{Events}
\end{subfigure}

\vspace{0.5cm}

\begin{subfigure}{0.4\linewidth}
\begin{tikzpicture}
\begin{axis}[
    xbar,
    width=\linewidth, height=4.5cm,
    xmin=1.8, xmax=4.2,
    xlabel={Appropriateness},
    symbolic y coords={
        Qwen2.5Omni,
        Q-CosyVoice2,
        G-CosyVoice2,
        GPT-4o-realtime,
        GLM4-Voice,
        KimiAudio,
        Doubao},
    ytick=data,
    nodes near coords,
    bar width=5pt,
    enlarge y limits=0.15,
    y dir=reverse,
    label style={font=\scriptsize},
    tick label style={font=\scriptsize},
    nodes near coords style={font=\scriptsize}
]
\addplot coordinates {
    (1.8133,Qwen2.5Omni)
    (3.0800,Q-CosyVoice2)
    (3.2200,G-CosyVoice2)
    (3.2933,GPT-4o-realtime)
    (3.4600,GLM4-Voice)
    (3.5333,KimiAudio)
    (3.7400,Doubao)
};
\end{axis}
\end{tikzpicture}
\caption{FantasyOccupation}
\end{subfigure}
\hspace{20pt}
\begin{subfigure}{0.4\linewidth}
\begin{tikzpicture}
\begin{axis}[
    xbar,
    width=\linewidth, height=4.5cm,
    xmin=2, xmax=4.2,
    xlabel={Appropriateness},
    symbolic y coords={
        Qwen2.5Omni,
        Q-CosyVoice2,
        GLM4-Voice,
        GPT-4o-realtime,
        G-CosyVoice2,
        KimiAudio,
        Doubao},
    ytick=data,
    nodes near coords,
    bar width=5pt,
    enlarge y limits=0.15,
    y dir=reverse,
    label style={font=\scriptsize},
    tick label style={font=\scriptsize},
    nodes near coords style={font=\scriptsize}
]
\addplot coordinates {
    (2.1200,Qwen2.5Omni)
    (2.8200,Q-CosyVoice2)
    (2.8500,GLM4-Voice)
    (2.8700,GPT-4o-realtime)
    (3.2900,G-CosyVoice2)
    (3.4400,KimiAudio)
    (3.8500,Doubao)
};
\end{axis}
\end{tikzpicture}
\caption{NamedCharacters}
\end{subfigure}

\vspace{0.5cm}

\begin{subfigure}{0.4\linewidth}
\begin{tikzpicture}
\begin{axis}[
    xbar,
    width=\linewidth, height=4.5cm,
    xmin=2, xmax=4.2,
    xlabel={Appropriateness},
    symbolic y coords={
        GLM4-Voice,
        GPT-4o-realtime,
        Qwen2.5Omni,
        G-CosyVoice2,
        Q-CosyVoice2,
        KimiAudio,
        Doubao},
    ytick=data,
    nodes near coords,
    bar width=5pt,
    enlarge y limits=0.15,
    y dir=reverse,
    label style={font=\scriptsize},
    tick label style={font=\scriptsize},
    nodes near coords style={font=\scriptsize}
]
\addplot coordinates {
    (2.8958,GLM4-Voice)
    (2.9375,GPT-4o-realtime)
    (2.9750,Qwen2.5Omni)
    (3.3792,G-CosyVoice2)
    (3.3917,Q-CosyVoice2)
    (3.4792,KimiAudio)
    (3.8750,Doubao)
};
\end{axis}
\end{tikzpicture}
\caption{RelativeRoles}
\end{subfigure}
\hspace{20pt}
\begin{subfigure}{0.4\linewidth}
\begin{tikzpicture}
\begin{axis}[
    xbar,
    width=\linewidth, height=4.5cm,
    xmin=2, xmax=4.2,
    xlabel={Appropriateness},
    symbolic y coords={
        Qwen2.5Omni,
        GPT-4o-realtime,
        GLM4-Voice,
        G-CosyVoice2,
        Q-CosyVoice2,
        KimiAudio,
        Doubao},
    ytick=data,
    nodes near coords,
    bar width=5pt,
    enlarge y limits=0.15,
    y dir=reverse,
    label style={font=\scriptsize},
    tick label style={font=\scriptsize},
    nodes near coords style={font=\scriptsize}
]
\addplot coordinates {
    (2.7400,Qwen2.5Omni)
    (2.9730,GPT-4o-realtime)
    (3.2780,GLM4-Voice)
    (3.4190,G-CosyVoice2)
    (3.4640,Q-CosyVoice2)
    (3.5760,KimiAudio)
    (3.8340,Doubao)
};
\end{axis}
\end{tikzpicture}
\caption{SocialIdentity}
\end{subfigure}

\vspace{0.5cm}

\begin{subfigure}{0.4\linewidth}
\begin{tikzpicture}
\begin{axis}[
    xbar,
    width=\linewidth, height=4.5cm,
    xmin=1.8, xmax=4.2,
    xlabel={Appropriateness},
    symbolic y coords={
        Qwen2.5Omni,
        GPT-4o-realtime,
        GLM4-Voice,
        KimiAudio,
        Q-CosyVoice2,
        G-CosyVoice2,
        Doubao},
    ytick=data,
    nodes near coords,
    bar width=5pt,
    enlarge y limits=0.15,
    y dir=reverse,
    label style={font=\scriptsize},
    tick label style={font=\scriptsize},
    nodes near coords style={font=\scriptsize}
]
\addplot coordinates {
    (1.9447,Qwen2.5Omni)
    (2.9854,GPT-4o-realtime)
    (3.1431,GLM4-Voice)
    (3.1545,KimiAudio)
    (3.5089,Q-CosyVoice2)
    (3.5740,G-CosyVoice2)
    (3.8585,Doubao)
};
\end{axis}
\end{tikzpicture}
\caption{Traits}
\end{subfigure}

\caption{Model rankings within domains for Appropriateness dimension of \textbf{Mandarin} archetype evaluation (higher is better). Each bar shows the mean score for that domain.}
\label{fig:mandarin-archetype}
\end{figure}

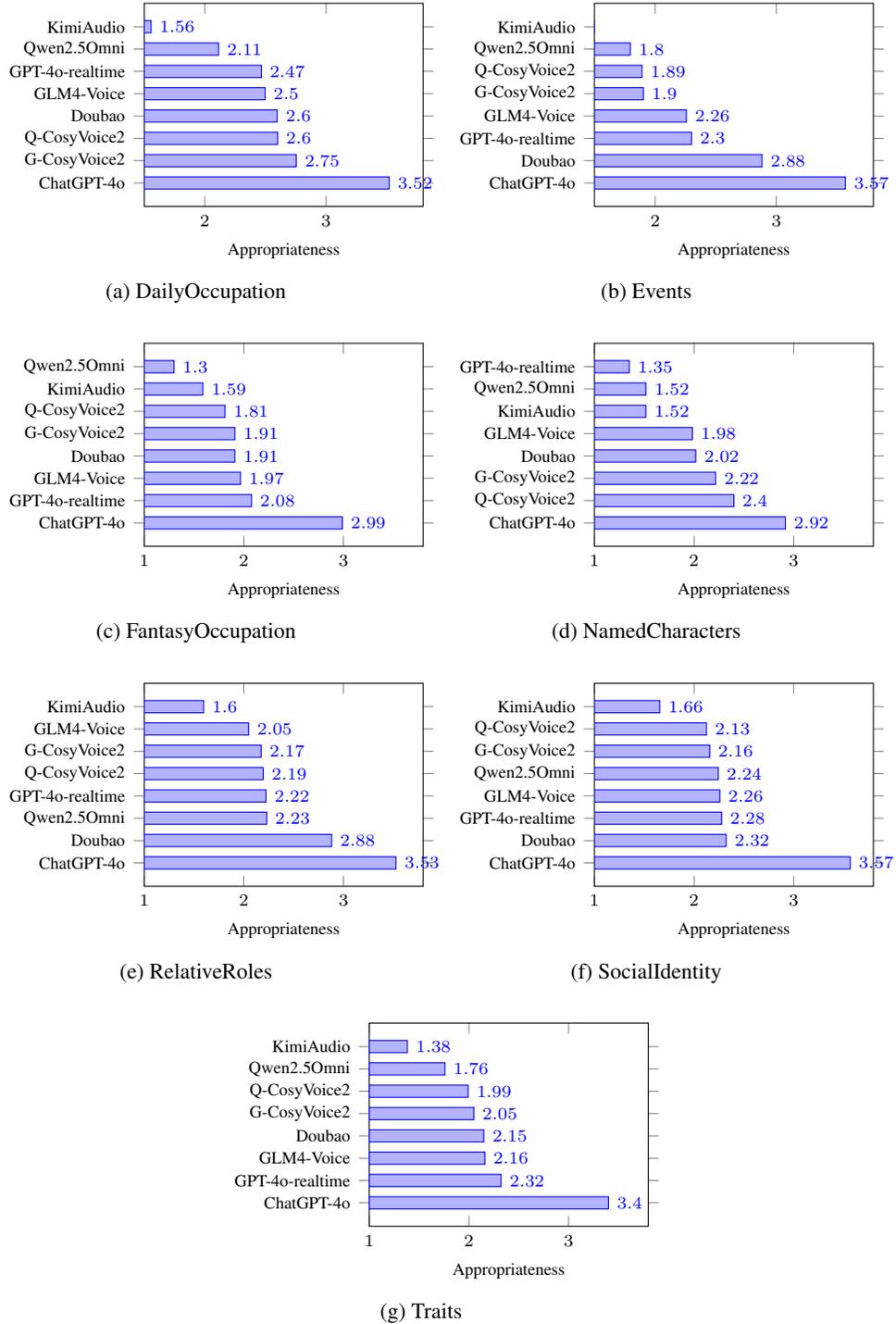
\begin{figure}[tp]
\centering


\begin{subfigure}{0.4\linewidth}
\begin{tikzpicture}
\begin{axis}[
    xbar,
    width=\linewidth, height=4.5cm,
    xmin=1.5, xmax=3.8,
    xlabel={Appropriateness},
    symbolic y coords={
        KimiAudio,
        Qwen2.5Omni,
        GPT-4o-realtime,
        GLM4-Voice,
        Doubao,
        Q-CosyVoice2,
        G-CosyVoice2,
        ChatGPT-4o},
    ytick=data,
    nodes near coords,
    bar width=5pt,
    enlarge y limits=0.15,
    y dir=reverse,
    label style={font=\scriptsize},
    tick label style={font=\scriptsize},
    nodes near coords style={font=\scriptsize}
]
\addplot coordinates {
    (1.5567,KimiAudio)
    (2.1133,Qwen2.5Omni)
    (2.4667,GPT-4o-realtime)
    (2.4967,GLM4-Voice)
    (2.5967,Doubao)
    (2.6000,Q-CosyVoice2)
    (2.7533,G-CosyVoice2)
    (3.5200,ChatGPT-4o)
};
\end{axis}
\end{tikzpicture}
\caption{DailyOccupation}
\end{subfigure}
\hspace{20pt}
\begin{subfigure}{0.4\linewidth}
\begin{tikzpicture}
\begin{axis}[
    xbar,
    width=\linewidth, height=4.5cm,
    xmin=1, xmax=3.8,
    xlabel={Appropriateness},
    symbolic y coords={
        KimiAudio,
        Qwen2.5Omni,
        Q-CosyVoice2,
        G-CosyVoice2,
        GLM4-Voice,
        GPT-4o-realtime,
        Doubao,
        ChatGPT-4o},
    ytick=data,
    nodes near coords,
    bar width=5pt,
    enlarge y limits=0.15,
    y dir=reverse,
    label style={font=\scriptsize},
    tick label style={font=\scriptsize},
    nodes near coords style={font=\scriptsize}
]
\addplot coordinates {
    (1.4839,KimiAudio)
    (1.7957,Qwen2.5Omni)
    (1.8925,Q-CosyVoice2)
    (1.9032,G-CosyVoice2)
    (2.2581,GLM4-Voice)
    (2.3011,GPT-4o-realtime)
    (2.8817,Doubao)
    (3.5699,ChatGPT-4o)
};
\end{axis}
\end{tikzpicture}
\caption{Events}
\end{subfigure}

\vspace{0.5cm}

\begin{subfigure}{0.4\linewidth}
\begin{tikzpicture}
\begin{axis}[
    xbar,
    width=\linewidth, height=4.5cm,
    xmin=1, xmax=3.8,
    xlabel={Appropriateness},
    symbolic y coords={
        Qwen2.5Omni,
        KimiAudio,
        Q-CosyVoice2,
        G-CosyVoice2,
        Doubao,
        GLM4-Voice,
        GPT-4o-realtime,
        ChatGPT-4o},
    ytick=data,
    nodes near coords,
    bar width=5pt,
    enlarge y limits=0.15,
    y dir=reverse,
    label style={font=\scriptsize},
    tick label style={font=\scriptsize},
    nodes near coords style={font=\scriptsize}
]
\addplot coordinates {
    (1.3000,Qwen2.5Omni)
    (1.5889,KimiAudio)
    (1.8111,Q-CosyVoice2)
    (1.9111,G-CosyVoice2)
    (1.9111,Doubao)
    (1.9667,GLM4-Voice)
    (2.0778,GPT-4o-realtime)
    (2.9889,ChatGPT-4o)
};
\end{axis}
\end{tikzpicture}
\caption{FantasyOccupation}
\end{subfigure}
\hspace{20pt}
\begin{subfigure}{0.4\linewidth}
\begin{tikzpicture}
\begin{axis}[
    xbar,
    width=\linewidth, height=4.5cm,
    xmin=1, xmax=3.8,
    xlabel={Appropriateness},
    symbolic y coords={
        GPT-4o-realtime,
        Qwen2.5Omni,
        KimiAudio,
        GLM4-Voice,
        Doubao,
        G-CosyVoice2,
        Q-CosyVoice2,
        ChatGPT-4o},
    ytick=data,
    nodes near coords,
    bar width=5pt,
    enlarge y limits=0.15,
    y dir=reverse,
    label style={font=\scriptsize},
    tick label style={font=\scriptsize},
    nodes near coords style={font=\scriptsize}
]
\addplot coordinates {
    (1.3500,GPT-4o-realtime)
    (1.5167,Qwen2.5Omni)
    (1.5167,KimiAudio)
    (1.9833,GLM4-Voice)
    (2.0167,Doubao)
    (2.2167,G-CosyVoice2)
    (2.4000,Q-CosyVoice2)
    (2.9167,ChatGPT-4o)
};
\end{axis}
\end{tikzpicture}
\caption{NamedCharacters}
\end{subfigure}

\vspace{0.5cm}

\begin{subfigure}{0.4\linewidth}
\begin{tikzpicture}
\begin{axis}[
    xbar,
    width=\linewidth, height=4.5cm,
    xmin=1, xmax=3.8,
    xlabel={Appropriateness},
    symbolic y coords={
        KimiAudio,
        GLM4-Voice,
        G-CosyVoice2,
        Q-CosyVoice2,
        GPT-4o-realtime,
        Qwen2.5Omni,
        Doubao,
        ChatGPT-4o},
    ytick=data,
    nodes near coords,
    bar width=5pt,
    enlarge y limits=0.15,
    y dir=reverse,
    label style={font=\scriptsize},
    tick label style={font=\scriptsize},
    nodes near coords style={font=\scriptsize}
]
\addplot coordinates {
    (1.5972,KimiAudio)
    (2.0486,GLM4-Voice)
    (2.1736,G-CosyVoice2)
    (2.1944,Q-CosyVoice2)
    (2.2222,GPT-4o-realtime)
    (2.2292,Qwen2.5Omni)
    (2.8819,Doubao)
    (3.5278,ChatGPT-4o)
};
\end{axis}
\end{tikzpicture}
\caption{RelativeRoles}
\end{subfigure}
\hspace{20pt}
\begin{subfigure}{0.4\linewidth}
\begin{tikzpicture}
\begin{axis}[
    xbar,
    width=\linewidth, height=4.5cm,
    xmin=1, xmax=3.8,
    xlabel={Appropriateness},
    symbolic y coords={
        KimiAudio,
        Q-CosyVoice2,
        G-CosyVoice2,
        Qwen2.5Omni,
        GLM4-Voice,
        GPT-4o-realtime,
        Doubao,
        ChatGPT-4o},
    ytick=data,
    nodes near coords,
    bar width=5pt,
    enlarge y limits=0.15,
    y dir=reverse,
    label style={font=\scriptsize},
    tick label style={font=\scriptsize},
    nodes near coords style={font=\scriptsize}
]
\addplot coordinates {
    (1.6550,KimiAudio)
    (2.1250,Q-CosyVoice2)
    (2.1567,G-CosyVoice2)
    (2.2433,Qwen2.5Omni)
    (2.2583,GLM4-Voice)
    (2.2767,GPT-4o-realtime)
    (2.3217,Doubao)
    (3.5683,ChatGPT-4o)
};
\end{axis}
\end{tikzpicture}
\caption{SocialIdentity}
\end{subfigure}

\vspace{0.5cm}

\begin{subfigure}{0.4\linewidth}
\begin{tikzpicture}
\begin{axis}[
    xbar,
    width=\linewidth, height=4.5cm,
    xmin=1, xmax=3.8,
    xlabel={Appropriateness},
    symbolic y coords={
        KimiAudio,
        Qwen2.5Omni,
        Q-CosyVoice2,
        G-CosyVoice2,
        Doubao,
        GLM4-Voice,
        GPT-4o-realtime,
        ChatGPT-4o},
    ytick=data,
    nodes near coords,
    bar width=5pt,
    enlarge y limits=0.15,
    y dir=reverse,
    label style={font=\scriptsize},
    tick label style={font=\scriptsize},
    nodes near coords style={font=\scriptsize}
]
\addplot coordinates {
    (1.3821,KimiAudio)
    (1.7588,Qwen2.5Omni)
    (1.9919,Q-CosyVoice2)
    (2.0488,G-CosyVoice2)
    (2.1491,Doubao)
    (2.1599,GLM4-Voice)
    (2.3225,GPT-4o-realtime)
    (3.4011,ChatGPT-4o)
};
\end{axis}
\end{tikzpicture}
\caption{Traits}
\end{subfigure}

\caption{Model rankings within domains for Appropriateness dimension of \textbf{English} archetype evaluation (higher is better). Each bar shows the mean score for that domain.}
\label{fig:english-archetype}
\end{figure}

\paragraph{Domain-Level Trends.}
Table~\ref{tab:domain-annotation} and Figures~\ref{fig:mandarin-archetype}–\ref{fig:english-archetype} further highlight differences across domains. In Mandarin, performance is relatively consistent across categories, with means clustering around 3.1–3.3. The highest mean is in \texttt{SocialIdentity}, with \texttt{RelativeRoles} and \texttt{DailyOccupation} close behind, suggesting that models handle culturally grounded archetypes more reliably. In English, however, domain variation is much greater. Categories such as \texttt{FantasyOccupation} and \texttt{NamedCharacters} receive notably lower scores (means below 2.0), reflecting the difficulty of producing expressive, role-appropriate speech in imaginative or culturally specific contexts. Occupational roles remain relatively easier for models in both languages, likely because these scenarios are closer to training data distributions.

\paragraph{Interpretation and Broader Implications.}
Although cross-lingual comparisons are invalid, this does not diminish the utility of archetype evaluation for role-play benchmarking. In fact, the observed inconsistencies reinforce a central point: successful role-play requires models to adapt to the cultural and linguistic norms of the target language community. A system that performs well in Mandarin but poorly in English cannot be considered universally capable, since end users will evaluate appropriateness and naturalness according to their native expectations. Thus, archetype evaluation highlights not only model-specific strengths and weaknesses but also the need for culturally grounded evaluation pipelines.

Another notable pattern is the trade-off between acoustic quality and role embodiment. CosyVoice2 variants achieve strong audio quality but weaker human likeness and appropriateness, whereas \texttt{ChatGPT-4o} prioritizes role coherence at the expense of audio fidelity. This suggests that future systems must better integrate high-quality synthesis with nuanced role adherence. Similarly, the consistently lower scores for imaginative and identity-driven roles underscore the limitations of current SFMs in capturing subtle sociocultural expectations and narrative context.

\paragraph{Summary.}
In sum, archetype evaluation reveals three major findings:
(1) Mandarin and English annotations cannot be directly compared, due to linguistic and cultural annotation inconsistencies, consistent with prior evidence from SVCC 2023.
(2) Within each language, clear model rankings emerge, \texttt{Doubao} dominates in Mandarin, while \texttt{ChatGPT-4o} excels in English.
(3) Domain sensitivity remains a challenge, particularly for imaginative or culturally rich roles, where scores are substantially lower, with substantial variation among systems.




\FloatBarrier
\section{Detailed Data Analysis (Realism Evaluation)}
\label{app:realism-data-analysis}

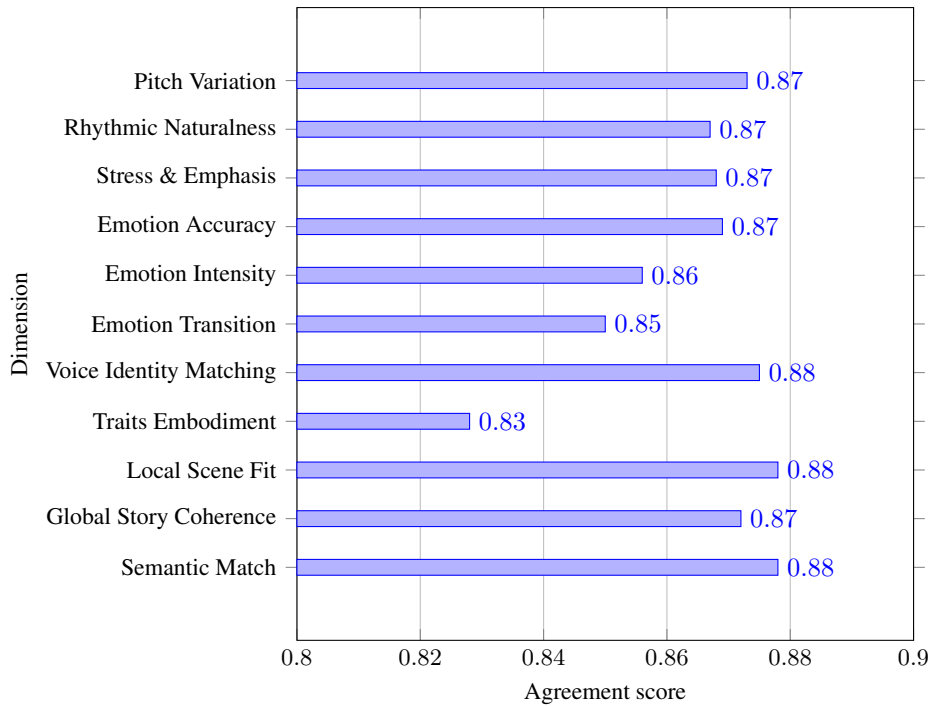
\begin{figure}[tp]
\centering
\begin{tikzpicture}
\begin{axis}[
    xbar,
    width=0.7\linewidth, height=10cm,
    xmin=0.80, xmax=0.9,
    xlabel={Agreement score},
    ylabel={Dimension},
    y dir=reverse,
    ytick={1,2,3,4,5,6,7,8,9,10,11},
    yticklabels={
        Pitch Variation,
        Rhythmic Naturalness,
        Stress \& Emphasis,
        Emotion Accuracy,
        Emotion Intensity,
        Emotion Transition,
        Voice Identity Matching,
        Traits Embodiment,
        Local Scene Fit,
        Global Story Coherence,
        Semantic Match
    },
    bar width=6pt,
    nodes near coords,
    nodes near coords align={horizontal},
    enlarge y limits=0.15,
    xmajorgrids,
    yticklabel style={font=\footnotesize},
    tick label style={font=\footnotesize},
    xlabel style={font=\footnotesize},
    ylabel style={font=\footnotesize},
]
\addplot coordinates {
    (0.873,1)   
    (0.867,2)   
    (0.868,3)   
    (0.869,4)   
    (0.856,5)     
    (0.850,6)     
    (0.875,7)   
    (0.828,8)   
    (0.878,9)   
    (0.872,10)  
    (0.878,11)  
};
\end{axis}
\end{tikzpicture}
\caption{Agreement scores across evaluation dimensions. Higher values indicate stronger annotator consistency.}
\label{fig:realism-agreement}
\end{figure}

\paragraph{Overview.}
Although realism evaluation targets nuanced paralinguistic behaviors, \rev{human annotators reach moderate-to-substantial agreement on most dimensions once the rubric disentangles the aspects to be judged, though agreement is lower on some real-recording dimensions (e.g.\ \emph{Accuracy} $\alpha\approx0.18$; see Tab.~\ref{tab:audit-kappa}).}
As shown in the agreement summary in Fig.~\ref{fig:realism-agreement}, most dimensions fall in the $0.86$--$0.88$ range, with the lowest agreement on \emph{Traits Embodiment} ($\approx 0.83$) and the highest on \emph{Local Scene Fit} / \emph{Semantic Match} ($\approx 0.88$). 
Emotion-related, progressive metrics (Intensity and Transition) are slightly lower ($\approx 0.85$), reflecting the added difficulty of judging graded expressivity.

\paragraph{Agreement definition.}
Let $R$ denote the possible score range (e.g., $R=4$ for a $1$--$5$ scale). 
For a sample $i$ with annotator scores $\{s_{ij}\}_j$, we compute the sample standard deviation $\sigma_i$ and define the per-sample agreement
\[
a_i \;=\; \max\!\bigl(0,\;\min(1,\;1 - \sigma_i/R)\bigr).
\]
We then report the overall agreement $A = 1 - \bar{\sigma}/R$, where $\bar{\sigma}$ is the mean of $\{\sigma_i\}$ over samples with at least two annotations.
We also clamp $A$ to $[0,1]$ and ignore single-annotator cases.
A minimal, reproducible implementation is:

\paragraph{\rev{Relationship to chance-corrected agreement and the human reference level.}}
The metric $A = 1-\bar{\sigma}/R$ above is a deliberately simple, scale-aware summary; it is \emph{not} Krippendorff's $\alpha$, Fleiss' $\kappa$, or any standard ordinal inter-rater agreement (IAA). Because $A$ collapses any constant-offset disagreement into the same number as a tight cluster on a single bin, it should be read as a coarse, descriptive \emph{raw-concentration score} rather than as a demonstrated lower bound or a chance-corrected agreement. \rev{We therefore complement it with the standard estimators, all reported per dimension in App.~\ref{app:tier1}: ordinal Krippendorff's $\alpha$ (and quadratic-weighted $\kappa$, Sec.~\ref{app:audit}) as chance-corrected inter-rater agreement, and the single-rater-vs-consensus Pearson as a human \emph{reference level} (Tab.~\ref{tab:tier1-ceiling}) against which the model--human Pearson values in Tabs.~\ref{tab:role-play-exp-archetype}--\ref{tab:role-play-exp-realism-real} should be read. We call this a reference level, not a formal upper bound, since a model could in principle exceed a single rater. One reading caveat carries over to those tables: mapping gated \emph{Emotion Intensity}/\emph{Emotion Transition} entries to the score floor (the floor-imputation convention) inflates both $A$ and the correlations on those two dimensions, because annotators who all hit the floor look perfectly consistent under $\sigma$-based agreement; the main tables and the reference-level column in Tab.~\ref{tab:tier1-ceiling} instead adopt \emph{masking} (dropping gated entries) as the primary convention, with floor-imputation reported only as a secondary sensitivity analysis.} The $A$ values are provided as a simple reading of the rubric's internal consistency.

\begin{lstlisting}[language=Python,basicstyle=\ttfamily\small,caption={Agreement computation (concise).}]
import numpy as np

def compute_agreement(values_by_sample, max_possible_range):
    stds, per_sample = [], {}
    for i, scores in enumerate(values_by_sample):
        x = [float(s) for s in scores if s is not None]
        if len(x) < 2: continue
        x = np.asarray(x, dtype=float)
        s = x.std(ddof=1)
        a = 1 - (s / max_possible_range) if max_possible_range > 0 else 1
        a = max(0, min(1, a))
        per_sample[f"sample_{i}"] = {
            "scores": x.tolist(), "mean": round(x.mean(), 3),
            "std": round(s, 3), "range": round(x.max()-x.min(), 3),
            "agreement_score": round(a, 3), "annotator_count": len(x)
        }
        stds.append(s)
    mean_std = np.mean(stds) if stds else np.nan
    A = 1 - (mean_std / max_possible_range) if stds else np.nan
    if not np.isnan(A): A = max(0, min(1, A))
    return {"mean_intra_sample_std": mean_std, 
            "sample_agreements": per_sample,
            "overall_agreement": A}
\end{lstlisting}

\begin{figure}[tp]
\centering

\begin{subfigure}[t]{0.48\linewidth}
  \centering
  \includegraphics[width=\linewidth]{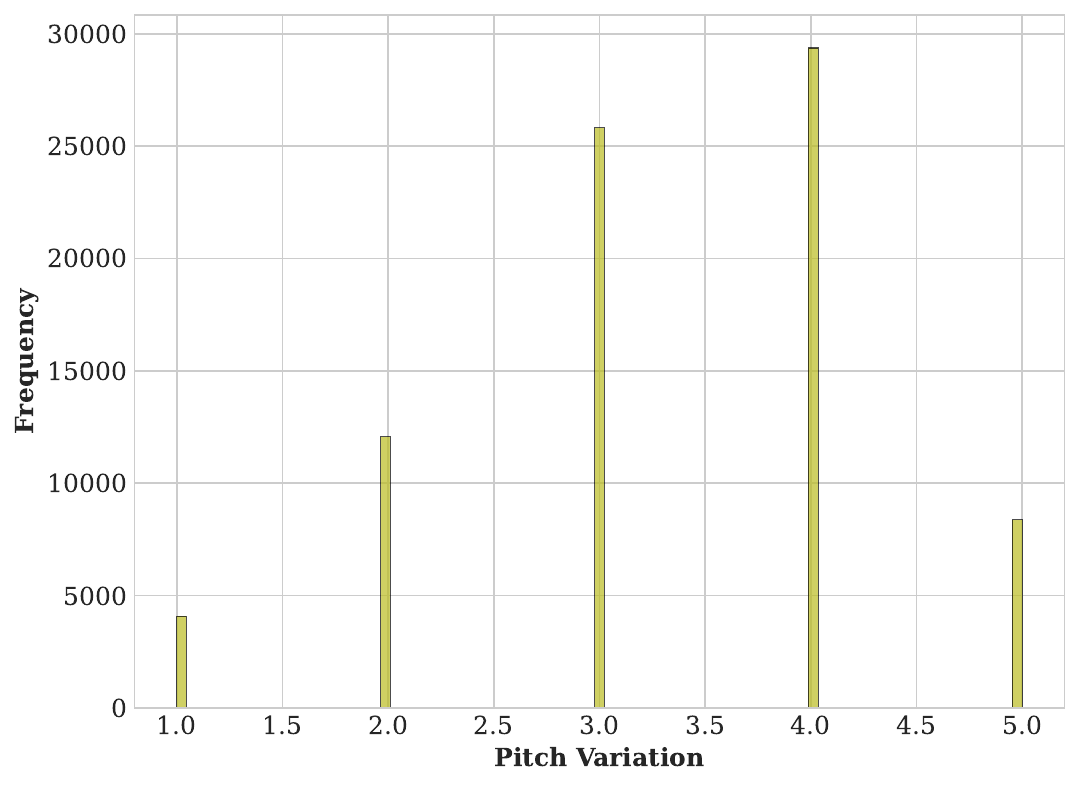}
  \subcaption{Distribution of Pitch Variation}
\end{subfigure}
\hfill
\begin{subfigure}[t]{0.48\linewidth}
  \centering
  \includegraphics[width=\linewidth]{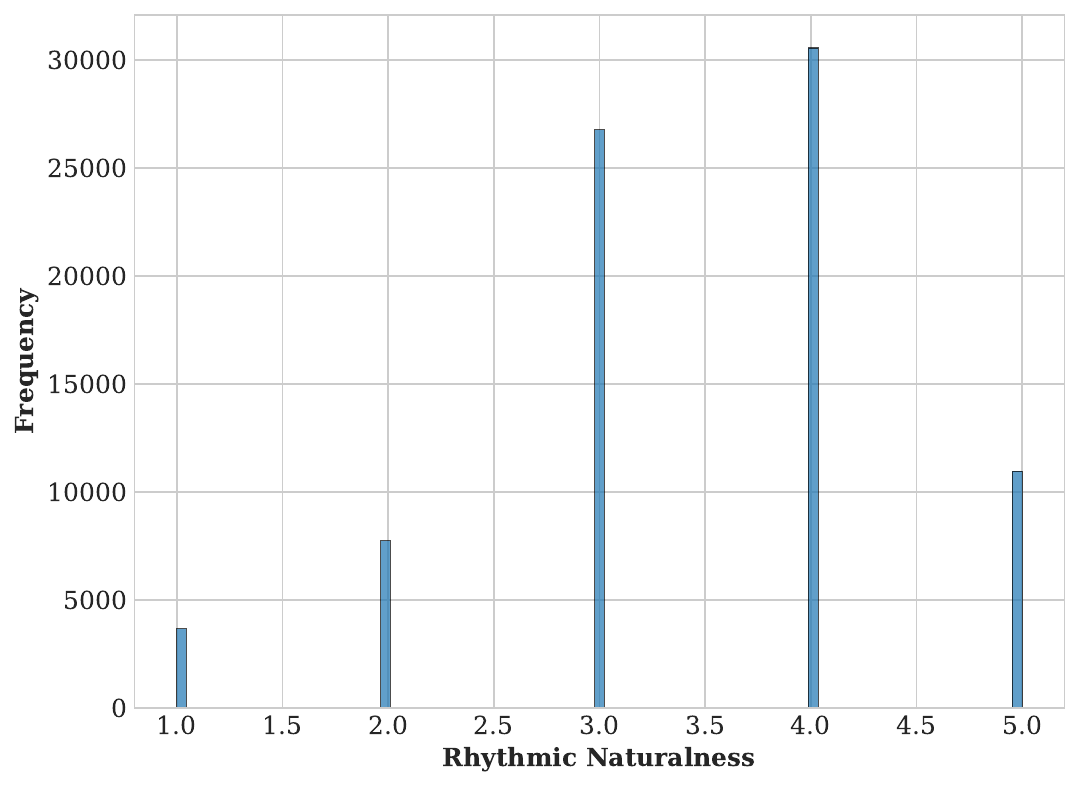}
  \subcaption{Distribution of Rhythmic Naturalness}
\end{subfigure}

\vspace{0.8em}

\begin{subfigure}[t]{0.48\linewidth}
  \centering
  \includegraphics[width=\linewidth]{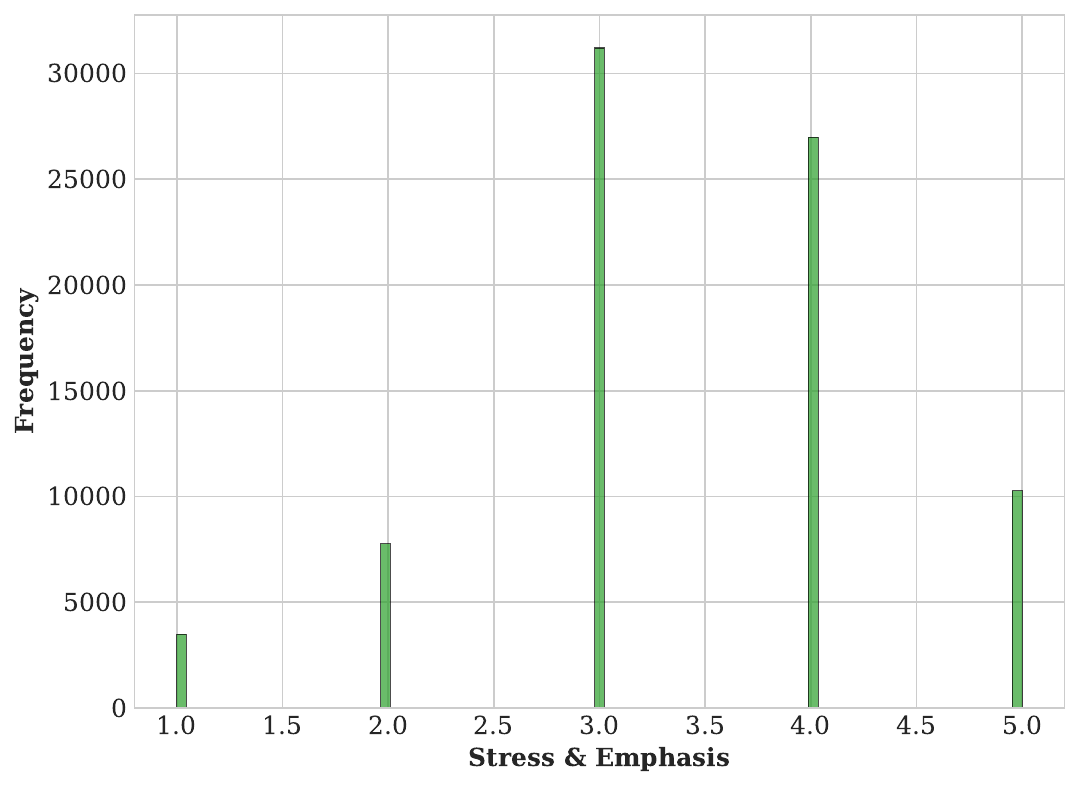}
  \subcaption{Distribution of Stress \& Emphasis}
\end{subfigure}
\hfill
\begin{subfigure}[t]{0.48\linewidth}
  \centering
  \includegraphics[width=\linewidth]{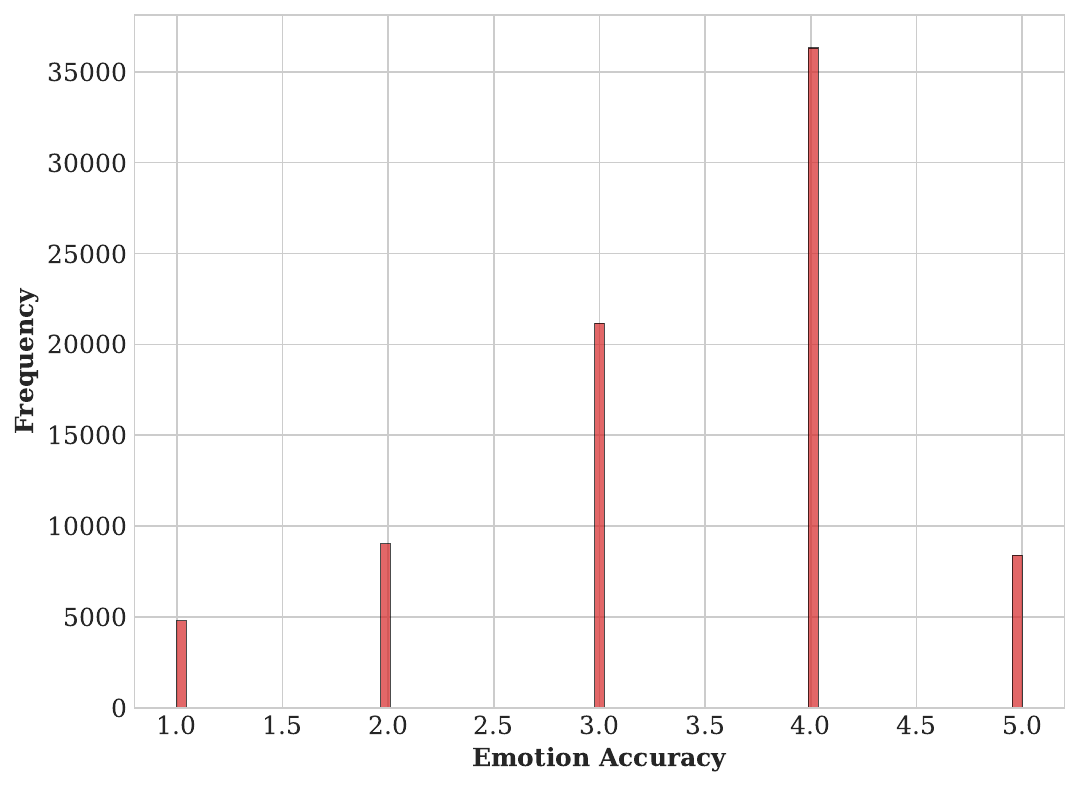}
  \subcaption{Distribution of Emotion Accuracy}
\end{subfigure}

\vspace{0.8em}

\begin{subfigure}[t]{0.48\linewidth}
  \centering
  \includegraphics[width=\linewidth]{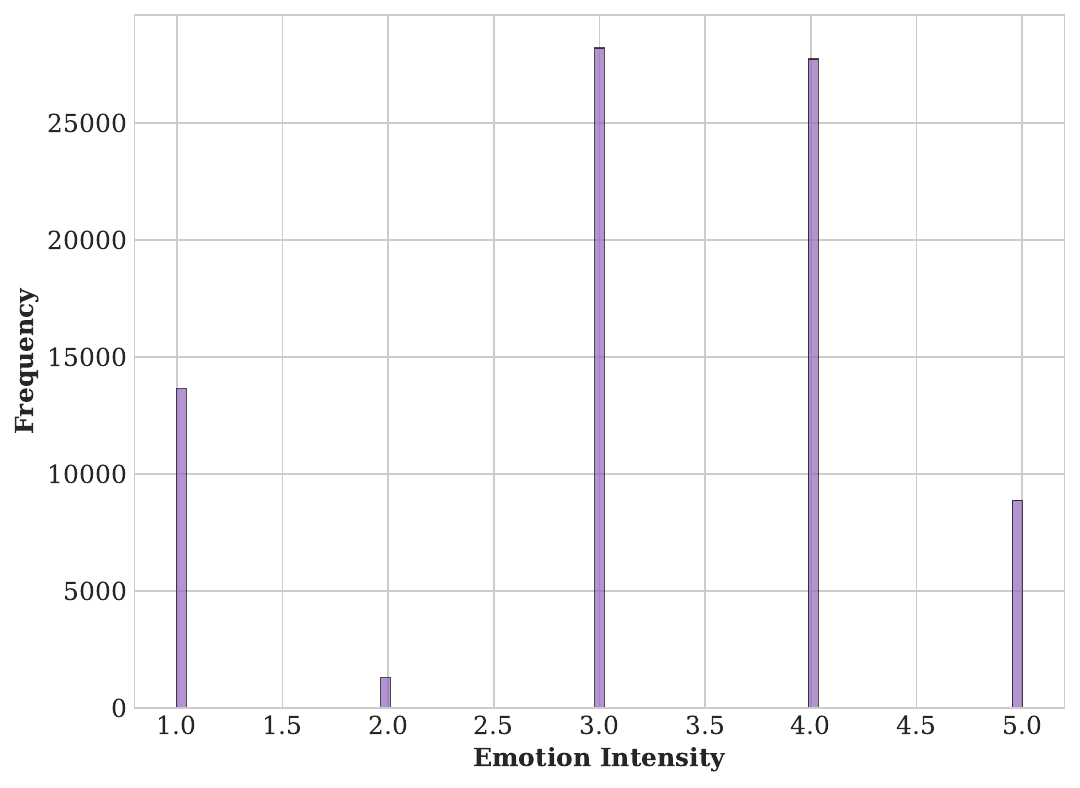}
  \subcaption{Distribution of Emotion Intensity}
\end{subfigure}
\hfill
\begin{subfigure}[t]{0.48\linewidth}
  \centering
  \includegraphics[width=\linewidth]{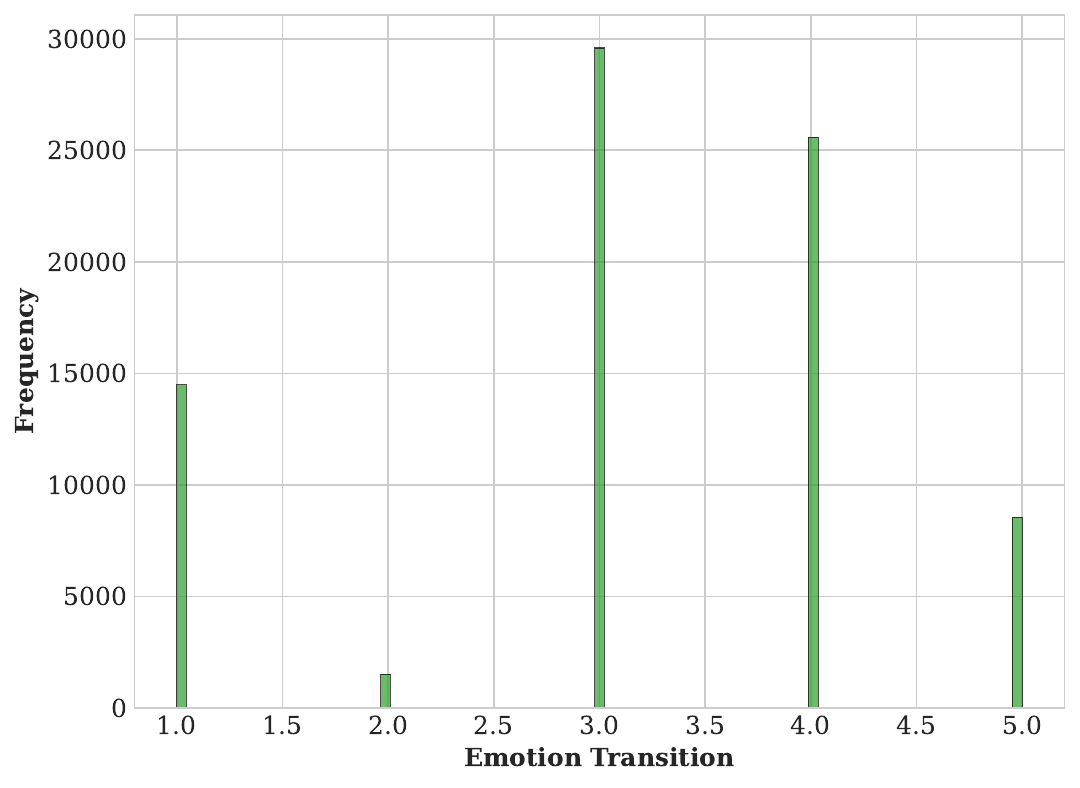}
  \subcaption{Distribution of Emotion Transition}
\end{subfigure}

\caption{Distributions of evaluation dimensions in Realism Evaluation (1).}
\label{fig:realism-histograms-1}
\end{figure}

\begin{figure}[tp]
\centering
\begin{subfigure}[t]{0.48\linewidth}
  \centering
  \includegraphics[width=\linewidth]{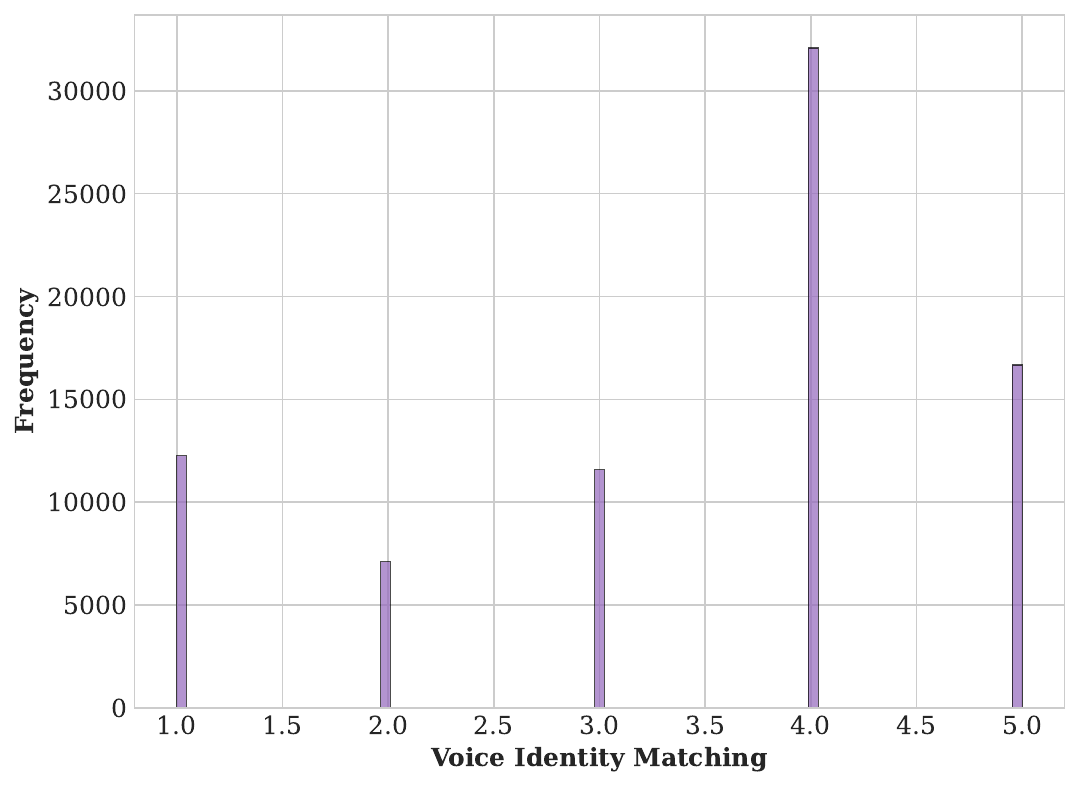}
  \subcaption{Distribution of Voice Identity Matching}
\end{subfigure}
\hfill
\begin{subfigure}[t]{0.48\linewidth}
  \centering
  \includegraphics[width=\linewidth]{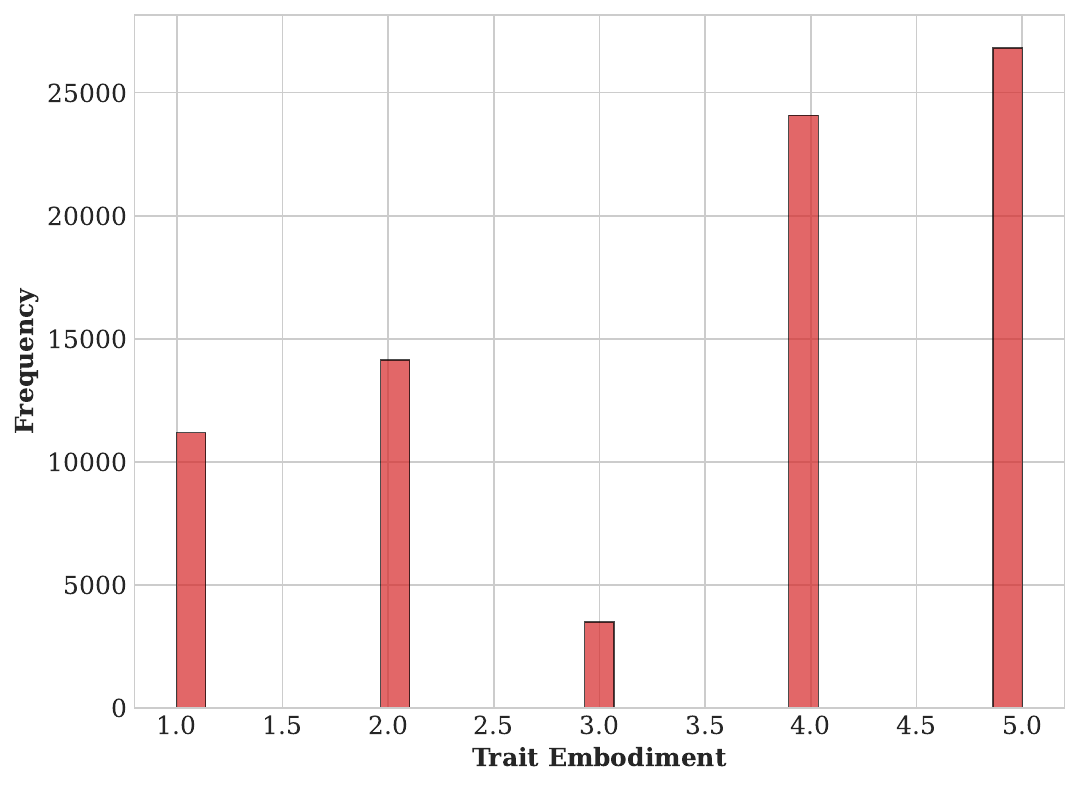}
  \subcaption{Distribution of Trait Embodiment}
\end{subfigure}

\vspace{0.8em}

\begin{subfigure}[t]{0.48\linewidth}
  \centering
  \includegraphics[width=\linewidth]{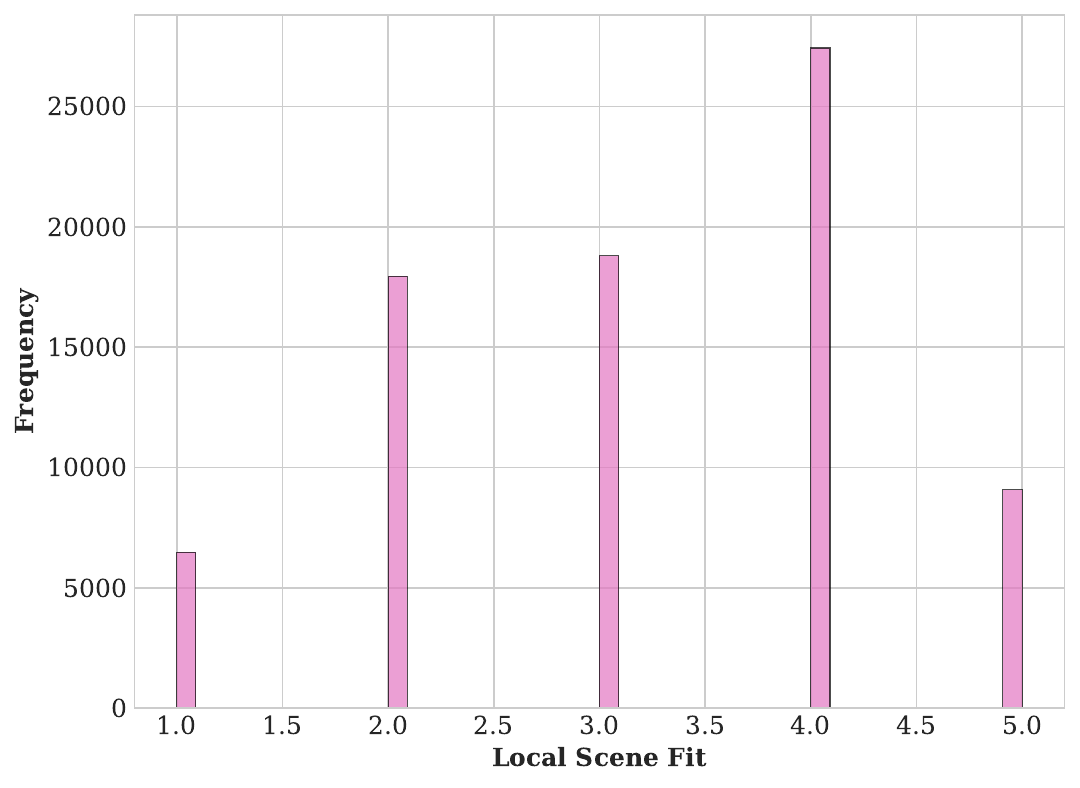}
  \subcaption{Distribution of Local Scene Fit}
\end{subfigure}
\hfill
\begin{subfigure}[t]{0.48\linewidth}
  \centering
  \includegraphics[width=\linewidth]{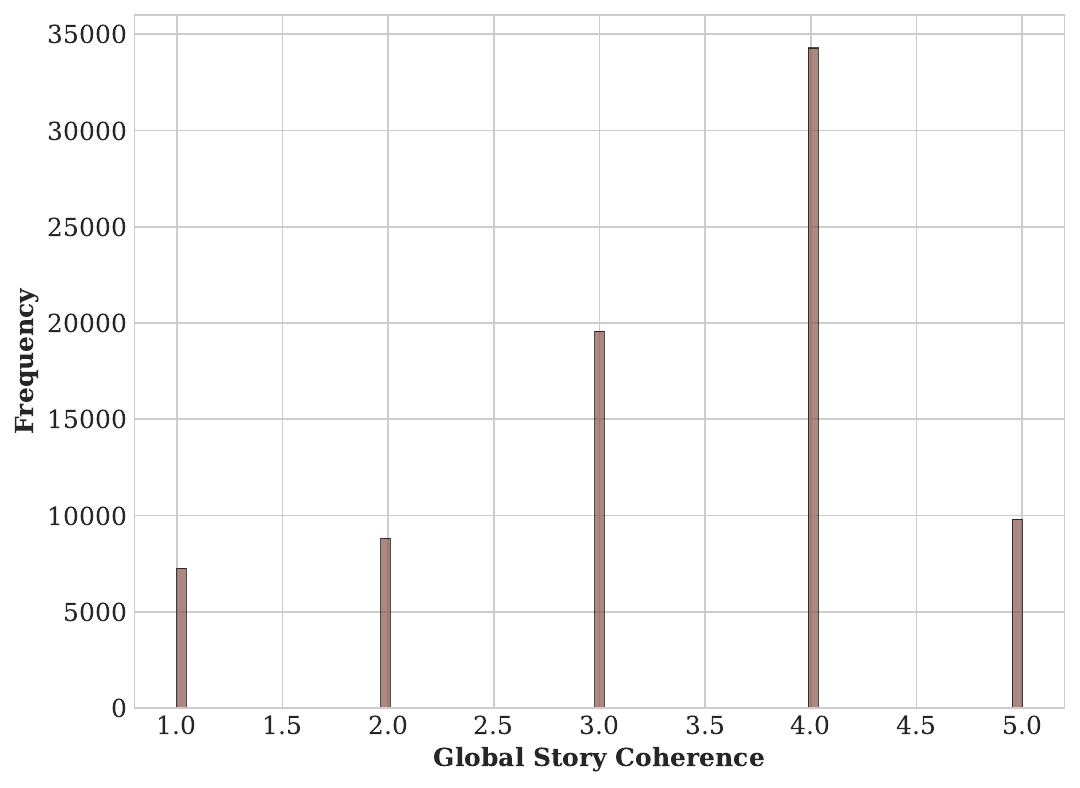}
  \subcaption{Distribution of Global Story Coherence}
\end{subfigure}

\vspace{0.8em}

\begin{subfigure}[t]{0.48\linewidth}
  \centering
  \includegraphics[width=\linewidth]{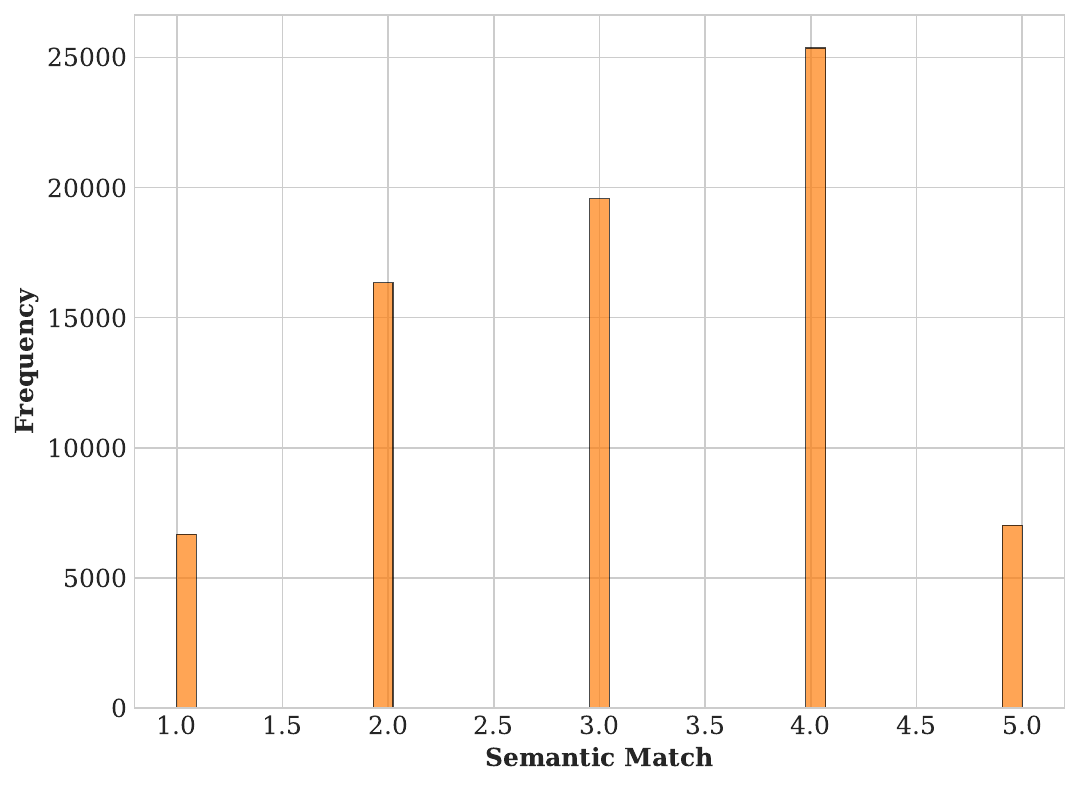}
  \subcaption{Distribution of Semantic Match}
\end{subfigure}

\caption{Distributions of evaluation dimensions in Realism Evaluation (2).}
\label{fig:realism-histograms-2}
\end{figure}

\begin{figure}[tp]
\centering
\begin{subfigure}[t]{0.48\linewidth}
  \centering
  \includegraphics[width=\linewidth]{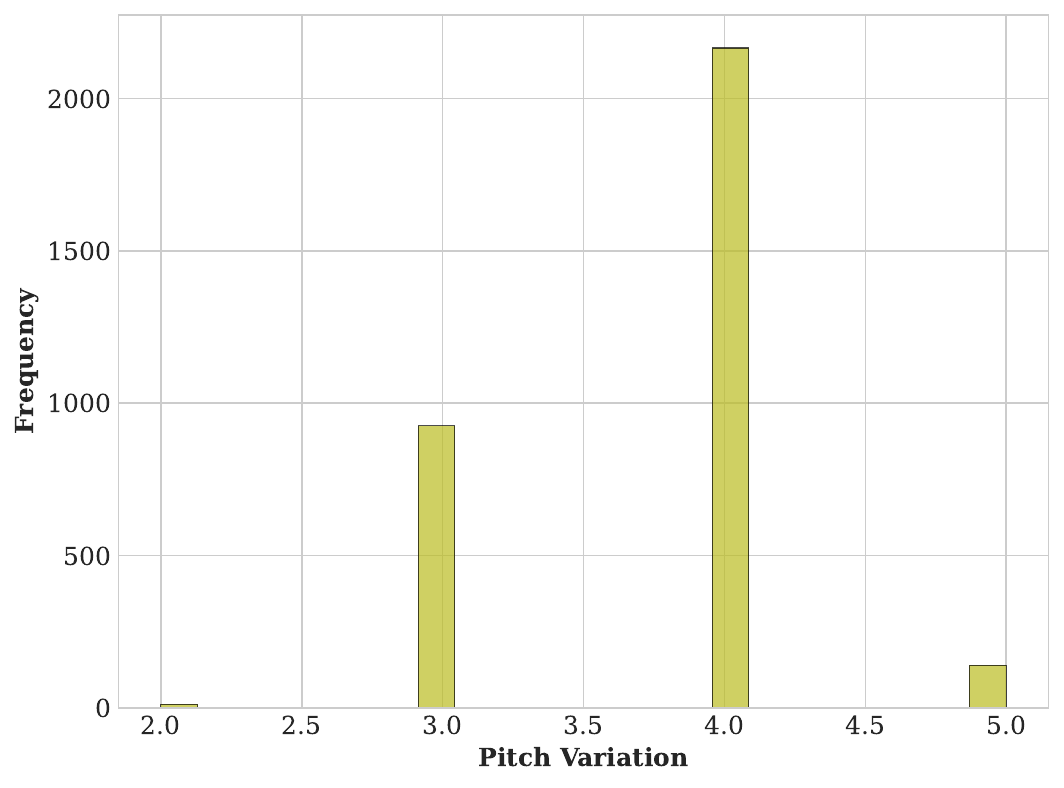}
  \subcaption{Professional Voice Actor}
\end{subfigure}\hfill
\begin{subfigure}[t]{0.48\linewidth}
  \centering
  \includegraphics[width=\linewidth]{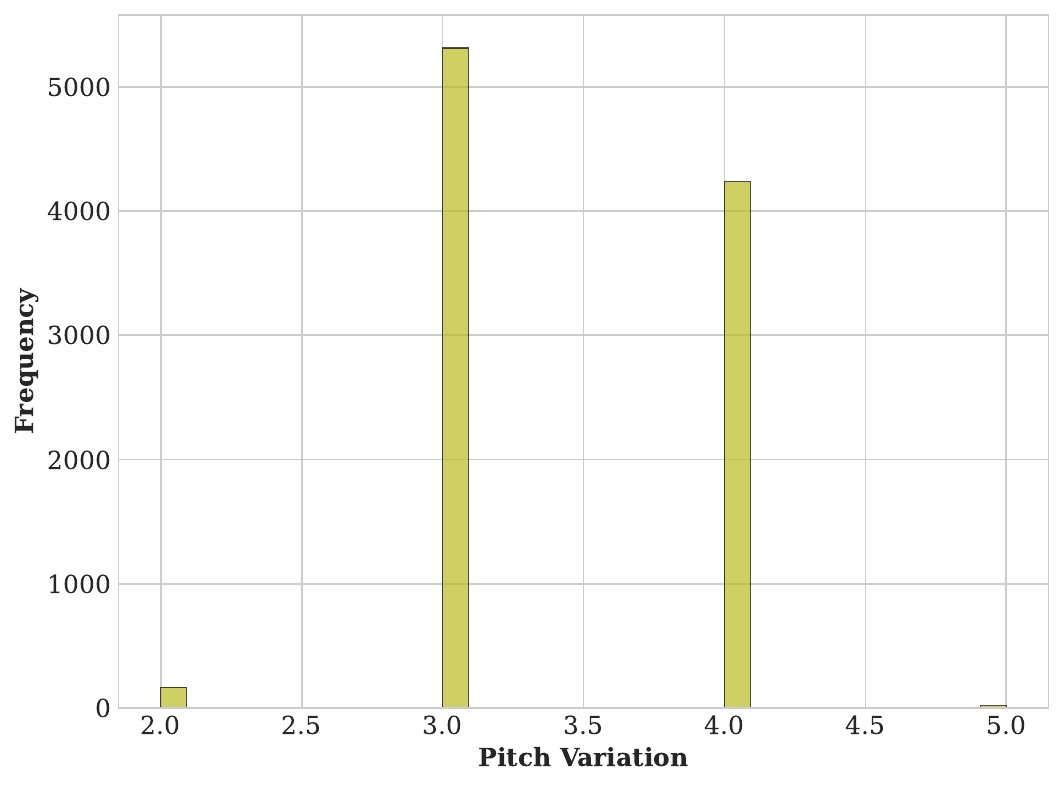}
  \subcaption{Amateur Voice Actor}
\end{subfigure}
\caption{Distribution of Pitch Variation in Realism Evaluation: Professional vs. Amateur.}
\label{fig:cmp-pitch-variation}
\end{figure}

\begin{figure}[tp]
\centering
\begin{subfigure}[t]{0.48\linewidth}
  \centering
  \includegraphics[width=\linewidth]{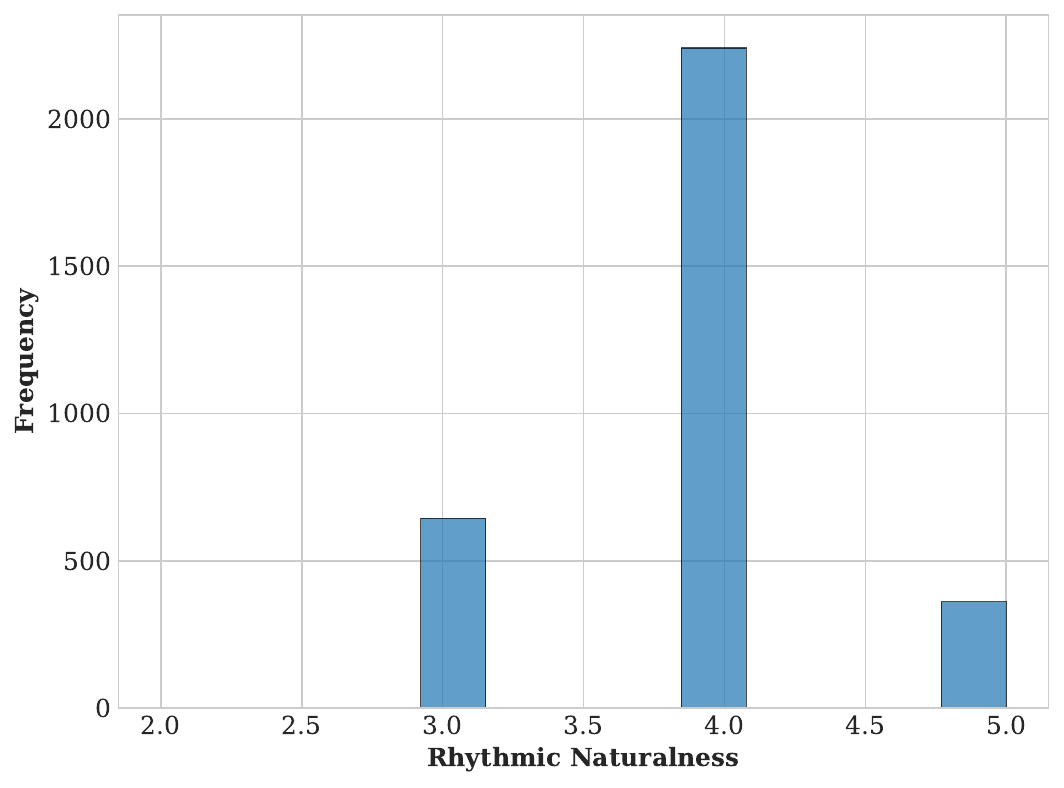}
  \subcaption{Professional Voice Actor}
\end{subfigure}\hfill
\begin{subfigure}[t]{0.48\linewidth}
  \centering
  \includegraphics[width=\linewidth]{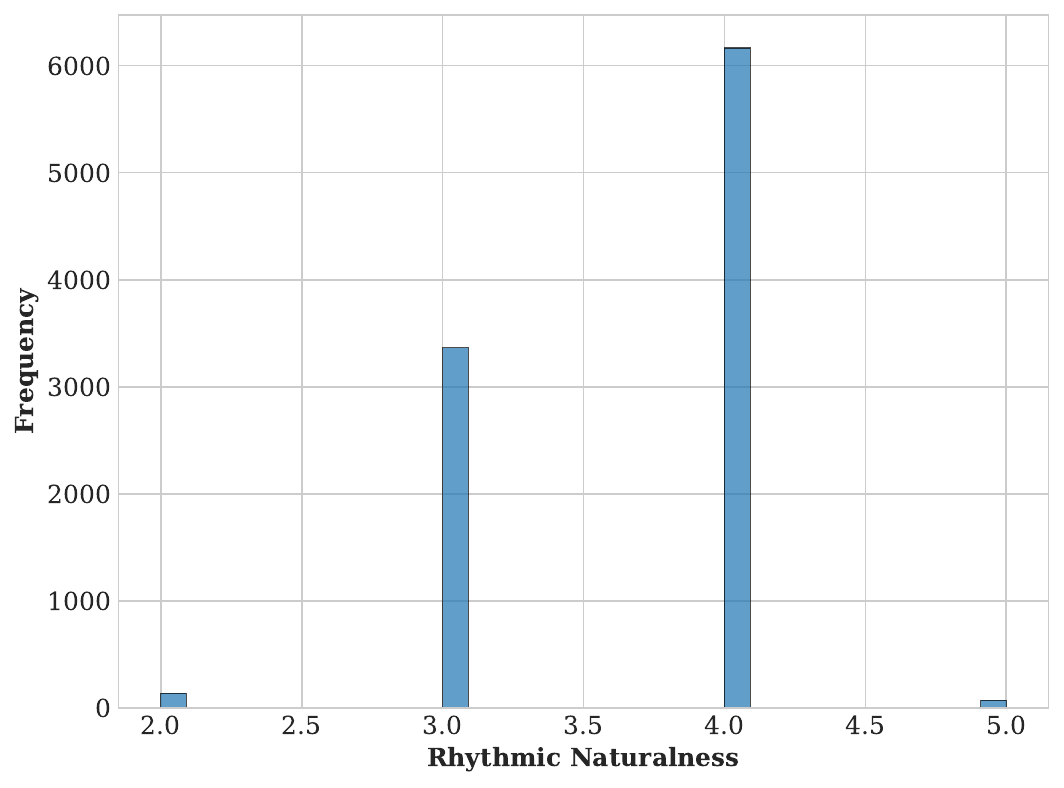}
  \subcaption{Amateur Voice Actor}
\end{subfigure}
\caption{Distribution of Rhythmic Naturalness in Realism Evaluation: Professional vs. Amateur.}
\label{fig:cmp-rhythmic-naturalness}
\end{figure}

\begin{figure}[tp]
\centering
\begin{subfigure}[t]{0.48\linewidth}
  \centering
  \includegraphics[width=\linewidth]{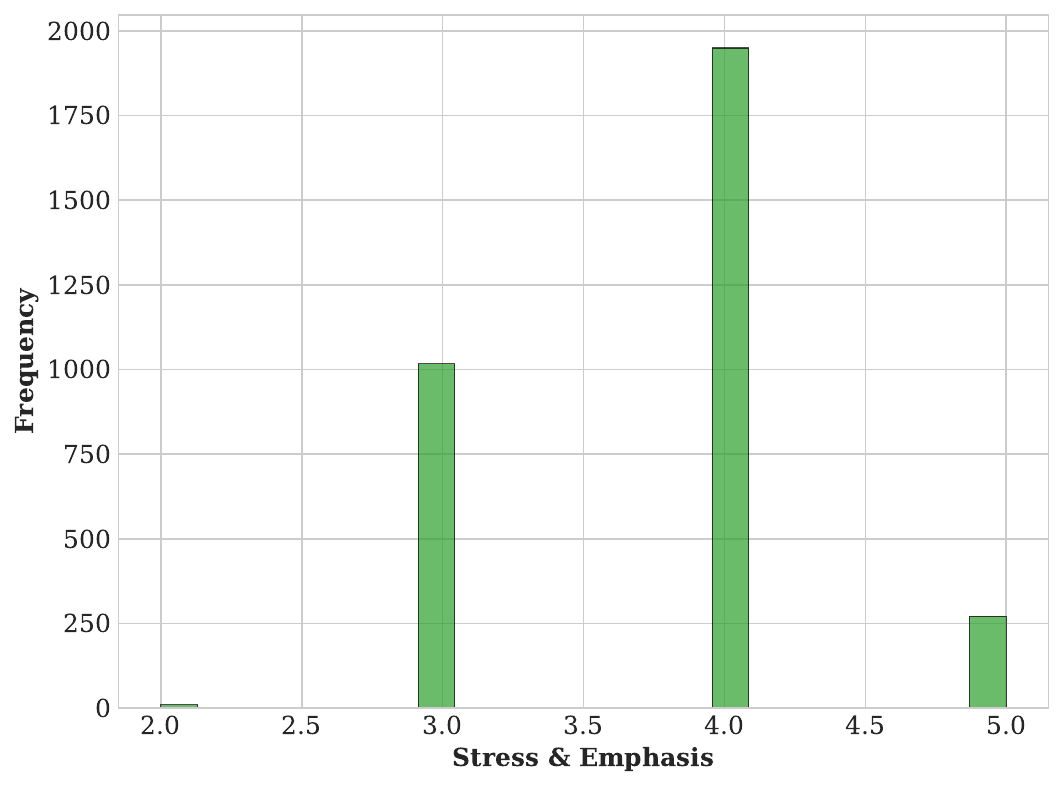}
  \subcaption{Professional Voice Actor}
\end{subfigure}\hfill
\begin{subfigure}[t]{0.48\linewidth}
  \centering
  \includegraphics[width=\linewidth]{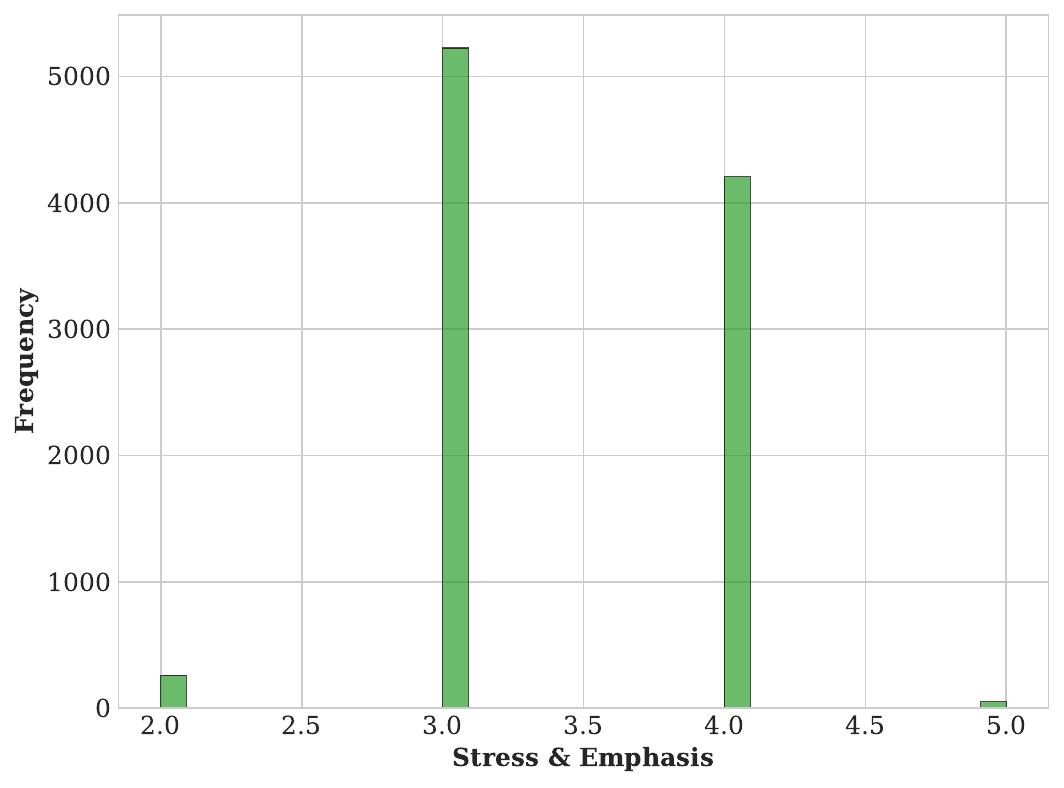}
  \subcaption{Amateur Voice Actor}
\end{subfigure}
\caption{Distribution of Stress \& Emphasis in Realism Evaluation: Professional vs. Amateur.}
\label{fig:cmp-stress-emphasis}
\end{figure}

\begin{figure}[tp]
\centering
\begin{subfigure}[t]{0.48\linewidth}
  \centering
  \includegraphics[width=\linewidth]{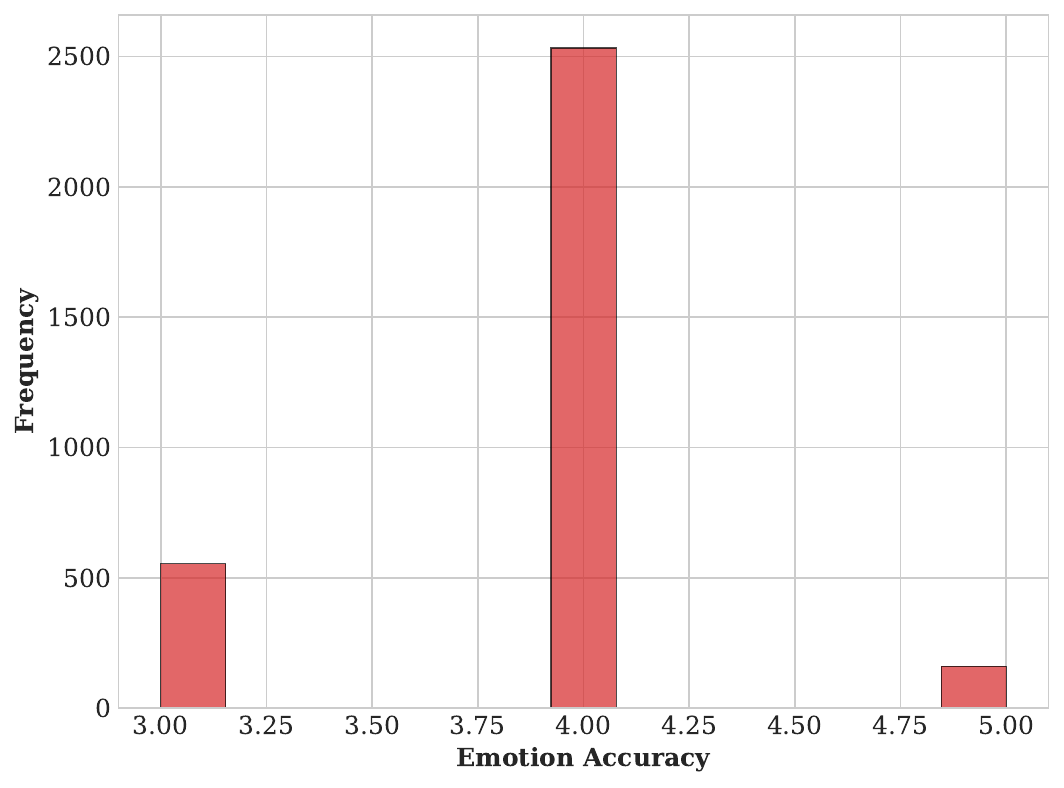}
  \subcaption{Professional Voice Actor}
\end{subfigure}\hfill
\begin{subfigure}[t]{0.48\linewidth}
  \centering
  \includegraphics[width=\linewidth]{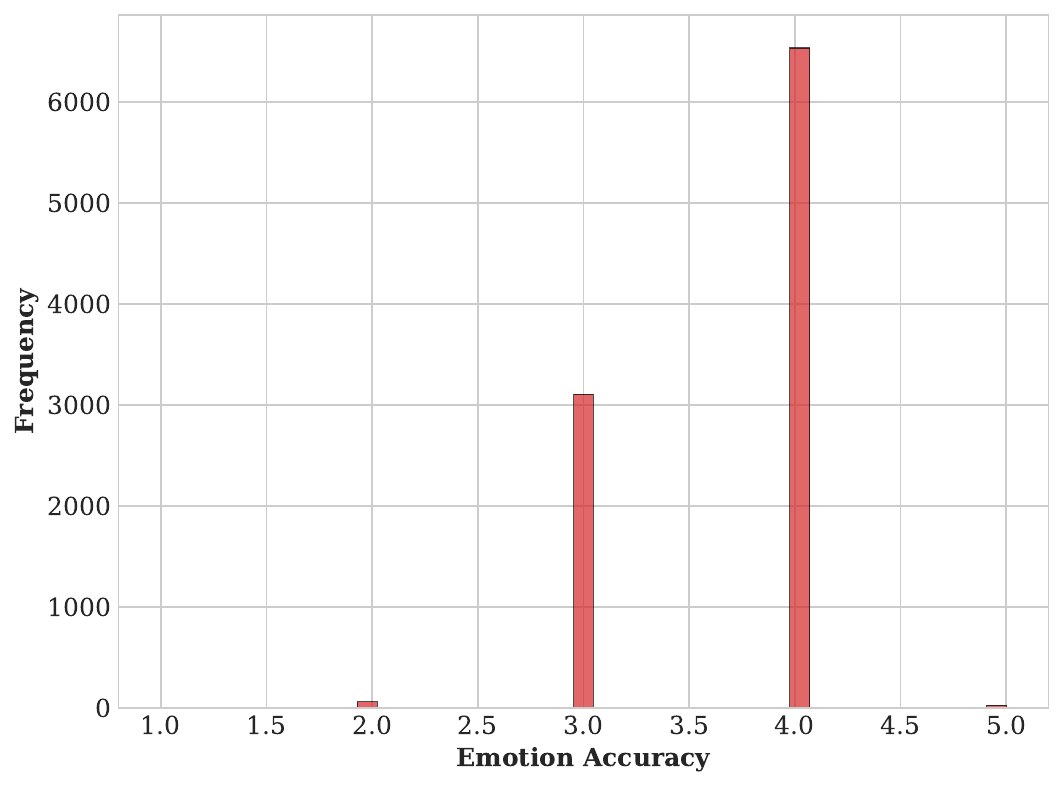}
  \subcaption{Amateur Voice Actor}
\end{subfigure}
\caption{Distribution of Emotion Accuracy in Realism Evaluation: Professional vs. Amateur.}
\label{fig:cmp-emotion-accuracy}
\end{figure}

\begin{figure}[tp]
\centering
\begin{subfigure}[t]{0.48\linewidth}
  \centering
  \includegraphics[width=\linewidth]{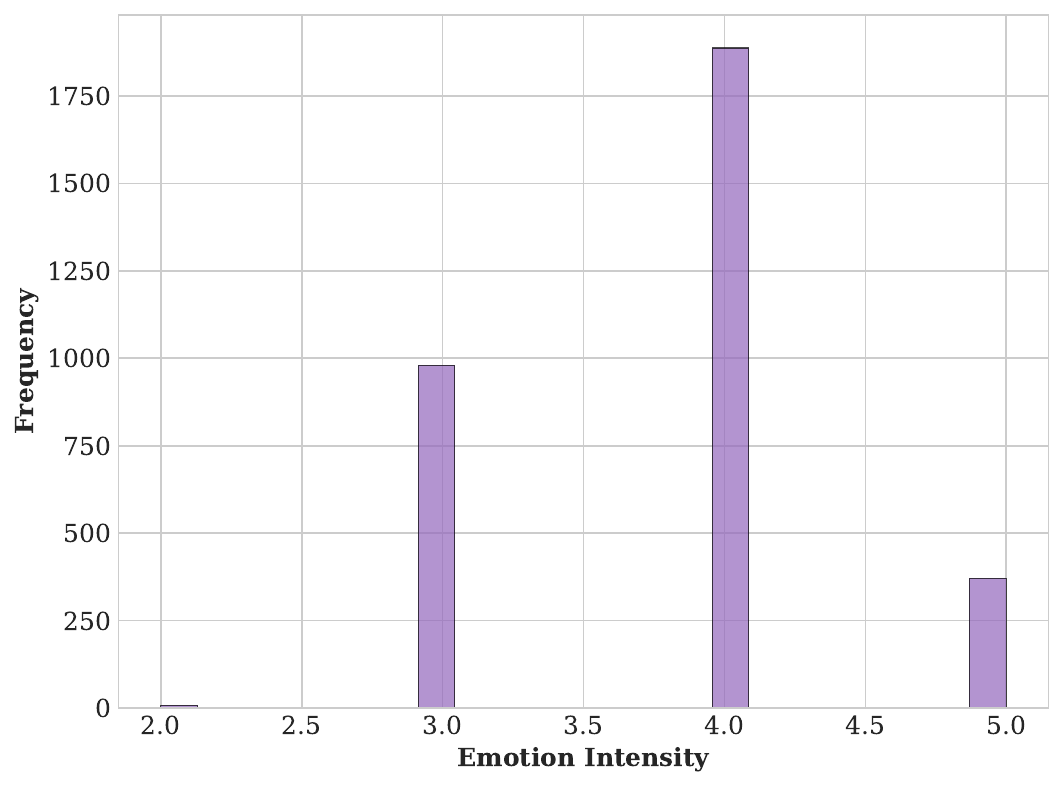}
  \subcaption{Professional Voice Actor}
\end{subfigure}\hfill
\begin{subfigure}[t]{0.48\linewidth}
  \centering
  \includegraphics[width=\linewidth]{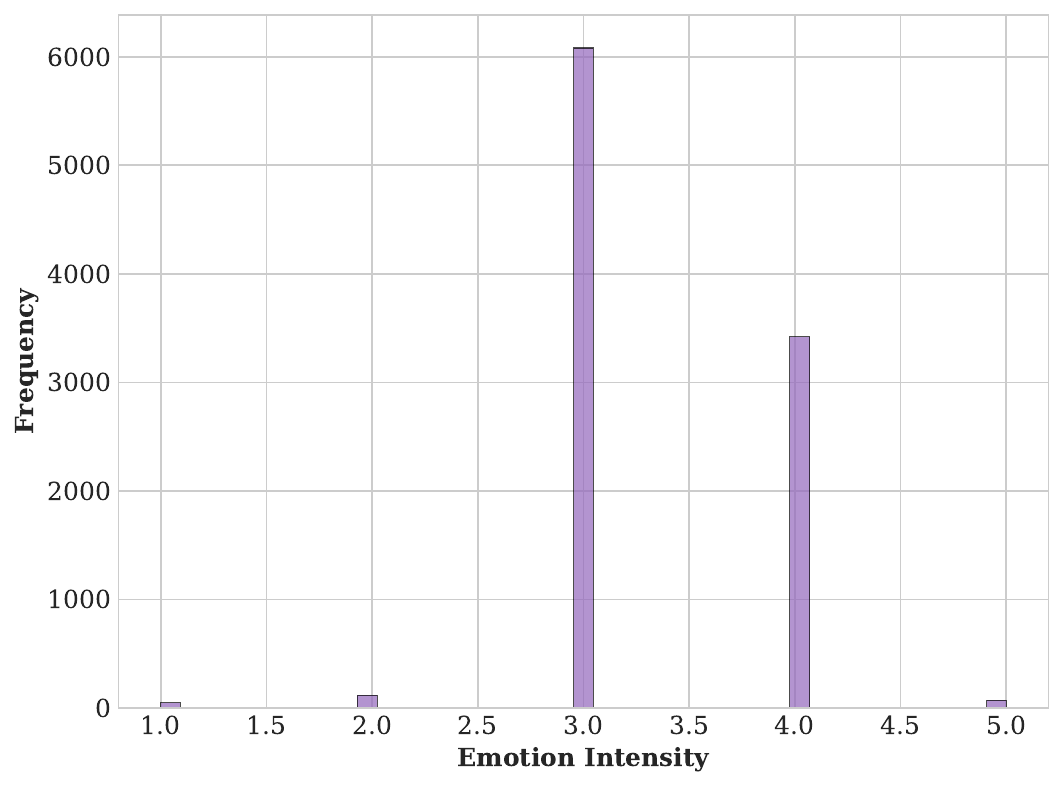}
  \subcaption{Amateur Voice Actor}
\end{subfigure}
\caption{Distribution of Emotion Intensity in Realism Evaluation: Professional vs. Amateur.}
\label{fig:cmp-emotion-intensity}
\end{figure}

\begin{figure}[tp]
\centering
\begin{subfigure}[t]{0.48\linewidth}
  \centering
  \includegraphics[width=\linewidth]{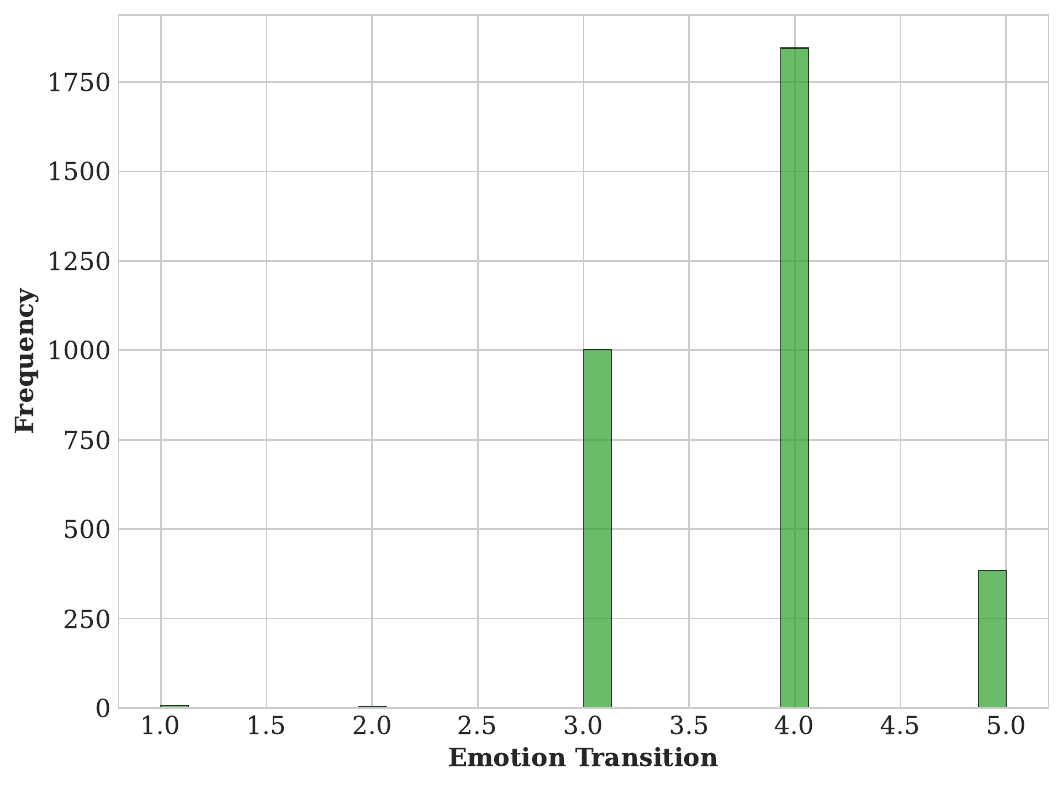}
  \subcaption{Professional Voice Actor}
\end{subfigure}\hfill
\begin{subfigure}[t]{0.48\linewidth}
  \centering
  \includegraphics[width=\linewidth]{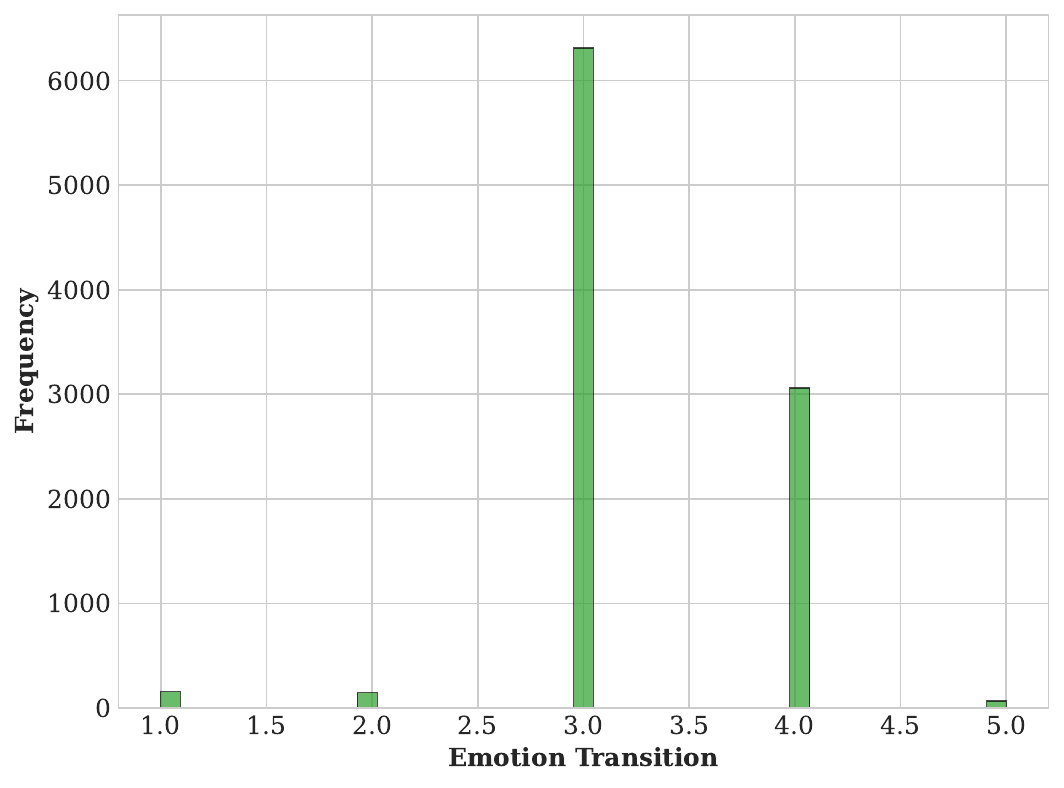}
  \subcaption{Amateur Voice Actor}
\end{subfigure}
\caption{Distribution of Emotion Transition in Realism Evaluation: Professional vs. Amateur.}
\label{fig:cmp-emotion-transition}
\end{figure}

\begin{figure}[tp]
\centering
\begin{subfigure}[t]{0.48\linewidth}
  \centering
  \includegraphics[width=\linewidth]{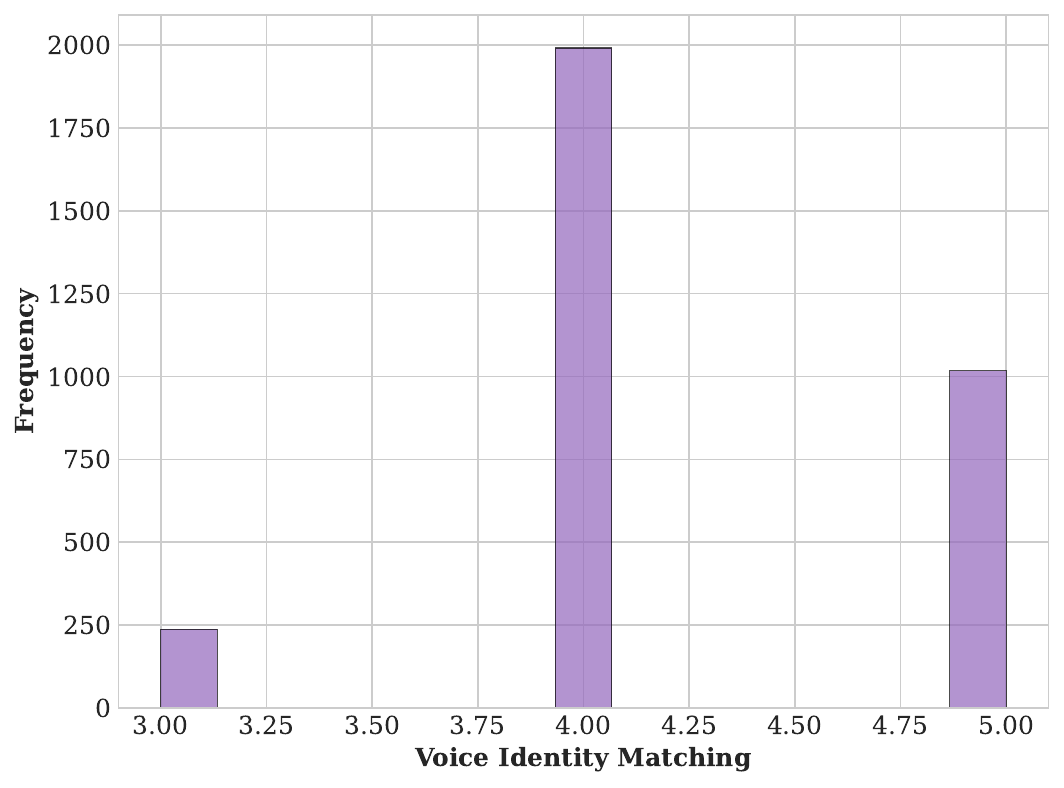}
  \subcaption{Professional Voice Actor}
\end{subfigure}\hfill
\begin{subfigure}[t]{0.48\linewidth}
  \centering
  \includegraphics[width=\linewidth]{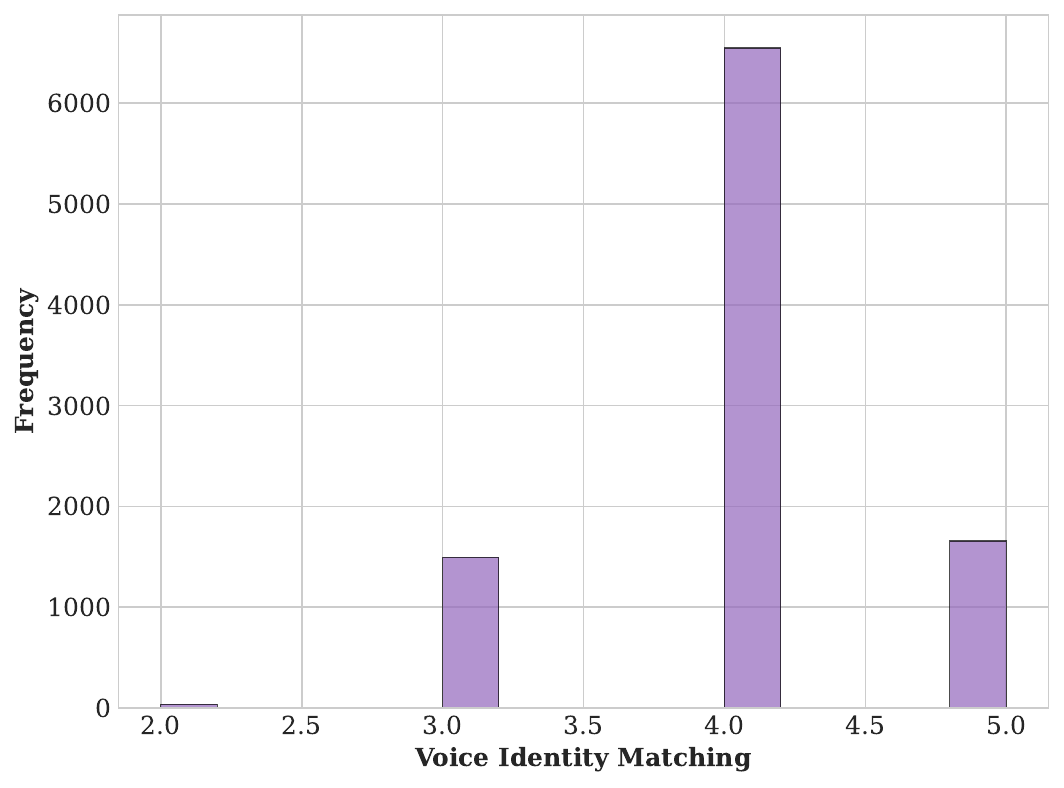}
  \subcaption{Amateur Voice Actor}
\end{subfigure}
\caption{Distribution of Voice Identity Matching in Realism Evaluation: Professional vs. Amateur.}
\label{fig:cmp-voice-identity}
\end{figure}

\begin{figure}[tp]
\centering
\begin{subfigure}[t]{0.48\linewidth}
  \centering
  \includegraphics[width=\linewidth]{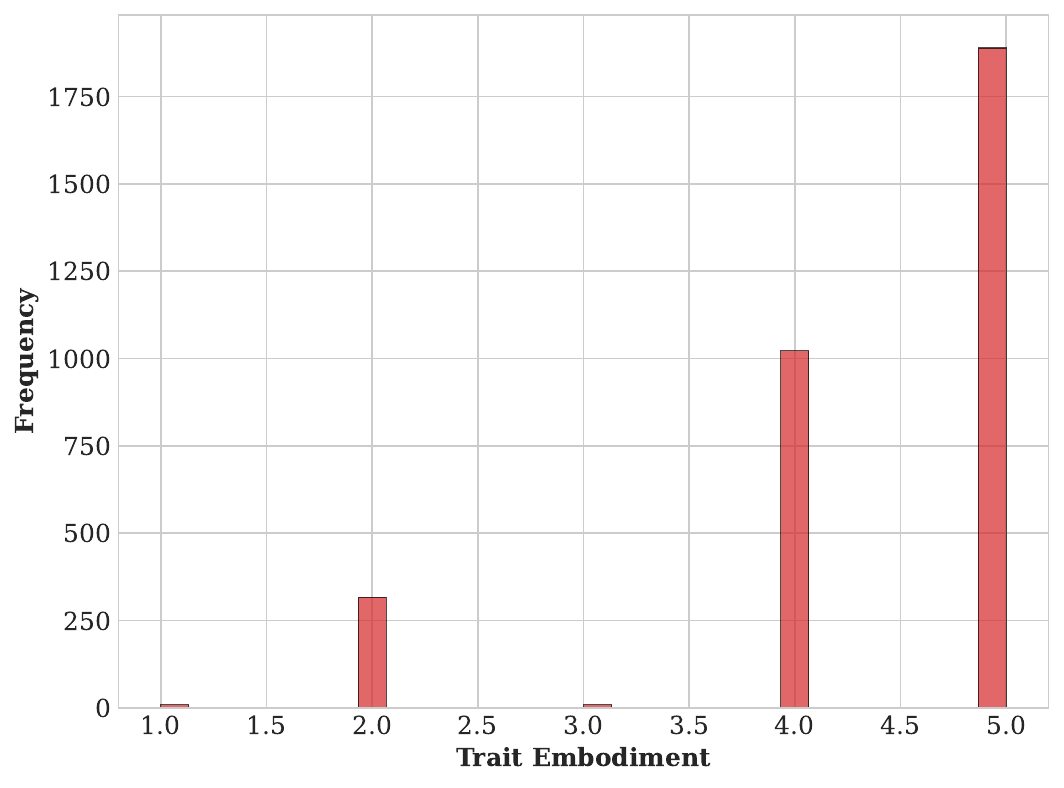}
  \subcaption{Professional Voice Actor}
\end{subfigure}\hfill
\begin{subfigure}[t]{0.48\linewidth}
  \centering
  \includegraphics[width=\linewidth]{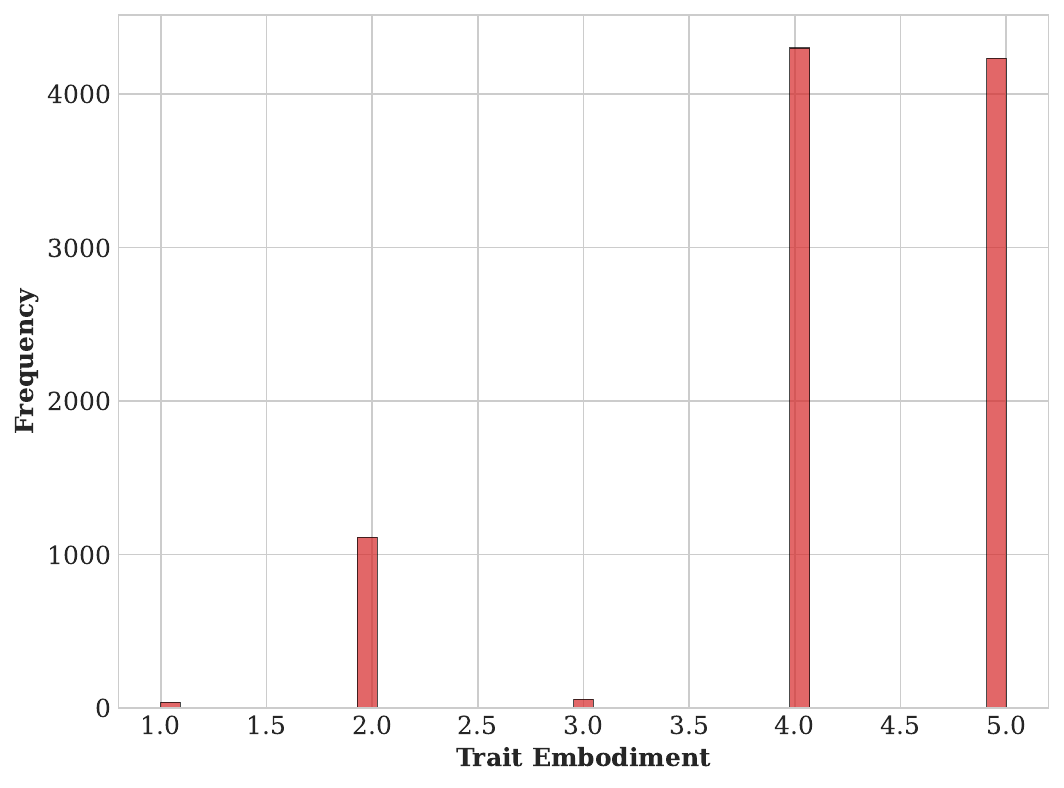}
  \subcaption{Amateur Voice Actor}
\end{subfigure}
\caption{Distribution of Traits Embodiment in Realism Evaluation: Professional vs. Amateur.}
\label{fig:cmp-trait-embodiment}
\end{figure}

\begin{figure}[tp]
\centering
\begin{subfigure}[t]{0.48\linewidth}
  \centering
  \includegraphics[width=\linewidth]{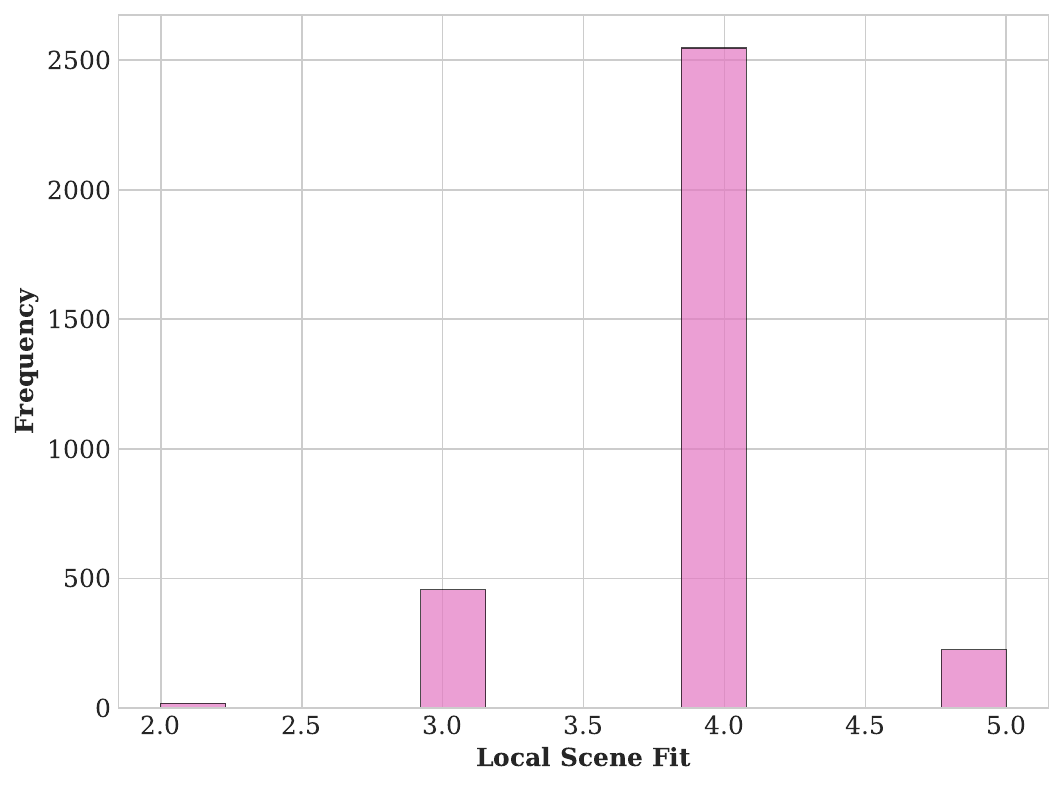}
  \subcaption{Professional Voice Actor}
\end{subfigure}\hfill
\begin{subfigure}[t]{0.48\linewidth}
  \centering
  \includegraphics[width=\linewidth]{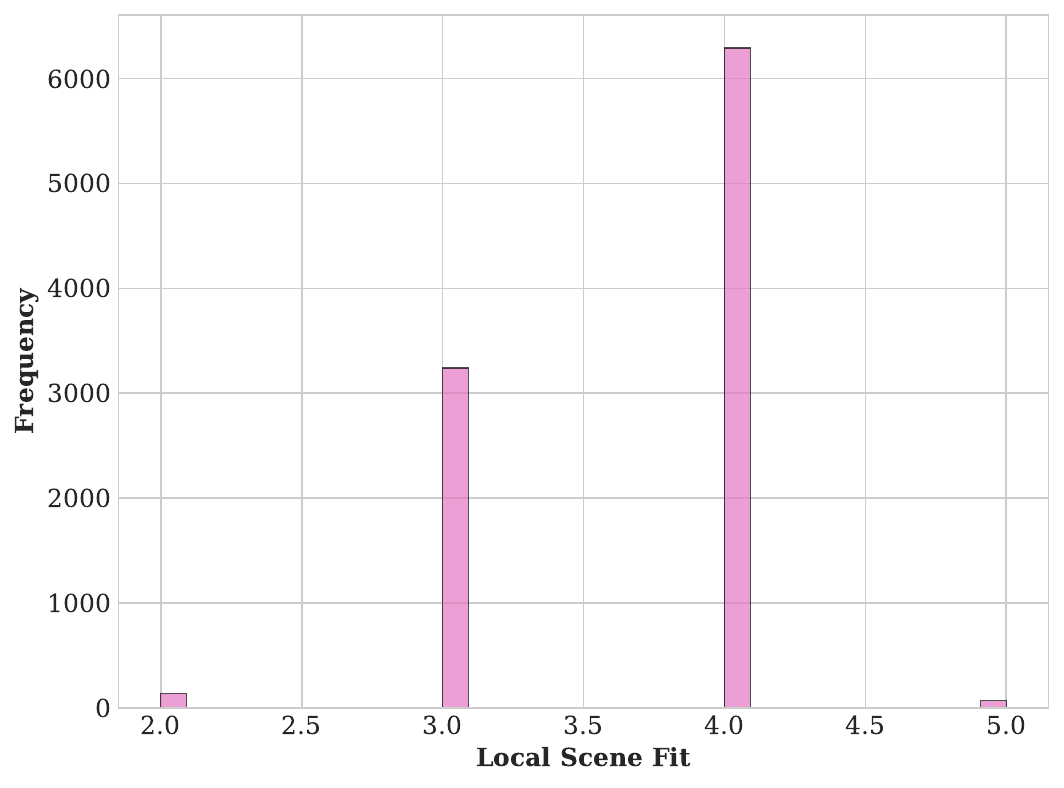}
  \subcaption{Amateur Voice Actor}
\end{subfigure}
\caption{Distribution of Local Scene Fit in Realism Evaluation: Professional vs. Amateur.}
\label{fig:cmp-local-scene-fit}
\end{figure}

\begin{figure}[tp]
\centering
\begin{subfigure}[t]{0.48\linewidth}
  \centering
  \includegraphics[width=\linewidth]{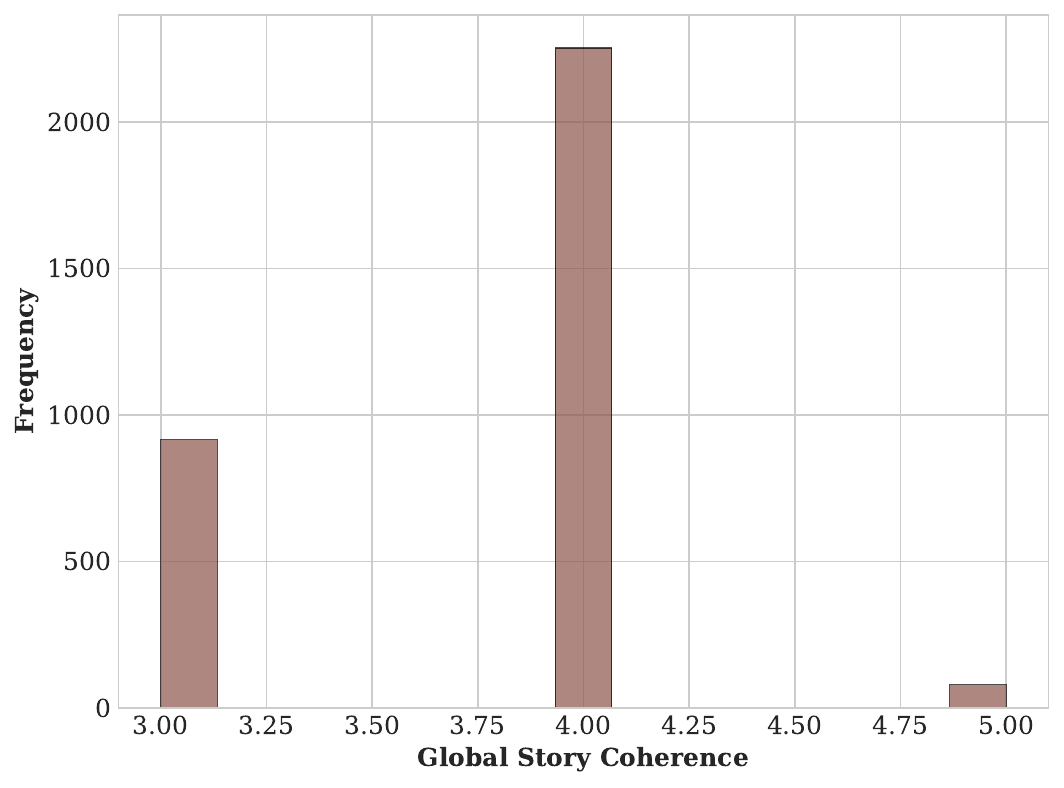}
  \subcaption{Professional Voice Actor}
\end{subfigure}\hfill
\begin{subfigure}[t]{0.48\linewidth}
  \centering
  \includegraphics[width=\linewidth]{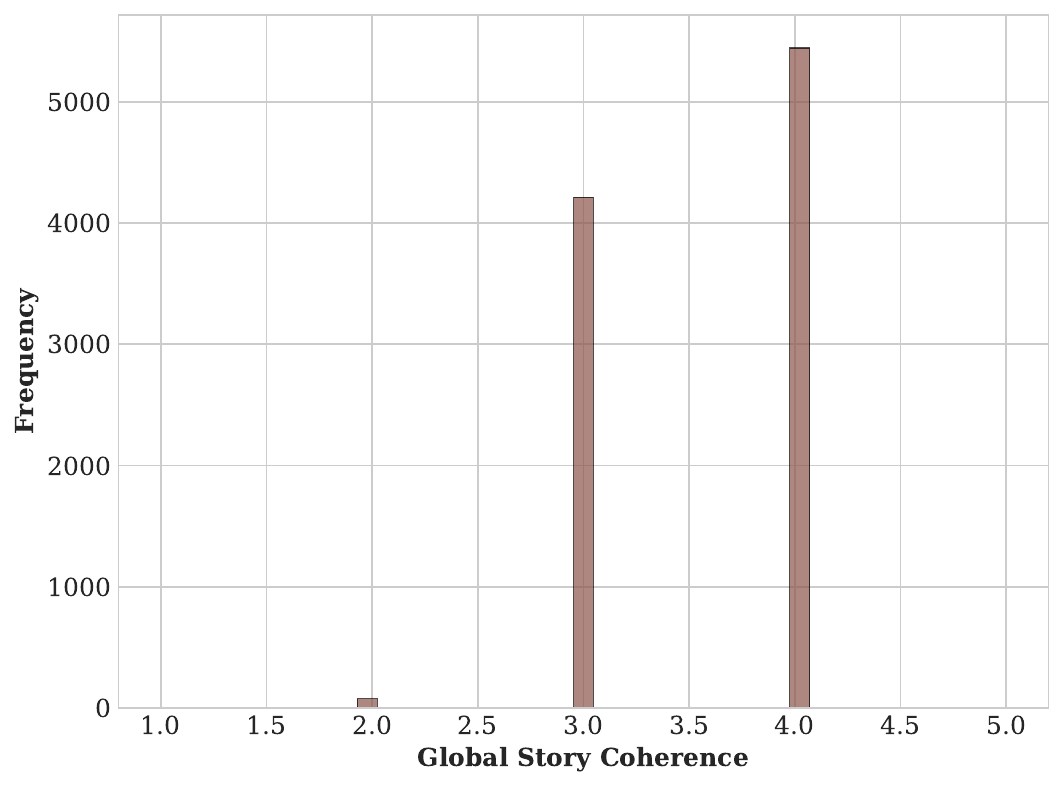}
  \subcaption{Amateur Voice Actor}
\end{subfigure}
\caption{Distribution of Global Story Coherence in Realism Evaluation: Professional vs. Amateur.}
\label{fig:cmp-global-story-coherence}
\end{figure}

\begin{figure}[tp]
\centering
\begin{subfigure}[t]{0.48\linewidth}
  \centering
  \includegraphics[width=\linewidth]{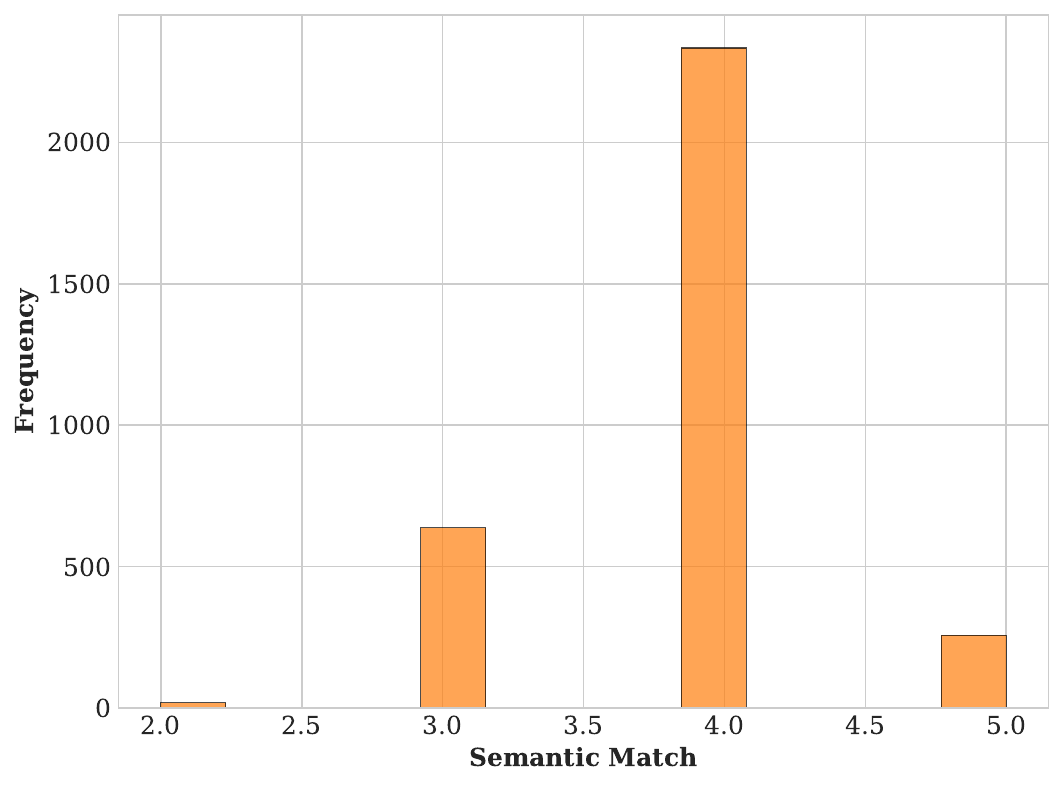}
  \subcaption{Professional Voice Actor}
\end{subfigure}\hfill
\begin{subfigure}[t]{0.48\linewidth}
  \centering
  \includegraphics[width=\linewidth]{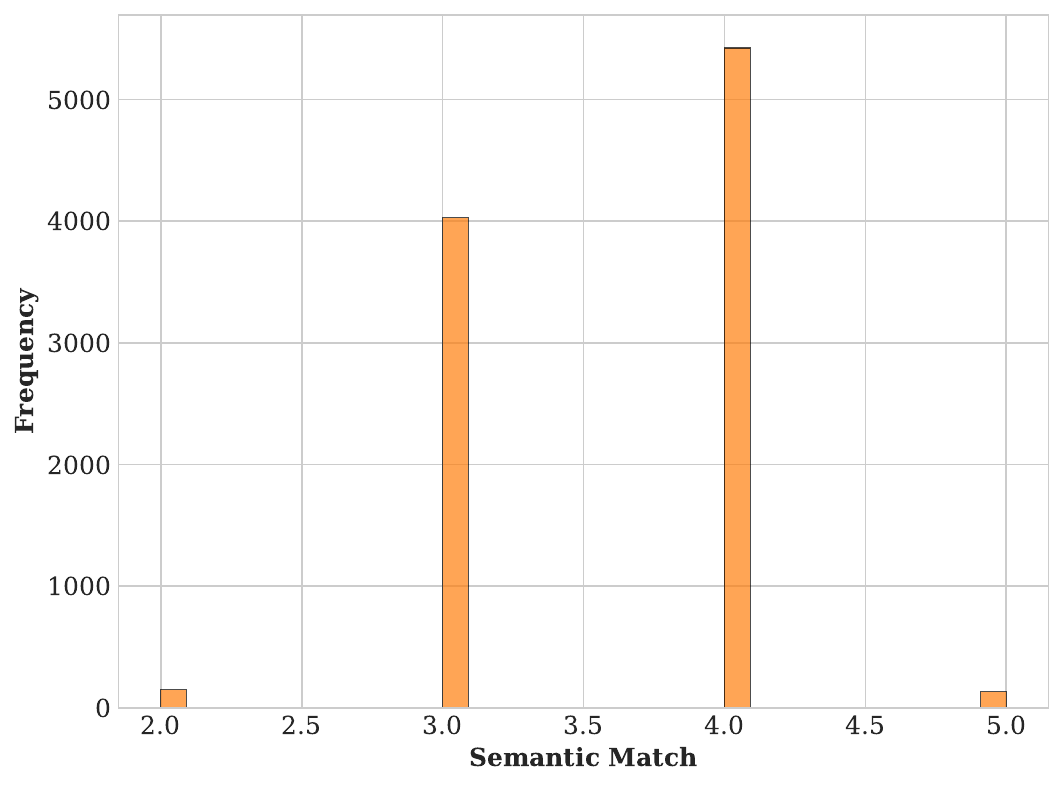}
  \subcaption{Amateur Voice Actor}
\end{subfigure}
\caption{Distribution of Semantic Match in Realism Evaluation: Professional vs. Amateur.}
\label{fig:cmp-semantic-match}
\end{figure}

\paragraph{Distributional characteristics.}
Fig.~\ref{fig:realism-histograms-1}--\ref{fig:realism-histograms-2} visualize the score distributions per dimension.
We highlight three patterns:
(i) \textbf{Progressive/gated emotion metrics.} For \emph{Emotion Intensity} and \emph{Emotion Transition}, we assign \texttt{1} when the scene is marked \emph{N/A} (i.e., emotion accuracy is insufficient to meaningfully judge intensity/transition). This produces a visible mass at $1$, a sparse bin at $2$, and slightly lower overall agreement; downstream analyses conditioning on adequate \emph{Emotion Accuracy} (e.g., mean~$\geq 3$) recover more bell-shaped distributions. This floor-imputation is used only for these descriptive histograms; the headline correlations in Sec.~\ref{sec:exp} use \emph{masking} (dropping gated entries), as noted in the caveats there.
(ii) \textbf{Traits Embodiment.} Scores tend to avoid the middle ($\sim\!3$), concentrating on low values when traits clearly mismatch and high values when traits are strongly realized, reflecting the inherently compositional and role-specific nature of trait judgments. (Its inter-rater agreement is nonetheless mid-range rather than the lowest; the two gated emotion dimensions have the lowest agreement on this split, Tab.~\ref{tab:audit-kappa}.)
(iii) \textbf{Other dimensions.} \emph{Pitch Variation}, \emph{Rhythmic Naturalness}, \emph{Stress \& Emphasis}, \emph{Voice Identity Matching}, \emph{Local Scene Fit}, \emph{Global Story Coherence}, and \emph{Semantic Match} are approximately unimodal with modes at $3$--$4$, consistent with moderate–strong realization in the dataset and with the high agreement reported.

\paragraph{Professional vs.\ Amateur voice actors.}
The paired histograms in Figs.~\ref{fig:cmp-pitch-variation}--\ref{fig:cmp-semantic-match} compare professional (left) and amateur (right) recordings:
\begin{itemize}\itemsep3pt
    \item \textbf{Prosody control (Pitch, Rhythm, Stress).} Professional speech is shifted toward higher bins with tighter spread, indicating steadier control and fewer low-end tails.
    \item \textbf{Emotion rendering (Accuracy, Intensity, Transition).} Professionals show higher centers and reduced mass at $1$ (fewer gated \texttt{N/A}s), suggesting more consistent emotional grounding and smoother dynamics; amateurs exhibit wider variance and more mass at the extremes.
    \item \textbf{Identity and traits (Voice Identity, Traits Embodiment).} Professionals present clearer peaks in upper bins, while amateurs are broader or bimodal, reflecting inconsistent character anchoring.
    \item \textbf{Scene and semantics (Local Scene Fit, Global Coherence, Semantic Match).} Differences are present but smaller; both groups cluster in the mid–high range, with professionals exhibiting slightly narrower distributions.
\end{itemize}
Overall, professionals outperform amateurs on expressivity and control, while narrative/semantic alignment shows a milder gap.

\paragraph{Implications for modeling.}
First, the \rev{generally moderate-to-substantial agreement (with some lower-agreement dimensions on real recordings)} suggests the rubric isolates perceptually coherent factors on most dimensions, so learned evaluators can target them reliably.
Second, the gated behavior of \emph{Intensity/Transition} motivates a \emph{hierarchical} evaluator: verify \emph{Emotion Accuracy} before scoring graded expressivity, or model the joint structure with conditional heads.
Third, \emph{Traits Embodiment} remains challenging; models may benefit from explicit character-trait priors, long-term context aggregation, or contrastive supervision between matched and mismatched traits.
Finally, prosody-focused objectives (e.g., pitch/rhythm/stress control) are likely to yield the largest quality gains for amateur-style speech.

\FloatBarrier
\section{Evaluation Model Experiments}
\subsection{Zeroshot/Fewshot Setup for Evaluation Models}
\label{app:zeroshot-evaluation-setup}

For both zero-shot and few-shot experiments, we adopt a version of the annotation guidelines to ensure consistency with the human annotation process. The exact rubric prompts for archetype and realism evaluation used are provided below:

\begin{tcolorbox}[colback=gray!5!white, colframe=black!75!white,
                  title={Rubrics Prompts (Archetype Evaluation)},
                  fonttitle=\bfseries, fontupper=\ttfamily, breakable]

Follow the rubric EXACTLY and return ONLY a minified JSON object (valid for json.loads) — no extra text. \\
Set \texttt{content\_pass} to \texttt{false} (i.e., reject) if ANY apply: \\
- Clip $\sim$1–2s (too short to judge). \\
- Obvious breakdown/loop/restart. \\
- Any multilingual word/phrase (except extremely common one-offs like ``croissant''). \\
- Irrelevant to prompt’s role/scene. \\
- Instruction failure: breaks persona/scene, 4th-wall narration, multi-party narration, long placeholder scaffolding. \\
Otherwise: PASS. \\\\

\textbf{WHAT TO DO} \\
1) Apply the rubrics below strictly. If torn between two scores, choose the LOWER one. \\
2) Return ONLY a minified JSON object (valid for \texttt{json.loads}). Integers 1–5. No extra text. \\
3) If not rejected, judge performance (HOW it’s said), not wording. \\
4) If rejected, set all scores to 1. \\\\

\textbf{0) Content-Pass} \\
If rejected, set to \texttt{false}, otherwise set to \texttt{true}. \\\\

\textbf{1) AudioQuality (Context-Free)} \\
Focus ONLY on digital/synthetic artifacts (noise, hiss, metallic timbre, clipping, glitchy transitions, pitch collapse, static). \\
After you record this score, PRETEND these issues do not exist for later dimensions. \\
\emph{Scale:} \\
- 5 (Excellent): Clean/clear; no noticeable artifacts. \\
- 4 (Good): Minor imperfections (slight hiss/sharpness) under close listen. \\
- 3 (Fair): Noticeable artifacts (buzz/echo/synthetic texture) that impact clarity. \\
- 2 (Poor): Frequent glitches/clipping/metallic resonance/harsh transitions. \\
- 1 (Very Poor): Dominated by artifacts (static/crackling/synthesis failure/severe clipping or pitch collapse). \\\\

\textbf{2) HumanLikeness (Context-Free)} \\
How human-like is the delivery? Ignore AudioQuality issues (already captured) and ignore contextual fit. \\
\emph{Scale:} \\
- 5 (Definitely Human): Indistinguishable; rich intonation, natural real-time pacing, consistent identity; organic breaths/pauses. \\
- 4 (Most Likely Human): Generally convincing; minor stiffness or overly even timing/pauses. \\
- 3 (Could be Human or AI): Mixed traits; some mechanical pauses/inflection, other parts human-like. \\
- 2 (Mostly Likely AI): Clearly artificial; choppy rhythm, flat/illogical inflections, awkward or forced breaths. \\
- 1 (Definitely AI): Rigid read-aloud monotony; no natural rhythm/pauses; entirely machine-like. \\\\

\textbf{3) Appropriateness (Context-Dependent)} \\
Tonal fit to the prompt’s role \& scene (stereotype-based). If ANY portion before 30s is ill-fitting, choose the LOWER score. \\
Ignore AudioQuality and generic naturalness here—judge only contextual tone/demeanor/status/urgency/emotion. \\
\emph{Scale:} \\
- 5 (Completely Appropriate): Fits both role AND scene perfectly; fully aligned with emotion, social dynamics, status, and events; immersive and consistent. \\
- 4 (Mostly Appropriate): Strong match with minor flaws (e.g., slight intensity/timing/subtlety issues) that do not break alignment. \\
- 3 (Adequately Appropriate): Broad fit to role/scene but lacks depth or shows mild inconsistencies in key elements. \\
- 2 (Slightly Appropriate): Only role OR scene fit is acceptable; superficial/partial alignment; missing/mismatched cues. \\
- 1 (Completely Inappropriate): Fails role AND scene; tone/status/urgency clearly mismatched. \\\\

\textbf{OUTPUT SCHEMA (minified JSON)} \\
\{\,"content\_pass":true/false,"audio\_quality":1-5,\\"human\_likeness":1-5,"appropriateness":1-5\,\}

\end{tcolorbox}

\begin{tcolorbox}[colback=gray!5!white, colframe=black!75!white,
                  title={Rubrics Prompts (Realism Evaluation)},
                  fonttitle=\bfseries, fontupper=\ttfamily, breakable]

\textbf{WHAT TO DO} \\
1. \textbf{Read first:} Global Profile $\Rightarrow$ Local Scene. \\
2. \textbf{Play} the audio \textbf{while skimming} the transcript (transcript may be imperfect). \\
3. \textbf{Focus only on the target speaker.} \\
\hspace*{1em} -- Ignore background SFX / music / other voices except where they influence the target character’s timing or tone. \\
4. \textbf{Apply the rubric below EXACTLY} and return \textbf{only} a minified JSON object – no prose, line-breaks, or extra keys (valid for \texttt{json.loads}). \\
5. \textbf{Scale is 1 (poor) … 5 (excellent).} Use integers only. \\\\

\textbf{DIMENSIONS \& RULES} \\
For every dimension, 5 = best possible, 1 = worst. Use these anchor descriptions to choose the closest integer. \\\\

\textbf{1) PitchDynamics} \\
5: Natural, situation-appropriate melodic movement; subtle, believable inflections. \\
4: Generally varied and fitting, minor stiffness or flat spots. \\
3: Neutral or limited inflection; neither distracting nor expressive. \\
2: Forced or sudden shifts that feel “performed.” \\
1: Monotone or blatantly mis-pitched for the scene. \\\\

\textbf{2) RhythmicNaturalness} \\
5: Flowing, lifelike rhythm; natural pauses. \\
4: Mostly smooth; slight hiccups. \\
3: Even pacing; read-aloud style. \\
2: Halting, mechanical, broken phrases. \\
1: Robotic timing, destroys immersion. \\\\

\textbf{3) StressEmphasis} \\
5: Perfect stress on key words; enhances meaning. \\
4: Generally correct with few mistakes. \\
3: Mostly flat; generic sentential stress. \\
2: Over-emphasis or misplaced stress. \\
1: Flat OR every word equally stressed. \\\\

\textbf{4) EmotionAccuracy} \\
5: Emotion fits precisely; subtle, authentic. \\
4: Appropriate but slightly under/overstated. \\
3: Neutral, misses clear opportunity. \\
2: Noticeably off (e.g., joking tone in crisis). \\
1: Jarring or nonsensical emotion. \\\\

\textbf{5) EmotionIntensity (only if 4 $\geq$ 3)} \\
5: Perfectly calibrated. \\
4: Minor overshoot/undershoot. \\
3: Mildly too strong/weak. \\
2: Clearly over- or under-acted. \\
1: Wildly inappropriate. \\\\

\textbf{6) EmotionalDynamicRange (only if 5 $\geq$ 3)} \\
5: Nuanced, organic shifts. \\
4: Variation present, somewhat constrained. \\
3: Mostly steady; slight modulation. \\
2: Flat OR abrupt jumps. \\
1: Chaotic, inconsistent emotion. \\\\

\textbf{7) VoiceIdentityMatching} \\
5: Perfectly matches profile. \\
4: Acceptable, minor mismatch. \\
3: Generic voice. \\
2: Noticeable mismatch. \\
1: Obviously wrong identity. \\\\

\textbf{8) TraitEmbodiment} \\
5: Core traits audible, believable. \\
4: Most traits present, one weak. \\
3: Traits faint, neutral delivery. \\
2: Traits clash or unclear. \\
1: No traits or fully off. \\\\

\textbf{9) LocalSceneFit} \\
5: Perfect alignment with objective \& mood. \\
4: Minor mismatch. \\
3: Neutral fit. \\
2: Somewhat confusing. \\
1: Contradicts scene. \\\\

\textbf{10) GlobalStoryFit} \\
5: Consistent with character arc. \\
4: Slightly atypical but plausible. \\
3: Generic filler. \\
2: Conflicts with backstory. \\
1: Breaks identity entirely. \\\\

\textbf{11) SemanticMatchness} \\
5: Perfectly advances scene. \\
4: Minor mismatch but acceptable. \\
3: Neutral filler. \\
2: Partially conflicts with scene. \\
1: Directly contradicts/derails. \\\\

\textbf{Penalty Rule:} If the speaker is clearly the wrong person, set \texttt{voice\_identity\_matching = 1} and \texttt{trait\_embodiment = 1}. \\\\

\textbf{OUTPUT SCHEMA (minified JSON)} \\
\{\, "pitch\_dynamics":1-5, "rhythmic\_naturalness":1-5, "stress\_emphasis":1-5, "emotion\_accuracy":1-5, "emotion\_intensity":1-5, "emotional\_dynamic\_range":1-5, "voice\_identity\_matching":1-5, "trait\_embodiment":1-5, "local\_scene\_fit":1-5, "global\_story\_fit":1-5, "semantic\_matchness":1-5 \}

\end{tcolorbox}

For zero-shot and few-shot settings, we condition the model with the rubric prompt and then append few-shot demonstration triplets (input, audio, gold JSON) to mirror the human annotation protocol. As shown in Listing~\ref{list:zeroshot-fewshot}, we use \texttt{Gemini2.5Pro} for these few-shot exemplars and require a minified JSON output consistent with the rubric schema.

\lstdefinestyle{pythonstyle}{
  language=Python,
  basicstyle=\ttfamily\small,
  numbers=left,
  numberstyle=\tiny,
  stepnumber=1,
  numbersep=5pt,
  frame=single,
  breaklines=true,
  breakatwhitespace=true,
  tabsize=2,
  showstringspaces=false,
  keywordstyle=\color{blue},
  commentstyle=\color{gray},
  stringstyle=\color{orange},
}

\begin{lstlisting}[style=pythonstyle, caption={Prompt construction for zeroshot and fewshot experiments (Gemini 2.5 Pro).},label=list:zeroshot-fewshot]
import json
from typing import List, Dict, Literal

Mode = Literal["archetype", "realism"]

def _minify_json(d: Dict) -> str:
    return json.dumps(d, separators=(",", ":"))

def _format_input_desc(sample: Dict, mode: Mode) -> str:
    if mode == "archetype":
        q = sample.get("question", "")
        return f"Prompt's role and scene: {q}"
    gp = sample.get("char_profile", "")
    ls = sample.get("local_scene", "")
    cs = sample.get("char_style", "")
    return f"Global Profile: {gp}; Local Scene: {ls}; Traits: {cs}"

def _audio_part_from_wav(path: str, wav_to_base64) -> Dict:
    b64 = wav_to_base64(path)
    return {
        "type": "input_audio",
        "input_audio": {"data": "data:audio/wav;base64," + b64, "format": "wav"},
    }

def build_prompt_with_examples(
    rubric_prompt: str,
    few_shot_examples: List[Dict],
    current_item: Dict,
    mode: Mode,
    wav_to_base64,
    model_name: str = "gemini-2.5-pro",
) -> List[Dict]:
    """Constructs rubric + few-shot + current eval into a single prompt."""
    parts: List[Dict] = []

    # 1) Rubric
    parts.append({"type": "text", "text": rubric_prompt.strip() + "\n\n"})

    # 2) Few-shot demonstrations
    for i, ex in enumerate(few_shot_examples, start=1):
        parts.append({"type": "text", "text": f"**Few-shot Example {i}**\n\n"})
        parts.append({"type": "text", "text": f"**Example Input:** {_format_input_desc(ex, mode)}\n\n"})
        parts.append(_audio_part_from_wav(ex["wav_path"], wav_to_base64))
        if "annotation_prompt" in ex:
            gold_json = _minify_json(ex["annotation_prompt"])
            parts.append({"type": "text", "text": f"\n\n**Example Output:** {gold_json}\n\n"})

    # 3) Current evaluation
    parts.append({"type": "text", "text": "**Current Evaluation**\n\n"})
    parts.append({"type": "text", "text": f"**Current Input:** {_format_input_desc(current_item, mode)}\n\n"})
    parts.append(_audio_part_from_wav(current_item["wav_path"], wav_to_base64))
    parts.append({"type": "text",
                  "text": "\n\nPlease evaluate the current audio and return ONLY a minified JSON."})

    return [
        {"role": "system", "content": f"You are a helpful assistant using {model_name}."},
        {"role": "user", "content": parts},
    ]
\end{lstlisting}

\subsection{Zeroshot/Fewshot Ablation Experiments}
\label{app:zeroshot-ablation}

Table~\ref{tab:role-play-fewshot-ablation} shows how different prompting strategies affect realism evaluation performance with \texttt{Gemini2.5Pro}. \rev{The averages in this prompting-ablation table ($0.353$/$0.390$/$0.433$/$0.459$/$0.469$) are reported under the earlier floor-\emph{imputation} convention and are \emph{not} directly comparable to the masking-convention headline numbers in Sec.~\ref{ssec:main-results}; they are retained only to show the \emph{relative} effect of run-averaging and demonstration count within a fixed convention.} Several observations can be drawn:

\begin{itemize}
    \item Effect of repeated runs.
When comparing single-run decoding (\texttt{Zeroshot-1run}) against averaging over five runs (\texttt{Zeroshot}), we see consistent gains across nearly all dimensions (average score improves from 0.353 to 0.390). This suggests that sampling variance plays a substantial role in the reliability of large language model (LLM) evaluators, and averaging multiple outputs reduces this variance, producing more stable correlation with human annotations.
    \item Impact of few-shot conditioning.
Introducing even three examples (\texttt{3-Shot}) yields a clear improvement over zero-shot (0.433 vs. 0.390). This indicates that providing explicit annotated demonstrations helps the model better align with the rubric. Interestingly, when using only a single run with few-shot demonstrations (\texttt{3-Shot-1run}), the performance is only marginally better than pure zeroshot, highlighting that both averaging and demonstrations are complementary: demonstrations improve grounding, while multiple runs mitigate stochasticity.
    \item Scaling the number of demonstrations.
Increasing the few-shot set size from three to six and nine further boosts performance (0.459 and 0.469, respectively). The improvements are most pronounced in prosodic and emotional categories (e.g., \emph{Pitch}, \emph{Accuracy}, \emph{Intensity}), suggesting that more examples provide stronger guidance in modeling subtle paralinguistic cues. Gains in \emph{Character Consistency} and \emph{Contextual Relevance} are also noticeable, although smaller in magnitude, which may indicate that these higher-level judgments are less sensitive to prompt scaling once a few demonstrations are provided.
\end{itemize}

In general, the results demonstrate two complementary strategies for stabilizing LLM-based evaluators: (i) averaging multiple runs to reduce randomness, and (ii) enriching prompts with more few-shot exemplars to improve adherence to rubric-based evaluation. Combining both leads to the strongest overall performance, with the nine-shot condition achieving the highest average correlation~(0.469).

\begin{table}[!t]
    \centering
    \caption{Realism role-play evaluator performance with \texttt{Gemini2.5Pro} in different few-shot conditions. Here, ``1run'' denotes a single generation run. \rev{Averages are reported under the earlier floor-\emph{imputation} convention for the gated emotion columns and are therefore \emph{not} directly comparable to the masking-convention headline numbers in Sec.~\ref{ssec:main-results}; they are retained only to show the \emph{relative} effect of run-averaging and demonstration count within a fixed convention.}}
\resizebox{\linewidth}{!}{
    \begin{tabular}{l|c|c|c|c|c|c|c|c|c|c|c}
    \toprule
    \multirow{2}{*}{Config.} & \multicolumn{3}{|c|}{Prosodic Dynamics} & \multicolumn{3}{|c|}{Emotional Expressiveness} & \multicolumn{2}{|c|}{Character Consistency} & \multicolumn{2}{|c|}{Contextual Relevance} & \multirow{2}{*}{Avg.} \\
    \cmidrule{2-11}
    & Pitch & Rhythm & Stress & Accuracy & Intensity & Transition & Identity & Traits & Local & Global &  \\
    \midrule
        \texttt{Zeroshot-1run} & 0.277 & 0.325 & 0.303 & 0.355 & 0.286 & 0.295 & 0.480 & 0.475 & 0.308 & 0.424 & 0.353 \\
        \texttt{Zeroshot} & 0.319 & 0.352 & 0.336 & 0.391 & 0.316 & 0.336 & 0.529 & 0.509 & 0.349 & 0.465 & 0.390 \\
       \texttt{3-Shot-1run}& 0.316 & 0.335 & 0.323 & 0.358 & 0.279 & 0.294 & 0.530 & 0.476 & 0.335 & 0.428 & 0.367 \\
       \texttt{3-Shot} & 0.400 & 0.450 & 0.432 & 0.414 & 0.344 & 0.353 & 0.584 & 0.505 & 0.374 & 0.477 & 0.433 \\
        \texttt{6-Shot} & \textbf{0.419} & 0.431 & 0.427 & 0.455 & 0.393 & 0.391 & 0.606 & 0.540 & 0.404 & 0.519 & 0.459 \\
        \texttt{9-Shot} & \textbf{0.419} & \textbf{0.453} & \textbf{0.434} & \textbf{0.464} & \textbf{0.404} & \textbf{0.416} & \textbf{0.613} & \textbf{0.551} & \textbf{0.407} & \textbf{0.524} & \textbf{0.469} \\
        
    \bottomrule
    \end{tabular}}
    \label{tab:role-play-fewshot-ablation}
\end{table}

\FloatBarrier
\section{\rev{Additional Analyses and Controls}}
\label{app:tier1}

\rev{This appendix reports human-reference agreement, negative discrimination, negative-supervision ablations, system-level ranking, statistical analyses, and robustness controls that complement the headline correlations of Sec.~\ref{ssec:main-results}. Unless a step is explicitly identified as a retraining experiment, the analyses are computed from the already-collected human annotations and model predictions. For the realism benchmark-distribution test, each per-dimension correlation is computed over the records with a usable human label on that dimension: ${\approx}1{,}966$ of the $2{,}000$ records for the eight non-gated dimensions and ${\approx}1{,}796$ for \emph{Emotion Intensity}/\emph{Transition}, where the masking convention drops gated entries.}

\paragraph{\rev{Single-rater human agreement (a reference level).}} \rev{Because the Pearson values in Tabs.~\ref{tab:role-play-exp-archetype}--\ref{tab:role-play-exp-realism-real} are model-vs-consensus correlations, a natural human reference is how well a \emph{single human} agrees with the consensus of the remaining raters. We call this a reference level rather than a formal upper bound: a model could in principle exceed a single rater, so ``percentage of the reference'' below is a descriptive comparison, not a fraction of a proven ceiling. For every sample we hold out one rater and correlate that rater's score against the mean of the rest, pooling the (held-out, consensus) pairs over all raters and samples; this \emph{single-vs-rest} Pearson is reported per dimension in Tab.~\ref{tab:tier1-ceiling}, alongside ordinal Krippendorff's $\alpha$ (a chance-corrected inter-rater agreement), the final DRAME-Eval correlation, and DRAME-Eval relative to the single-rater level. On benchmark-distribution data, DRAME-Eval reaches $76$--$94\%$ of the single-rater level across the ten realism dimensions (median $\approx\!88\%$): it is already close to what a single human achieves. On the clean authentic-speech set, however, its correlation is near or below zero on most dimensions ($-0.041$ on \emph{Voice Identity} at single-rater level $0.48$, and negative on the prosodic/emotional dimensions), \emph{while single-rater agreement on those dimensions remains substantial}; the model retains signal only on \emph{Trait Embodiment} ($+0.419$ against $0.58$). Humans thus still agree on realism in genuine acted speech where the model does not, which reframes the synthetic-to-authentic gap as a model limitation rather than a task that is simply too hard. On the archetype set the single-vs-rest level is $0.761$/$0.756$/$0.724$ for Audio Quality / Human Likeness / Appropriateness, with DRAME-Eval at approximately $90\%$/$90\%$/$73\%$. The single-rater level and the DRAME-Eval column in Tab.~\ref{tab:tier1-ceiling} both use the released (masked) labels and coincide with the main realism table (Tab.~\ref{tab:role-play-exp-realism}), which adopts the same masking convention across all ten dimensions (including the two gated emotion dimensions); the floor-imputation variant is reported only as the secondary sensitivity analysis in this appendix.}

\begin{table}[!t]
    \centering
    \caption{\rev{\textbf{Single-rater human agreement for realism evaluation} (pooled over languages; the authentic-speech set is Mandarin-only). \emph{Single-rater} is the single-rater-vs-consensus Pearson, a human reference level (not a formal upper bound); \emph{K-$\alpha$} is ordinal Krippendorff's $\alpha$; \emph{DRAME-Eval} is the final Instruct-LoRA model--consensus Pearson; \emph{\%s.r.}\ is DRAME-Eval relative to the single-rater level. All cells are point estimates; representative $95\%$ sample-level bootstrap half-widths on the per-dimension single-vs-rest values are ${\approx}\pm0.02$, and per-model $95\%$ CIs on the within-language macro-averages are tabulated in Tab.~\ref{tab:tier1-perlang}, together with the EN/ZH point estimates.}}
    \label{tab:tier1-ceiling}
{\color{revcolor}\small
\begin{minipage}{\linewidth}
    \centering
    \textbf{(a) Benchmark distribution (\texttt{test\_base})}\\[2pt]
    \begin{tabular}{ll|cccc}
    \toprule
    Group & Dim. & \makecell{Single-\\rater} & K-$\alpha$ & \makecell{DRAME-\\Eval} & \%\,s.r. \\
    \midrule
    \multirow{3}{*}{\makecell[l]{Prosodic\\Dynamics}}
      & Pitch  & 0.693 & 0.580 & 0.621 & 90 \\
      & Rhythm & 0.687 & 0.565 & 0.627 & 91 \\
      & Stress & 0.672 & 0.547 & 0.631 & 94 \\
    \midrule
    \multirow{3}{*}{\makecell[l]{Emotional\\Express.}}
      & Accuracy   & 0.696 & 0.562 & 0.632 & 91 \\
      & Intensity  & 0.623 & 0.476 & 0.494 & 79 \\
      & Transition & 0.614 & 0.464 & 0.469 & 76 \\
    \midrule
    \multirow{2}{*}{\makecell[l]{Character\\Consist.}}
      & Identity & 0.845 & 0.735 & 0.660 & 78 \\
      & Traits   & 0.752 & 0.636 & 0.655 & 87 \\
    \midrule
    \multirow{2}{*}{\makecell[l]{Contextual\\Relevance}}
      & Local  & 0.734 & 0.644 & 0.563 & 77 \\
      & Global & 0.757 & 0.642 & 0.668 & 88 \\
    \midrule
    \multicolumn{2}{l|}{\textbf{Average}} & 0.707 & 0.585 & 0.602 & 85 \\
    \bottomrule
    \end{tabular}
\end{minipage}

\vspace{6pt}

\begin{minipage}{\linewidth}
    \centering
    \textbf{(b) Real recordings (Mandarin)}\\[2pt]
    \begin{tabular}{ll|cccc}
    \toprule
    Group & Dim. & \makecell{Single-\\rater} & K-$\alpha$ & \makecell{DRAME-\\Eval} & \%\,s.r. \\
    \midrule
    \multirow{3}{*}{\makecell[l]{Prosodic\\Dynamics}}
      & Pitch  & 0.603 & 0.446 & -0.357 & -59 \\
      & Rhythm & 0.453 & 0.287 & -0.051 & -11 \\
      & Stress & 0.554 & 0.385 & -0.228 & -41 \\
    \midrule
    \multirow{3}{*}{\makecell[l]{Emotional\\Express.}}
      & Accuracy   & 0.323 & 0.182 & -0.161 & -50 \\
      & Intensity  & 0.756 & 0.674 & -0.333 & -44 \\
      & Transition & 0.722 & 0.625 & -0.291 & -40 \\
    \midrule
    \multirow{2}{*}{\makecell[l]{Character\\Consist.}}
      & Identity & 0.480 & 0.315 & -0.041 & -9 \\
      & Traits   & 0.582 & 0.474 & 0.419 & 72 \\
    \midrule
    \multirow{2}{*}{\makecell[l]{Contextual\\Relevance}}
      & Local  & 0.636 & 0.472 & -0.262 & -41 \\
      & Global & 0.502 & 0.341 & -0.118 & -24 \\
    \midrule
    \multicolumn{2}{l|}{\textbf{Average}} & 0.561 & 0.420 & -0.142 & -25 \\
    \bottomrule
    \end{tabular}
\end{minipage}}
\end{table}

\paragraph{\rev{Positive-vs-negative discrimination.}} \rev{The contrastive negatives (Sec.~\ref{sec:realism-data}) are designed to test whether a judge attends to delivery, role, and scene cues, but Sec.~\ref{ssec:main-results} reports a single Pearson correlation over all scored records (positives and negatives alike) and does not isolate how sharply each judge separates the two classes. Tab.~\ref{tab:tier1-negdisc} adds that view on the benchmark-distribution realism set ($210$ positives, $832$ role-mismatch and $958$ context-mismatch negatives; each record is a distinct source clip---the negatives are not resampled from an in-set positive's audio but are separate clips paired with a perturbed \emph{text} component (profile/scene), the spoken audio left unmodified; the underlying audio pool is mixed real/TTS, not uniformly real). For each negative type we measure, on the dimension(s) the perturbation targets (\emph{Voice Identity}+\emph{Traits} for role, \emph{Local}+\emph{Global} scene fit for context), how well each model's score separates positives from negatives (AUROC) and the positive$-$negative score gap. Human raters clearly down-score the negatives (gap $1.304$ for role, $0.536$ for context; Mann--Whitney $p<10^{-25}$). The open-source judges are essentially at chance (role AUROC $0.51$/$0.55$ for \texttt{KimiAudio}/\texttt{Qwen2Audio}), i.e.\ they do not register the mismatch at all, whereas DRAME-Eval attains the highest role-mismatch AUROC ($0.872$). On context-mismatch, however, \texttt{Qwen2.5Omni} ($0.851$, open-source), \texttt{GPT4o-audio} ($0.815$), and \texttt{GPT4o-audio} few-shot ($0.709$) all discriminate better than DRAME-Eval ($0.679$), so DRAME-Eval is best on role but not on context discrimination. The delivery-mismatch negatives are not included here: saved predictions exist only for DRAME-Eval, and the released delivery set contains no in-set positive reference, so a controlled AUROC would conflate synthetic-vs-real with the perturbation. This analysis shows the negatives are learnable and graded; on its own it does \emph{not} isolate the marginal value of \emph{training} with negatives. We therefore ran the complementary ablation, retraining \texttt{DRAME-Eval} with the negatives withheld while holding the positive set and the optimizer budget fixed ($2{,}000$ LoRA steps for every configuration, so differences reflect training-data composition rather than training length; absolute correlations consequently sit below the final $10$k-step model, and only within-family deltas are interpretable). Tab.~\ref{tab:tier1-negablate} reports the result: withholding \emph{all} negatives (positives only) inverts positive-vs-negative discrimination, from AUROC $0.757$ to $0.404$ (below chance), and lowers the average benchmark-distribution correlation from $0.432$ to $0.249$; removing a single branch is intermediate, and the context-coherence (\emph{global-story}) negatives carry more of the discrimination signal than the local-story negatives ($0.597$ versus $0.696$ AUROC when the respective branch is dropped). Under this training setup, negative-containing training is important for learning the tested positive-versus-negative role-fit distinction. Because the positive-only condition also contains less training data, this experiment does not fully separate the effect of negative supervision from that of training-set size and diversity, so the below-chance AUROC (a qualitative sign inversion) is a more robust read than the per-branch Pearson differences.}

\begin{table}[!t]
    \centering
    \caption{\rev{\textbf{Positive-vs-negative discrimination} on the benchmark-distribution realism set. \emph{AUROC} uses each model's score on the targeted dimension(s) to separate positives ($n{=}210$) from negatives; $0.5$ is chance. \emph{Gap} is the positive$-$negative mean score on those dimensions. For reference, the human gaps are $0.536$ (context) and $1.304$ (role). Each benchmark-distribution test record is a distinct source clip: the $210$ positives and $1{,}790$ negatives are drawn from disjoint source segments (a served negative retains its own segment's original stored audio, which may be real or TTS; the matching-text positive counterpart of that exact segment is not itself a second record in this set), so the AUROC contrasts the $210$ positives against the negatives across distinct clips rather than within an audio family. We report point estimates only.}}
    \label{tab:tier1-negdisc}
{\small\color{revcolor}
    \begin{tabular}{l|l|c|c|c|c}
    \toprule
    \multirow{2}{*}{Config.} & \multirow{2}{*}{Model} & \multicolumn{2}{c|}{Context mismatch} & \multicolumn{2}{c}{Role mismatch} \\
    \cmidrule{3-6}
    & & AUROC & Gap (pos$-$neg) & AUROC & Gap (pos$-$neg) \\
    \midrule
    \multirow{5}{*}{Zeroshot}
      & \texttt{Gemini2.5Pro} & 0.665 & 0.725 & 0.821 & 1.681 \\
      & \texttt{GPT4o-audio}  & 0.815 & 0.560 & 0.847 & 1.483 \\
      & \texttt{KimiAudio}    & 0.493 & -0.002 & 0.512 & 0.010 \\
      & \texttt{Qwen2Audio}   & 0.557 & 0.058 & 0.555 & 0.056 \\
      & \texttt{Qwen2.5Omni}  & 0.851 & 0.845 & 0.782 & 0.683 \\
    \midrule
    \multirow{2}{*}{Fewshot}
      & \texttt{Gemini2.5Pro} & 0.645 & 0.602 & 0.839 & 1.682 \\
      & \texttt{GPT4o-audio}  & 0.709 & 0.472 & 0.859 & 1.213 \\
    \midrule
    Finetune & \texttt{DRAME-Eval} & 0.679 & 0.312 & \textbf{0.872} & 1.000 \\
    \bottomrule
    \end{tabular}}
\end{table}

\begin{table}[!t]
    \centering
    \caption{\rev{\textbf{Contribution of negative supervision during training.} Each row retrains \texttt{DRAME-Eval} from the same base under an identical $2{,}000$-step LoRA budget, differing only in which negative branches are retained; the positive set is held constant. \emph{pos/neg AUROC} separates the held-out positives from the \texttt{weak\_*} negatives ($0.5$ is chance); \emph{Avg.\ Pearson} is the mean over the $10$ realism dimensions on the benchmark-distribution test set. Because the budget is matched, absolute levels sit below the final $10$k-step model ($0.602$) and only within-family deltas are meaningful; the positives-only split is also smaller, so the qualitative below-chance AUROC is a more robust read than the per-branch Pearson differences. Point estimates only.}}
    \label{tab:tier1-negablate}
{\small\color{revcolor}
    \begin{tabular}{l|c|c|c}
    \toprule
    Training data & \# train & pos/neg AUROC & Avg.\ Pearson \\
    \midrule
    Full (positives $+$ both negative branches) & $12{,}070$ & \textbf{0.757} & \textbf{0.432} \\
    \quad without local-story negatives         & $7{,}894$  & 0.696 & 0.400 \\
    \quad without global-story negatives        & $7{,}760$  & 0.597 & 0.328 \\
    Positives only (no negatives)               & $3{,}584$  & 0.404 & 0.249 \\
    \bottomrule
    \end{tabular}}
\end{table}

\paragraph{\rev{System-level ranking.}} \rev{Used as an automatic judge, a model's practical value is whether it ranks \texttt{SFnd} systems the way humans do. We aggregate predictions to per-system means over the eight archetype systems and correlate them with the human per-system means; Tab.~\ref{tab:tier1-sysrank} reports the system-level Pearson, Spearman, and Kendall-$\tau$ averaged over the three archetype dimensions. Despite non-negative \emph{sample}-level Pearson (${\approx}0.23$ for the zero-shot proprietary judges under masking, Tab.~\ref{tab:role-play-exp-archetype}), the zero-/few-shot judges do not produce a reliable \emph{system} ranking: their coefficients span roughly $-0.32$ to $0.40$ with inconsistent sign across the three measures (e.g.\ \texttt{Gemini2.5Pro} zero-shot Spearman $-0.32$; \texttt{Qwen2Audio} Pearson $0.39$ but Spearman/Kendall near or below zero). \texttt{DRAME-Eval} shows the strongest system-level agreement across all three coefficients (Pearson $0.74$, Spearman $0.50$, Kendall $0.41$); \texttt{KimiAudio} is also positive on all three but markedly weaker ($0.23$/$0.40$/$0.21$), while the remaining judges are inconsistent in sign. With only eight systems this is a small-$n$, exploratory comparison, so we report all three coefficients rather than rely on any single one or on an interval estimate; the qualitative separation between DRAME-Eval and the judges is consistent across them. DRAME-Eval has seen these systems' style during training (only role$\times$scene pairs are held out across splits), so part of this advantage is benchmark-distribution familiarity at the system level.}

\begin{table}[!t]
    \centering
    \caption{\rev{\textbf{System-level ranking} on the archetype test set ($8$ \texttt{SFnd} systems). Each value is the correlation between a model's per-system mean and the human per-system mean, averaged over Audio Quality / Human Likeness / Appropriateness. \emph{P/S/K} = Pearson / Spearman / Kendall-$\tau$. Sample-level archetype averages (Tab.~\ref{tab:role-play-exp-archetype}) are shown for contrast. $^\dagger$\texttt{Qwen2Audio} returns parseable continuous scores on only ${\sim}13/1{,}499$ archetype items, so its per-system means---and hence these ranking coefficients---rest on very few items and are unreliable; its sample-level average is omitted for the same reason.}}
    \label{tab:tier1-sysrank}
{\small\color{revcolor}
    \begin{tabular}{l|l|c|c|c|c}
    \toprule
    Config. & Model & Sys.\ P & Sys.\ S & Sys.\ K & Sample-level Avg. \\
    \midrule
    \multirow{5}{*}{Zeroshot}
      & \texttt{Gemini2.5Pro} & -0.061 & -0.317 & -0.190 & \rev{0.236} \\
      & \texttt{GPT4o-audio}  & -0.153 & 0.040 & 0.119 & \rev{0.233} \\
      & \texttt{KimiAudio}    & 0.233 & 0.397 & 0.214 & \rev{0.070} \\
      & \texttt{Qwen2Audio}$^\dagger$   & 0.393 & -0.012 & -0.079 & \rev{--} \\
      & \texttt{Qwen2.5Omni}  & -0.110 & -0.103 & -0.048 & \rev{-0.008} \\
    \midrule
    \multirow{2}{*}{Fewshot}
      & \texttt{Gemini2.5Pro} & 0.045 & -0.103 & -0.167 & \rev{0.158} \\
      & \texttt{GPT4o-audio}  & -0.106 & -0.190 & -0.167 & \rev{-0.102} \\
    \midrule
    Finetune & \texttt{DRAME-Eval} & \textbf{0.739} & \textbf{0.500} & \textbf{0.405} & 0.629 \\
    \bottomrule
    \end{tabular}}
\end{table}

{\color{revcolor}
\subsection{Statistical analyses and additional controls}
\label{app:audit}
This appendix reports confidence intervals and paired tests, metric-consistency checks, per-language breakdowns, shortcut controls, and negative-sample validation. All are offline recomputations from the experiment annotations and model predictions unless a GPU re-run is explicitly noted; our loaders reproduce the reported point estimates.

\paragraph{Bootstrap confidence intervals and paired tests.} Using $2{,}000$ \emph{sample-level} bootstrap resamples (under the masking sensitivity), the benchmark-distribution gains are significant: on the realism base set \texttt{DRAME-Eval} scores $0.602$ $[0.573,0.627]$, and the paired advantage over the strongest few-shot baseline, \texttt{Gemini2.5Pro} few-shot ($0.387$), is $\Delta=+0.215$ $[0.178,0.250]$; over the weaker \texttt{GPT4o-audio} few-shot ($0.305$) it is larger, $\Delta=+0.297$ $[0.256,0.336]$, and over \texttt{Gemini2.5Pro} zero-shot $+0.249$ (all $p\le0.001$, i.e.\ no resample reversed the sign). On the clean authentic-speech set the ordering is different: \texttt{GPT4o-audio} few-shot has the highest pooled point estimate ($0.226$ $[0.190,0.262]$), and \texttt{DRAME-Eval} ($-0.142$ $[-0.160,-0.125]$ sample-level) does not improve on it; under the sample-level paired bootstrap on the common $1{,}084$ samples \texttt{DRAME-Eval} is lower by $\Delta=-0.369$ $[-0.397,-0.343]$ ($p\le0.001$), while the four-speaker clustered interval on that difference is reported below, so the sample-level and speaker-level comparisons must be read separately. With only four held-out speakers, the clustered bootstrap is necessarily wide: \texttt{DRAME-Eval}'s Avg CI becomes $[-0.29,+0.07]$, and the paired-difference interval becomes $[-0.56,+0.04]$. The latter includes zero, making the speaker-level comparison inconclusive; thus, neither analysis provides evidence of a positive \texttt{DRAME-Eval} advantage. Because samples can share a source series (and thus speakers and production style), we also ran a clustered bootstrap that resamples by source series rather than by sample. It widens the interval only modestly and leaves the benchmark-distribution gain strongly significant: the paired advantage over the strongest few-shot baseline (\texttt{Gemini2.5Pro}) is $\Delta=+0.213$ $[0.174,0.239]$ ($p\le0.001$; \texttt{DRAME-Eval} Avg $[0.560,0.627]$), versus the sample-level $[0.178,0.250]$ (the corresponding advantage over \texttt{GPT4o-audio} few-shot is larger, $\Delta=+0.294$ $[0.225,0.337]$). The realism base test assigns a distinct source clip to every record ($2{,}000$ records over $2{,}000$ unique source clips; the negatives are separate clips paired with perturbed \emph{text}, not audio twins of an in-set positive), so there is no positive/negative audio family finer than the source series, and the series-clustered bootstrap above is the appropriate coarser check (App.~\ref{app:audit} leakage analysis). For the authentic-speech set the relevant clustering axis is the four held-out speakers, addressed directly by the between-speaker decomposition above (within-speaker $r\approx+0.03$). Per-language estimates and macro-average CIs are reported for every realism model and for \texttt{DRAME-Eval} on archetype (Tab.~\ref{tab:tier1-perlang}); Tab.~\ref{tab:tier1-ceiling} reports per-dimension point estimates and the representative bootstrap spread (${\approx}\pm0.02$). With $2{,}000$ resamples the finest resolvable two-sided $p$ is ${\approx}0.001$, so we report empirical $p\le0.001$ throughout and note that no resample reversed the sign.

\paragraph{Why a held-out set is required, and speaker decomposition.} The final \texttt{DRAME-Eval} is an in-domain-adapted model whose training data includes $1{,}080$ in-house real recordings, so it must be evaluated on data with no training overlap; its proper held-out set is \texttt{test\_real\_v2} ($1{,}084$ samples, zero training overlap), on which the average Pearson is $-0.142$ (a base-only model with no real-recording training data scores $-0.024$ over the same pool). The human labels moreover carry substantial \emph{between-speaker} structure: a speaker-mean oracle (predicting each speaker's mean human score) attains $0.538$, so a positive pooled correlation alone cannot establish within-speaker discrimination. For \texttt{DRAME-Eval}, the measured mean within-speaker correlation is only $+0.029$ (per speaker $-0.09$ to $+0.12$). To make the human reference comparable, we recompute single-rater-vs-consensus agreement under the \emph{same} within-speaker protocol (per-speaker single-vs-rest Pearson, averaged over the four held-out speakers with ${\ge}270$ utterances): humans reach $+0.34$ within speaker (per speaker $+0.20$ to $+0.50$), well above the model's $+0.029$. The pooled single-rater figure ($0.561$) is higher because it also absorbs between-speaker variance, so the like-for-like human--model gap is the within-speaker $+0.34$ vs $+0.029$, and the pooled $0.561$ should be read only against the pooled model number ($-0.142$). A matched control that varies only train/test speaker overlap (retraining with disjoint vs.\ overlapping speaker sets) changes held-out correlation by $-0.01$ to $+0.03$, i.e.\ speaker overlap has essentially no effect, so we do not describe the model as ``speaker-calibrated''.

\paragraph{Metric consistency: masking vs.\ imputation.} The main tables (Tabs.~\ref{tab:role-play-exp-archetype}--\ref{tab:role-play-exp-realism}) report every row under masking, recomputed from the saved predictions. Table~\ref{tab:audit-archmask} places those masked archetype averages next to the floor-\emph{imputation} averages as a sensitivity check: under imputation the baselines rise sharply (e.g.\ \texttt{Gemini2.5Pro} zero-shot $0.236\!\to\!0.456$) and \texttt{GPT4o-audio} few-shot flips from \emph{negative} ($-0.102$) to $0.331$, i.e.\ imputation to the score floor hides a real baseline failure. Masking is therefore both the reproducible convention and the one \emph{less} favorable to \texttt{DRAME-Eval}'s lead, so the reported gap is conservative. For the realism base set, the release-reproducible sensitivity floors missing human labels to $1$ while leaving model predictions unchanged: \texttt{DRAME-Eval} has Avg $0.602$ under masking and $0.614$ under human-only imputation (Intensity $0.494\!\to\!0.552$; Transition $0.469\!\to\!0.536$). The earlier internal ablation in Tab.~\ref{tab:role-play-realism-basic-ablation} also floored the model side and reports Avg $0.625$; that legacy convention is not recoverable from the release-native predictions alone and is not the same sensitivity analysis. \rev{The archetype baseline predictions in Tab.~\ref{tab:audit-archmask} reproduce these masked averages exactly from the saved files, so the \texttt{DRAME-Eval} lead carries a paired bootstrap test on the common $1{,}312$ archetype items (shared resample indices on per-dimension common ids): $\Delta=+0.393$ $[0.338,0.445]$ over \texttt{Gemini2.5Pro} zero-shot ($0.236$), $+0.396$ $[0.352,0.439]$ over \texttt{GPT4o-audio} zero-shot ($0.233$), and larger against the few-shot judges, all with $p\le0.001$ (no resample of $2{,}000$ reversed the sign). The archetype benchmark-distribution advantage is thus significant under the same paired procedure used for realism, not merely a point-estimate gap.}

\begin{table}[!t]
\centering\small
\caption{\textbf{Archetype under a consistent masking convention} (recomputed from saved predictions). Imputed averages are shown for reference. Baseline intervals use $1{,}000$ bootstrap resamples; the final-model interval uses the $2{,}000$-resample primary bootstrap analysis. $^\dagger$\texttt{Qwen2Audio} zero-shot has only ${\sim}13$ usable masked items (${\sim}15$ under imputation), so both averages are omitted as unreliable.}
\label{tab:audit-archmask}
\begin{tabular}{l|c|c}
\toprule
Model & Avg.\ (mask) [95\% CI] & Avg.\ (impute) \\
\midrule
\texttt{DRAME-Eval} (Instruct-LoRA) & $\mathbf{0.629}$ $[0.593,0.663]$ & $0.694$ \\
\texttt{Gemini2.5Pro} (ZS) & $0.236$ $[0.194,0.280]$ & $0.456$ \\
\texttt{GPT4o-audio} (ZS) & $0.233$ $[0.201,0.266]$ & $0.427$ \\
\texttt{KimiAudio} (ZS) & $0.070$ $[0.029,0.108]$ & $0.105$ \\
\texttt{Qwen2.5Omni} (ZS) & $-0.008$ $[-0.053,0.038]$ & $0.013$ \\
\texttt{Gemini2.5Pro} (FS) & $0.158$ $[0.116,0.196]$ & $0.464$ \\
\texttt{GPT4o-audio} (FS) & $-0.102$ $[-0.146,-0.059]$ & $0.331$ \\
\texttt{Qwen2Audio} (ZS)$^\dagger$ & --- & --- \\
\bottomrule
\end{tabular}
\end{table}

\paragraph{Per-language breakdowns.} Pooled numbers can hide language-specific behavior, so we report English/Mandarin separately with a macro-average. On the realism base set \texttt{DRAME-Eval} scores EN $0.611$ / ZH $0.570$ (macro $0.591$, close to the $0.602$ pool), and it leads every baseline in both languages; the proprietary judges lean Mandarin (e.g.\ \texttt{Gemini2.5Pro} zero-shot EN $0.292$ / ZH $0.431$). The archetype set shows a pooling effect worth flagging: \texttt{DRAME-Eval} pools to $0.629$ but is EN $0.474$ / ZH $0.305$ (macro $0.389$), because an aligned between-language mean shift inflates the pooled correlation; the per-language numbers are the cleaner measure of within-language discrimination. The authentic-speech set is Mandarin-only, so its numbers carry no English component. The per-model EN and ZH averages (point estimates) together with the macro-average and its $95\%$ sample-level bootstrap CI are given in Tab.~\ref{tab:tier1-perlang} for the realism set (all models) and for the DRAME-Eval archetype row (we report the CI on the macro-average only, not on the individual EN/ZH point estimates); the per-language archetype \emph{baseline} predictions were not reconstructable from the saved files, so the archetype per-language comparison is limited to DRAME-Eval rather than being a full all-model comparison.

\begin{table}[t]
\centering
\caption{\rev{\textbf{Per-language breakdown} of the pooled realism base-set averages (Tab.~\ref{tab:role-play-exp-realism}) and the \texttt{DRAME-Eval} archetype average (Tab.~\ref{tab:role-play-exp-archetype}), under the masking convention. Each cell is the unweighted mean Pearson over the ten realism (three archetype) dimensions; \emph{Macro} is the mean of EN and ZH with its $95\%$ sample-level bootstrap CI. The archetype baseline rows are omitted because their saved predictions do not reproduce cleanly per language (see the metric-consistency note); the authentic-speech set is Mandarin-only and has no English component.}}
\label{tab:tier1-perlang}
{\small\color{revcolor}
\begin{tabular}{l|l|c|c|c}
\toprule
Task & Model (config) & EN Avg & ZH Avg & Macro Avg [95\% CI] \\
\midrule
\multirow{8}{*}{\makecell[l]{Realism\\base}}
  & \texttt{DRAME-Eval} (FT)        & \textbf{0.611} & \textbf{0.570} & \textbf{0.591 [0.568, 0.613]} \\
  & \texttt{Gemini2.5Pro} (ZS)      & 0.292 & 0.431 & 0.362 [0.332, 0.389] \\
  & \texttt{GPT4o-audio} (ZS)       & 0.220 & 0.348 & 0.284 [0.253, 0.313] \\
  & \texttt{KimiAudio} (ZS)         & 0.074 & 0.060 & 0.067 [0.034, 0.100] \\
  & \texttt{Qwen2Audio} (ZS)        & 0.077 & $-0.014$ & 0.032 [$-0.005$, 0.066] \\
  & \texttt{Qwen2.5Omni} (ZS)       & 0.086 & 0.246 & 0.166 [0.134, 0.196] \\
  & \texttt{Gemini2.5Pro} (FS)      & 0.348 & 0.430 & 0.389 [0.362, 0.416] \\
  & \texttt{GPT4o-audio} (FS)       & 0.246 & 0.396 & 0.321 [0.291, 0.351] \\
\midrule
Archetype & \texttt{DRAME-Eval} (FT) & 0.474 & 0.305 & 0.389 [0.348, 0.432] \\
\bottomrule
\end{tabular}}
\end{table}

\paragraph{Chance-corrected agreement (weighted $\kappa$).} Alongside the ordinal Krippendorff's $\alpha$ already in Tab.~\ref{tab:tier1-ceiling}, we computed average-pairwise quadratic-weighted $\kappa$ per dimension; the full per-dimension values for all three splits (including the archetype dimensions) are in Tab.~\ref{tab:audit-kappa}. The two estimators show broadly similar per-dimension patterns (e.g.\ \emph{Voice Identity} highest, the two emotion dimensions lowest on the base set) and give similar macro values --- realism base $\alpha=0.585$ / $\kappa_w=0.605$, real recordings $0.420$ / $0.417$, archetype $0.618$ / $0.643$ --- so the reliability interpretation does not depend on one estimator (the fine-grained rank of adjacent, near-tied dimensions does differ between them). All evaluation samples have at least three raters (most samples have four to six), and the label is the rater mean. The base realism training corpus additionally retains $31$ two-rater records; this exception does not affect the evaluation-split agreement estimates.

\begin{table}[!t]
\centering\small
\caption{\rev{\textbf{Per-dimension chance-corrected inter-rater agreement} for all three splits: ordinal Krippendorff's $\alpha$ and average-pairwise quadratic-weighted $\kappa$ ($\kappa_w$). Archetype has three scored dimensions. The two estimators show broadly similar per-dimension patterns and yield similar macro conclusions, so the reliability interpretation does not depend on one estimator.}}
\label{tab:audit-kappa}
{\small\color{revcolor}
\begin{tabular}{l|cc|cc|cc}
\toprule
& \multicolumn{2}{c|}{Realism base} & \multicolumn{2}{c|}{Real recordings} & \multicolumn{2}{c}{Archetype} \\
Dimension & $\alpha$ & $\kappa_w$ & $\alpha$ & $\kappa_w$ & $\alpha$ & $\kappa_w$ \\
\midrule
Pitch          & 0.580 & 0.584 & 0.446 & 0.452 & --- & --- \\
Rhythm         & 0.565 & 0.577 & 0.287 & 0.301 & --- & --- \\
Stress         & 0.547 & 0.560 & 0.385 & 0.399 & --- & --- \\
Accuracy       & 0.562 & 0.585 & 0.182 & 0.192 & --- & --- \\
Intensity      & 0.476 & 0.518 & 0.674 & 0.638 & --- & --- \\
Transition     & 0.464 & 0.506 & 0.625 & 0.595 & --- & --- \\
Identity       & 0.735 & 0.774 & 0.315 & 0.326 & --- & --- \\
Traits         & 0.636 & 0.647 & 0.474 & 0.429 & --- & --- \\
Local          & 0.644 & 0.634 & 0.472 & 0.489 & --- & --- \\
Global         & 0.642 & 0.661 & 0.341 & 0.347 & --- & --- \\
Audio Quality  & --- & --- & --- & --- & 0.614 & 0.659 \\
Human Likeness & --- & --- & --- & --- & 0.649 & 0.655 \\
Appropriateness& --- & --- & --- & --- & 0.589 & 0.614 \\
\midrule
\textbf{Average} & \textbf{0.585} & \textbf{0.605} & \textbf{0.420} & \textbf{0.417} & \textbf{0.618} & \textbf{0.643} \\
\bottomrule
\end{tabular}}
\end{table}

\paragraph{Shortcut controls: does the model use audio, and is it a real/synthetic or naturalness detector?} Several controls rule out the obvious confounds. \emph{(i) Audio is necessary:} replacing the audio with silence collapses the benchmark-distribution realism correlation from $0.602$ to $-0.035$, and the evaluated text-only judge (given gold transcript, profile, and scene but no audio) reaches $0.186$, compared with $0.602$ for the full model (we report these as separate operating points and do not decompose the correlation into additive audio/text shares). \emph{(ii) Both modalities improve role fit:} blanking the profile/scene text leaves an audio-only correlation of $0.392$ that is delivery-quality variance; on the positive-vs-negative role-fit axis of this matched modality ablation, the audio-only, text-only, and full-model AUROCs are $0.408$, $0.599$, and $0.711$ respectively, under the intended score direction, which is why an audio-only score is not itself evidence of role-play evaluation. (This $0.711$ is the full model on the balanced role-fit ablation subset, with audio-only/text-only obtained by blanking a modality at inference; it is a different measurement from the shipped evaluator's role-mismatch AUROC of $0.872$ in Tab.~\ref{tab:tier1-negdisc}, which applies the unmodified model to the full role-mismatch negative set, so the two are not directly comparable.) \emph{(iii) Source identity alone does not explain the score:} a WavLM probe separates real from TTS audio at AUROC $0.959$ (strong source cues clearly exist) yet linearly explains only $3\%$ ($R^2{=}0.03$) of \texttt{DRAME-Eval}'s score variance, making a pure source-identity shortcut unlikely; we cannot, however, rule out nonlinear or interaction-based use of those cues. \emph{(iv) Not a naturalness proxy:} UTMOS naturalness \emph{anti}-correlates with the human construct (mean $-0.26$). \emph{(v) Not a metadata shortcut:} baselines that see only source-show identity ($0.172$) or the construction oracle ($0.278$) fall far short of $0.602$; a frozen WavLM+ridge audio baseline reaches $0.426$ benchmark-distribution---above the evaluated prompted judges but below \texttt{DRAME-Eval}'s $0.602$---and falls to $-0.157$ on authentic speech (a fairly simple acoustic model is thus a strong benchmark-distribution baseline, but it too fails to transfer). \emph{(vi) Mel/vocoder audio reconstruction fidelity:} feeding mel-reconstructed audio (BigVGAN) to \texttt{DRAME-Eval} instead of the original drifts its score by only $0.073$ (correlation $0.989$), so mel-domain storage does not materially change \texttt{DRAME-Eval}'s scores. We did not re-collect human ratings on the reconstructed audio, so this establishes score stability, not that reconstruction preserves human-perceived acoustic properties.

\paragraph{Held-out synthesis-system generalization.} To test transfer to an unseen synthesis system, we retrained the archetype evaluator with all \texttt{Doubao}-system data removed and evaluated on \texttt{Doubao}, against a matched control that keeps it. The unseen-system model retains $87\%$ of control performance ($0.759\!\to\!0.661$); \emph{Audio Quality} and \emph{Human Likeness} transfer almost fully ($-0.03$/$-0.04$), while \emph{Appropriateness} --- the most role-conditioned dimension --- drops most ($0.71\!\to\!0.48$). The judge thus generalizes to an unseen system, with the role-fit dimension the least transferable and the clearest open target.

\paragraph{\rev{Cross-lingual (monolingual-training) generalization.}} \rev{To probe transfer across languages we retrain the realism evaluator on a single training language and evaluate on both, under the same matched $2{,}000$-step budget (so absolute levels again sit below the final model and only within-family deltas are read). Tab.~\ref{tab:tier1-xling} gives the $2\times2$ train-language$\times$test-language matrix. Transfer is limited and asymmetric: a Mandarin-only evaluator retains about $35\%$ of the English-only model's English performance ($0.156$ vs.\ $0.441$), while an English-only evaluator retains about $62\%$ of the Mandarin-only topline on Mandarin ($0.360$ vs.\ $0.579$). Under the matched budget the bilingual mixture reduces English correlation from $0.441$ to $0.353$ while leaving Mandarin essentially unchanged ($0.563$ vs.\ $0.579$). This pattern is consistent with negative transfer or with insufficient English exposure in the Mandarin-dominant mixture, which is larger and includes the Mandarin-only real recordings; the present experiment does not distinguish these explanations. A language-balanced or longer-trained control would separate them, which we leave to future work.}

\begin{table}[!t]
    \centering
    \caption{\rev{\textbf{Cross-lingual generalization} of the realism evaluator (avg-$10$ Pearson on the benchmark-distribution test set). Each model is retrained on a single training language (or the bilingual mixture) under a matched $2{,}000$-step LoRA budget and evaluated on the English ($n{=}1{,}002$) and Mandarin ($n{=}998$) test slices. This retraining probe assigns language by the presence of Chinese characters in the transcript; two records tagged Mandarin in the canonical metadata are consequently grouped with English here. The canonical metadata and the main per-language analysis instead use the stored language tags ($1{,}000$ records per language). Because the budget is matched, absolute levels sit below the final $10$k-step model and only within-family deltas are meaningful; the monolingual splits also differ in size (the Mandarin split is larger and contains the Mandarin-only real recordings), so we read the direction of the effect, not the precise magnitudes. Point estimates only.}}
    \label{tab:tier1-xling}
{\small\color{revcolor}
    \begin{tabular}{l|c|c}
    \toprule
    Training data & English test & Mandarin test \\
    \midrule
    English only        & \textbf{0.441} & 0.360 \\
    Mandarin only       & 0.156 & \textbf{0.579} \\
    Bilingual mixture   & 0.353 & 0.563 \\
    \bottomrule
    \end{tabular}}
\end{table}

\paragraph{Negative-sample validation and leakage.} Negatives are counted by type and severity (train $2{,}504$ positive / $4{,}310$ role-mismatch / $4{,}176$ context-mismatch, severity $L0/L1/L2 = 2{,}866/2{,}823/2{,}797$; benchmark-distribution test $210/832/958$). Human scores decrease monotonically with intended severity (role-negative targeted-dimension mean $3.04\!\to\!2.44$ from $L0$ to $L2$, versus $3.69$ for positives), confirming the perturbations are graded. Mismatch type was hidden from annotators and used only for internal QA; empirically, rater confidence is essentially flat across positive/role/context samples (spread $0.169$), consistent with blind annotation. For leakage, there are zero exact (audio$+$profile) duplicates across splits; the realism train/test split is \emph{clip}-disjoint but not fully character-disjoint ($75/210$ test positives reuse a training character with a different utterance); the archetype split has zero role$\times$scene prompt overlap. To check whether the character reuse inflates results, a seen-vs-unseen source control (using source domain as a proxy, since an exact character-level familiarity flag is not reproducible from the release) finds no evidence of inflation in this proxy: \texttt{DRAME-Eval} scores $0.570$ on the seen (games) domain versus $0.611$ on the unseen (TV) domain, a difference of $\approx-0.04$ in the direction opposite to familiarity gain.

\paragraph{Bottom-up labeling reliability and Semantic Match.} The realism ``media context'' is reconstructed automatically (a profile via \texttt{Gemini-2.0-flash} from turn-by-turn dialogue and captions, a scene via \texttt{DeepSeek-V3}; the transcript is left unmodified), then human-rated. Positives receive high human scores on the reconstructed-context dimensions (Local $3.71$ / Global $3.60$ / Semantic $3.64$, ${\sim}98\%\!\ge\!3$), and $97.3\%$ of prompts are judged sufficient for the rating, indicating the reconstruction is sufficient for \emph{ratability} (annotators can score against it, and flag it when it is not). \rev{Because annotators evaluate the audio against the provided profile and scene, the directly measured criterion is whether the reconstructed context is coherent and sufficient for annotation, which the high positive scores and $97.3\%$ prompt-sufficiency rate support. These figures do not measure exact agreement with the original media, and we did not conduct a separate manual source-fidelity audit.} \emph{Semantic Match} is near-redundant with \emph{Local Scene Fit} (per-source $r=0.86$--$0.99$) and is therefore retained as an annotation but excluded as a separate (eleventh) model output; it is \emph{not} mathematically merged into any of the ten scored dimensions, and the experiment annotations retain it for transparency.

\paragraph{Qualitative error analysis.} On the clean authentic-speech set, \texttt{DRAME-Eval}'s largest errors occur on samples where human raters agree perfectly (zero rater variance) yet the model is off by two or more points, concentrated on \emph{Voice Identity} ($r=-0.041$) and scene-fit dimensions; \emph{Trait Embodiment} is the exception, with a pooled per-dimension correlation of $r=+0.419$ on this set. Public qualitative examples and score comparisons will be restricted to materials authorized for redistribution.
} 

\end{document}